%% file: main.tex
\PassOptionsToPackage{table,prologue}{xcolor}
\documentclass
{llncs}

\usepackage{etoolbox}
\newtoggle{shortpaper} 

\usepackage{subcaption} 

\usepackage{multirow}
\usepackage{multicol}

\usepackage{thmtools}
\usepackage{thm-restate}

\usepackage{amsmath}
\usepackage{amssymb}
\usepackage{cite}
\usepackage{mathpartir} 

\usepackage{adjustbox}
\usepackage{refinement}
\usepackage{bm}
\usepackage{hhline}

\usepackage{xspace}

\usepackage[inline]{enumitem}
\setlist[itemize]{leftmargin=*,label={$\bullet$}}
\setlist[enumerate]{label={\arabic*)}}

\usepackage{hyperref}

\iftoggle{shortpaper}
{} 
{ 
  \hypersetup{
    colorlinks,
    linkcolor=blue,
    citecolor=blue,
    urlcolor=blue
  }
  \renewcommand{\orcidID}[1]{\href{http://orcid.org/#1}
    {\raisebox{-1.25pt}{\includegraphics{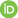}}}\xspace}
}

\usepackage{stmaryrd} 

\usepackage{tikz}
\usetikzlibrary{trees}
\usetikzlibrary{patterns.meta,shapes.multipart,arrows,positioning,calc}
\usetikzlibrary{fit,arrows,snakes,backgrounds,automata,shapes,chains}

\usepackage{wrapfig}

\usepackage{framed}

\usepackage[capitalize]{cleveref} 

\include{macros-specific} 
\include{macros-generic}
\include{macros-authors}

\allowdisplaybreaks

\begin{document}

\title{Specifying and Verifying RDMA Synchronisation (Extended Version)
}

\author{
Guillaume Ambal\inst{1}\orcidID{0000-0002-4667-7266}
\and
Max {Stupple} \inst{1}\orcidID{0009-0002-5603-7585}
\and
Brijesh Dongol \inst{2}\orcidID{0000-0003-0446-3507}
\and
 Azalea {Raad}\inst{1}\orcidID{0000-0002-2319-3242}
}
\institute{Imperial College London, London, UK\\
  \email{\{g.ambal,max.stupple21,azalea.raad\}@imperial.ac.uk}
  \and University of Surrey, Guildford, UK\\
  \email{b.dongol@surrey.ac.uk}}

\maketitle

\begin{abstract}
\input{abstract}
\end{abstract}

\input{intro}

\input{overview}

\input{rdmawait}

\input{locklib}

\input{strongrdma}

\input{related}

\subsubsection*{Acknowledgements.}
\begin{credits}
  Ambal is supported by the EPSRC grant EP/X037029/1. Raad is supported by a
  UKRI fellowship MR/V024299/1, by the EPSRC grant EP/X037029/1, and by VeTSS.
  Dongol is supported by EPSRC grants EP/Y036425/1, EP/X037142/1, EP/V038915/1,
  and EP/X015149/1; and Royal Society grant IES$\backslash$R1$\backslash$221226;
  and VeTSS.
\end{credits}

\bibliographystyle{splncs04}
\bibliography{refs}

\iftoggle{shortpaper}
  {
    \nocite{extended-version}
  }
  {
    \appendix
    \newpage

    \input{brl}

    \input{proofs}

    \input{rdmatsormw}

    \input{rmws}

    \input{op-decl-proof}
  }

\end{document}

%% file: macros-specific.tex
\usepackage{pifont}
\newcommand{\newxspacedcommand}[2]{\newcommand{#1}{\xspace{#2}\xspace}}

\newcommand{\img}[1]{\ensuremath{\mathtt{img}(#1)}\xspace}

\newcommand{\queue}[1]{\ensuremath{#1^\mathbf{*}}}

\newcommand{\assign}{:=}
\newcommand{\assignbrl}{\assign_{\brl}}

\newcommand{\x}{x}
\newcommand{\y}{y}
\newcommand{\z}{z}

\newcommand{\doneE}{\ensuremath{\mathtt{ack_p}}\xspace}
\newcommand\doneEL\doneE
\newcommand\doneER\doneE
\newcommand{\cn}{\ensuremath{\mathtt{cn}}\xspace}

\newcommand{\Local}[1][]{\ensuremath{\mathbf{L}}\xspace}
\newcommand{\Remote}[1][]{\ensuremath{\mathbf{R}}\xspace}

\newcommand{\wbR}[1][]{\ensuremath{\mathbf{wb_R#1}}\xspace}
\newcommand{\wbL}[1][]{\ensuremath{\mathbf{wb_L#1}}\xspace}

\newcommand{\code}[1]{\ensuremath{\mathtt{#1}}\xspace}
\newcommand{\clcas}{\code{CAS}}

\newcommand{\cskip}{\code{skip}}

\newcommand{\cassume}{\code{assume}}

\newcommand{\cmfence}{\code{mfence}}
\newcommand{\cnfence}[1]{\code{rfence}(#1)}

\newcommand{\textget}{\text{get}\xspace}
\newcommand{\textgets}{\text{gets}\xspace}
\newcommand{\textput}{\text{put}\xspace}
\newcommand{\textputs}{\text{puts}\xspace}

\newcommand{\Progs}{\ensuremath{\mathsf{Prog}}\xspace}
\newcommand{\prog}{\ensuremath{\mathsf{P}}\xspace}
\newcommand{\Program}{\prog}
\newcommand{\Nodes}{{\sf Node}}
\newcommand{\Threads}{{\sf Tid}}
\newcommand{\Var}{{\sf Loc}}
\renewcommand{\Loc}{{\sf Loc}}
\newcommand{\Val}{{\sf Val}}
\newcommand{\Method}{{\sf Method}}
\newcommand{\Vals}{{\sf Val}}
\newcommand{\Exps}{{\sf Exp}}

\newcommand{\ev}{\action} 

\newcommand{\RCAS}{\mathtt{RCAS}}
\newcommand{\RFAA}{\texttt{RFAA}}
\newcommand{\rRMW}{\texttt{RRMW}}

\newcommand{\abexp}{e\xspace}
\newcommand{\fv}[1]{\mathsf{elocs}(#1)}

\newcommand{\node}{n\xspace}
\newcommand{\nodefun}[1]{{\tt \node}(#1)\xspace}

\newcommand{\threadt}{t\xspace}
\newcommand{\threadfun}[1]{{\tt \threadt}(#1)\xspace}
\newcommand{\methodfun}[1]{{\tt m}(#1)\xspace}
\newcommand{\val}{\ensuremath{v}\xspace}

\newcommand{\upd}{\ensuremath{u}\xspace}
\newcommand{\valr}{v_{\text{r}}\xspace}
\newcommand{\valw}{v_{\text{w}}\xspace}
\newcommand{\vale}{v_{\text{e}}\xspace}
\newcommand{\valu}{v_{\text{u}}\xspace}

\newcommand{\Wid}{{\sf Wid}}
\newcommand{\wid}{\ensuremath{d}\xspace}

\newcommand{\loc}{\ensuremath{\mathtt{loc}}\xspace}

\newcommand{\neqn}[1]{\overline{#1}}

\newcommand{\poll}{\ensuremath{\mathtt{poll}}\xspace}

\newcommand{\ms}{\\[3pt]}

\newcommand{\aident}{\iota} 

\newcommand{\Alab}{{\sf ELab}}
\newcommand{\set}[1]{\left\{{#1}\right\}}
\newcommand{\tup}[1]{\langle{#1}\rangle}
\newcommand{\st}{\; | \;}

\newcommand{\thread}{\mathsf{thread}}

\newcommand{\EId}{{\sf EventId}}

\newcommand{\action}{\ensuremath{\mathsf{e}}}
\newcommand{\saction}{\ensuremath{\mathsf{s}}}
\newcommand{\events}{\ensuremath{E}}
\newcommand{\Events}{\mathsf{Event}}
\newcommand{\SEvents}{\mathsf{SEvent}}
\newcommand{\Actions}{\Events}
\newcommand{\ActionsExt}{\events^{\text{ext}}}

\newcommand{\CPUactions}{\events^{\text{cpu}}}
\newcommand{\NICactions}{\ensuremath{\events^{\text{nic}}}}

\newcommand{\Labs}{\ensuremath{\mathsf{Lab}}\xspace}

\newcommand{\typ}{\mathsf{type}}
\newcommand{\lab}{l}

\newcommand{\rtsowrite}{\ensuremath{\mathtt{Write}^{\rm\sc TSO}}\xspace}
\newcommand{\rtsoread}{\ensuremath{\mathtt{Read}^{\rm\sc TSO}}\xspace}
\newcommand{\rtsopoll}{\ensuremath{\mathtt{Poll}}\xspace}
\newcommand{\rtsomf}{\ensuremath{\mathtt{Mfence}^{\rm\sc TSO}}\xspace}
\newcommand{\rtsocas}{\ensuremath{\mathtt{CAS}^{\rm\sc TSO}}\xspace}
\newcommand{\rtsoget}{\ensuremath{\mathtt{Get}^{\rm\sc TSO}}\xspace}
\newcommand{\rtsoput}{\ensuremath{\mathtt{Put}^{\rm\sc TSO}}\xspace}
\newcommand{\rtsorf}{\ensuremath{\mathtt{Rfence}^{\rm\sc TSO}}\xspace}
\newcommand{\rtsoadd}{\ensuremath{\mathtt{SetAdd}}\xspace}
\newcommand{\rtsorm}{\ensuremath{\mathtt{SetRemove}}\xspace}
\newcommand{\rtsoempty}{\ensuremath{\mathtt{SetIsEmpty}}\xspace}
\newcommand{\rtsorcas}{\ensuremath{\mathtt{RCAS}^{\rm\sc TSO}}\xspace}
\newcommand{\rtsorfaa}{\ensuremath{\mathtt{RFAA}^{\rm\sc TSO}}\xspace}

\newcommand{\rlwrite}{\ensuremath{\mathtt{Write}}\xspace}
\newcommand{\rlread}{\ensuremath{\mathtt{Read}}\xspace}
\newcommand{\rlmf}{\ensuremath{\mathtt{Mfence}}\xspace}
\newcommand{\rlcas}{\ensuremath{\mathtt{CAS}}\xspace}
\newcommand{\rlfaa}{\ensuremath{\mathtt{FAA}}\xspace} 
\newcommand{\rlwait}{\ensuremath{\mathtt{Wait}}\xspace}
\newcommand{\rlget}{\ensuremath{\mathtt{Get}}\xspace}
\newcommand{\rlput}{\ensuremath{\mathtt{Put}}\xspace}
\newcommand{\rlrf}{\ensuremath{\mathtt{Rfence}}\xspace}
\newcommand{\rlrcas}{\ensuremath{\mathtt{RCAS}}\xspace}
\newcommand{\rlrfaa}{\ensuremath{\mathtt{RFAA}}\xspace}

\newcommand{\rget}{\ensuremath{\mathtt{Get}}\xspace}
\newcommand{\rput}{\ensuremath{\mathtt{Put}}\xspace}
\newcommand{\rcas}{\ensuremath{\mathtt{RCAS}}\xspace}
\newcommand{\rfaa}{\ensuremath{\mathtt{RFAA}}\xspace}
\newcommand{\rrmw}{\ensuremath{\mathtt{rRMW}}\xspace}

\newcommand{\nR}{\ensuremath{\mathtt{nR}}\xspace}

\newcommand{\nlR}{\ensuremath{\mathtt{nlR}}\xspace}
\newcommand{\nlW}{\ensuremath{\mathtt{nlW}}\xspace}
\newcommand{\nrR}{\ensuremath{\mathtt{nrR}}\xspace}
\newcommand{\nrW}{\ensuremath{\mathtt{nrW}}\xspace}
\newcommand{\nEX}{\ensuremath{\mathtt{nEX}}\xspace}
\newcommand{\nlEX}{\ensuremath{\mathtt{nlEX}}\xspace}
\newcommand{\nrEX}{\ensuremath{\mathtt{nrEX}}\xspace}

\newcommand{\lF}[1][]{\ensuremath{\mathtt{F}}#1\xspace}
\newcommand{\lP}{\ensuremath{\mathtt{P}}\xspace}
\newcommand{\nF}{\ensuremath{\mathtt{nF}}\xspace}
\newcommand{\lR}{\ensuremath{\mathtt{lR}}\xspace}
\newcommand{\lW}{\ensuremath{\mathtt{lW}}\xspace}
\newcommand{\RMW}{\ensuremath{\mathtt{CAS}}\xspace}
\newcommand{\RMWF}{\ensuremath{\mathtt{CASF}}\xspace}

\newcommand{\narR}{\ensuremath{\mathtt{narR}}\xspace}
\newcommand{\narW}{\ensuremath{\mathtt{narW}}\xspace}

\newcommand{\naF}{\ensuremath{\mathtt{naF}}\xspace}

\newcommand{\lX}{\ensuremath{\mathtt{X}}\xspace}

\newcommand{\brlwrite}{\ensuremath{\mathtt{Write}_{\brl}}\xspace}
\newcommand{\brlread}{\ensuremath{\mathtt{Read}_{\brl}}\xspace}
\newcommand{\brlwait}{\ensuremath{\mathtt{Wait}_{\brl}}\xspace}
\newcommand{\brlgf}{\ensuremath{\mathtt{GFence}}\xspace}
\newcommand{\brlbr}{\ensuremath{\mathtt{Bcast}_{\brl}}\xspace}

\newcommand{\lklgena}{\ensuremath{\mathtt{Acq}}\xspace}
\newcommand{\lklgenr}{\ensuremath{\mathtt{Rel}}\xspace}

\newcommand{\lklwa}{\ensuremath{\mathtt{Acq}_{\textsc{wl}}}\xspace}
\newcommand{\lklwr}{\ensuremath{\mathtt{Rel}_{\textsc{wl}}}\xspace}

\newcommand{\lklna}{\ensuremath{\mathtt{Acq}_{\textsc{nl}}}\xspace}
\newcommand{\lklnr}{\ensuremath{\mathtt{Rel}_{\textsc{nl}}}\xspace}

\newcommand{\SAG}{\mathcal{G}\xspace}
\newcommand{\EG}{\ensuremath{G}\xspace}

\newcommand{\lklsa}{\ensuremath{\mathtt{Acq}_{\textsc{sl}}}\xspace}
\newcommand{\lklsr}{\ensuremath{\mathtt{Rel}_{\textsc{sl}}}\xspace}

\newcommand{\strlw}{\ensuremath{\mathtt{Write}_\textsc{sc}}\xspace}
\newcommand{\strlr}{\ensuremath{\mathtt{Read}_\textsc{sc}}\xspace}
\newcommand{\strlcas}{\ensuremath{\mathtt{CAS}_\textsc{sc}}\xspace}
\newcommand{\strlfaa}{\ensuremath{\mathtt{FAA}_\textsc{sc}}\xspace}

\newxspacedcommand{\POdef}{\sc POdef}
\newxspacedcommand{\HBdef}{\sc HBdef}
\newxspacedcommand{\MOdef}{\sc MOdef}
\newxspacedcommand{\SCdef}{\sc SCdef}
\newxspacedcommand{\SWdef}{\sc SWdef}
\newxspacedcommand{\Acyclicity}{\sc Acyclicity}

\newcommand{\arr}[2][]{\xrightarrow[#1]{#2}}

\newcommand{\mytextsf}[1]{\textnormal{\textsf{#1}}}
\newcommand{\makerel}[2][.]{\ensuremath{{\color{#1}\mytextsf{#2}}}\xspace}
\newcommand{\makerelext}[2][]{\ensuremath{#2_{\mytextsf{e}}\xspace}}
\newcommand{\makerelint}[2][]{\ensuremath{#2_{\mytextsf{i}}}\xspace}

\definecolor{darkgrey}{rgb}{0.4,0.4,0.4}
\definecolor{darkturquoise}{rgb}{0.012, 0.502, 0.486}
\definecolor{raocolor}{HTML}{006600}
\definecolor{rocolor}{HTML}{4B0082}
\colorlet{colorSC}{red!80!black}
\definecolor{raocolor}{HTML}{006600}
\definecolor{arcolor}{HTML}{FB607F}
\colorlet{colorPO}{darkgray!80!black}
\colorlet{colorPPO}{magenta}
\colorlet{colorOPPO}{magenta}
\colorlet{colorIPPO}{cyan}
\colorlet{colorRF}{green!60!black}
\colorlet{colorMO}{orange}
\colorlet{colorRO}{rocolor}
\colorlet{colorFR}{purple}
\colorlet{colorSO}{red}
\colorlet{colorHB}{blue}
\colorlet{colorPF}{brown!80!black}
\colorlet{colorRAO}{raocolor}
\colorlet{colorAR}{arcolor}
\colorlet{colorOB}{teal}
\colorlet{colorIB}{blue}

\newcommand{\po}{\makerel[colorPO]{po}}
\newcommand{\ppo}{\makerel[colorPPO]{ppo}}
\newcommand{\ippo}{\makerel[colorIPPO]{ippo}}
\newcommand{\oppo}{\makerel[colorOPPO]{oppo}}
\newcommand{\tagppo}{\makerel[colorPPO]{sto}}
\newcommand{\rf}{\makerel[colorRF]{rf}}
\newcommand{\rfe}{\makerelext[colorRF]{\rf}}

\newcommand{\rfi}{\makerelint[colorRF]{\rf}}
\newcommand{\mo}{\makerel[colorMO]{mo}}

\newcommand{\ro}{\makerel[colorRO]{nfo}}
\newcommand{\rao}{\makerel[colorRAO]{rao}}
\newcommand{\fr}{\makerel[colorFR]{rb}}

\newcommand{\ar}{\makerel[colorAR]{ar}}

\newcommand{\fri}{\makerelint[colorFR]{\fr}}
\newcommand{\hb}{\makerel[colorHB]{hb}}
\newcommand{\pf}{\makerel[colorPF]{pf}}
\newcommand{\pfget}{\makerel[colorPF]{pfg}}
\newcommand{\pfput}{\makerel[colorPF]{pfp}}

\newcommand{\ib}{\makerel[colorIB]{ib}}
\newcommand{\ob}{\makerel[colorOB]{ob}}
\newcommand{\IB}{\makerel[colorIB]{IB}}
\newcommand{\OB}{\makerel[colorOB]{OB}}

\newcommand{\lo}{\makerel[colorRAO]{lo}}

\newcommand{\iso}{\makerel[colorSO]{iso}}
\newcommand{\so}{\makerel[colorSO]{so}}

\newcommand{\sloc}{\makerel{sloc}}
\newcommand{\sthd}{\makerel{sthd}}
\newcommand{\sqp}{\ensuremath{\mathtt{sqp}}\xspace}
\newcommand{\Inst}{\ensuremath{\mathtt{Inst}}\xspace}

\tikzset{
   every path/.style={>=stealth},
   po/.style={->,color=colorPO,thin,shorten >=-0.5mm,shorten <=-0.5mm},
   rf/.style={->,color=colorRF,dashed,shorten >=-0.5mm,shorten <=-0.5mm},
   mo/.style={->,color=colorMO,dotted,very thick,shorten >=-0.5mm,shorten <=-0.5mm},
   fr/.style={->,color=colorFR,dotted,thick,shorten >=-0.5mm,shorten <=-0.5mm},
}

\newcommand{\vr}{\ensuremath{\mathtt{v}_\mathtt{R}}\xspace}
\newcommand{\vw}{\ensuremath{\mathtt{v}_\mathtt{W}}\xspace}

\newxspacedcommand{\Read}{\mathcal{R}}
\newxspacedcommand{\Write}{\mathcal{W}}
\newxspacedcommand{\ReadSC}{\ensuremath{\mathcal{R}_\textsc{sc}}}
\newxspacedcommand{\WriteSC}{\ensuremath{\mathcal{W}_\textsc{sc}}}

\newcommand{\updmem}{\lhd}

\newxspacedcommand{\modelRDMA}{{RDMA}\xspace}
\newxspacedcommand{\modelTSO}{{TSO}\xspace}
\newxspacedcommand{\modelPSO}{{PSO}\xspace}
\newxspacedcommand{\modelSC}{{SC}\xspace}
\newxspacedcommand{\modelRDT}{RDT}
\newxspacedcommand{\modelSCw}{\emph{The quite nice and fairly SC execution}}

\newxspacedcommand{\disipClServ}{{\it \textbf{client-server discipline}}}
\newxspacedcommand{\TSOva}{\sc TSOva}
\newxspacedcommand{\SWva}{\sc SWva}
\newxspacedcommand{\MOva}{\sc MOva}

\newxspacedcommand{\compNodes}{\Node}
\newxspacedcommand{\compCPU}{CPU}
\newxspacedcommand{\compNIC}{NIC}
\newxspacedcommand{\compProcs}{Proc}
\newxspacedcommand{\compRegisters}{Register}
\newxspacedcommand{\compOpId}{Id}
\newxspacedcommand{\compMemLoc}{Loc}
\newxspacedcommand{\compVal}{Val}
\newxspacedcommand{\compOp}{Op}
\newxspacedcommand{\compQP}{queue pair}

\newxspacedcommand{\compActions}{Action}

\newxspacedcommand{\emptylist}{[\thinspace]}

\newxspacedcommand{\nicOutQueue}{nic\_out\_que}
\newxspacedcommand{\nicInQueue}{nic\_in\_que}
\newxspacedcommand{\nicCompleted}{rem\_nic\_acks}
\newxspacedcommand{\lastRDMAop}{last\_rdma\_op}
\newxspacedcommand{\nicLastAcked}{cq\_last}
\newxspacedcommand{\nicCQ}{cq}
\newxspacedcommand{\nicBuffer}{nic\_buffer}
\newxspacedcommand{\nicBufferLoc}{nic\_buff\_loc}
\newxspacedcommand{\nicBufferRem}{nic\_buff\_rem}
\newxspacedcommand{\buffer}{buffer}
\newxspacedcommand{\memory}{memory}

\newxspacedcommand{\procNode}{node}

\newxspacedcommand{\typeStatus}{Status}

\newxspacedcommand{\valueCompleted}{``compl"}
\newxspacedcommand{\valuePending}{``pend"}
\newxspacedcommand{\valueACK}{ACK}

\newxspacedcommand{\execProcess}{exec}
\newxspacedcommand{\bufferFlush}{bufferFlush}
\newxspacedcommand{\nicBufferFlush}{nicBufferFlush}
\newxspacedcommand{\execNICsecondStep}{execLocalPreps}
\newxspacedcommand{\execNICthirdStep}{execRemRecieve}
\newxspacedcommand{\execNICfourthStep}{execRemRespond}
\newxspacedcommand{\execNICfifthStep}{handleResponse}

\newxspacedcommand{\execNICsingleStep}{{nicExec}\xspace}

\newxspacedcommand{\hlprIsRemRead}{isRemoteRead}
\newxspacedcommand{\hlprIsRemWrite}{isRemoteWrite}
\newxspacedcommand{\hlprMakeRemWrite}{makeRemoteWrite}
\newxspacedcommand{\hlprIsAtomic}{isAtomic}
\newxspacedcommand{\hlprPerformAtomic}{performAtomic}
\newxspacedcommand{\hlprIsCompletion}{isCompletions}
\newxspacedcommand{\hlprMakecompletion}{makeCompletion}
\newxspacedcommand{\hlprGetQueuePair}{getQueuePair}

\newcommand{\tso}{\stso}
\newcommand{\Atm}{\ensuremath{\mathsf{A}}\xspace}
\newcommand{\Mem}{\ensuremath{\mathsf{M}}\xspace}
\newcommand{\AMem}{\Mem}

\newcommand{\mfence}{\cmfence}
\newcommand{\CAS}{\texttt{CAS}}

\newcommand{\upvar}[2]{[#1 \mapsto #2]}

\newxspacedcommand{\ParTSO}{ParTSO}

\newcommand{\defeq}{\triangleq}

\renewcommand{\new}{\text{new}}

\newcommand{\ctrtc}[1][\threadt]{\xrightarrow{#1}_c}

\newcommand{\trp}{\Rightarrow}
\newcommand{\RAMap}{\mathsf{RAMap}}

\newcommand{\Mems}{\mathsf{Mem}}
\newcommand{\AMems}{\mathsf{AMem}}

\newcommand{\SQPairs}{\mathsf{SQPair}}
\newcommand{\SQPmaps}{\mathsf{SQPMap}}
\newcommand{\AQPairs}{\mathsf{AQPair}}
\newcommand{\qp}{\ensuremath{\mathsf{qp}}\xspace}
\newcommand{\simqp}{\qpbis}

\newcommand{\actseq}{\rightarrowtail}

\newcommand{\ATSOs}{\mathsf{ASBuff}}

\newcommand\partso{\tso}
\newcommand\ntso\partso

\newcommand{\old}{\text{old}}

\newcommand{\aLabels}{\ensuremath{\mathsf{ALabel}}\xspace}

\newcommand{\paths}{\ensuremath{\mathsf{Path}}\xspace}

\newcommand{\amtr}[1]{\xrightarrow{#1}}

\newcommand{\flB}{\ensuremath{\mathsf{B}}\xspace}
\newcommand{\wfp}{\ensuremath{\mathsf{wfp}}\xspace}
\newcommand{\wf}{\ensuremath{\mathsf{wf}}\xspace}
\newcommand{\complete}{\ensuremath{\mathsf{complete}}\xspace}
\newcommand{\fresh}{\ensuremath{\mathsf{fresh}}\xspace}%
\newcommand{\wfrd}{\ensuremath{\mathsf{wfrd}}\xspace}

\newcommand{\bufFlushOrd}{\ensuremath{\mathsf{bufFlushOrd}}\xspace}%
\newcommand{\nicActOrder}{\ensuremath{\mathsf{nicActOrder}}\xspace}%
\newcommand{\remoteAtomic}{\ensuremath{\mathsf{nicAtomicity}}\xspace}%
\newcommand{\pollOrder}{\ensuremath{\mathsf{pollOrder}}\xspace}%

\newcommand{\sameqp}{\ensuremath{\mathsf{sameqp}}\xspace}%

\newcommand{\getG}{\ensuremath{\mathsf{getEG}}\xspace}%
\newcommand{\getA}{\ensuremath{\mathsf{getA}}}

\newcommand{\getIL}{\ensuremath{\mathsf{getI{\lambda}}}}
\newcommand{\getOL}{\ensuremath{\mathsf{getO{\lambda}}}}
\newcommand{\getCPU}{\ensuremath{\mathsf{getCPU}}}
\newcommand{\getTSO}{\ensuremath{\mathsf{getTSO}}}

\newcommand{\precpi}{\ensuremath{\mathsf{\prec_{\pi}}}}
\newcommand{\locveq}{\ensuremath{\mathsf{eq_{\loc\&\val}}}}
\newcommand{\pathread}{\ensuremath{\mathsf{read}}}
\newcommand{\bigpar}[1]{ \left( #1 \right) }
\newcommand{\setpred}[2]{ \left\{ #1 \;\middle|\; #2 \right\} }

\newcommand{\rst}[1]{|_{#1}}
\newcommand{\imm}[1]{{#1}{\rst{\text{imm}}}}

\newcommand{\cmark}{\textcolor{green!60!black}{\ding{51}}\xspace}
\newcommand{\xmark}{\textcolor{red!80!black}{\ding{55}}\xspace}
\newcommand{\checkyes}{\cmark}
\newcommand{\checkno}{\xmark}

\newcommand{\cyes}{\checkyes}
\newcommand{\cno}{\checkno}
\newcommand{\cqp}{{\sc sn}\xspace}

\newcommand{\kstar}{{\displaystyle{*}}}
\newcommand{\choice}{\operatorname{\displaystyle{+}}}

\newcommand{\pipe}[1][]{\ensuremath{\mathbf{pipe#1}}\xspace}

\newcommand{\qpbis}[1][]{\ensuremath{\mathbf{sqp#1}}\xspace}
\newcommand{\queueset}[1]{\queue{\setarray{#1}}}

\newcommand{\pipearrow}{\ensuremath{\rightarrow_\mathtt{sqp}}}
\newcommand{\pipealab}[1]{\ensuremath{\xrightarrow{#1}_\mathtt{sqp}}}

\newcommand{\TSO}{\text{TSO}\xspace}

\newcommand{\temp}{r}

\newcommand{\intabT}[1]{\begin{tabular}[t]{@{}c@{}}#1\end{tabular}}

\newcommand{\inarr}[1]{\begin{array}{@{}l@{}}#1\end{array}}
\newcommand{\inarrT}[1]{\begin{array}[t]{@{}l@{}}#1\end{array}}

\newcommand{\inarrII}[2]{\begin{array}{@{}l@{~~}|@{~~}l@{}}\inarr{#1}&\inarr{#2}\end{array}}

\newcommand{\rdmatso}{\text{\sc rdma}\textsuperscript{{\sc tso}}\xspace}
\newcommand{\rdmasc}{\text{\sc rdma}\textsuperscript{{\sc sc}}\xspace}
\newcommand{\rdmawait}{\text{\sc rdma}\textsuperscript{{\sc wait}}\xspace}
\newcommand{\rdmatsormw}{\ensuremath{\text{\sc rdma}^{\text{\sc tso}}_{\text{\sc rmw}}}\xspace}
\newcommand{\rdmascrmw}{\ensuremath{\text{\sc rdma}^{\text{\sc sc}}_{\text{\sc rmw}}}\xspace}
\newcommand{\rdmawaitrmw}{\ensuremath{\text{\sc rdma}^{\text{\sc wait}}_{\text{\sc rmw}}}\xspace}
\newcommand{\rdmasync}{\ensuremath{\text{\sc rdma}^{\text{\sc wait}}_{\text{\sc rmw}}}\xspace} 
\newcommand{\loco}{\text{\sc loco}\xspace}

\newcommand{\brl}{\text{\sc sv}\xspace} 
\newcommand{\strl}{\rdmascrmw} 
\newcommand{\lklw}{\text{\sc wlock}\xspace}
\newcommand{\lkln}{\text{\sc nlock}\xspace}
\newcommand{\lkls}{\text{\sc slock}\xspace}

\newcommand{\Comms}{\ensuremath{\mathsf{Comm}}}
\renewcommand{\comm}[1][\node]{\ensuremath{\mathsf{C}^{#1}}}

\newcommand{\PComms}{\ensuremath{\mathsf{PComm}}}
\newcommand{\pcomm}[1][\node]{\ensuremath{\mathsf{c}^{#1}}}

\newcommand{\LComms}{\ensuremath{\mathsf{CComm}}}
\newcommand{\lcomm}[1][\node]{\ensuremath{\mathsf{cc}^{#1}}}

\newcommand{\RComms}{\ensuremath{\mathsf{RComm}}}
\newcommand{\rcomm}[1][\node]{\ensuremath{\mathsf{rc}^{#1}}}

\newcommand{\corerma}{\text{coreRMA}\xspace}
\newcommand{\stso}{\ensuremath{\mathsf{b}}\xspace}
\newcommand{\STSOs}{\mathsf{SBuff}}
\newcommand{\TSOmap}{\ensuremath{\mathsf{B}}\xspace}
\newcommand{\TSOmaps}{\mathsf{SBMap}}
\newcommand{\NQPmap}{\ensuremath{\mathsf{QP}}\xspace}
\newcommand{\NQPmaps}{\mathsf{QPMap}}

\newcommand{\framework}{{\sc mowgli}\xspace}
\newcommand{\arrcomm}[1]{\ensuremath{\xrightarrow{#1}}}
\newcommand{\arrprog}[2][\threadt]{\ensuremath{\xrightarrow{#1:#2}}}
\newcommand{\arrlab}[2][\threadt]{\xrightarrow{#1:#2}}
\newcommand{\arreps}{\arrlab{\varepsilon}}

\newcommand{\slW}{\lW}
\newcommand{\slR}{\lR}
\newcommand{\slP}{\lP}
\newcommand{\slF}{\lF}
\newcommand{\srget}{\rget}
\newcommand{\srput}{\rput}
\newcommand{\snF}{\ensuremath{\mathtt{rF}}\xspace}
\newcommand{\CASFail}{\ensuremath{\mathtt{CASF}}\xspace}
\newcommand{\CASSucc}{\ensuremath{\mathtt{CASS}}\xspace}

\newcommand{\lb}{\ensuremath{l}}
\newcommand{\empL}{\ensuremath{\varepsilon}}

\definecolor{shadecolor}{HTML}{F0E0F0}

\newcommand{\WBRs}{\mathsf{WBR}}

\newcommand{\WBLs}{\mathsf{WBL}}
\newcommand{\Pipes}{\mathsf{Pipe}}

\definecolor{highlightcolor}{HTML}{F0E0F0}
\definecolor{highlightcoloralt}{HTML}{FFFFE3}

\newcommand{\shade}[2][highlightcolor]{\colorbox{#1}{#2}}
\newcommand{\tightshade}[2][highlightcolor]{\setlength{\fboxsep}{0pt}\colorbox{#1}{#2}}
\newcommand{\shademath}[2][highlightcolor]{\colorbox{#1}{\ensuremath{#2}}}
\newcommand{\tightshademath}[2][highlightcolor]{\setlength{\fboxsep}{0pt}\colorbox{#1}{\ensuremath{#2}}}

\renewcommand{\red}[1]{{\color{red}#1}}

\newcommand{\myblue}[1]{{\color{blue}#1}}
\newcounter{rowcounter}
\newcounter{colcounter}
\newcommand{\APipes}{\mathsf{APipe}}
\newcommand{\AWBRs}{\mathsf{AWBR}}
\newcommand{\AWBLs}{\mathsf{AWBL}}

\newcommand{\vsep}{\vspace*{-5pt}}
\newcommand{\ASTSOs}{\mathsf{ASBuff}}
\newcommand{\ATSOmap}{\TSOmap}
\newcommand{\ATSOmaps}{\mathsf{ASBMap}}
\newcommand{\ANQPmap}{\NQPmap}
\newcommand{\ANQPmaps}{\mathsf{AQPMap}}

\newcommand{\rowc}{\Alph{rowcounter}\refstepcounter{rowcounter}}
\newcommand{\colc}{\thecolcounter\refstepcounter{colcounter}}
\definecolor{cellname}{rgb}{0.5,0.5,0.5}
\newcommand\colheader{\cellcolor{cellname}\color{white}\textbf{\colc}}
\newcommand\rowheader[1]{\cellcolor{cellname}\color{white}\textbf{\rowc}&#1}

\newcommand{\chlm}{\cellcolor{highlightcolor}}

\newcommand{\Tags}{{\sf Stamp}}

\newcommand{\cons}{\mathcal{C}}

\newcommand{\exec}{\mathcal{G}}

\newcommand{\tagt}{\ensuremath{a}\xspace}
\newcommand{\gettags}{\ensuremath{\mathtt{stmp}}\xspace}

\newcommand{\gettagsrl}{\ensuremath{\mathtt{stmp}_\mathtt{RW}}\xspace} 
\newcommand{\gettagsbrl}{\ensuremath{\mathtt{stmp}_\mathtt{SV}}\xspace}

\newcommand{\gettagslklw}{\ensuremath{\mathtt{stmp}_\mathtt{WL}}\xspace}
\newcommand{\gettagslkln}{\ensuremath{\mathtt{stmp}_\mathtt{NL}}\xspace}
\newcommand{\gettagslkls}{\ensuremath{\mathtt{stmp}_\mathtt{SL}}\xspace}
\newcommand{\gettagsstrl}{\ensuremath{\mathtt{stmp}_\mathtt{SC}}\xspace} 

\newcommand{\gettagsrtso}{\ensuremath{\mathtt{stmp}_\mathtt{TSO}}\xspace}

\newcommand{\tagcread}{\ensuremath{\mathtt{aCR}}\xspace}
\newcommand{\tagcwrite}{\ensuremath{\mathtt{aCW}}\xspace}
\newcommand{\tagcas}{\ensuremath{\mathtt{aCAS}}\xspace}
\newcommand{\tagmf}{\ensuremath{\mathtt{aMF}}\xspace}
\newcommand{\tagwait}{\ensuremath{\mathtt{aWT}}\xspace}

\newcommand{\tagnlr}[1][\node]{\ensuremath{\mathtt{aNLR}_{#1}}\xspace}
\newcommand{\tagnlw}[1][\node]{\ensuremath{\mathtt{aNLW}_{#1}}\xspace}
\newcommand{\tagnrr}[1][\node]{\ensuremath{\mathtt{aNRR}_{#1}}\xspace}
\newcommand{\tagnarr}[1][\node]{\ensuremath{\mathtt{aNAR}_{#1}}\xspace}
\newcommand{\tagnrw}[1][\node]{\ensuremath{\mathtt{aNRW}_{#1}}\xspace}
\newcommand{\tagnf}[1][\node]{\ensuremath{\mathtt{aRF}_{#1}}\xspace}
\newcommand{\taggf}[1][\node]{\ensuremath{\mathtt{aGF}_{#1}}\xspace}

\newcommand{\implwait}{\ensuremath{I_\mathtt{W}}\xspace}

\newcommand{\impllklw}{\ensuremath{I_\mathtt{WL}}\xspace}
\newcommand{\impllkln}{\ensuremath{I_\mathtt{NL}}\xspace}
\newcommand{\impllkls}{\ensuremath{I_\mathtt{SL}}\xspace}
\newcommand{\implstrl}{\ensuremath{I_\mathtt{SC}}\xspace} 

\newcommand{\true}{\code{true}}

\newcommand{\vect}[1]{\widetilde{#1}}
\newcommand{\outcome}{\ensuremath{\mathtt{outcome}}\xspace}
\newcommand{\abs}{\ensuremath{\mathtt{abs}}\xspace}
\newcommand{\absf}[5][f]{\ensuremath{\abs_{#3,#2}^{#1}(#4, #5)}}

\newcommand{\interp}[1]{\llbracket #1 \rrbracket}
\newcommand{\interpt}[1]{\llbracket #1 \rrbracket_{\threadt}}
\newcommand{\impl}[1]{\llfloor #1 \rrfloor}
\newcommand{\impli}[1]{\llfloor #1 \rrfloor_{I}}
\newcommand{\implti}[1]{\llfloor #1 \rrfloor_{\threadt,I}}

\newcommand{\impliw}[1]{\llfloor #1 \rrfloor_{\implwait}}

\newcommand{\progf}{\ensuremath{\mathsf{f}}}
\newcommand{\progp}{\ensuremath{\mathsf{p}}}
\newcommand{\Progp}{\ensuremath{\mathsf{SeqProg}}}
\newcommand{\progps}{\vect{\progp}}

\newcommand{\cardinal}[1]{\ensuremath{\#(#1)}}

\newcommand{\SVar}{{\rm SVar} \xspace}

\newcommand{\LetC}[3]{\code{let}\, #1  =  #2 \, \code{in} \, #3}
\newcommand{\LetF}[2]{\code{let}\, #1 \  #2}

\newcommand{\paperbreak}{\mathtt{break}}

\newcommand{\namedinferrule}[3]{\inferrule{#1}{#2}}

\newcommand{\xx}{\texttt{x}\xspace}
\newcommand{\yy}{\texttt{y}\xspace}

\newcommand{\evlink}{[\hyperlink{cite.extended-version}{EV}]}


%% file: macros-generic.tex
\newcommand{\DONE}[1]{ }

\renewcommand{\paragraph}[1]{\smallskip\noindent{\textbf{#1.}}}

\renewcommand{\ie}{\text{i.e.}\xspace}
\renewcommand{\Ie}{\text{I.e.}\xspace}
\renewcommand{\eg}{\text{e.g.}\xspace}
\renewcommand{\Eg}{\text{E.g.}\xspace}

\newcommand{\resp}{\text{resp.}\xspace}

\renewcommand{\cf}{\textit{cf.}\xspace}
\newcommand{\etal}{\text{et al.}\xspace}

\Crefname{figure}{\text{Fig.}}{\text{Figs.}}
\crefname{figure}{\text{Fig.}}{\text{Figs.}}

\crefname{section}{\S{}}{Section}
\crefformat{section}{#2\S{}#1#3}
\Crefformat{section}{Section~#2#1#3}
\crefname{theorem}{Theorem}{Theorems}
\crefformat{theorem}{Theorem~#2#1#3}
\Crefformat{theorem}{Theorem~#2#1#3}
\crefname{corollary}{Corollary}{Corollaries}
\crefformat{corollary}{Corollary~#2#1#3}
\Crefformat{corollary}{Corollary~#2#1#3}

\crefformat{appendix}{#2\S{}#1#3}
\Crefname{appendix}{Appendix}{Appendix}
\Crefformat{appendix}{Appendix~#2#1#3}

\crefname{definition}{\text{Def.\xspace}}{\text{Definitions}}
\Crefname{definition}{\text{Definition}}{\text{Definitions}}

\crefname{lemma}{\text{Lem.\xspace}}{\text{Lemmas}}
\Crefname{lemma}{\text{Lemma}}{\text{Lemmas}}

\newcommand{\setarray}[1]{
\ensuremath{
	\left\{
		\begin{array}{@{} l @{}}
			#1
		\end{array}
	\right\}
}
}

\newcounter{mylabelcounter}

\makeatletter
\newcommand{\labelAxiom}[2]{%
\hfill{\textsc{(#1)}}\refstepcounter{mylabelcounter}
\immediate\write\@auxout{%
  \string\newlabel{#2}{{\unexpanded{\textsc{#1}}}{\thepage}{{\unexpanded{\textsc{#1}}}}{mylabelcounter.\number\value{mylabelcounter}}{}}
}%
}
\makeatother

\newcommand{\suq}{\subseteq}
\newcommand{\inv}[1]{\ensuremath{#1^{-1}}}

\newlength{\Oldarrayrulewidth}

\newcolumntype{?}{!{\vrule width 1pt}}


%% file: macros-authors.tex
\newcommand{\ignore}[1]{}

\newcommand{\orix}[1]{}

\newcommand{\gax}[1]{}

\newcommand{\azaleax}[1]{}

\newcommand{\haggaix}[1]{}

\definecolor{bdcolor}{HTML}{e83858}

%% file: abstract.tex

Remote direct memory access (RDMA) allows a machine to directly read
from and write to the memory of remote machine, enabling
high-throughput, low-latency data transfer. Ensuring correctness of
RDMA programs has only recently become possible with the formalisation
of \rdmatso semantics (describing the behaviour of RDMA networking
over a TSO CPU). However, this semantics currently lacks a
formalisation of remote synchronisation, meaning that the
implementations of common abstractions such as locks cannot be
verified. In this paper, we close this gap by presenting \rdmatsormw,
the first semantics for remote `read-modify-write' (RMW) instructions
over TSO. It turns out that remote RMW operations are weak and only ensure
atomicity against other remote RMWs.  We therefore build a
set of composable synchronisation abstractions starting with the
\rdmasync library. Underpinned by \rdmasync, we then specify, 
implement and verify three classes of remote locks that are
suitable for different scenarios. Additionally, we develop the notion
of a strong RDMA model, \rdmascrmw, which is akin to sequential
consistency in shared memory architectures. Our libraries are built to
be compatible with an existing set of high-performance libraries called \loco, which
ensures compositionality and verifiability. 


%% file: intro.tex

\section{Introduction}
\label{sec:intro}

{\em Remote Direct Memory Access} (RDMA), as implemented by RoCE and
Infiniband, is a high-performance networking technology that enables
low-latency wire-speed data transmission. Specifically, an RDMA device
can directly read and write from the memory of a remote (network) \emph{node} (machine),
bypassing the remote CPU and operating system. RDMA technology has been
used in high-performance computing applications (including
supercomputers) since the early 2000s, and is being branched out
to support a much wider range of applications, ranging from
production-grade data
centres~\cite{zhu2015congestion,lu2018multi,wang2023srnic} to
distributed AI training~\cite{gangidi2024rdma}. Thus, there is
currently a push towards developing programmer-friendly libraries to
improve the reliability and robustness of such applications.


To enable rigorous development and verification, there is ongoing work
aimed at formalising the semantics of RDMA architectures, primarily
the RDMA memory model. Dan et al~\cite{rdma-sc} proposed an early
model, called \corerma, which was used to formalise the behaviours of
remote read/write operations, assuming a sequentially consistent CPU.
Ambal et al~\cite{OOPSLA-24} have presented a more realistic \rdmatso
specification, which assumes a total-store-order (TSO) CPU (\eg as
implemented by Intel processors) that (unlike \corerma) has been validated
against real RoCE and Infiniband hardware. \rdmatso precisely
describes the interaction between the CPU and NIC (Network Interface
Card) and the reorderings that they allow. Their formalisation
comprises both declarative and operational models (which are proved
equivalent). However, \rdmatso only covers a \emph{subset} of RDMA
instructions. In particular it only covers local (\ie CPU-level)
`read-modify-write' (RMW) synchronisation, relegating remote (\ie RDMA)
RMWs to future work. This means that \rdmatso cannot be used to
specify and verify locks and other related high-level mechanisms that
require synchronisation at the network level.

\begin{wrapfigure}[11]{r}{0.47\textwidth}
  \vspace{-25pt}
  \centering\scalebox{0.8}{
  \begin{tikzpicture}
  \node [color=teal](rdmatso) at (3.5,-1.2) {\rdmatso \cite{OOPSLA-24}};
  \node [color=purple] (rdmawait) at (3.5,0) {\rdmawait};
  \node (rdmatsormw) at (0,-1.2) {\rdmatsormw \iftoggle{shortpaper}{\evlink}{\cref{sec:rdmatsormw}}};
  \node (rdmasync) at (0,0) {\rdmasync (\cref{sec:rdmasync})};
  \node [color=purple] 
    (sv) at (3.5,1) {\brl };
  \node(lklw) at (1.2,1.9) {\lklw (\cref{sec:locklib-weak})};
  \node(lkln) at (-1.2,1.9) {\lkln (\cref{sec:locklib-node})};
  \node(lkls) at (1.2,3.1) {\lkls (\cref{sec:locklib-strong})};
  \node(strl) at (-1.2,3.1) {\strl (\cref{sec:strl})};

  \draw[-latex,dashed,thick] (rdmasync) to node[above]{\footnotesize extends} (rdmawait);
  \draw[-latex,dashed,thick] (rdmatsormw) to node[above]{\footnotesize extends} (rdmatso);

  \draw[-latex,color=purple] (sv) to (rdmawait);
  \draw[-latex] (lklw) to node[below]{\footnotesize \iftoggle{shortpaper}{\evlink}{\cref{sec:lklw-library}}} (sv);
  \draw[-latex] (lklw) to node[right]{\footnotesize \iftoggle{shortpaper}{\evlink}{\cref{sec:lklw-library}}} (rdmasync);
  \draw[-latex] (lkln) to node[left]{\footnotesize \iftoggle{shortpaper}{\evlink}{\cref{sec:lkln-library}}}  (rdmasync);
  \draw[-latex] (lkls) to node[right]{\footnotesize \iftoggle{shortpaper}{\evlink}{\cref{sec:lkls-library}}} (lklw);
  \draw[-latex] (strl) to node[right]{\footnotesize \iftoggle{shortpaper}{\evlink}{\cref{sec:strl-library}}} (lkln);
  \draw[-latex,color=purple] (rdmawait) to 
  (rdmatso);
  \node[draw,
  fit=(sv) (rdmawait),color=purple,fill=gray,opacity=0.1] (fit) {};
  \node[above=0mm of fit,color=purple]  {\footnotesize LOCO~\cite{POPL-LOCO}};
  \draw[-latex] (rdmasync) to node[right]{\footnotesize \iftoggle{shortpaper}{\evlink}{\cref{sec:rdmatsormw}}} (rdmatsormw);
\end{tikzpicture}}
\vspace{-20pt}
\caption{Development overview}
\label{fig:overview}
\vspace{-20pt}
\end{wrapfigure}
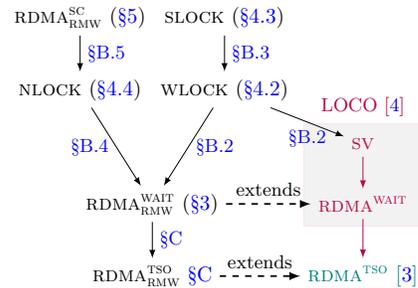

In this work, we address this gap and extend the existing efforts
with a notion of remote (RDMA) synchronisation. 
Specifically, we develop the \rdmatsormw model by extending \rdmatso to account for remote RMWs. 
To ensure the fidelity of our extension, we developed \rdmatsormw by careful inspection of the Infiniband technical manual~\cite{rdmaspec} and in close consultation with engineers at NVIDIA, the largest manufacturer of RDMA products worldwide (after acquiring Mellanox in 2019).
We then build a series of synchronisation libraries and prove them correct (as
we discuss below). An overview of our development is given in
\cref{fig:overview}. \iftoggle{shortpaper}{The full development is available in the arXiv extended version~\evlink.}{}

Remote RMW instructions are surprisingly \emph{weak} in that they
only guarantee a weak form of isolation: remote RMWs
are atomic \emph{only} with respect to other remote RMWs and not CPU accesses or remote read and write operations
(\cf~weak transactional
isolation~\cite{rpsi,rsi,4069174,DBLP:journals/pacmpl/DongolJR18,DBLP:conf/pldi/ChongSW18}). 
We provide a set of litmus tests that exemplify these behaviours in two-
and three-node configurations. 
A second challenge is that (like \rdmatso) \rdmatsormw is {\em not} compositional: the semantics of a certain remote operation, $\rtsopoll$, directly depends on the \emph{exact number of remote operations} in the program up to that point! As such, one cannot specify the behaviour of $\rtsopoll$ \emph{modularly} (in isolation). 

To address both issues, we build on the {\em Library of Composable
  Objects} (\loco) framework~\cite{rdmaloco,POPL-LOCO}, which is a
modular set of objects for constructing RDMA libraries. We start at
the lowest level of \loco, called \rdmawait, which is a compositional analogue of \rdmatso (\ie also does not support remote RMWs). 
As shown in \cref{fig:overview}, the \rdmawait library itself is implemented using \rdmatso.

Importantly, \rdmawait abstracts \rdmatso by replacing its non-modular operation ($\rtsopoll$) with a modular analogue ($\rlwait$, see \cref{sec:ov-background}). 
As such, unlike \rdmatso, \rdmawait is \emph{modular} and can be \emph{composed} with other \loco libraries (thanks to its \rlwait operation).  
Accordingly, we develop \rdmasync by extending \rdmawait with RMW operations. 
Specifically,  in \rdmasync we specify two remote RMWs: \rlrcas (remote
compare-and-swap) and \rlrfaa (remote fetch-and-add). 
In doing so, we also ensure that our extensions are compatible with \rdmawait and the modular design of \loco, thus guaranteeing that \rdmasync is also modular and can be composed with other \loco libraries.

We next use \rdmasync to develop several RDMA libraries (\cref{fig:overview}). 
First,  we combine \rdmasync with the \emph{shared variable} (\brl) library (that provides a mechanism for broadcasting to many nodes) of \loco to develop three lock libraries with varying synchronisation
guarantees (\cref{sec:locklib}), each offering a different trade-off between intuitive behaviours and efficiency. 
Second,  we develop an RDMA library with strong \emph{sequential consistency} (SC)~\cite{sc} semantics (\cref{sec:strl}).

Our first lock library is a \emph{weak lock}, \lklw, that provides \emph{mutual
exclusion} across multiple threads over the network, but does not
provide any ordering guarantees on RDMA instructions enclosed within critical sections.
Nevertheless, it is possible to recover such strong ordering guarantees on RDMA operations within
a \lklw critical section by inserting a \emph{global fence} immediately before the
lock is released. 
To capture this, we thus develop a \emph{strong lock}, \lkls, that guarantees the desired strong guarantees by executing a global fence before releasing the lock.
The most novel aspect of our library is the notion of a \emph{node lock}, \lkln,
that takes a node $n$ as a parameter, and only guarantees synchronisation
on RDMA operations specific to $n$, while operations within a critical
section acting on a different node $n' \ne n$ are left
unsynchronised. 

Interestingly, we show that it is possible to build a
novel, strong model for RDMA using \lkln. 
Specifically, we develop the \strl library, which, unlike \rdmawait, provides support
for strong isolation of remote RMW instructions, with strong synchronisation
akin to SC.
\footnote{In related work, Ambal \etal~\cite{ESOP-25} write
  \text{\sc rdma}\textsuperscript{{\sc sc}} for an RDMA model where the underlying CPU is SC
  (instead of TSO). This is unrelated to \rdmascrmw. }

For each library $L$ in our development (\cref{fig:overview}), we 
\begin{enumerate*}
	\item \emph{formally specify} $L$; 
	\item develop a \emph{reference implementation} of $L$ using lower-level
libraries; and
	\item \emph{prove} our implementation is \emph{correct} against its specification.
\end{enumerate*}
For (1) and (3), we use \framework~\cite{POPL-LOCO}, a declarative
framework previously used to verify a subset of \loco
 (those \emph{without} RMWs). 
\framework is a compositional framework for specification and verification of very weak libraries where program order is not preserved (\eg RDMA programs). 
However, previous definitions~\cite{POPL-LOCO} are not sufficient to specify
remote RMWs out of the box, and we extend them with the features needed~(\cref{sec:rdmasync}).


\paragraph{Contributions}
Our core contributions are as follows.
\begin{enumerate*}[label=\bfseries(\arabic*)]
\item We develop the \emph{first formal semantics of remote RMWs} through the
  \rdmatsormw and \rdmasync models by carefully inspecting the (informal)
  technical specification~\cite{rdmaspec}. Our models have further been
  validated by NVIDIA engineers.
\item We extend the definitions of \framework to support RMW operations, and use
  it to develop \emph{several programmer-friendly and composable RDMA
    libraries}. Specifically,
  we specify, implement and verify \emph{three lock libraries} offering varying
  degrees of synchronisation guarantees and efficiency; and
\item we develop a novel, strong RDMA model, \strl, ensuring strong isolation of
  RDMA instructions with strong synchronisation guarantees of SC.
\end{enumerate*}

\paragraph{Outline}
The remainder of this article is organised as follows. 
In \cref{sec:ov} we discuss the necessary background and present an intuitive overview of our contributions.  
In \cref{sec:rdmasync} we describe how we extend the \framework framework and present our \rdmasync model.
In \cref{sec:locklib} we present our three lock libraries (including their specification,
implementation, and verification), which we build on top of \rdmasync.
In \cref{sec:strl} we specify, implement, and verify our \strl library (simulating SC in RDMA programs). 
Finally, we discuss related work in \cref{sec:related-work}.


%% file: overview.tex

\section{Background and Overview}
\label{sec:ov}


We present an intuitive account of our contributions via a series of litmus tests.  
We begin with a summary of necessary background (\cref{sec:ov-background} and \cref{sec:backgr-fram}).
We discuss the behaviour of remote RMW (`read-modify-write') synchronisation, culminating in our formal \rdmawaitrmw
model (\cref{sec:ov-rmw}).
We then describe our RDMA libraries (\cref{sec:ov-locks}), including locks and
a library for sequential consistency we build from it.

\paragraph{Terminology and Litmus Test Notation}
Throughout this article, we present small examples (litmus tests) to
highlight particular behaviours. A single vertical bar (\eg in
\cref{subfig:lklw-sync}) separates threads on the \emph{same}
(network) node, while a double vertical bar (\eg in
\cref{subfig:rempollcpu}) separates \emph{distinct} nodes.  For each
annotated outcome, \checkyes denotes that the outcome is
\emph{allowed} by the semantics, while \checkno states that the
outcome is \emph{disallowed}. To distinguish local and remote (memory)
locations, we write $x^\node$ for a location on a remote node $n$, and
write $x$ for a location on the current local node.  We number nodes
from left to right, starting at $1$. The statement on the top line of
each column denotes where locations reside as well as their initial
values; \eg $x\!=\!0$ and $z\!=\!0$ on top of \cref{subfig:rempollcpu}
denote that $x$ and $z$ respectively reside on nodes $1$ and $2$ with
initial value $0$. When a thread on local node $n$ issues a remote
operation to be executed on remote node $n'$, we denote this by
stating that the operation is by $n$ \emph{towards} $n'$.



\input{overview-background}

\input{overview-rmw}

\input{overview-libraries}


%% file: overview-background.tex

\subsection{Background: \rdmatso, \rdmawait, and \loco}
\label{sec:ov-background}



\paragraph{The \rdmatso Model}
Ambal \etal~\cite{OOPSLA-24} developed \rdmatso, the first formal
model of RDMA programs where the underlying CPUs are assumed to follow
the x86-TSO memory model~\cite{DBLP:conf/tphol/OwensSS09}.
\rdmatso formalises the semantics of \emph{RDMA Writes} (referred to as \emph{\textputs}),  
\emph{RDMA Reads} (referred to as \emph{\textgets}) and \emph{polling} instructions, executed by the \emph{network interface card} (NIC).
A put operation towards $n$, written $x^\node \assign y$,  reads from local location $y$ (referred to as a \emph{NIC local read}) and writes to remote location $x$ on node $n$ (a \emph{NIC remote write}). 
Similarly, a get operation towards $n$, written $y \assign x^\node$ reads from remote location $x$ (a \emph{NIC remote read}) and writes to local location $y$ (a \emph{NIC local write}). 
The \rdmatso semantics is  unintuitive as remote operations are executed by NIC \emph{independently} from later CPU operations,  \emph{as if} run in parallel to them. 
For instance, the program $z^2 \assign x ; x \assign 1$ (comprising a put towards node $2$, followed by a standard CPU store) can result in $z$ containing value $1$ as
follows:
 \begin{enumerate*}[label=\arabic*)]
 \item CPU offloads the put instruction to the NIC;
 \item CPU executes $x \assign 1$;
 \item NIC executes the put, fetching the {\em new} value $1$ of $x$ and
   updating the remote location $z$ in node $2$ to this new value.
 \end{enumerate*}
To prevent this weak behaviour, a programmer can \emph{poll} the remote instruction (towards node $2$) by executing $\rtsopoll(2)$, as shown in \cref{subfig:rempollcpu}: this blocks the CPU until the NIC confirms that the put has been executed, thereby preventing the above scenario. 


\input{poll_vs_wait}

The polling system on RDMA hardware (and thus \rdmatso) is highly brittle in that it synchronises with the \emph{earliest} (in program order) unpolled remote operation. 
For instance, in \cref{subfig:remrempollcpu} the single poll only acknowledges the first put, and the second put can be arbitrarily
delayed, once again enabling the outcome $z=1$. 
Preventing unintended weak behaviours therefore often relies on  \emph{counting} remote operations and polling them accordingly; \eg in this case we must use two polls to prevent the weak outcome, as in \cref{subfig:remrempollpollcpu}.

\paragraph{The \rdmawait Model}
The non-local semantics of polls does not lend itself to compositional programming and verification. 
That is,  the polling semantics depends on the \emph{exact number} of earlier remote operations towards the same node.
To address this, recent work developed \loco~\cite{POPL-LOCO} as an RDMA library for composable  objects with a more abstract completion system that ensures modularity and compositionality
through a \emph{waiting} instruction that is analogous to polling but is compositional. 
Specifically,  in \loco each remote operation is associated with a \emph{work identifier}, $\wid \in \Wid$, and the wait operation $\rlwait(\wid)$ ensures the acknowledgement of all previous operations with this identifier (multiple remote operations may have the same identifier). 
This is illustrated in \cref{subfig:wait1,subfig:wait2} (obtained from \cref{subfig:rempollcpu,subfig:remrempollcpu} by replacing polls with waits), where $z^2 \assign^d x $ denotes a put (as before) with work id $d$.
Unlike previously in \cref{subfig:remrempollcpu}, adding an earlier put in
\cref{subfig:wait2} (with different work id $e$) towards the same remote node
does not alter the behaviour of $\rlwait(\wid)$. Since, in this case, the
ordering between the two puts is also preserved, the weak outcome $z\!=\!1$
remains prohibited.

From a reordering perspective, \rdmawait is still quite
permissive. For example, because a remote NIC sends an acknowledgement
for a put as soon as it is received (but before the put takes effect
in memory), \rdmawait permits the store-buffering behaviour in \cref{fig:wait-SB}.  
Therefore, using \rdmawait,
\loco additionally implements a \emph{global-fence operation} towards a node
$n$, written $\brlgf(\set{\node})$, that blocks until all previous
remote operations towards $\node$ are \emph{fully} completed
(see~\cref{sec:brl}). 
Replacing $\rlwait(d)$ and $\rlwait(e)$ in
\cref{fig:wait-SB} respectively with fences $\brlgf(\set{2})$ and
$\brlgf(\set{1})$ would prevent the store-buffering
behaviour.





\subsection{Background: \framework}
\label{sec:backgr-fram}
To support compositional specification and verification, Ambal \etal have developed the
\framework framework~\cite{POPL-LOCO}. 
They have specified the \rdmawait formal model (obtained from \rdmatso by replacing the \code{poll} instruction with
\code{Wait}) in \framework and subsequently used it as a foundation for developing 
and verifying a suite
of RDMA libraries. The principal one is a \emph{shared variable} (\brl)
library (see~\cref{sec:brl}), where each node possesses a local copy of each
variable $x$. The methods include store ($x \assignbrl v$) and load
($a \assignbrl x$) operations to access the local copy, as well as a broadcast
($\brlbr(x)$) operation to forward the local value to other nodes.

\paragraph{Specification}
\framework~\cite{POPL-LOCO} is a declarative framework for modularly specifying and
verifying libraries in the context of (very) weak concurrency models. 
Unlike other  declarative frameworks in the literature~\cite{DBLP:journals/pacmpl/RaadDRLV19,DBLP:conf/esop/StefanescoRV24}, \framework can handle the 
behaviours allowed by RDMA programs. 
The key novelty in \framework enabling this is the use of a fixed set of
\emph{stamps}, $\Tags = \set{\tagt_1, \ldots}$, and the \emph{stamp-order} relation,
$\tagppo \suq \Tags \times \Tags$, defined as a subset of the program order that is \emph{preserved}. This then allows one to define weak libraries where the program order is not fully preserved, as is the case in RDMA.

We present the stamps and their ordering in~\cref{fig:to} (assuming
that the underlying CPUs follow the TSO model). Intuitively, each
stamp denotes a behaviour category, such as a CPU write (\tagcwrite),
a CPU read (\tagcread), a NIC remote read (\tagnrr) or write (\tagnrw)
towards $n$, or a NIC local read (\tagnlr) or write (\tagnlw) towards
$\node$. Compared to~\cite{POPL-LOCO}, we also introduce a new stamp
\tagnarr to represent the ordering guarantees of remote RMWs
(see~\cref{sec:ov-rmw}).




\begin{figure}[t]
  \begin{minipage}[b]{.3\textwidth}
    \begin{subfigure}[b]{\textwidth}
      \centering
      \vspace{0pt} \small
      \scalebox{0.9}{  \begin{tabular}{|@{\hspace{3pt}}c @{\hspace{3pt}} || @{\hspace{3pt}}c@{\hspace{3pt}}|}
    \hline
    $y=0$
    & $x=0$\\
    \hline
    $\inarr{\phantom{a} \vspace{-6pt}\\
    x^2 \assign^d 1 \\
    \rlwait{(d)} \\
    a \assign y\\\phantom{a} \vspace{-6pt}
    }$ &
         $\inarr{\phantom{a} \vspace{-6pt}\\
         y^1 \assign^e 1 \\
    \rlwait{(e)} \\
    b \assign x\\\phantom{a} \vspace{-6pt}
    }$ \\
    \hline
      \end{tabular}}
    \vspace{-5pt}
    $$(a,b)=(0,0)\ \text{\checkyes}$$
    \vspace{-20pt}
  \end{subfigure}
  \caption{Store buffering}
  \label{fig:wait-SB}
  \end{minipage}
  \hfill
  \begin{minipage}[b]{.65\textwidth}
    \begin{subfigure}[b]{.32\textwidth}
      \centering
      \vspace{0pt} \small
      \scalebox{0.9}{\begin{tabular}{|@{\hspace{2pt}}c @{\hspace{2pt}} || @{\hspace{2pt}}c @{\hspace{2pt}}|}
        \hline
        \multicolumn{2}{|c|}{$\myblue{\SVar \ x = 0}$} \\
        \hline
        & $z = 0$ \\
        \hline
        $\inarr{\phantom{a} \vspace{-6pt} \\
        z^2 \assign 1 \\ \myblue{x \assignbrl 1} \\ \myblue{\brlbr(x)} \\
        \phantom{a} \vspace{-6pt}}$
        & $\inarr{\myblue{a \assignbrl x} \\ b \assign z}$
        \\
        \hline
      \end{tabular}}
        \vspace{-5pt}
    \caption{$(a,b)\!=\!(1,0)$ \text{\checkno}}
    \label{subfig:bcast1}
  \end{subfigure}
  \qquad
  \begin{subfigure}[b]{.55\textwidth}
    \centering
    \vspace{0pt} \small
    \scalebox{0.9}{\begin{tabular}{|@{\hspace{2pt}}c @{\hspace{2pt}} || @{\hspace{2pt}}c @{\hspace{2pt}} || @{\hspace{2pt}}c @{\hspace{2pt}}|}
      \hline
      \multicolumn{3}{|c|}{$\myblue{\SVar \ x = 0}$} \\
      \hline
      & $y, z = 0, 0$ &   \\
      \hline
      $\inarr{\phantom{a} \vspace{-6pt} \\
      z^2 \assign 1 \\ \myblue{x \assignbrl 1} \\ \myblue{\brlbr(x)} \\
      \phantom{a} \vspace{-6pt}}$
      & $\inarr{a \assign y \\ b \assign z}$
      & $\inarr{\myblue{c \assignbrl x} \\ y^2 \assign 1}$
      \\
      \hline
    \end{tabular}}
  \caption{$(a,b,c)=(1,0,1)\ \text{\checkyes}$}
  \label{subfig:bcast2}
\end{subfigure}
    \vspace{-5pt}
\caption{Shared variable examples}
\label{fig:broadcast}
\end{minipage}
\end{figure}

This stamp mechanism addresses two problems. The first is the reordering of methods
of different libraries. As libraries are defined \emph{independently}, the exact
interaction between pairs of methods of different libraries cannot be explicit.
Instead, libraries can associate their method calls with generic behaviour categories (stamps), so that their interactions can be implicitly deduced.
For instance, in \cref{subfig:bcast1}, $z^2 \!\assign\! 1$ and $b \!\assign\! z$ are
part of \rdmawaitrmw, while $\myblue{\brlbr(x)}$ and $\myblue{a \!\assignbrl\! x}$
are part of the \brl library. To determine if the outcome
$(a,b) \!=\! (1,0)$ is allowed, we need to check if $z^2 \!\assign\! 1$ and
$\myblue{\brlbr(x)}$ can be reordered on node $1$, and if
$\myblue{a \assignbrl x}$ and $b \!\assign\! z$ can be reordered on node~$2$. The
semantics of the two libraries (\cref{sec:rdmasync} and
\iftoggle{shortpaper}{\evlink}{\cref{sec:app-brl}})
ensure that $z^2 \!\assign\! 1$ and $\myblue{\brlbr(x)}$ behave as remote writes towards
node $2$ (stamp $\tagnrw[2]$) and that $\myblue{a \assignbrl x}$ and
$b \assign z$ behave as CPU reads (stamp \tagcread).
This enforces their respective program orders as
$\tup{z^2 \assign 1, \tagnrw[2]} \allowbreak \arr{\ppo} \myblue{\tup{\brlbr(x),
    \tagnrw[2]}}$ and
$\myblue{\tup{a \assignbrl x, \tagcread}} \arr{\ppo} \tup{b \assign z,
  \tagcread}$, where $\ppo$ is the \emph{preserved program order}, \ie they cannot be reordered. 
  Moreover, if $a\!=\!1$, then we have the \emph{happens-before} ($\hb$) relation 
$\myblue{\tup{\brlbr(x), \tagnrw[2]}} \arr{\hb} \myblue{\tup{a \assignbrl x,
    \tagcread}}$, and as $\ppo \suq \hb$, by transitivity we have $\tup{z^2 \assign 1, \tagnrw[2]} \arr{\hb} \tup{b \assign z,
  \tagcread}$, \ie the weak outcome $(a,b) \!=\! (1,0)$ is prohibited.


The second problem stamps address is the \emph{partial} execution of methods. A
method call may have multiple visible effects, and observing one does not
necessarily imply that others are also observed. In \cref{subfig:bcast2} the
shared variable $\myblue{x}$ is read by the \emph{third} node, which then sends
a message to node~$2$ (through $y^2 \assign 1$). 
As such, when $(a,c)\!=\!(1,1)$, we
have a $\arr\hb$ chain from $\myblue{\brlbr(x)}$ to $b \assign z$ and may naturally expect $b\!=\!1$.
However, this is \emph{not} the case. 
Specifically,  as per the semantics of \brl, $\myblue{\brlbr(x)}$ is associated with (at least) two
stamps, $\tagnrw[2]$ (remote write towards node $2$) and $\tagnrw[3]$ (remote
write towards node $3$), where the latter is observed but is \emph{not}
ordered with the earlier $z^2 \assign 1$ operation (as they are toward different nodes).
That is, we have the $\hb$ orders 
$\tup{z^2 \assign 1, \tagnrw[2]} \arr{\ppo \suq \hb} \myblue{\tup{\brlbr(x),
    \tagnrw[2]}}$ (as in example \cref{subfig:bcast1}) and
$\myblue{\tup{\brlbr(x), \tagnrw[3]}} \arr{\hb} \tup{b \assign z, \tagcread}$,
and when put together they do \emph{not} imply
$\tup{z^2 \assign 1, \tagnrw[2]} \arr{\hb} \tup{b \assign z, \tagcread}$,
allowing the weak outcome $b\!=\!0$.
In other words, $z^2 \assign 1$ and $\myblue{\brlbr(x)}$ can be \emph{partially}
reordered: although their respective updates (on $z$ and $x$) towards node $2$ stay ordered,  the update on $x$ towards \emph{other} nodes (\ie node $3$) may take place before $z^2 \assign 1$ is executed.
A pair formed by a method call and a stamp is called a subevent. Associating a
method call with multiple stamps generates multiple subevents and allows us to
express such nuances. 



\paragraph{Implementation and Soundness}
Within the \framework framework, Ambal \etal~\cite{POPL-LOCO} also
formalise the notion of a \emph{library implementation} and what it
means for an implementation $I$ to be \emph{sound} against its
specification, \ie that the behaviours of the implementation are
contained in those of its specification.  To enable proving
implementation soundness \emph{compositionally}, they establish a
\emph{local soundness} theorem. Specifically, to
show that an implementation $I$ of library $L$ is correct, one must
show that for \emph{all} client programs $P$ with calls to $L$ (where
$P$ may in general contain calls to libraries other than $L$),
replacing the calls to $L$ with their corresponding (inlined)
implementation yields the same outcomes. Intuitively, as the only
calls being replaced (inlined) are those of $L$, the calls to
libraries other than $L$ should not affect the outcome.  That is, it
should be sufficient to show that the implementation is \emph{locally
  sound} by considering client programs that only constitute calls to
$L$.  Ambal \etal then prove 
that local
soundness implies soundness: if $I$ is a locally sound implementation
of $L$ (\ie for all client programs that only comprise calls to $L$),
then $I$ is a sound implementation of $L$ (\ie for all client
programs).

As we discuss below, we use \framework to specify several RDMA libraries and verify their implementations, as shown in \cref{fig:overview}.




%% file: poll_vs_wait.tex

\begin{figure}[t]
  \begin{minipage}[b]{0.62\textwidth}
  \begin{subfigure}[b]{0.26\textwidth}\small
  \vspace{0pt}
  \scalebox{0.95}{\begin{tabular}{|@{\hspace{3pt}}c @{\hspace{3pt}} || @{\hspace{2pt}}c@{\hspace{2pt}}|}
  \hline
  $x\!=\!0$
  & $z\!=\!0$\\
  \hline
  $\inarr{
     \phantom{a} \\
    z^2 \assign x \\ 

    \rtsopoll{(2)} \\
    x \assign 1 \\ 
    \phantom{a}
  }$
  & \\
  \hline
  \end{tabular}}
  \caption{
  $z=0$\; \checkyes\\ \phantom{(a)} $z=1$\; \checkno}
  \label{subfig:rempollcpu}
  \end{subfigure}
  \quad
  \begin{subfigure}[b]{0.26\textwidth}\small
  \vspace{0pt}
  \scalebox{0.95}{\begin{tabular}{|@{\hspace{3pt}}c @{\hspace{3pt}} || @{\hspace{2pt}}c@{\hspace{2pt}}|}
  \hline
  $x\!=\!0$
  & $z\!=\!0$\\
  \hline
  $\inarr{
    z^2 \assign x \\ 
    z^2 \assign x \\ 
    \rtsopoll{(2)} \\
        x \assign 1\\
        \phantom{a}
  }$
  & \\
  \hline
  \end{tabular}}
  \caption{$z=0$\;\checkyes\\
  \phantom{(b)} $z=1$\,\checkyes}
  \label{subfig:remrempollcpu}
\end{subfigure} \quad
  \begin{subfigure}[b]{0.26\textwidth} \small
  \vspace{0pt}
  \scalebox{0.95}{\begin{tabular}{|@{\hspace{3pt}}c @{\hspace{3pt}} || @{\hspace{2pt}}c@{\hspace{2pt}}|}
  \hline
  $x\!=\!0$
  & $z\!=\!0$\\
  \hline
  $\inarr{
    z^2 \assign x \\ 
    z^2 \assign x \\ 
    \rtsopoll{(2)} \\
    \rtsopoll{(2)} \\
    x \assign 1
  }$
  & \\
  \hline
  \end{tabular}}
   \caption{$z=0$\;\checkyes\\
  \phantom{(c)} $z=1$\,\checkno}
  \label{subfig:remrempollpollcpu}
\end{subfigure}
\vspace{-5pt}
\caption{Polling on \rdmatso}
\label{fig:ex-rdmatso}
\end{minipage}
\hspace{-1em}
\begin{minipage}[b]{0.37\textwidth}
    \begin{subfigure}[b]{0.46\textwidth} \small
      \vspace{0pt}
      \scalebox{0.95}{
  \begin{tabular}{|@{\hspace{3pt}}c @{\hspace{3pt}} || @{\hspace{2pt}}c@{\hspace{2pt}}|}
  \hline
  $x\!=\!0$
  & $z\!=\!0$\\
  \hline
  $\inarr{
    \phantom{a} \\
    z^2 \assign^d x \\ 

    \rlwait{(d)} \\
    x \assign 1 \\ 
    \phantom{a}
  }$
  & \\
  \hline
  \end{tabular}}
   \caption{$z=0$\;\checkyes\\
  \phantom{(a)} $z=1$\,\checkno}
  \label{subfig:wait1}
  \end{subfigure}
  \quad
  \begin{subfigure}[b]{0.45\textwidth}\small
  \vspace{0pt}\scalebox{0.95}{
  \begin{tabular}{|@{\hspace{3pt}}c @{\hspace{3pt}} || @{\hspace{2pt}}c@{\hspace{2pt}}|}
  \hline
  $x\!=\!0$
  & $z\!=\!0$\\
  \hline
  $\inarr{
    z^2 \assign^e x \\ 
    z^2 \assign^d x \\ 
    \rlwait{(d)} \\
        x \assign 1\\
        \phantom{a}

  }$
  & \\
  \hline
  \end{tabular}}
   \caption{$z=0$\;\checkyes\\
  \phantom{(b)} $z=1$\,\checkno}
  \label{subfig:wait2}
\end{subfigure} 

\vspace{-5pt}
\caption{Waiting on \rdmawait}
\label{fig:ex-rdmawait}
\end{minipage}
\end{figure}


%% file: overview-rmw.tex

\subsection{Remote Read-Modify-Write Operations}
\label{sec:ov-rmw}



\paragraph{CPU Read-Modify-Writes}
Read-modify-writes (RMW) are a category of synchronisation operations that simultaneously read the value $v$ of a location and update (modify-write) it in place. 
Examples of common RMWs include the compare-and-swap, $\rlcas(x,v_1,v_2)$, instruction (it reads the current value $v$ of $x$ and updates it to $v_2$ if $v=v_1$ and otherwise leaves it unchanged); and the fetch-and-add, $\rlfaa(x,v)$, instruction (it increments the value of $x$ by $v$ unconditionally). 
Both operations return the old value of $x$. 
These operations are 
useful for ensuring inter-thread synchronisation and are often used to implement strong synchronisation mechanisms such as locks (mutexes).

\begin{wrapfigure}[5]{r}{0.32\textwidth}\footnotesize
  \vspace{-20pt}
  \begin{tabular}{|@{\hspace{3pt}}c @{\hspace{3pt}}|}
  \hline
  $x = 0$ \\
  \hline
  $\inarrII{
    \phantom{a} \vspace{-5.5pt} \\
    a \assign \rlcas(x,0,2) \\
    \phantom{a} \vspace{-5.5pt}
  }{
    x \assign 1
  }$
  \\
    \hline
  \multicolumn{1}{c}{$x = 2$ \checkno }
  \end{tabular}
\end{wrapfigure}

CPU RMWs behave \emph{atomically}: their `read' and `modify-write' phases cannot be interleaved by concurrent instructions. 
As such, RMWs are commonly referred to as `atomic operations'. 
This is illustrated in the example across where the outcome $x\!=\!2$ is disallowed. 
If the right thread executes first,  then $x$ is updated to $1$ and subsequently the \rlcas fails. 
If the left thread executes first, then the right thread overwrites $x$ to $1$.




\paragraph{Remote RMWs}
The RDMA hardware specification~\cite{rdmaspec} optionally supports two remote RMW instructions, referred to as `atomics\footnote{Although RMWs are commonly referred to as `atomics' in the RDMA specification,  they do \emph{not} always behave atomically.}':
$\rlrcas(a,x,v_1,v_2)$,  analogous to $a \assign \rlcas(x,v_1,v_2)$ on CPUs,  and $\rlrfaa(a,x,v)$,  analogous to $a \assign \rlfaa(x,v)$ on CPUs, where $x$ is a remote location in both cases. %

\begin{figure}[t]
\begin{subfigure}[b]{0.24\textwidth}\small
  \begin{tabular}{|@{\hspace{3pt}}c @{\hspace{3pt}} || @{\hspace{3pt}}c@{\hspace{3pt}}|}
  \hline
  & $x = 0$ \\
  \hline
  $\inarr{
    \phantom{a} \vspace{-5.5pt} \\
    \rlrcas\\(a,x^2,0,2) \\
    \phantom{a} \vspace{-5.5pt}
  }$
  &
  $\inarr{
    x \assign 1
  }$
  \\
    \hline
  \end{tabular}
  \caption{$x = 2$ \checkyes}
  \label{subfig:rcas-atomicity-1}
\end{subfigure}
\quad
\begin{subfigure}[b]{0.34\textwidth} \small
  \begin{tabular}{|@{\hspace{3pt}}c @{\hspace{3pt}} || @{\hspace{3pt}}c@{\hspace{3pt}} || @{\hspace{3pt}}c@{\hspace{3pt}}|}
  \hline
  & & $x = 0$ \\
  \hline
  $\inarr{
    \phantom{a} \vspace{-5.5pt} \\
    \rlrcas\\(a,x^3,0,2) \\
    \phantom{a} \vspace{-5.5pt}
  }$
  &
  $\inarr{
    x^3 \assign 1
  }$ &
  \\
  \hline
  \end{tabular}
  \caption{$x = 2$ \checkyes}
  \label{subfig:rcas-atomicity-2}
\end{subfigure}
\quad
\begin{subfigure}[b]{0.31\textwidth}
  \begin{tabular}{|@{\hspace{3pt}}c @{\hspace{3pt}} || @{\hspace{3pt}}c@{\hspace{3pt}} || @{\hspace{3pt}}c@{\hspace{3pt}}|}
  \hline
  & & $x = 0$ \\
  \hline
  $\inarr{
    \phantom{a} \vspace{-5.5pt} \\
    \rlrcas\\(a,x^3,0,2) \\
    \phantom{a} \vspace{-5.5pt}
  }$
  &
  $\inarr{
    \rlrfaa\\(b,x^3,1) \\
  }$ &
  \\
    \hline
  \end{tabular}
  \caption{$x = 2$ \checkno}
  \label{subfig:rcas-atomicity-3}
\end{subfigure}
\vspace{-5pt}
\caption{Examples showcasing the limited atomicity of remote RMW operations}
\label{fig:rcas-atomicity}
\end{figure}

Unlike CPU RMWs, remote RMWs do \emph{not} always behave atomically: 
their `read' and `modify-write' phases may be interleaved by other CPU or (remote) put/get instructions. 
This is illustrated in the examples of \cref{subfig:rcas-atomicity-1,subfig:rcas-atomicity-2}, executing a remote CAS in parallel with a CPU store (\cref{subfig:rcas-atomicity-1}) and a put (\cref{subfig:rcas-atomicity-2}), where the remote CAS can first read $0$ from $x$,  be interleaved with the concurrent CPU store/put writing $1$ to $x$, and then update $x$ to $2$. 

This weakness is due to an inherent hardware limitation.  
Atomicity is possible on CPUs because a CPU core can:
\begin{enumerate*}[label=\arabic*)]
\item request exclusive access to a cache line;
\item read the cache line;
\item write to the cache line;
\item release the cache line.
\end{enumerate*}
During periods of exclusive access, other components (\eg other CPU cores or the NIC) cannot access the cache line in-between the `read' and
`modify-write'.
This, however, is not feasible over RDMA since NICs cannot lock 
a cache line; they can only submit read and write operations to their PCIe root complex. 
As such, it is not possible to block accesses by other components (\eg the CPU) interleaving
between the NIC's `read' and the `modify-write'.

Nevertheless, remote RMWs do behave atomically with respect to other remote RMWs. 
For instance, as shown in \cref{subfig:rcas-atomicity-3}, a remote FAA cannot interleave between the `read' and `modify-write' phases of a remote CAS.

In practice, one can ensure atomicity of accesses to a
location $x$ by ensuring $x$ is \emph{only} ever accessed through
remote RMWs.  As such, it is common for RDMA programs to access {\em
  local} locations (\ie those residing on their node) through remote
RMWs (via loop-back). To provide atomicity between remote RMWs and
other operations, we require software solutions, as supported
by \rdmascrmw. 



\paragraph{Extending \rdmatso with RMWs}
The \rdmatso and \rdmawait models do not include the semantics of remote RMWs; 
we close this gap in this work.
Specifically, starting from \rdmatso~\cite{OOPSLA-24}, we formulate
\rdmatsormw both declaratively and operationally and prove the two
characterisations are equivalent (see \iftoggle{shortpaper}{\evlink}{\cref{sec:rmw}}).

Our main reference for modelling the semantics of remote RMWs is the Infiniband technical
specification~\cite{rdmaspec}. However, as the specification is often
ambiguous, we developed our model in close collaboration with NVIDIA experts
specialising in RDMA hardware who confirmed the expected behaviours of RMWs and that our model captures them faithfully. 



Compared to \rdmatso, our \rdmatsormw declarative model brings two
important changes. The first is a new relation, the
`\emph{remote-atomic-order}' $\rao$, capturing the mutual exclusion of
remote RMWs. We require a total order $\rao_\node$ on
remote RMWs towards each node $\node$, such that $\rao_\node \suq \hb$ (\ie it induces synchronisation). 
%
The second is a new stamp, \tagnarr (the `NIC atomic read'), encoding
the new ordering guarantees of the read phase of a remote RMW (see
\cref{fig:to}). 
Recall that a \rlget performs a NIC remote read (stamp \tagnrr) followed by a NIC local write (\tagnlw), while a \rlput performs a NIC local read (\tagnlr) followed by a NIC remote write (\tagnrw). 
Analogously, a remote RMW, \eg $\rlrfaa(x,y,v)$, performs (up to) three NIC accesses: 
\begin{enumerate*}[label=\arabic*)]
	\item it remotely reads $y$ (\tagnarr); 
	\item remotely updates $y$ (\tagnrw); and
	\item locally writes (the return value) to $x$ (\tagnlw).  
\end{enumerate*}


\begin{figure}[t]
\begin{subfigure}[b]{0.25\textwidth}\small
  \begin{tabular}{|@{\hspace{3pt}}c @{\hspace{3pt}} || @{\hspace{3pt}}c@{\hspace{3pt}}|}
  \hline
  & $x,y = 0,0$ \\
  \hline
  $\inarr{
    \phantom{a} \vspace{-5.5pt} \\
    a \assign x^2 \\
    y^2 \assign 1 \\
    \phantom{a} \vspace{-5.5pt}
  }$
  &
  $\inarr{
    b \assign y \\
    x \assign 1
  }$
  \\
    \hline
  \end{tabular}
  \caption{$(a,b) = (1,1)$ \checkyes}
  \label{fig:rcas-reordering-1}
\end{subfigure}
\;\;
\begin{subfigure}[b]{0.34\textwidth}\small
  \begin{tabular}{|@{\hspace{3pt}}c @{\hspace{3pt}} || @{\hspace{3pt}}c@{\hspace{3pt}}|}
  \hline
  & $x,y = 0,0$ \\
  \hline
  $\inarr{
    \phantom{a} \vspace{-5.5pt} \\
    \rlrcas(a,x^2,8,9) \\
    y^2 \assign 1 \\
    \phantom{a} \vspace{-5.5pt}
  }$
  &
  $\inarr{
    b \assign y \\
    x \assign 1
  }$
  \\
    \hline
  \end{tabular}
  \caption{$(a,b) = (1,1)$ \checkno}
  \label{fig:rcas-reordering-2}
\end{subfigure}
\;\;
\begin{subfigure}[b]{0.34\textwidth}\small
  \begin{tabular}{|@{\hspace{3pt}}c @{\hspace{3pt}} || @{\hspace{3pt}}c@{\hspace{3pt}}|}
  \hline
  $y = 0$ & $x = 0$ \\
  \hline
  $\inarr{
    \phantom{a} \vspace{-5.5pt} \\
    \rlrfaa(\_,x^2,1) \\
    \rtsopoll(2) \\
    a \assign y \\
    \phantom{a} \vspace{-5.5pt}
  }$
  &
  $\inarr{
    \rlrfaa(\_,y^1,1) \\
    \rtsopoll(1) \\
    b \assign x
  }$
  \\
    \hline
  \end{tabular}
  \caption{$(a,b) = (0,0)$ \checkyes}
  \label{fig:rcas-reordering-3}
\end{subfigure}
\vspace{-5pt}
\caption{Examples of remote RMW behaviours and how they compare from \rlput{s}.}
\label{fig:rcas-reordering}
\end{figure}

Note that the stamp \tagnarr is required because a remote read stamp
(\tagnrr) is insufficient for modelling the stronger ordering
guarantees of the `read' phase of an RMW.  We show an
example of this in
\cref{fig:rcas-reordering-1,fig:rcas-reordering-2}.  The (remote) read
phase of a \rlget (\tagnrr) may be delayed (reordered) after a later
remote write (\tagnrw of a \rlput or RMW).
As such,  the weak `load-buffering' behaviour in \cref{fig:rcas-reordering-1} is allowed.
By contrast, the read phase of a remote RMW (\tagnarr) cannot be delayed, and thus the analogous behaviour is prohibited in \cref{fig:rcas-reordering-2}.

Finally, the example in \cref{fig:rcas-reordering-3} shows that the remote write
(`modify') phase (\tagnrw) of an RMW behaves similarly to that of a \rlput. In
particular, a poll does \emph{not} enforce the full completion of the remote write and
thus the weak `store-buffering' behaviour presented is allowed, similarly
to~\Cref{fig:wait-SB}.

\paragraph{Supporting Modularity with \rdmawaitrmw} As \rdmatsormw is not modular, we develop \rdmawaitrmw by
adapting \rdmawait~\cite{POPL-LOCO,rdmaloco}.
We then implement
\rdmawaitrmw using \rdmatsormw and prove that it is correct
(\iftoggle{shortpaper}{\evlink}{\cref{sec:rdmatsormw}}) against its specification.


%% file: overview-libraries.tex



\subsection{Modular RDMA Synchronisation Libraries}
\label{sec:ov-locks}

We 
now 
implement RDMA libraries \emph{modularly}
and specify and verify them using \framework.  A key use case of our
remote RMWs is for implementing network-wide locks that ensure mutual
exclusion of critical sections.  A lock library provides two main
operations, $\lklgena(l)$ and $\lklgenr(l)$, for acquiring and
releasing a lock $l$, respectively.  When specifying such a network
lock, there are several choices for defining its semantics as there
are trade-offs between the guarantees (strength) of a lock and the
efficiency of its implementation. 

\Cref{fig:overview-lock} presents several variants of an example where two threads use a lock $l$ to  access locations $x$ and $y$ in a critical section (the first thread writing to $x$ and $y$ and the last thread reading from $x$ and $y$). 
As the locks are expected to ensure atomicity of the critical sections (enclosed within the lock acquisition and release blocks), the expected outcomes are either $a\!=\!b\!=\!0$ or $a\!=\!b\!=\!1$, \ie not $a\!\ne\!b$. 
However, ensuring this strong guarantee for locks is not straightforward over RDMA. 
Specifically, while ensuring mutual exclusion is necessary for prohibiting the weak $a\!\ne\!b$ behaviour, it is not sufficient. We must additionally ensure that the operations enclosed in a critical section are   \emph{completed} before the end of the critical section (and hence are not \emph{reordered} past the lock release). 
However, as we demonstrated above,  meeting these latter constraints are not always straightforward due to the weak ordering guarantees on remote operations.


\paragraph{Weak Lock Library}
The weakest network lock that we  consider 
ensures mutual exclusion \emph{only}, but does not prohibit the enclosed operations from being reordered. 
As shown in \cref{subfig:lklw-sync}, when the operations enclosed in a critical section are CPU loads and stores, the weak outcome $a\!\ne\!b$ is prohibited.  
By contrast, as shown in \cref{subfig:lklw}, when the enclosed operations are over RDMA (two \rlput{s} in \cref{subfig:lklw}), then a weak lock is insufficient and we may observe $a\!\ne\!b$.
This is because the remote operations may not complete before the critical section ends.  

Thus, we require an operation akin to a global fence (see
\cref{sec:ov-background}) to ensure that these remote operations are
completed.  Note that as shown in \cref{subfig:nolklwgf}, the global
fence in isolation (without the protection provided by a weak lock) is
also insufficient for prohibiting the weak behaviour as it only
provides intra-thread synchronisation (and does not ensure mutual
exclusion).
However, as shown in \cref{subfig:lklwgf}, if we combine a weak lock with a global fence, we can attain the desired strong guarantees and prohibit $a\!\ne\!b$.

\begin{figure}[t]
  \vspace{-10pt}
  \centering
  \noindent
    \begin{subfigure}[b]{0.28\textwidth}\small\centering
  \vspace{0pt}
  \scalebox{0.9}{\begin{tabular}{|@{\hspace{2pt}}c @{\hspace{2pt}}|}
  \hline
  $x,y = 0,0$\\
  \hline
  \phantom{a} \vspace{-8pt} \\
  $\inarrII{
    \lklwa(l) \\
    x \assign 1 \\
    y \assign 1 \\
    \lklwr(l)
  }{
    \lklwa(l) \\
    a \assign x \\
    b \assign y \\
    \lklwr(l)
  }$
    \vspace{-8pt} \\
    \phantom{a} \\
    \hline
  \end{tabular}} 
\caption{$a \neq b$ \checkno}
  \label{subfig:lklw-sync}
\end{subfigure}
\quad
  \begin{subfigure}[b]{0.26\textwidth}\small\centering
  \vspace{0pt}
  \scalebox{0.9}{
  \begin{tabular}{|@{\hspace{3pt}}c @{\hspace{3pt}} || @{\hspace{3pt}}c@{\hspace{3pt}}|}
  \hline
  & $x,y = 0,0$\\
  \hline
  $\inarr{
    \phantom{a} \vspace{-8pt} \\
    \lklwa(l) \\
    x^2 \assign 1 \\
    y^2 \assign 1 \\
    \lklwr(l) \\
    \phantom{a} \vspace{-8pt}
  }$
  &
  $\inarr{
    \lklwa(l) \\
    a \assign x \\
    b \assign y \\
    \lklwr(l)
  }$
  \\
  \hline
  \end{tabular}} 
  \caption{$a \neq b$ \checkyes}
  \label{subfig:lklw}
  \end{subfigure}
\quad
  \begin{subfigure}[b]{0.3\textwidth}\small\centering
  \vspace{0pt}
  \scalebox{0.9}{
  \begin{tabular}{|@{\hspace{3pt}}c @{\hspace{3pt}} || @{\hspace{3pt}}c@{\hspace{3pt}}|}
  \hline
  & $x,y = 0,0$\\
  \hline
  $\inarr{
    x^2 \assign 1 \\
    y^2 \assign 1 \\
    \brlgf(\set{2}) \\
  }$
  &
  $\inarr{
    a \assign x \\
    b \assign y \\
  }$
  \\
  \hline
  \end{tabular}} 
  \caption{$a \neq b$ \checkyes}
  \label{subfig:nolklwgf}
  \end{subfigure}

  \medskip\noindent
  \begin{subfigure}[b]{0.28\textwidth}\small\centering
  \vspace{0pt}
  \scalebox{0.9}{
  \begin{tabular}{@{}|@{\hspace{3pt}}c @{\hspace{3pt}} || @{\hspace{3pt}}c@{\hspace{3pt}}|}
  \hline
  & $x,y = 0,0$\\
  \hline
  $\inarr{
    \lklwa(l) \\
    x^2 \assign 1 \\
    y^2 \assign 1 \\
    \brlgf(\set{2}) \\
    \lklwr(l)
  }$
  &
  $\inarr{
    \lklwa(l) \\
    a \assign x \\
    b \assign y \\
    \lklwr(l)
  }$
  \\
  \hline
  \end{tabular}} 
  \caption{$a \neq b$ \checkno}
  \label{subfig:lklwgf}
  \end{subfigure}
\quad
  \begin{subfigure}[b]{0.26\textwidth}\small\centering
  \vspace{0pt}
  \scalebox{0.9}{
  \begin{tabular}{|@{\hspace{3pt}}c @{\hspace{3pt}} || @{\hspace{3pt}}c@{\hspace{3pt}}|}
  \hline
  $x = 0$ & $y = 0$\\
  \hline
  $\inarr{
    \phantom{a} \vspace{-5.5pt} \\
    \lklsa(l) \\
    y^2 \assign 1 \\
    x \assign 1 \\
    \lklsr(l) \\
    \phantom{a} \vspace{-5.5pt}
  }$
  &
  $\inarr{
    \lklsa(l) \\
    a \assign x^1 \\
    b \assign y \\
    \lklsr(l)
  }$
  \\
  \hline
  \end{tabular}} 
  \caption{$a \neq b$ \checkno}
  \label{subfig:lkls}
  \end{subfigure}
\quad 
  \begin{subfigure}[b]{0.3\textwidth}\small\centering
    \scalebox{0.85}{
      \begin{tabular}{|@{\hspace{3pt}}c @{\hspace{3pt}} || @{\hspace{3pt}}c @{\hspace{3pt}} || @{\hspace{3pt}}c@{\hspace{3pt}}|}
  \hline
  & $x,y = 0,0$ &\\
  \hline
  & \lkln $l$ & \\
  \hline
  $\inarr{
    \phantom{a} \vspace{-8pt} \\
    \lklna(l^2) \\
    x^2 \assign 1 \\
    y^2 \assign 1 \\
    \lklnr(l^2) \\
    \phantom{a} \vspace{-8pt}
  }$
  & &
  $\inarr{
    \lklna(l^2) \\
    a \assign x^2 \\
    b \assign y^2 \\
    \lklnr(l^2)
  }$
  \\
  \hline
\end{tabular}
}
\caption{$a \neq b$ \checkno}
\label{subfig:overview-lkln}
\end{subfigure}
\vspace{-5pt}
\caption{Examples of  weak, strong, and node lock behaviours}
\label{fig:overview-lock}
\end{figure}



\paragraph{Strong Lock Library}
The weak lock library discussed above is efficient and gives programmers full control over synchronisation. 
However, if not used correctly, without the relevant global fences, its guarantees are not as strong as one may expect. 
That is, in designing the weak lock library, we opted for better performance over the strength of guarantees. 
We next develop a \emph{strong} lock library that achieves the desired strong guarantees 
(without the need for additional synchronisation via fences). 
Specifically,  on releasing a strong lock \emph{all} earlier operations are guaranteed to have \emph{fully} completed. 

This is illustrated in \cref{subfig:lkls}, where the outcome $a\!\ne\!b$ is once again prohibited. 
However, the strong guarantees of strong locks come at the cost of their implementation efficiency. 
Intuitively, an implementation of a strong lock release issues a global fence towards \emph{every} node on the network to ensure that there are no pending remote operations. 
This is in contrast to a weak lock release implementation that issues no fence by default, and developers have full control over fencing the relevant nodes. 



\paragraph{Node Lock Library}
As a midway between the efficient weak locks, requiring manual synchronisation, and the
inefficient strong locks, we develop the concept of (more fine-grained) \emph{node} locks.
Intuitively, as shown in \cref{subfig:overview-lkln}, a node lock $l$ is associated with a specific
node $\node$ (node $2$ in \cref{subfig:overview-lkln}) and provides strong guarantees only for locations on $\node$. 

A node lock is stronger than a weak lock: as shown in
\cref{subfig:overview-lkln} the weak behaviour $a\!\ne\!b$ is prohibited without
the need for additional synchronisation. Moreover, a node lock is weaker than a
strong lock in two ways. First, it only provides guarantees for operations
towards \emph{one} node. For instance, consider a variant of the example in
\cref{subfig:overview-lkln} where the lock $l$ is associated with node $1$
(instead of $2$); the outcome $a \neq b$ would once again be allowed.
Second, it does \emph{not} provide any intra-thread ordering guarantees in that
releasing a node lock does not guarantee that previous
operations (even towards the associated node) have completed. For instance, in
the program $ \lklna(l^2) ; z^2 \assign x ; \lklnr(l^2) ; x \assign 1$ the
outcome $z=1$ is \emph{allowed}: $z^2 \assign x$ and the lock release may not
have fully completed before the CPU runs the subsequent $x \assign 1$ store; 
\ie, while $z^2 \assign x$ cannot be reordered past $\lklnr(l^2)$,
the $x \assign 1$ can be reordered \emph{before} both of them. 
More concretely, our implementation of $\lklnr$ (\cref{sec:locklib-node-implem}) comprises RDMA operations that can be delayed after later CPU operations.


Nevertheless, as a common usage of a lock is to protect a specific object that is likely to
reside on a single node, this level of guarantee is sufficient for many
applications, while enabling efficient implementations.



\paragraph{The \rdmascrmw Library}
Lastly, to simplify RDMA programming, we specify and implement the \rdmascrmw library that fully abstracts away the notion of nodes and provides strong \emph{sequentially consistent} (SC)~\cite{sc} semantics via four (per-location) instructions, $\strlw$, $\strlr$, $\strlcas$, and $\strlfaa$ analogous to stores, loads, and RMWs on CPUs with strong SC semantics. 
Our implementation uses node locks to wrap RDMA operations and ensure they become visible in the order they are submitted. 
Indeed, as we discuss later in \cref{sec:strl}, we can use the same approach to implement \emph{any concurrent data structure} over RDMA, and show that it is correct in that it is \emph{linearisable}~\cite{DBLP:journals/toplas/HerlihyW90}.




%% file: rdmawait.tex

\section{Extending \rdmawait to \rdmasync}
\label{sec:rdmasync}
We present \emph{\rdmasync model}, an extension of \rdmawait~\cite{POPL-LOCO} with remote RMW instructions. 
Our definitions naturally extend those of \rdmawait. To underline the distinction between the two, we have \tightshade{highlighted} our extensions from \rdmawait to \rdmawaitrmw. 
We specify \rdmasync in \framework~\cite{POPL-LOCO}, yielding a \emph{modular} semantics that enables \emph{compositional} reasoning.  
In particular, as we show below, since \loco libraries can be freely composed together, this allows us to use the \emph{locality} result of \framework to verify each library modularly (in isolation).
We proceed with an account of \framework preliminaries (\cref{subsec:rdmasync_prelim}) and present \rdmawaitrmw in \cref{subsec:rdmasync_model}.



\subsection{The \framework Framework Preliminaries}
\label{subsec:rdmasync_prelim}

\framework assumes a type $\Val$ of values and a type $\Loc \suq \Val$ of
locations.
We also assume two sets for threads $\threadt \in \Threads$ and nodes
$\node \in \Nodes$, where each thread $\threadt$ is associated with a node
$\nodefun{\threadt}$. Recall from \cref{sec:backgr-fram} that \framework can be
instantiated with a set of \emph{stamps} $\Tags$ and a relation
$\tagppo \suq \Tags \times \Tags$.
In the case of \rdmawait and \rdmawaitrmw, the stamps and their associated
$\tagppo$ are as presented in \cref{fig:to}. Note that certain stamps, \eg
\tagnlr, are associated with a node $\node$, and each induce a \emph{family} of
stamps, \eg $\tagnlr[] \defeq \bigcup_{\node \in \Nodes} \set{\tagnlr}$. The
\shade{highlighted} sections (row H and column 8) denote our extensions from
\rdmawait to \rdmawaitrmw and are associated with the new stamp family
\tagnarr[] used to specify RMWs (see~\cref{sec:backgr-fram}).
The \cyes (\eg in cell A2) denotes that the corresponding stamps (\eg \tagcread
and \tagcwrite) are \emph{ordered}. This means that the program order between
relevent subevents (pair of a function call and a stamp, see \cref{def:events})
is \emph{preserved} and thus their effects are observed in order. Conversely,
\cno denotes that the stamps are \emph{not ordered} (they may be reordered) and
thus the effects of subevents with these stamps may be observed \emph{out of
  order}. The \cqp denotes the stamps are ordered if and only if they are
associated with the \emph{same node}.

\input{tags}



\paragraph{Libraries}
Intuitively,  a library $L$ specification identifies its associated \emph{methods} as well as the semantics of these methods.
A \emph{method call} is of the form $m(\vect{v})$, where $m$ denotes the method name and $\vect{v}$ denote its arguments. 
Ambal \etal capture the method semantics in \framework by identifying the set of \emph{executions} that are \emph{$L$-consistent} in that they uphold the guarantees promised by $L$. 
To this end, they associate $L$ with a set $\cons$ of \emph{$L$-consistent executions}.
A \emph{library} is then formally defined as a triple $L\!=\! \tup{M,\loc,\cons}$, where $M$ is its set of \emph{method names} (\eg
$\rlwrite$ or $\rlput$); $\loc$ associates each method call with its set
of accessed locations (within the method call arguments); and $\cons$ is its set of {\em $L$-consistent executions}.
(\framework further requires $\cons$ to adhere to some basic 
properties to ensure modularity~\cite{POPL-LOCO}, which we elide here.)
We use the prefix `$L.$' to project the components
of a library $L$, \eg $L.M$.



\paragraph{Events and Executions}
In the literature of declarative models, traces of a program are represented as a set of \emph{executions}. 
An execution is a graph comprising:
\begin{enumerate*}
	\item a set of \emph{events} (graph nodes), where each event is associated with the execution of a method call; and 
	\item a number of relations on events (graph edges). 
\end{enumerate*}
For instance, if thread $\threadt$ executes a $\rlread(x)$ and reads value $v$, the corresponding event is of the form $\tup{\threadt, \aident, \tup{\rlread, (x), v}}$, where $\aident$ denotes its (unique) event identifier.  
Identifiers serve to distinguish calls to the same method (with same arguments and output) by the same thread in an execution. 
For an event $\action$, we write $\threadfun{\action}$ and $\methodfun{\action}$ to extract its thread and  method name, respectively.
%
%

\begin{definition}[Events and Executions]
\label{def:execs}
\label{def:events}
An \emph{event} is a tuple $\tup{\threadt, \aident, \tup{m, \vect{v}, v'}}$,
where $\threadt \in \Threads$ denotes the executing thread, $\aident$ is the
(unique) event identifier, $m$ denotes the method being executed,
$\vect v \in \Val^*$ is the method \emph{input} (arguments) and $v' \in \Val$ is
its \emph{output} (return value, which may be unit $()$). An \emph{execution}
$\exec$ is a tuple $\tup{E, \po, \gettags, \so, \hb}$ where:
\begin{itemize}
\item $E$ is the set of events; 
\item $\po \suq E \times E$ is the (strict) program order, total for each thread;
\item $\gettags : E \rightarrow \mathcal{P}(\Tags)$ associates each event with a
  \emph{non-empty} set of stamps and induces a set of \emph{subevents},
  $\SEvents \defeq \setpred{\tup{\action,\tagt}}{\action \in E \land \tagt \in
    \gettags(\action)}$;
\item $\so \suq \SEvents \times \SEvents$ is the \emph{synchronisation order},
  representing the intra-library dependencies exported by each library;
\item $\hb \suq \SEvents \times \SEvents$ is the \emph{happens-before order},
  a strict partial order
  such that $\so \cup \ppo \suq \hb$, where
  $\ppo \suq \SEvents \times \SEvents$ denotes the
  \emph{preserved program order} capturing inter-library dependencies and is
  defined as follows\vspace{5pt}:
\centerline{$\ppo \defeq
{\setpred{\tup{\tup{\action_1,\tagt_1},\tup{\action_2,\tagt_2}}}{
    \tup{\action_1,\action_2} \in \po \land \tagt_i \in \gettags(\action_i)
    \land \tup{\tagt_1,\tagt_2} \in \tagppo}}$\vspace{-5pt}}
\end{itemize}
\end{definition}



\paragraph{Notations}
Given a set $A$ and a relation $\mathsf r \subseteq A \times A$, we write $\mathsf{r}^+$ for the
transitive closure of $\mathsf r$; $\mathsf r^*$ for its reflexive transitive closure; $\inv{\mathsf r}$
for the inverse of $\mathsf r$; and $[A]$ for the identity relation on $A$, \ie
$\set{\tup{a, a} \mid a \!\in\! A}$. We write $\mathsf{r}_1; \mathsf{r}_2$ for the relational
composition of $\mathsf{r}_1$ and $\mathsf{r}_2$:
$\{\tup{a, b} \mid \exists c.\, \tup{a, c} \!\in\! \mathsf{r}_1 \land \tup{c, b} \!\in\! \mathsf{r}_2\}$.
We write $A\rst{c}$ to restrict $A$ with condition $c$.
For instance, given a set of events $E$, we define $E\rst{L} \defeq \setpred{\action \in E}{\methodfun{\action} \!\in\! L.M}$,
$E\rst{\threadt} \defeq \setpred{\action \in E}{\threadfun{\action} = \threadt}$, and we write $E\rst{\wid}$ for the set
of events in $E$ with work identifier $\wid$. 
We define $E_x \defeq \setpred{\action \in E}{x \in \loc(\action)}$.
Similarly, we define 
$\mathsf r\rst{c} \defeq [A\rst{c}];\mathsf r;[A\rst{c}]$ (\eg $\po\rst{\threadt}$) 
and $\mathsf{r}_x \defeq [E_x]; \mathsf r; [E_x]$ (\eg $\po_x$).
Given a subset $A' \subseteq A$, we define $\mathsf r\rst{A'} \defeq [A'];\mathsf r;[A']$. 
When $\mathsf r$ is a strict partial order, we write $\imm{\mathsf r}$ for its immediate edges, \ie
$\mathsf r \setminus (\mathsf r;\mathsf r)$.

Given execution $\exec \!=\! \tup{E, \po, \gettags, \allowbreak\so, \hb}$, we write $\exec\rst{L}$ for
$\langle E\rst{L}, \po\rst{L}, \gettags\rst{L}, \allowbreak\so\rst{L}, \hb\rst{L}\rangle$, where
$\gettags\rst{L}$ denotes the function obtained by restricting the domain of
$\gettags$ (\ie $E$) to $E\rst{L}$.
%
We use the prefix `$\exec.$' to project the components of $\exec$ (\eg
$\exec.\po$), including its derived ones (\eg $\exec.\SEvents$). Given a stamp
$\tagt$, we write $\exec.\tagt$ for
$\setpred{\saction \in \exec.\SEvents}{\saction = \tup{\_, \tagt}}$; analogously
for a stamp family, \eg $\exec.\tagnrr[]$.
%
We define the set of \emph{read subevents} as
$\exec.\Read \defeq \exec.\tagcread \cup \exec.\tagcas \cup \exec.\tagnlr[] \cup
\shademath{\exec.\tagnarr[]} \cup \exec.\tagnrr[]$, and \emph{write subevents}
as
$\exec.\Write \defeq \exec.\tagcwrite \cup \exec.\tagcas \cup
\exec.\tagnlw[]\cup \exec.\tagnrw[]$.
%
Given a set of subevents $A$, we define
$A_x \defeq \setpred{\saction \in A}{\loc(\saction) = \set{x}}$; \eg
$\exec.\Write_x$ is the set of write subevents on $x$.
When the choice of $\exec$ is clear, we omit `$\exec.$', \eg we simply write $\Write$ for $\exec.\Write$ and $[\tagcwrite]$ for $[\exec.\tagcwrite]$.



\paragraph{Consistency}
An execution is \emph{consistent} against a set of libraries $\Lambda$ iff 
\begin{enumerate*}
	\item $\exec\rst L$ is $L$-consistent for each $L \in \Lambda$; 
	\item its events and their synchronisation are those of the libraries in $\Lambda$; and 
	\item its happens-before relation is irreflexive. 
\end{enumerate*}
Note that the first condition ensures \emph{modularity} as each library can specify independently the
visible behaviours of its functions (stamps), its allowed outcomes (consistency) and the synchronisation (guarantees) it offers (\so).

\begin{definition}[Consistency]
\label{def:lambda-cons}
Let $\Lambda$ be a set of libraries where $L_1.M \cap L_2.M \!=\! \emptyset$ for
distinct $L_1, L_2$. An execution $\exec \!=\! \tup{E, \po, \gettags, \so, \hb}$
is {\em $\Lambda$-consistent} iff:
\begin{itemize}
\item For all $L \in \Lambda$: $\exec \rst{L} \in L.\cons$ (\ie $\exec$ is
  $L$-consistent for each $L \in \Lambda$);
\item $E = \bigcup_{L \in \Lambda} E\rst{L}$ and
  $\so = \bigcup_{L \in \Lambda} \so\rst{L}$; and
\item $\hb$ is irreflexive (\ie $\hb$ is a strict partial order). 
\end{itemize}
\end{definition}



\subsection{The Declarative \rdmawaitrmw Model}
\label{sec:rdmasync-sem}

We present \rdmawaitrmw as an extension of \rdmawait~\cite{POPL-LOCO} with remote RMWs. 
Our definitions naturally extend those of \rdmawait. To underline the distinction between the two, we have \tightshade{highlighted} our extensions from \rdmawait to \rdmawaitrmw. 

\label{subsec:rdmasync_model}

\paragraph{The \rdmawaitrmw Methods}
\rdmawaitrmw methods extend those of \rdmawait with remote RMWs as defined by the following grammar, where \rdmawait methods comprise local (CPU) operations on TSO machines and remote operations.  
\vspace{-10pt}
\begin{align*}
m(\vect{v}) & ::=
\rlwrite(x,v)
\mid \rlread(x)
\mid \rlcas(x,v_1,v_2)
\mid \rlmf() && {\color{teal} \text{\small/\!/ \rdmawait: Local}}\\
& \quad \
\mid \rlget(x,y,\wid)
\mid \rlput(x,y,\wid)
\mid \rlwait(\wid)
\mid \rlrf(\node) 
&& {\color{teal} \text{\small/\!/ \rdmawait: Remote}}
 \\
& \quad \
\mid \shademath{\rlrcas(x,y,v_1,v_2,\wid)
  \mid \rlrfaa(x,y,v,\wid)}
  && {\color{teal} \text{\small/\!/ Remote RMWs}}
  \vspace{-10pt}
\end{align*}
The remote operations comprise \rlget, \rlput, \rlwait (as described in \cref{sec:ov-background}), and \rlrf instructions. 
Note that for 
readability in our examples we write $x \assign^\wid y^n$ (\resp $x^n \assign^\wid y$) for $\rlget(x,y,\wid)$ (\resp $\rlput(x,y,\wid)$). 
Similarly, we write $x \assign v$ (\resp $a \assign x$) for $\rlwrite(x,v)$ (\resp $\LetC{a}{\rlread(x)}{\ldots}$). 
%
The $\rlrf(\node)$ denotes a \emph{remote fence} that strongly
orders all operations towards $\node$ without blocking the (local) CPU.
That is, given a (sequential) program of the form $C; \rlrf(\node); C'$, all remote operations towards $n$ in $C$ are ordered before those in $C'$. 
The $\rlrcas(x,y,v_1,v_2,\wid)$ is the remote analogue of writing
$\LetC{v\!}{\!\rlcas(y,v_1,v_2)}{\,\rlwrite(x,v)}$ with work identifier $\wid$, where
the RMW is run on remote location $y$ and the result is written to local location $x$. 
Similarly, $\rlrfaa(x,y,v,\wid)$ increments (remote) $y$ by $v$ and
writes its old value to $x$.



\paragraph{Well-addressed \rdmawaitrmw Executions}
We assume each location $x$ is associated with exactly one node denoted by $\nodefun{x}$. 
We write $\nodefun\threadt$ to denote the node on which $\threadt$ is run.
An execution $\exec$ is \emph{well-addressed} iff it comprises method calls (in $\exec.E$) with appropriate local locations when expected; \eg for each $\rlwrite(x,\_)$ or $\rlput(\_,x,\_)$ call by thread $\threadt$ in $\exec$,  $\nodefun{x} {=} \nodefun{\threadt}$.
%
%
%
We define $\loc$ for \rdmawaitrmw as expected; \eg $\loc(\rlwrite(x,\_)) {=} \{x\}$,  $\loc(\rlput(x,y,\_)) {=} \{x,y\}$ and $\loc(\rlmf){=} \emptyset$.

\paragraph{Well-stamped \rdmawaitrmw Executions}
An execution $\exec$ is \emph{well-stamped} if for all $\action = \tup{\_, \_, \tup{m, (\vect{v}), v'}} \in \exec.E$: $\exec.\gettags(\action) \in \gettagsrl(m(\vect v), v')$, with $\gettagsrl$ defined as follows. 
Note that depending on whether $\rlrcas$ calls succeed, they may have multiple valid sets of stamps; as such, the $\gettagsrl$ function returns a set of stamp sets (set of set of stamps), though in all cases but for $\rlrcas$ this set is a singleton. 
\small
\[
\begin{array}{@{} l @{}}
\begin{array}{@{} r @{\hspace{1pt}} l @{}}
	\gettagsrl(\rlcas(x,v_1,\_), v_2) \!\defeq 
	& \begin{cases}
		\!\big\{\!\!\set{\tagmf, \tagcread}\!\!\big\} & \!\!\text{if } v_1 \neq v_2 \\
		\!\big\{\!\!\set{\tagcas}\!\!\big\} & \!\!\text{if } v_1 = v_2
	\end{cases} \\
	\gettagsrl(\rlget(x,y^\node,\_), \_) \!\defeq 
	& \big\{\!\!\set{\tagnrr[\node], \tagnlw[\node]}\!\!\big\} 
	\\
	\gettagsrl(\rlput(x^\node,y,\_), \_) \!\defeq 
	& \big\{\!\!\set{\tagnlr[\node], \tagnrw[\node]}\!\!\big\} 
	\\
	 {\chlm \gettagsrl(\rlrfaa(x,\!y^\node\!\!,\_,\_), \_) \!\defeq} 
	 & {\chlm \big\{\!\!\set{\tagnarr, \tagnrw, \tagnlw}\!\!\big\}}	
\end{array}
\quad
\begin{array}{@{} r @{\hspace{1pt}} l @{}}
	\gettagsrl( \rlwrite(x,v), \_) \!\defeq 
	& \big\{\!\!\set{\tagcwrite}\!\!\big\} 
	\\
	\gettagsrl(\rlread(x), \_) \!\defeq 
	&  \big\{\!\!\set{\tagcread}\!\!\big\} 
	\\
	 \gettagsrl(\rlmf(), \_) \!\defeq 
	 & \big\{\!\!\set{\tagmf}\!\!\big\} 
	 \\
	 \gettagsrl(\rlwait(\wid), \_) \!\defeq 
	 & \big\{\!\!\set{\tagwait}\!\!\big\}	 
	 \\
	 \gettagsrl(\rlrf(\node), \_) \!\defeq 
	 & \big\{\!\!\set{\tagnf}\!\!\big\}	 	 
\end{array} \\
	{\chlm 
	\gettagsrl(\rlrcas(x,y^\node,\_,\_,\_), \_) 
	\defeq
	\big\{\!\!\set{\tagnarr[\node], \tagnlw[\node]}, \set{\tagnarr[\node], \tagnrw[\node], \tagnlw[\node]}\!\!\big\}
	}
\end{array}	
\]
\normalsize

A successful remote RMW has three stamps for reading the remote location, modifying it, and writing it to the local location, while a failed \rlrcas does not modify the remote location. 
Recall that the remote read of a remote RMW yields stamp 
\tagnarr, which offers more guarantees than the stamp \tagnrr of \rlget{s}.



\medskip We extend the location function (\loc, defined above for \rdmawaitrmw) to subevents. 
For method calls corresponding to \emph{local operations} (with one or zero locations) their subevents have the same locations. 
The subevents of \rlget, \rlput, \rlrcas, and \rlrfaa are associated with the relevant location as expected.
For instance, if $\action \!=\! \tup{\_, \_, \tup{\rlget, (x,y,\wid), \_}}$ (with subevents $\tagnrr$ and $\tagnlw$), then
$\loc(\tup{\action, \tagnrr}) \!=\! \set{y}$ and
$\loc(\tup{\action, \tagnlw}) \!=\! \set{x}$; whereas if
$\action \!=\! \langle\_, \_, \langle\rlrfaa, (x,y,v,\allowbreak\wid),\_\rangle\rangle$, then
$\loc(\tup{\action, \tagnarr}) {=} \!\set{y}$,
$\loc(\tup{\action, \tagnrw}) {=} \!\set{y}$ and
$\loc(\tup{\action, \tagnlw}) {=} \!\set{x}$.



\paragraph{Well-formed \rdmawaitrmw Executions}
We shortly define the notion of \rdmawaitrmw-consistency for an execution $\exec$. 
To do this, we need a few auxiliary functions and relations as follows.
We assume functions $\vr : \exec.\Read \rightarrow \Val$ and
$\vw : \exec.\Write \rightarrow \Val$, which associate each read
(\resp write) subevent with the value returned
(\resp written). 
We define the `\emph{reads-from}' relation, $\rf \suq \exec.\Write \times \exec.\Read$, on subevents of the same location with matching values (formalised below);
the `\emph{modification-order}' relation, $\mo \suq \exec.\Write \times \exec.\Write$,  describing a (total) order in which writes reach the memory; and 
the `\emph{NIC flush order}', $\ro$,  capturing the PCIe guarantee that NIC reads flush previous NIC writes.
For remote RMWs, we define the `\emph{remote-atomic-order}', \shademath{\phantom{'}\rao\phantom{,}}, 
describing the (total) order in which (remote read parts of) remote RMWs towards each node are executed.
A tuple $\tup{\vr, \vw, \rf, \mo, \ro, \shademath{\phantom{'}\rao\phantom{,}}}$ is {\em well-formed} if the following hold for all $\action, v, v', v_1,v_2,\saction_1, \saction_2, \node, x, y$.
\begin{itemize}
\item If $\action$ is of the form $\tup{\rlread, \_, v}$ or $\tup{\rlcas, \_, v}$, 
then $\vr(\action) = v$.
\item  If $\action$ is of the form $\tup{\rlwrite, (\_, v), \_}$ or $\tup{\rlcas, (\_, v', v), v'}$, 
then $\vw(\action) = v$  
\item  If $\saction_1 {=} \tup{\action, \tagnlr} \wedge \saction_2 {=}
  \tup{\action, \tagnrw}$, then $\vr(\saction_1) {=}
  \vw(\saction_2)$; \textit{mutatis mutandis} for $\saction_1 = \tup{\action, \tagnrr} , \saction_2 =
  \tup{\action, \tagnlw}$ and $\shademath{\saction_1 = \tup{\action, \tagnarr} , \saction_2 =
  \tup{\action, \tagnlw}}$.
%
\item 
  $\tup{\saction_1, \saction_2} \in \rf
  \Rightarrow \loc(\saction_1) = \loc(\saction_2) \land \vw(\saction_1) =
  \vr(\saction_2)$.  
\item $\rf^{-1}$ is a function, \ie every
  read is related to at most one write. If a read is not related to a write, it returns zero:
  $
  \saction_2 \not\in \img{\rf}
  \Rightarrow \vr(\saction_2) = 0$.
\item $\mo \defeq \bigcup_{x \in \Loc} \mo_x$, where each $\mo_x$ is a strict total order on
  $\exec.\Write_x$. 
\item if $\tup{\saction_1, \saction_2} \in (\tagnlr \times \tagnlw)
  \cup ((\tagnrr \cup \tightshademath{\tagnarr}) \times \tagnrw)$
   and 
  $\threadfun{\saction_1} {=} \threadfun{\saction_2}$\\
  then
  $\tup{\saction_1, \saction_2} \in \ro \cup \inv\ro$. \vspace{-10pt}
\end{itemize}
\begin{snugshade}
\begin{itemize}
\item \rlrcas succeeds iff it reads the expected value, in which case it
  overwrites with the given value. 
  That is, given $\action = \tup{\_, \_, \tup{\rlrcas, (x,y,v_1,v_2,\_), \_}}$:\\
  if $\gettags(\action) {=} \set{\tagnarr[\nodefun{y}], \tagnlw[\nodefun{y}]}$,
  then $\vr(\tup{\action,\tagnarr[\nodefun{y}]}) \neq v_1$; and\\
  if 
  	$
		\gettags(\action){=} \set{\!\tagnarr[\nodefun{y}], \!\tagnrw[\nodefun{y}], \!\tagnlw[\nodefun{y}]}
	$, 
  then 
  $ \vr(\tup{\action,\tagnarr[\nodefun{y}]}) = v_1$ and\\
  $\vw(\tup{\action,\tagnrw[\nodefun{y}]}) = v_2$.
\item 
  If
  $\action = \tup{\_, \_, \tup{\rlrfaa, (x,y,v,\_), \_}}$, then
  $\vw(\tup{\action,\tagnrw[\nodefun{y}]}) =
  \vr(\tup{\action,\tagnarr[\nodefun{y}]}) + v$.
\item $\rao \defeq \bigcup_{n \in \Nodes} \rao_{n}$, where
  $\rao_{n}$ is a strict total order on the set of subevents
  $\setpred{\tup{\action, \tagnarr}}{\action = \tup{\_, \_, \tup{m, (x,y,\ldots),
    \_}} \land m \in \set{\rlrfaa,\rlrcas} \land \nodefun{y} = \node}$
\end{itemize}
\end{snugshade}



\noindent We distinguish the point subevents \emph{start} executing (point of `issue') from when they \emph{complete}. 
We define the \emph{issued-before} relation, $\ib$, to record dependencies between the starts of subevents, while $\so$ records dependencies between their ends.
Note that $\ib$ and $\so$ are incomparable: $\tup{\saction_1, \saction_2} \!\in\! \ib$ does \emph{not} imply $\tup{\saction_1, \saction_2} \!\in\! \so$ and vice versa. 
We define \emph{instantaneous subevents}, 
$\exec.\Inst \defeq \exec.\SEvents \!\setminus\! (\exec.\tagcwrite \cup
\exec.\tagnlw[] \cup \exec.\tagnrw[])$, as those that start and end at the same time.

Given an execution $\exec$ and well-formed $\tup{\vr, \vw, \rf, \mo, \ro, \rao}$, we further define the following relations that will help us define $\ib$ and $\so$ for \rdmawaitrmw:
\small
\[
\begin{array}{@{} c @{}}
	\fr \!\defeq\! \setpred{\tup{r,w} \in \exec.\Read \times \exec.\Write} 
{\begin{array}{@{} l @{}}
	 \left(\tup{r,w} \in (\inv\rf; \mo) \lor r \not\in \img{\rf}\right)\\
  	\land\ \loc(r) = \loc(w) 
\end{array}} \setminus [\exec.\SEvents]
\\
\fri \!\defeq\! [\tagcread] ; ((\po \cup \inv{\po}) \cap \fr) ; [\tagcwrite]
\qquad\;
\pfget \!\defeq\! \!\setpred{\tup{\tup{\action_1, \!\tagnlw},\!\tup{\action_2, \!\tagwait}}}
{\exists \wid.\tup{\action_1,\!\action_2} \!\in\! \po\rst{\wid}}
\\
\rfi \!\defeq\! [\tagcwrite] ; (\po \cap \rf) ; [\tagcread]
\quad\;
\rfe \!\defeq\! \rf \!\setminus\! \rfi
\quad
\pfput \!\defeq\! \!\setpred{\tup{\tup{\action_1, \!\tagnrw},\!\tup{\action_2, \!\tagwait}}}
{\exists \wid.\tup{\action_1,\!\action_2} \!\in\! \po\rst{\wid}}
\\
\iso \defeq  \inarrT{
  \phantom{\cup} \setpred{\tup{\tup{\action, \tagmf},\tup{\action, \tagcread}}} {\methodfun{\action} = \rlcas} \\
  \cup \setpred{\tup{\tup{\action, \tagnrr},\tup{\action, \tagnlw}}} {\methodfun{\action} = \rlget} 
  \cup \setpred{\tup{\tup{\action, \tagnlr},\tup{\action, \tagnrw}}} {\methodfun{\action} = \rlput} \\
  \shademath{\cup \setpred{\tup{\tup{\action, \tagnarr},\tup{\action, \tagnlw}}} {\methodfun{\action} \in \set{\rlrcas, \rlrfaa}}} \\
  \shademath{\cup \setpred{\tup{\tup{\action, \tagnarr},\tup{\action, \tagnrw}}} {\methodfun{\action} \in \set{ \rlrcas, \rlrfaa} \land \tagnrw \in \gettags(\action)}}
}
\end{array}
\]
\normalsize
The $\fr$ denotes the `\emph{reads-before}' relation: given a read $r$ that reads from a write $w_r$,  \ie $\tup{w_r, r} \in \rf$, then $\fr$ relates $r$ to all writes $w$ (on the same location) that are $\mo$-\emph{later} than $w_r$. 
The \emph{internal} $\fr$ relation, $\fri$, restricts $\fr$ to CPU reads and writes on the same thread; similarly for $\rfi$ (internal $\rf$). 
The \emph{external} $\rf$, $\rfe$, is defined as $\rf$ edges that are not internal.
The \pfget (\resp\pfput) relation captures the synchronisation between the
local write subevent of a \rlget or remote RMW (\resp remote write
subevent of a \rlput or remote RMW) and a later \rlwait with the same work identifier.
As we describe shortly, while both are included in $\ib$, only \pfget is included in $\so$ as waiting
for a \rlput (or remote RMW) does not guarantee that the NIC remote write has completed.
The `\emph{internal synchronisation order}', $\iso$, captures
ordering between subevents of the same event and ensures that a failing CPU \rlcas performs a memory fence before reading;
RDMA operations (\rlget, \rlput, and \shade{remote RMW}) read before copying the value; 
and \tightshade{a successful remote RMW reads before updating the remote value}.
%
%

Finally, we define $\ib$ as follows and it includes a superset $\ippo$ of $\ppo$. 
Specifically, 
while a later CPU read might finish before an earlier CPU write or wait (cells B1 and B5, in \cref{fig:to}),  they start (are issued) in order; 
and while a remote fence does not guarantee previous NIC writes have completed (cells G11 and J11, in \cref{fig:to}), it guarantees they have at least started. \vspace{5pt}\\
\centerline{
	$\ib \defeq (\ippo \,\cup\, \iso \,\cup\, \rf \,\cup\, \pfget \,\cup\, \pfput \,\cup\, \ro \,\cup\, \fri)^+$
}\\
\centerline{
	with $
	\ippo \defeq \ppo \cup ([\tagcwrite] ; \po ; [\tagcread \cup \tagwait])
\cup\ \bigcup_{\node \in \Nodes} ([\tagnrw \cup \tagnlw] ; \po ; [\tagnf])
	$
}\vspace{-5pt}\\
%
%


We next define \emph{consistency} for \rdmawaitrmw. 
We require that \ib and \so be irreflexive (the latter is implied by irreflexivity
of $\hb$ in \cref{def:lambda-cons} as $\so \suq \hb$ (\cref{def:execs})).

\begin{definition}[\rdmawaitrmw-consistency]
  \label{def:rl-consistency}
  An execution $\exec = \tup{E, \po, \gettags,\so,\hb}$ is \emph{\rdmawaitrmw-consistent} iff it
  is well-addressed, well-stamped, and there exists a well-formed tuple $\tup{\vr, \vw, \rf,
  \mo, \ro, \shade{\phantom{'}\!\rao\!\!\!\phantom{,}}}$ such that:
  \begin{enumerate}
  \item 
    $\ib$ is irreflexive; and 
  \item 
    $\so = \iso \cup \rfe \cup \pfget \cup \ro \cup \fr \cup \mo \cup
    \tightshademath{\rao \cup ([\tagnrw[]]; \inv{\iso} ; \rao)} \cup
    ([\Inst];\ib)$.
  \end{enumerate}
\end{definition}

As described above, $\rao$ captures the order in which remote read parts of remote RMWs towards a node is executed.  
The extension $([\tagnrw[]]; \inv{\iso} ; \rao)$ ensures that remote RMWs towards the same
node do not overlap: if a remote RMW succeeds, then its remote write completes before the next RMW can read.


%% file: tags.tex

\begin{figure}[t]

\setcounter{rowcounter}{1}
\setcounter{colcounter}{1}
\vspace{3pt}
\begin{center} \footnotesize
\scalebox{0.95}{
\begin{tabular}{c|c|c|c|c|c|c|c|c?c|c|c|c|c|c|c|}
\multicolumn{4}{c}{} & \multicolumn{12}{c}{Later (in Program Order) Stamp} \\
  \hhline{~|-|-|-|-|-|-|-|-|-|-|-|-|-|-|-|}
\multicolumn{1}{c|}{} & \multicolumn{3}{c|}{\multirow{3}{*}{\LARGE \tagppo}} & \multicolumn{5}{c?}{single} & \multicolumn{7}{c|}{families} \\
  \hhline{~|~~~|-|-|-|-|-|-|-|-|-|-|-|-|}
\multicolumn{1}{c|}{} & \multicolumn{3}{c|}{} & \colheader & \colheader & \colheader & \colheader
   & \colheader & \colheader & \colheader & \colheader & \colheader & \colheader & \colheader & \colheader \\
\multicolumn{1}{c|}{} & \multicolumn{3}{c|}{}
                      & \tagcread & \tagcwrite & \tagcas & \tagmf & \tagwait & \tagnlr & \tagnrw & \chlm\tagnarr & \tagnrr & \tagnlw & \tagnf & \taggf \\
  \hhline{~|-|-|-|-|-|-|-|-|-|-|-|-|-|-|-|}
\multirow{10}{*}{\rotatebox{90}{Earlier (in Program Order) Stamp\hspace*{-10pt}}} & \multirow{5}{*}{\rotatebox{90}{single}}
   & \rowheader{\tagcread}& \cyes& \cyes & \cyes & \cyes & \cyes & \cyes & \cyes & \chlm\cyes & \cyes & \cyes & \cyes & \cyes \\
  \hhline{~|~|-|-|-|-|-|-|-|-|-|-|-|-|-|-|}
&  & \rowheader{\tagcwrite}& \cno& \cyes & \cyes & \cyes & \cno  & \cyes & \cyes & \chlm\cyes & \cyes & \cyes & \cyes & \cyes \\
  \hhline{~|~|-|-|-|-|-|-|-|-|-|-|-|-|-|-|}
&  & \rowheader{\tagcas} & \cyes & \cyes & \cyes & \cyes & \cyes & \cyes & \cyes & \chlm\cyes & \cyes & \cyes & \cyes & \cyes \\
  \hhline{~|~|-|-|-|-|-|-|-|-|-|-|-|-|-|-|}
&  & \rowheader{\tagmf}  & \cyes & \cyes & \cyes & \cyes & \cyes & \cyes & \cyes & \chlm\cyes & \cyes & \cyes & \cyes & \cyes \\
  \hhline{~|~|-|-|-|-|-|-|-|-|-|-|-|-|-|-|}
&  & \rowheader{\tagwait}& \cyes & \cyes & \cyes & \cyes & \cyes & \cyes & \cyes & \chlm\cyes & \cyes & \cyes & \cyes & \cyes \\
  \hhline{~|-|-|-|-|-|-|-|-|-|-|-|-|-|-|-|}
  \hhline{~|-|-|-|-|-|-|-|-|-|-|-|-|-|-|-|}
& \multirow{7}{*}{\rotatebox{90}{families}}
   & \rowheader{\tagnlr} & \cno  & \cno  & \cno  & \cno  & \cno  & \cqp  & \cqp  & \chlm\cqp  & \cqp  & \cqp  & \cqp  & \cqp \\
  \hhline{~|~|-|-|-|-|-|-|-|-|-|-|-|-|-|-|}
&  & \rowheader{\tagnrw} & \cno  & \cno  & \cno  & \cno  & \cno  & \cno  & \cqp  & \chlm\cqp  & \cqp  & \cqp  & \cno  & \cqp \\
  \hhline{~|~|-|-|-|-|-|-|-|-|-|-|-|-|-|-|}
&  & \rowheader{\chlm\tagnarr}& \chlm\cno  & \chlm\cno  & \chlm\cno  & \chlm\cno  & \chlm\cno  & \chlm\cno  & \chlm\cqp & \chlm\cqp  & \chlm\cqp  & \chlm\cqp  & \chlm\cqp  & \chlm\cqp \\
  \hhline{~|~|-|-|-|-|-|-|-|-|-|-|-|-|-|-|}
&  & \rowheader{\tagnrr} & \cno  & \cno  & \cno  & \cno  & \cno  & \cno  & \cno  & \chlm\cno  & \cno  & \cqp  & \cqp  & \cqp \\
  \hhline{~|~|-|-|-|-|-|-|-|-|-|-|-|-|-|-|}
&  & \rowheader{\tagnlw} & \cno  & \cno  & \cno  & \cno  & \cno  & \cno  & \cno  & \chlm\cno  & \cno  & \cqp  & \cno  & \cqp \\
  \hhline{~|~|-|-|-|-|-|-|-|-|-|-|-|-|-|-|}
&  & \rowheader{\tagnf}  & \cno  & \cno  & \cno  & \cno  & \cno  & \cqp  & \cqp  & \chlm\cqp  & \cqp  & \cqp  & \cqp  & \cqp \\
  \hhline{~|~|-|-|-|-|-|-|-|-|-|-|-|-|-|-|}
&  & \rowheader{\taggf}  & \cyes & \cyes & \cyes & \cyes & \cyes & \cyes & \cyes & \chlm\cyes & \cyes & \cyes & \cyes & \cyes \\
  \hhline{~|-|-|-|-|-|-|-|-|-|-|-|-|-|-|-|}
\end{tabular}}
\end{center}
%
%
%
\vspace*{-15pt}
\caption{The \tagppo order in \rdmawait and \rdmawaitrmw, where \shade{highlighted} cells denote our extensions from \rdmawait to \rdmawaitrmw.
The \cyes denotes that the (program-order-related) stamps are \emph{ordered}; 
the \cno denotes that the stamps are \emph{not ordered}; 
the \cqp denotes the stamps are ordered iff they are associated with the \emph{same node}.
}
\label{fig:to}
\end{figure}


%% file: locklib.tex

\section{Specifying and Verifying RDMA Lock Libraries}
\label{sec:locklib}
We use the \rdmasync library to \emph{specify, implement, and verify} three RDMA lock libraries.
As discussed in \cref{sec:ov-locks}, designing an RDMA lock presents a trade-off between
strong, intuitive behaviours and efficient implementations. 
As such, after introducing the required preliminaries (\cref{sec:basic-lock-spec}), we develop a weak (\lklw), strong (\lkls), and node
(\lkln) lock library. 




\subsection{Preliminaries}
\label{sec:basic-lock-spec}
\label{sec:brl}

\paragraph{Well-formed Locks}
A lock library typically  provides two methods $\lklgena(x)$ and $\lklgenr(x)$ for acquiring and releasing a (network-shared) lock $x$, ensuring mutual exclusion; \ie two threads cannot hold the lock on $x$ simultaneously. 
We assume the existence of a \emph{location function} $\loc$ such that $\loc(\lklgena(x)) \!=\! \loc(\lklgenr(x)) \!=\!~\set{x}$.
We further assume that locks are used in a \emph{well-formed} fashion: a thread only acquires (\resp releases) lock $x$ if it has not (\resp has) already acquired $x$.  
We formalise this in \cref{def:locklib-respect} below, requiring that each $\lklgena(x)$ (\resp $\lklgenr(x)$) is followed (\resp preceded) by $\lklgenr(x)$ (\resp $\lklgena(x)$) in program order.
\begin{definition}
  \label{def:locklib-respect}
  An execution $\tup{E, \po, \_, \_, \_}$ is {\em lock-well-formed} iff for all $x$:
  \begin{enumerate}
  \item for all $\action_a \in E_x$ there exists an $\action_r \in E_x$ such that
    $\tup{\action_a, \action_r} \in \imm{\po_x}$; and 
  \item for all $\action_r \in E_x$ there exists an $\action_a \in E_x$ such that
    $\tup{\action_a, \action_r} \in \imm{\po_x}$
  \end{enumerate}
  where 
  $\action_a,\action_r$ are
  acquire and release events: $\methodfun{\action_a} = \lklgena$ and
  $\methodfun{\action_r} = \lklgenr$. 
%
\end{definition}
Library guarantees only hold for programs that adhere to this well-formedness requirement.  For those that do not, 
\emph{any} behaviour is allowed.



\paragraph{Background: \brl Library}
Ambal \etal~\cite{POPL-LOCO} use \rdmawait to define higher-level
libraries such as a \emph{shared-variable} library (\brl) where each node maintains its own \emph{copy} for each location $x$. 
A thread then accesses (reads/writes) its own local copies, and can broadcast its local value to other nodes.
The \brl library comprises these methods: 
$M = \set{\brlwrite,\brlread,\brlbr,\brlwait,\brlgf}$.
The $\brlwrite(x,v)$ (\resp $\brlread(x)$) writes (\resp reads) value $v$ to the local copy of $x$ on the current node.
The $\brlbr(x,d,\set{\node_1,\ldots,\node_k})$ broadcasts the local value of
$x$ and overwrites $x$ on nodes $\node_1,\ldots,\node_k$, which may include the local node itself (where $d$ is the work id).
The $\brlwait(d)$ waits for previous broadcasts of the thread associated with
work id $\wid \in \Wid$. 
Finally, the global fence $\brlgf(\set{\node_1,\ldots,\node_k})$ ensures every previous operation of the thread towards nodes $\node_1,\ldots,\node_k$ is fully completed.
We repeat the formal semantics of $\brl$ in
\iftoggle{shortpaper}{the extended version~\evlink}{\cref{sec:app-brl}}.
In the remainder of this article we use \brl to implement several libraries.



\input{locklibweak}

\input{locklibstrong}

\input{locklibnode}


%% file: locklibweak.tex
\subsection{The Weak Lock Library}
\label{sec:locklib-weak}

We present our \lklw library, which only guarantees {\em mutual exclusion}, without any guarantees on the completion order of submitted RDMA operations.



\paragraph{The \lklw Specification}
\label{sec:locklib-weak-spec}
The stamps for \lklw are defined through the $\gettagslklw$ function as follows.
That is, acquiring a weak lock behaves as a memory fence (stamp \tagmf) on \TSO ,
while releasing it behaves merely as a write (\tagcwrite). 
\small
\begin{align*}
  \gettagslklw(\tup{\threadt, \_, \tup{\lklwa, (x), ()}}) & \defeq \set{\tagmf}
  &&&
  \gettagslklw(\tup{\threadt, \_, \tup{\lklwr, (x), ()}}) & \defeq \set{\tagcwrite}
\end{align*}
\normalsize
As we formulate in \cref{def:locklib-weak-cons} below (the second condition), \lklw provides synchronisation between lock releases and acquisitions of each lock.
\begin{definition}[\lklw-consistency]
  \label{def:locklib-weak-cons}
  A lock-well-formed execution $\exec = \langle E, \po, \allowbreak\gettags, \so, \hb\rangle$ is
  \emph{$\lklw$-consistent} iff:
  \begin{enumerate}
  \item $\gettags = \gettagslklw$ (where $\gettagslklw$ is as defined above); and 
  \item $\so = \bigcup_x \setpred{\tup{\tup{\action_1, \tagcwrite}, \tup{\action_2,
        \tagmf}}} {\tup{\action_1, \action_2} \in \inv{(\imm{\po_x})} ; \lo_{x}}$, where 
  $\lo_x$ is a total order on acquisition events on $x$, \ie on
  $\setpred{\action \in E_x}{\methodfun{\action} = \lklwa}$. 
  \end{enumerate}
\end{definition}

Given a release event $\action_1$ on $x$ (in a lock-well-formed execution), the $\inv{(\imm{\po_x})}$ component identifies an acquire event $\action_3$ that is the
latest corresponding acquire event on $x$ preceding $\action_1$ (in $\po$).
As such, $\so$ induces synchronisation between $\action_1$ and all later (in $\lo_x$) acquisition events $\action_2$.
Note that $\lo_x$ is also indirectly included in \hb, since the acquire and
release operations stay in order.

The release stamp (\tagcwrite) does not synchronise with previous RDMA-specific
stamps (bottom-left part of \Cref{fig:to}). As such, reacquiring a lock does not
guarantee that previous RDMA operations submitted with the lock are completed.







\begin{figure}[t]
\[
\begin{array}{@{} l @{\hspace{30pt}} l@{}}
  \begin{array}{@{} l @{}}
    \impllklw(\threadt, \lklwa, (x)) \defeq \\
      \quad \rlrfaa(p^\threadt_x,x_a,1,\wid) ; \ \rlwait(\wid) ; \\
      \quad \LetC{v}{\rlread(p^\threadt_x)}{} \\
      \quad \texttt{loop } \{ \texttt{if } \brlread(x_{1}) \!=\! v \texttt{ then } \paperbreak \texttt{ else } \\
      \qquad \qquad \ldots \\
      \qquad \qquad \texttt{if } \brlread(x_{T}) \!=\! v \texttt{ then } \paperbreak \ \} \\
  \end{array}
  & 
  \begin{array}{@{} l @{}}
     \impllklw(\threadt, \lklwr, (x)) \defeq \\
      \quad \LetC{v}{\rlread(p^\threadt_x)}{} \\
      \quad \brlwrite(x_\threadt, v+1) ; \\
      \quad \brlbr(x_\threadt, \_, \Nodes \setminus \set{\nodefun{\threadt}})
  \end{array}
\end{array}
\]
\vspace{-10pt}
\caption{The \lklw implementation using \rdmasync and \brl libraries.}
\label{fig:impllklw}
\end{figure}

\paragraph{The \lklw (Distributed) Implementation}
\label{sec:locklib-weak-implem}
%
We present our \lklw implementation in \cref{fig:impllklw} (via the \impllklw
function), inspired by the well-known ticket lock implementation.
For each lock location $x$, we create a ticket dispenser $x_a$ (on some
arbitrary node) that records the value of the next \emph{available} ticket,
thread-local locations ($p^\threadt_x$ for each
$\threadt \in \Threads = \set{1, \ldots, T}$) to track the ticket allocated to
$\threadt$ (\ie its turn), and shared variables $x_\threadt$ (for each
$\threadt \in \Threads$) to signal releasing the lock.

To release the lock on $x$, thread $\threadt$ writes the \emph{next} turn, \ie
$v {+} 1$ when $\threadt$ holds ticket $v$ (obtained by reading $p^\threadt_x$),
to its release location $x_\threadt$ and subsequently broadcasts it to all nodes
other than itself ($\nodefun{\threadt}$). To acquire the lock on $x$, thread
$\threadt$ calls a fetch-and-add on $x_a$ to fetch the next available ticket
(\ie its turn) in $p^\threadt_x$ and increments $x_a$. It then records its turn
in $v$ and repeatedly examines the release location $x_{\threadt'}$ of each
thread $\threadt' \in \set{1,\ldots,T}$ until one has value $v$, indicating that
its turn has come and thus $\threadt$ holds the lock.
Note that $\threadt'$ may be $\threadt$ itself, \ie $\threadt = \threadt'$, if
it was the last thread to release the lock.

At the cost of more network messages (through broadcasts), our implementation provides lower latency than centralised systems (\eg in \cref{fig:impllkln}) as messages are transmitted
directly from the thread releasing the lock to the next thread acquiring the
lock. 
We next prove (\cref{thm:wlock-sounndness}) that our implementation is correct against the \lklw specification with the full proof given in
\iftoggle{shortpaper}{the extended version~\evlink}{\cref{sec:lklw-library}}.




\begin{restatable}{theorem}{lklwlocsound}
  \label{thm:wlock-sounndness}
  The implementation \impllklw is sound.
\end{restatable}



%% file: locklibstrong.tex



\subsection{The Strong Lock Library}
\label{sec:locklib-strong}
%
%
%
%
We present our strong lock library \lkls that, as well as ensuring mutual exclusion of critical sections, additionally guarantees that \emph{all} earlier operations have \emph{fully} completed on releasing a strong lock. 
We present several examples of the `message-passing' behaviour in \cref{fig:ex-lklw-lkls-2} contrasting the behaviour of weak and strong locks when interacting with \rlget{s} and whether the weak outcome $a \ne b$ is allowed. 
In particular, we may observe $a \ne b$ when using a weak lock (\cref{subfig:lklw-2}) and this can be prohibited by explicitly waiting (using $ \rlwait(\wid)$) on the completion of the \rlget{s} before releasing the weak lock (\cref{subfig:lklwgf-2}). 
By contrast, when using a strong lock we no longer need to wait for their completion as this is guaranteed by the strong lock release (\cref{subfig:lkls-2}). 



\begin{figure}[t]
  \begin{subfigure}[b]{0.31\textwidth}\small
  \vspace{0pt}
  \scalebox{0.95}{
  \begin{tabular}{|@{\hspace{3pt}}c @{\hspace{3pt}} || @{\hspace{8pt}}c@{\hspace{8pt}}|}
  \hline
  $x,y = 0,0$ & \\
  \hline
  $\inarr{
    \phantom{a} \vspace{-5.5pt} \\
    \lklwa(l) \\
    x \assign 1 \\
    y \assign 1 \\
    \lklwr(l) \\
    \phantom{a} \vspace{-5.5pt}
  }$
  &
  $\inarr{
    \lklwa(l) \\
    a \assign x^1 \\
    b \assign y^1 \\
    \lklwr(l)
  }$
  \\
  \hline
  \end{tabular}} 
  \caption{$a \neq b$ 
    \checkyes}
  \label{subfig:lklw-2}
  \end{subfigure}
  \quad
  \begin{subfigure}[b]{0.31\textwidth}\small
  \vspace{0pt}
  \scalebox{0.95}{
  \begin{tabular}{|@{\hspace{3pt}}c @{\hspace{3pt}} || @{\hspace{8pt}}c@{\hspace{8pt}}|}
  \hline
  $x,y = 0,0$ & \\
  \hline
  $\inarr{
    \lklwa(l) \\
    x \assign 1 \\
    y \assign 1 \\
    \lklwr(l)
  }$
  &
  $\inarr{
    \lklwa(l) \\
    a \assign^\wid x^1 \\
    b \assign^\wid y^1 \\
    \rlwait(\wid) \\
    \lklwr(l)
  }$
  \\
  \hline
  \end{tabular}} 
  \caption{$a \neq b$ \checkno}
  \label{subfig:lklwgf-2}
  \end{subfigure}
  \quad
  \begin{subfigure}[b]{0.31\textwidth}\small
  \vspace{0pt}
  \scalebox{0.95}{
  \begin{tabular}{|@{\hspace{3pt}}c @{\hspace{3pt}} || @{\hspace{8pt}}c@{\hspace{8pt}}|}
  \hline
  $x,y = 0,0$ & \\
  \hline
  $\inarr{
    \phantom{a} \vspace{-5.5pt} \\
    \lklsa(l) \\
    x \assign 1 \\
    y \assign 1 \\
    \lklsr(l) \\
    \phantom{a} \vspace{-5.5pt}
  }$
  &
  $\inarr{
    \lklsa(l) \\
    a \assign x^1 \\
    b \assign y^1 \\
    \lklsr(l)
  }$
  \\
  \hline
  \end{tabular}} 
  \caption{$a \neq b$ 
    \checkno}
  \label{subfig:lkls-2}
  \end{subfigure}
\vspace*{-15pt}
\caption{Weak versus strong locks when interacting with \rlget instructions.}
\label{fig:ex-lklw-lkls-2}
\end{figure}








\paragraph{The \lkls Specification}
\label{sec:locklib-strong-spec}
The \lkls stamps are defined (via $\gettagslkls$) as:
\begin{align*}
  \gettagslkls(\tup{\threadt, \_, \tup{\lklsa, \!(x), \!()}}) & \!\defeq\! \set{\tagmf}
  &
  \gettagslkls(\tup{\threadt, \_, \tup{\lklsr, \!(x), \!()}}) & \!\defeq\!\!\!\! \bigcup_{\node \in \Nodes}\!\!\! \set{\taggf[\node]}
\end{align*}

As with \lklw, acquiring a strong lock behaves as a memory fence (\tagmf), while releasing it behaves as a global fence ($\taggf[]$), ensuring that all previous remote operations are completed.
\begin{definition}[\lkls-consistency]
  \label{def:locklib-strong-cons}
  A lock-well-formed execution $\exec = \langle E, \po, \allowbreak\gettags, \so, \hb\rangle$ is
  \emph{$\lkls$-consistent}
  iff: 
  \begin{enumerate}
  \item   $\gettags = \gettagslkls$ (where $\gettagslkls$ is defined above); and
  \item
    $\so = \bigcup_{x \in \Loc, n \in \Nodes} \setpred{\tup{\tup{\action_1,
        \taggf}, \tup{\action_2, \tagmf}}} {\tup{\action_1, \action_2} \in
      \inv{(\imm{\po_x})} ; \lo_x}$, where $\lo_x$ is a total order on
    $\setpred{\action \in E_x}{\methodfun{\action} = \lklsa}$.
  \end{enumerate}
\end{definition}


\paragraph{Strong Lock Implementation}
\label{sec:locklib-strong-implem}
We implement \lkls (via \impllkls) simply by combining the weak locks and global fences (from the \brl library) as follows:
\begin{align*}
\impllkls(\threadt, \lklsa, (x)) & \defeq \lklwa(x) 
&&&
\impllkls(\threadt, \lklsr, (x)) & \defeq \brlgf(\Nodes) ; \lklwr(x)
\end{align*}

Finally, we prove (\cref{thm:slock-sounndness}) that our implementation is sound against the \lkls specification with the full proof given in
\iftoggle{shortpaper}{the extended version~\evlink}{\cref{sec:lkls-library}}.

\begin{restatable}{theorem}{lklslocsound}
  \label{thm:slock-sounndness}
  The implementation \impllkls is sound.
\end{restatable}


%% file: locklibnode.tex
\subsection{The Node Lock Library}
\label{sec:locklib-node}
A common use case of locks is to protect an object (set of locations) on a specific node. 
In such cases, neither weak nor strong locks are suitable as they either incur a high
programmer burden (weak locks) or a high performance overhead (strong locks).
To address this,  we develop \emph{node locks},  $\lkln$, a novel lock library that provides synchronisation on a specific node.
Given a node lock $x$ on node $n$, we write $\nodefun{x}$ for $n$. 
A node lock $x$ ensures that on re-acquiring it all previous remote operations (within a critical section of $x$) towards $n$ are observable.


\paragraph{The \lkln Specification}
\label{sec:locklib-node-spec}
The \lkln stamps are defined (via $\gettagslkln$) as:
\small
\begin{align*}
  \gettagslkln(\tup{\threadt, \_, \tup{\lklna, \!(x), \!()}}) & \!\defeq\! \set{\tagmf}
  &\;
  \gettagslkln(\tup{\threadt, \_, \tup{\lklnr, \!(x), \!()}}) & \!\defeq\! \set{\tagnf[\nodefun{x}], \tagnrw[\nodefun{x}]}
  \vspace{-10pt}
\end{align*}
\normalsize
\begin{figure}[t]
  \centering
  \noindent
  \begin{subfigure}[b]{0.27\textwidth}\small\centering
    \scalebox{0.85}{
\begin{tabular}{|@{\hspace{3pt}}c @{\hspace{3pt}} || @{\hspace{3pt}}c @{\hspace{3pt}} || @{\hspace{3pt}}c@{\hspace{3pt}}|}
  \hline
  & $x = 0$ & $y=0$\\
  \hline
  $\inarr{
    \phantom{a} \vspace{-8pt} \\
    \lklsa(l) \\
    x^2 \assign 1 \\
    \lklsr(l) \\
    y^3 \assign 1 \\
    \phantom{a} \vspace{-8pt}
  }$
  & &
  $\inarr{
    a \assign y \\
    b \assign x^2 \\
  }$
  \\
  \hline
\end{tabular}} 
\caption{\intabT{$(a,b) = (1,0)$ \checkno\\~~}}
\label{fig:nkl-1}
\end{subfigure}
\hfill 
\begin{subfigure}[b]{0.3\textwidth}\small\centering
\scalebox{0.85}{    \begin{tabular}{|@{\hspace{3pt}}c @{\hspace{3pt}} || @{\hspace{3pt}}c @{\hspace{3pt}} || @{\hspace{3pt}}c@{\hspace{3pt}}|}
      \hline
      & $x = 0$ & $y=0$\\
      \hline
      & \lkln $l$ & \\
      \hline
      $\inarr{
      \phantom{a} \vspace{-8pt} \\
      \lklna(l^2) \\
      x^2 \assign 1 \\
      \lklnr(l^2) \\
      y^3 \assign 1 \\
      \phantom{a} \vspace{-8pt}
      }$
      & &
          $\inarr{
          a \assign y \\
      b \assign x^2 \\
      }$
      \\
      \hline
    \end{tabular}}
  \caption{\intabT{$(a,b) = (1,0)$ \checkyes \\ ~~}}
  \label{fig:nkl-2}
\end{subfigure} 
%
%
%
%
%
\hfill\begin{subfigure}[b]{0.405\textwidth}\small
    \scalebox{0.85}{
\begin{tabular}{|@{\hspace{3pt}}c @{\hspace{3pt}} || @{\hspace{3pt}}c @{\hspace{3pt}} || @{\hspace{3pt}}c@{\hspace{3pt}}|| @{\hspace{3pt}}c@{\hspace{3pt}}|}
  \hline
  & $x = 0$ & $y=0$ & $z=0$\\
  \hline
  & \lkln $l$ & & \\
  \hline
  $\inarr{
    \phantom{a} \vspace{-8pt} \\
    \lklna(l^2) \\
    x^2 \assign 1 \\
    z^4 \assign 1 \\
    \lklnr(l^2) \\
    y^3 \assign 1 \\
    \phantom{a} \vspace{-8pt}
  }$
  & &
  $\inarr{
    a \assign y \\
    \lklna(l^2) \\
    b \assign x^2 \\
    c \assign z^4 \\
    \lklnr(l^2)
  }$ &
  \\
  \hline
\end{tabular}} 
\caption{
  \intabT{$(a,b,c) \!=\! (1,0,\_)$ \checkno \\
  $(a,b,c) \!=\! (1,\_,0)$ \checkyes}}
\label{fig:lkln-3}
\end{subfigure}
\\
\vspace*{-5pt}
\caption{Strong (left) versus node (middle and right) locks examples.}
\label{fig:strong-nodespecific}
\end{figure}
Note that unlike \lklw, the \lkln releases use $\tagnf$ and $\tagnrw$ stamps to synchronise with previous remote operations towards $\node$ (\ie those with stamps \tagnarr, \tagnrr, and \tagnrw).
Importantly, note that unlike in \lkls, the release \emph{should not} include a global fence stamp (\taggf)
as that would be too strong. 
By using \tagnf and \tagnrw, we ensure that previous operations towards $\node$ are completed only when the lock is \emph{later re-acquired}, and they may not have yet completed on release. 
This means that, when appropriate, using a node lock is
more efficient than combining a weak lock with a global fence.

To understand the difference between strong and node locks,
consider the examples in \cref{fig:nkl-1,fig:nkl-2}, 
where the $x^2 \assign 1$ \rlput by node $1$ is enclosed \emph{within} a lock, while the $b \assign x^2$ \rlget by node $3$ is {\em without} a lock.
In \cref{fig:nkl-1}, $\lklsr(l)$ ensures that the earlier $x^2 \assign 1$ has completed. 
As such, $a=1$ implies that $y^3 \assign 1$ has been executed and that $x$ (in node $2$) has been
modified, ensuring $b=1$. 
By contrast, the $\lklnr(l^2)$ in \cref{fig:nkl-2} does not wait for $x^2 \assign 1$ to complete, \ie $x^2 \assign 1$ may complete after $y^3 \assign 1$. 
We can prevent this by enclosing $b \assign x^2$ within the node lock, as shown in \cref{fig:lkln-3}.
Specifically, $y^3 \assign 1$ in \cref{fig:lkln-3} may still complete before earlier remote operations.
However, $a=1$ implies that $y^3 \assign 1$ is executed, and that thread $1$ has at least
acquired the lock. 
As such, when thread $3$ acquires the lock via $\lklna(l)$, it synchronises with $\lklnr(l)$ in thread $1$ and ensures that $x^2 \assign 1$ is completed on lock acquisition.
Note that the node lock $l$ protects the accesses towards locations on
\emph{node~$2$ only}. Thus, in \cref{fig:lkln-3}, it only guarantees that
$x^2 \assign 1$ is completed but not necessarily $z^4 \assign 1$ (towards node
$4$), and thus $(a,c){=}(1,0)$ is an allowed outcome.
By contrast, \cref{subfig:overview-lkln} in the overview showcases the lock
guarantees: as $x$ and $y$ both reside on node $2$, the lock ensures that their
accesses by threads $1$ and $3$ are mutually exclusive, \ie $a \ne b$ is
disallowed. Specifically, if thread $1$ acquires $l$ first, $x$ and $y$ are
modified before being read by thread $3$, \ie $a\!=\!b\!=\!1$. Conversely, if
thread $3$ acquires $l$ first, $x$ and $y$ are read before being modified by
thread $1$, \ie $a\!=\!b\!=\!0$.

\begin{definition}[\lkln-consistency]
  \label{def:locklib-strong-cons}
  A lock-well-formed execution $\exec = \langle E, \po, \allowbreak\gettags, \so, \hb \rangle$ is $\lkln$-consistent
  iff: 
  \begin{enumerate}
  \item   $\gettags = \gettagslkln$ (where $\gettagslkln$ is as defined above);  and 
  \item  
    $\so = \inarrT{ \setpred{\tup{\tup{\action, \tagnf[\nodefun{\loc(\action)}]}, \tup{\action, \tagnrw[\nodefun{\loc(\action)}]}}}{\methodfun{\action} = \lklnr} \\
      \bigcup_{x \in \Loc} \setpred{\tup{\tup{\action_1, \tagnrw[\nodefun{\loc(\action_1)}]}, \tup{\action_2, \tagmf}}} {\tup{\action_1, \action_2} \in \inv{(\imm{\po_x})} ; \lo_x}}$
    
    where 
$\lo_x$ is a total order on
  $\setpred{\action \in E_x}{\methodfun{\action} = \lklna}$.
  \end{enumerate}
\end{definition}

\paragraph{The \lkln Implementation}
\label{sec:locklib-node-implem}
%
%
We implement \lkln as a centralised ticket lock using remote RMWs (\Cref{fig:impllkln}).  
For each (node) lock $x$ associated with node $\nodefun{x}$, we create two remote locations $x_a$ and $x_r$ on $\nodefun{x}$. 
As before,  $x_a$ is the ticket dispenser and records the next available ticket.
The $x_r$ tracks the release counter and indicates which ticket currently holds the lock. 
Each thread also uses a local location $p^\threadt_x$ to hold the result of remote operations.

\begin{wrapfigure}[14]{r}{0.45\linewidth}
  \vspace{-15pt}
  \scalebox{0.9}{
\begin{minipage}{\linewidth}
\begin{align*}
  & \impllkln(\threadt, \lklna, (x)) \defeq \\[-2pt]
  & \quad \rlrfaa(p^\threadt_x,x_a,1,\wid) ; \rlwait(\wid) ; \\[-2pt]
  & \quad \LetC{v}{\rlread(p^\threadt_x)}{} \\[-2pt]
  & \quad \texttt{loop } \{ \\[-2pt]
  & \qquad \rlget(p^\threadt_x,x_r,\wid); \rlwait(\wid); \\[-2pt]
  & \qquad \texttt{if } \rlread(p^\threadt_x) = v \texttt{ then } \paperbreak \  \} ; \\[-2pt]
  & \quad \rlwrite(p^\threadt_x, v+1) \\[5pt]
  & \impllkln(\threadt, \lklnr, (x)) \defeq \\[-2pt]
  & \quad \rlrf(\nodefun{x}) ; \\[-2pt]
  & \quad \rlput(x_r,p^\threadt_x,\_) 
\end{align*}
\end{minipage}} 
\vspace{-5pt}
\caption{Node lock implementation (\impllkln) using \rdmasync}
\label{fig:impllkln}
\vspace{-10pt}
\end{wrapfigure}
Acquiring the lock on $x$ calls a fetch-and-add on $x_a$ to fetch the next available ticket in $p^\threadt_x$ and increments $x_a$. 
It then records the ticket value in $v$ and repeatedly examines $x_r$ until it has value $v$, indicating that its turn has come and thus $\threadt$ holds the lock. 
Finally, it increments its ticket value in $p^\threadt_x$ in preparation for later releasing the lock; \ie $p^\threadt_x$ now records the ticket whose turn is next. 
As such, releasing the lock simply updates $x_r$ to $p^\threadt_x$ using a \rlput rather than an RMW; this is because only the lock holder can write to $x_r$. 
Note that the preceding \rlrf ensures that earlier \rlget operations towards $\nodefun{x}$ have completed before the lock is released. 
We prove (\cref{thm:nlock-sounndness}) that our implementation is correct against the \lkln specification with the full proof given in
\iftoggle{shortpaper}{the extended version~\evlink}{\cref{sec:lkln-library}}.

%
%
\begin{restatable}{theorem}{lklnlocsound}
  \label{thm:nlock-sounndness}
  The implementation \impllkln is sound.
\end{restatable}




%% file: strongrdma.tex

\section{The \strl Library}
\label{sec:strl}
%

We specify (\cref{sec:strl-spec}), implement, and verify (\cref{sec:strl-impl}) the \strl library that provides \emph{intuitive} read, write, and RMW operations with the strong semantics of \emph{sequential consistency} (SC)~\cite{sc}.  That is,  as with SC, the instructions in each thread under \strl are always observed in (program) order.
Moreover, unlike in \rdmawaitrmw or the lock libraries in \cref{sec:locklib}, the users do not need to specify whether a location is local or remote and which node it resides on. 
For instance, a user can simply call $\strlw(\xx,v)$ to write (with SC semantics) to location $\xx$, regardless of whether $\xx$ is local (on the current node) or remote. 
As such, we use the \texttt{typewriter} font and write $\xx$ to denote an abstract \rdmascrmw location whose underlying memory address may be local (\ie $\xx {=} x$) or on a remote node $n$ (\ie $\xx {=} x^n$).

\subsection{The \strl Specification}
\label{sec:strl-spec}
\paragraph{The \strl Methods}
The \strl library has four methods: 
$\strlr(\xx)$, to read from \xx;
$\strlw(\xx, v)$ to write $v$ to $\xx$;
$\strlcas(\xx,v_1,v_2)$, a compare-and-swap on $\xx$; and
$\strlfaa(\xx,v)$, a fetch-and-add on $\xx$.
We define \loc as expected, \ie
$\loc(\strlw(\xx,v)) \!=\! \loc(\strlr(\xx)) \!=\! \loc(\strlcas(\xx,v_1,v_2)) \!=\!
\loc(\strlfaa(\xx,v)) \allowbreak\!=\!\set{\xx}$.
We extend $\po$ and $\loc$ to subevents as expected.

\paragraph{Well-formedness} Given an \strl execution
$\exec$, we define the sets of read subevents ($\Read$) to comprise all subevents except writes and the set of write subevents ($\Write$) to include all subevents except reads and failed RMWs. 
\begin{align*}
  \Read & \!\defeq\! \setpred{\tup{\action,\tagmf}}{ \action \in \exec.E \setminus\!
  \set{\tup{\_,\_,\tup{\strlw,\_,\_}}}} \\
  \Write& \!\defeq\! \setpred{\tup{\action,\tagmf}}{ \action \in \exec.E \setminus
  \set{\tup{\_,\_,\tup{\strlr,\_,\_}}} \setminus\!
  \setpred{\tup{\_,\_,\tup{\strlcas,(\_,v,\_),v'}}}{v \!\neq\! v'}} 
\end{align*}
As before, a tuple $\tup{\vr, \vw, \rf, \mo}$ is \emph{well-formed}
if the following holds:
\begin{itemize}
\item $\vr$/$\vw$ map each read/write subevent to the value read/written:
\begin{align*}
	\!\!\vr(\tup{\tup{\_,\_,\tup{\_,\_,v}}, \_}) \!\defeq & v
	& \;\;\vw(\tup{\tup{\_,\_,\tup{\strlcas,(\_,v_1,v_2),v_1}}, \_}) & \!\defeq\! v_2 \\
	\!\!\vw(\tup{\tup{\_,\!\_,\tup{\strlw,(\_,v),\!\_}}, \!\_}) \!\defeq & v 
	& \;\;\vw(\tup{\tup{\_,\_,\tup{\strlfaa,(\_,v),v'}}, \_}) & \!\defeq\! v{+}v'
  \end{align*}
\item $\rf$ and $\mo$ satisfy the same constraints as well-formedness
  of \rdmawaitrmw (\cref{subsec:rdmasync_model}).  
\end{itemize}

We next define \strl-consistency, which requires that 
\begin{enumerate*}
	\item each event be associated with (single) stamp $\tagmf$; and
	\item $\so = \po \cup \rf \cup \mo \cup \fr$.
\end{enumerate*}
The former ensures that \strl calls remain ordered with respect to other non-RDMA operations. 
The latter captures the standard notion of happens-before in SC~\cite{DBLP:journals/pacmpl/RaadDRLV19}.
\begin{definition}[\strl-consistency]%
\label{def:rl-consistency}%
Execution $\exec$ is \strl-consistent if:
  \begin{enumerate}
  \item $\forall \action \in E. \ \gettags(e) = \set{\tagmf}$, and 
  \item there exists a well-formed $\tup{\vr, \vw, \rf,\mo}$ such that
    $\exec.\so = \exec.\po \cup \rf \cup \mo \cup \fr$, where $\fr$ is defined as in \cref{subsec:rdmasync_model}.
  \end{enumerate}
\end{definition}



\subsection{The \strl Implementation}
\label{sec:strl-impl}
We implement \strl using node locks and \rdmasync operations, as shown in
\cref{fig:implstrl}. 
For each \rdmascrmw location $\xx$, we create an \rdmawaitrmw location $x$ on some arbitrary node. We assume each thread $\threadt$ has
access to a private location $\temp_\threadt$ for recording the remote data it
reads, and a private location $p_\xx^\threadt$ for recording the value to be put
to a remote location (\ie the second argument of a \rlput)%
\footnote{In practice, we can use a \rlput with `inlined data' and forgo
  temporary location $p_\xx^\threadt$.}. Moreover, each location $\xx$ is
associated with
a node lock $l_\xx$ hosted on the same node as $x$.
We implement \rdmascrmw writes, reads, and RMWs respectively using \rlput,
\rlget, and remote RMWs of \rdmawaitrmw while holding the $l_\xx$ lock.
\begin{figure}[t]
  \noindent
  \scalebox{0.85}{
    \fbox{\begin{minipage}[t]{.22\textwidth}\vspace{-10pt}
\begin{align*}
  \implstrl&(\threadt, \strlw, (\xx,v)) \defeq \\[-2pt]
  & \lklna(l_\xx) ; \\[-2pt]
  & \rlwrite(p_\xx^\threadt,v) ; \\[-2pt]
  & \rlput(x,p_\xx^\threadt,\_) ; \\[-2pt]
  & \lklnr(l_\xx) \\[-2pt]
\end{align*}
\end{minipage}}}
\hfill
\scalebox{0.85}{
  \fbox{\begin{minipage}[t]{.2\textwidth}
\vspace{-10pt}\begin{align*}
  \implstrl&(\threadt, \strlr, (\xx)) \defeq \\[-2pt]
  & \lklna(l_\xx) ; \\[-2pt]
  & \rlget(\temp_\threadt,x,\wid) ; \\[-2pt]
  & \lklnr(l_\xx) ; \\[-2pt]
  & \rlwait(\wid) ; \\[-2pt]
  & \rlread(\temp_\threadt) 
\end{align*}
\vspace{-13.5pt}
\end{minipage}}}
\hfill
  \scalebox{0.85}{\fbox{\begin{minipage}[t]{.2\textwidth}\vspace{-10pt}
\begin{align*}
  \implstrl&(\threadt, \strlcas, (\xx,v_1,v_2)) \defeq \\[-2pt]
  & \lklna(l_\xx) ; \\[-2pt]
  & \rlrcas(\temp_\threadt,x,v_1,v_2,\wid) ; \\[-2pt]
  & \lklnr(l_\xx) ; \\[-2pt]
  & \rlwait(\wid) ; \\[-2pt]
  & \rlread(\temp_\threadt) 
\end{align*}\vspace{-13.5pt}
\end{minipage}}}
\hfill
\scalebox{0.85}{\fbox{\begin{minipage}[t]{.24\textwidth}\vspace{-10pt}
\begin{align*}
  \implstrl&(\threadt, \strlfaa, (\xx,v)) \defeq \\[-2pt]
  & \lklna(l_\xx) ; \\[-2pt]
  & \rlrfaa(\temp_\threadt,x,v,\wid) ; \\[-2pt]
  & \lklnr(l_\xx) ; \\[-2pt]
  & \rlwait(\wid) ; \\[-2pt]
  & \rlread(\temp_\threadt) 
\end{align*}\vspace{-13.5pt}
\end{minipage}}}
\caption{The implementation of \strl (through the \implstrl function)}
\label{fig:implstrl}
\end{figure}

Note that the $\strlw(\xx,v)$ implementation does not wait for
$\rlput(x,p_\xx^\threadt,\_)$ to complete.
As such, when running $\strlw(\xx,1);\strlw(\yy,1)$ in \cref{fig:SC-3} with
$\nodefun{x} \!\neq\! \nodefun{y}$, location $y$ may be modified before $x$. However, this
out-of-order completion is \emph{not observable}, \ie the
(non-SC) outcome $(a,b){=}(1,0)$ is disallowed,  because re-acquiring a node
lock makes all previous operations towards its node visible (see
\cref{sec:locklib-node}).
Specifically, $a{=}1$ implies that $\rlput(x,p_\xx^\threadt,\_)$ has been issued. 
As the implementation of $\strlr(\xx)$ acquires $l_\xx$,
this enforces $\rlput(x,p_\xx^\threadt,\_) $ to become visible; \ie $\strlr(\xx)$ reads $1$ and $(a, b) {=} (1, 0)$ is disallowed. 

In contrast to $\strlw(\xx,v)$, the implementations of the other three
operations must wait (via $\rlwait(\wid)$) for their remote operations to
complete prior to reading the result via $\rlread(\temp_\threadt)$ to ensure
they observe the correct value. For instance, were we to remove $\rlwait(\wid)$
in the implementation of $\strlr(\xx)$, the $\rlread(\temp_\threadt)$ could read
a stale value from $\temp_\threadt$ \emph{before}
$\rlget(\temp_\threadt, x, \wid)$ completes and updates $\temp_\threadt$.
Nevertheless, it is sufficient to wait for the remote operation to complete {\em
  after} releasing the lock. That is, it is possible for another thread to
acquire $l_\xx$ (and modify $x$) before $\rlget(\temp_\threadt, x, d)$
completes; however, the semantics of \lkln ensures that
$\rlget(\temp_\threadt, x, d)$ reads the old value into $\temp_\threadt$.




\begin{figure}[t]
  \centering
  %
  %
  \begin{subfigure}[b]{0.3\textwidth}\small
  \vspace{0pt}\centering
  \scalebox{0.9}{
  \begin{tabular}{|@{\hspace{3pt}}c @{\hspace{3pt}} || @{\hspace{2pt}}c@{\hspace{2pt}}|}
  \hline
  \multicolumn{2}{|c|}{ $\xx,  \yy = 0, 0$} \\
  \hline
  $\inarr{
    \phantom{a} \vspace{-8pt} \\
    \strlw(\xx,1) \\
    \strlw( \yy,1) \\
    \phantom{a} \vspace{-8pt}
  }$
  &
  $\inarr{
    a \assign \strlr( \yy) \\
    b \assign \strlr(\xx)
  }$
  \\
  \hline
  \end{tabular}} 
  \caption{$(a,b) = (1,0)$ \checkno}
  \label{fig:SC-3}
  \end{subfigure}
  \quad
  \begin{subfigure}[b]{0.32\textwidth}\small
  \vspace{0pt}    \centering
  \scalebox{0.9}{
  \begin{tabular}{|@{\hspace{3pt}}c @{\hspace{3pt}} || @{\hspace{2pt}}c@{\hspace{2pt}}|}
  \hline
  \multicolumn{2}{|c|}{ $\xx, \yy = 0, 0$} \\
  \hline
  $\inarr{
    \phantom{a} \vspace{-8pt} \\
    \strlw(\xx,1) \\
    a \assign \strlr( \yy) \\
    \phantom{a} \vspace{-8pt}
  }$
  &
  $\inarr{
    \strlw( \yy,1) \\
    b \assign \strlr(\xx)
  }$
  \\
  \hline
  \end{tabular}} 
  \caption{$(a,b) = (0,0)$ \checkno}
  \label{subfig:strl-sb2}
\end{subfigure}
\quad
  \begin{subfigure}[b]{0.3\textwidth}\small
  \vspace{0pt}\centering
  \scalebox{0.9}{
  \begin{tabular}{|@{\hspace{3pt}}c @{\hspace{3pt}} || @{\hspace{2pt}}c@{\hspace{2pt}}|}
  \hline
  \multicolumn{2}{|c|}{ $\xx = 0$} \\
  \hline
  $\inarr{
    \phantom{a} \vspace{-8pt} \\
    \strlcas(\xx,0,2) \\
    \phantom{a} \vspace{-8pt}
  }$
  &
  $\inarr{
    \strlw(\xx,1)
  }$
  \\
  \hline
  \end{tabular}} 
  \caption{$\xx = 2$ \checkno}
  \label{subfig:strl-rmw}
\end{subfigure}
\\[10pt]
    \begin{subfigure}[b]{0.3\textwidth}\small
  \vspace{0pt}\centering
  \scalebox{0.9}{
  \begin{tabular}{|@{\hspace{3pt}}c @{\hspace{3pt}} || @{\hspace{2pt}}c@{\hspace{2pt}}|}
  \hline
  \multicolumn{2}{|c|}{ $\xx = 0$} \\
  \hline
  & $z = 0$ \\
  \hline
  $\inarr{
    \phantom{a} \vspace{-8pt} \\
    z^2 \assign 1 \\
    \strlw(\xx,1) \\
    \phantom{a} \vspace{-8pt}
  }$
  &
  $\inarr{
    a \assign \strlr(\xx) \\
    b \assign z
  }$
  \\
  \hline
  \end{tabular}}
\caption{$(a,b) = (1, 0)$ \checkyes}
\label{fig:SC-4}
\end{subfigure}
\quad
\begin{subfigure}[b]{0.3\textwidth}\small
  \vspace{0pt}\centering
  \scalebox{0.9}{
  \begin{tabular}{|@{\hspace{3pt}}c @{\hspace{3pt}} || @{\hspace{2pt}}c@{\hspace{2pt}}|}
  \hline
  \multicolumn{2}{|c|}{ $\xx = 0$} \\
  \hline
  $z = 0$ & \\
  \hline
  $\inarr{
    \phantom{a} \vspace{-8pt} \\
    \strlw(\xx,1) \\
    z \assign 1 \\
    \phantom{a} \vspace{-8pt}
  }$
  &
  $\inarr{
    a \assign z^1 \\
    b \assign \strlr(\xx)
  }$
  \\
  \hline
  \end{tabular} }
\caption{$(a,b) = (1, 0)$ \checkyes}
\label{fig:SC-5}
\end{subfigure}
\vspace{-5pt}
\caption{\rdmascrmw examples}
\label{fig:ex-strl-sync}
\end{figure}

The \strl library, when used in isolation (without calls to \eg \rdmawaitrmw),  ensures SC behaviour. 
As such, the weak behaviours of `message-passing' in \cref{fig:SC-3} and `store-buffering' in \cref{subfig:strl-sb2} are disallowed. 
Moreover, \strl RMW operations are strongly isolated with \strl reads and writes; \eg outcome $\xx {=} 2$ is disallowed in \cref{subfig:strl-rmw}. 
This is in contrast to remote RMWs of \rdmawaitrmw, where outcome $x{=}2$ is allowed in 
\cref{subfig:rcas-atomicity-2}.
However, \strl operations does not ensure that earlier remote operations by \emph{other libraries} are completed,  and thus outcome $(a, b) {=} (1, 0)$ is allowed in both \cref{fig:SC-4,fig:SC-5}.

More generally, we can use this strategy to \emph{linearise}~\cite{DBLP:journals/toplas/HerlihyW90} accesses to any sequential data structure $D$ by wrapping each call to $D$ inside a node lock.
This allows us to port existing sequential data structures to RDMA settings with minimal effort. 
Finally, we prove (\cref{thm:strl-sounndness}) that our implementation is correct against the \strl specification with the full proof given in
\iftoggle{shortpaper}{the extended version~\evlink}{\cref{sec:strl-library}}.

\begin{restatable}{theorem}{strllocsound}
  \label{thm:strl-sounndness}
  The implementation \implstrl is sound.
\end{restatable}


%% file: related.tex

\section{Related Work}
\label{sec:related-work}



\smallskip\noindent\textbf{RDMA Semantics.}
The \corerma model~\cite{rdma-sc} is an early attempt at formalising remote
memory accesses, but this semantics does not match the RDMA technical
specification. This gap is addressed by \rdmatso~\cite{OOPSLA-24},
which formalises the actual RDMA semantics over TSO, but the
formalisation did not cover remote RMWs. A later model,
\rdmasc~\cite{ESOP-25}, explored the semantics from
\rdmatso~\cite{OOPSLA-24} but over an SC CPU alongside programming
strategies to efficiently prevent weak behaviours. \rdmasc is
unrelated to our work, including \rdmascrmw.



\smallskip\noindent\textbf{RDMA-Based Distributed Systems.}
Besides \loco~\cite{POPL-LOCO,rdmaloco}, 
prior work has covered a range of distributed systems, \eg consensus
protocols~\cite{aguilera-osdi-2020},
databases~\cite{li-fast-2023,alquraan-nsdi-2024}, stand-alone data
structures~\cite{brock-icpp-2019,devarajan-cluster-2020}. However, unlike \loco (and our work),
these are bespoke systems rather than a programming methodology
or library.



\smallskip\noindent\textbf{Verification.}  Our proofs for the
soundness of library implementations have followed the declarative
style~\cite{DBLP:journals/pacmpl/RaadDRLV19,DBLP:conf/esop/StefanescoRV24,POPL-LOCO}. For
\rdmatsormw (like \rdmatso), we also provide an operational
model~(see~\iftoggle{shortpaper}{\evlink}{\cref{sec:rmw}}) which could ultimately form a basis for a
program logic~(e.g., \cite{pierogi,DBLP:conf/cav/LahavDW23}),
ultimately enabling operational abstractions and proofs of
refinement~\cite{DBLP:journals/pacmpl/DalvandiD22,DBLP:journals/pacmpl/SinghL23}. We
consider such extensions to be future work.



\smallskip\noindent\textbf{RDMA Locks.}
There are several implementations of network locks using RDMA operations, including centralised lock
managers~\cite{DBLP:journals/corr/ChungZ15}, decentralised
algorithms~\cite{DBLP:conf/sigmod/YoonCM18}, asymmetric implementations to
favour local accesses~\cite{alock}, and technology-agnostic designs that are more general
than RDMA~\cite{DBLP:conf/icpp/DevulapalliW05}.
Other stated objectives of these implementations can include fairness,
starvation freedom, low latency, load balancing, scalability, contention
mitigation, fault tolerance~\cite{DBLP:conf/fast/Gao0S25}, etc.

However, none of these existing implementations have been formally
verified (since \rdmatsormw is the first formal semantics of remote
RMWs).  These works, at most, have offered intuitive explanations to
support the correctness of their approach.
Moreover, these implementations lack an explicit description of the
interaction guarantees between locks and other RDMA operations, which
as we have seen can be subtle. In most cases, programmers are made
responsible to ensure relevant operations are completed before
releasing the lock, thus aligning with the weak lock semantics that we
have presented.


%% file: brl.tex

\section{\brl Library Semantics}
\label{sec:app-brl}



As per other RDMA libraries, we assume a set of nodes, $\Nodes$, of fixed size.
Each thread $\threadt$ is associated to a node $\nodefun{\threadt}$. The \brl
library uses the following 5 methods:
\begin{align*}
  m(\vect{v}) ::= & \ \phantom{\mid}
                \brlwrite(x,v)
                \mid \brlread(x)
                \mid \brlbr(x,d,\set{\node_1,\ldots,\node_k}) \\
  &
                \mid \brlwait(d)
                \mid \brlgf(\set{\node_1,\ldots,\node_k})
\end{align*}
\begin{multicols}{2}
  \begin{itemize}
  \item $\brlwrite : \Loc \times \Val \rightarrow ()$
  \item $\brlread : \Loc \rightarrow \Val$
  \item $\brlwait : \Wid \rightarrow ()$
  \item $\brlgf : \mathcal{P}(\Nodes) \rightarrow ()$
  \end{itemize}
\end{multicols}

$\brlwrite(x,v)$ writes a new value $v$ to the location $x$ of the current node.
$\brlread(x)$ reads the location $x$ of the current node and returns its value.\linebreak
$\brlbr(x,d,\set{\node_1,\ldots,\node_k})$ broadcasts the local value of $x$ and
overwrites the values of the copies of $x$ on the nodes
$\set{\node_1,\ldots,\node_k}$, which might include the local node.
$\brlwait(d)$ waits for previous broadcasts of the thread marked with the same
work identifier $\wid \in \Wid$. As is the case for \rlput, this operation only
guarantees that the local values of the broadcasts have been read, but not that
remote copies have been modified. Finally, the global fence operation
$\brlgf(\set{\node_1,\ldots,\node_k})$ ensures every previous operation of the
thread towards one of the nodes in the argument is fully finished, including the
writing part of broadcasts.



We then require the stamping function $\gettagsbrl$:
\begin{align*}
  \gettagsbrl(\tup{\_, \_, \tup{\brlwrite, \_, \_}}) & = \set{\tagcwrite}
  \\
  \gettagsbrl(\tup{\_, \_, \tup{\brlread, \_, \_}}) &= \set{\tagcread}
  \\
  \gettagsbrl(\tup{\_, \_, \tup{\brlwait, \_, \_}}) &= \set{\tagwait}
  \\
  \gettagsbrl(\tup{\_, \_, \tup{\brlgf, (\set{\node_1,\ldots,\node_k}), \_}}) &=
                                                                                \set{\taggf[\node_1] ,\ldots ,\taggf[\node_k]}
  \\
  \gettagsbrl(\tup{\_, \_, \tup{\brlbr, (\_, \_, \set{\node_1,\ldots,\node_k}), \_}}) &=
                                                                                        \set{\tagnlr[\node_1] , \tagnrw[\node_1] , \ldots ,\tagnlr[\node_k] , \tagnrw[\node_k]}
\end{align*}
Broadcasts are associated with a NIC local read and NIC remote write stamp for
each remote node they are broadcasting towards. Similarly, global fence
operations are associated with a global fence stamp for each node.

With this, the stamp order is enough to enforce the behaviour of the global
fence. If we have a program
$\brlbr(x,\wid,\set{\ldots,\node,\ldots}) ; \brlgf(\set{\ldots,\node,\ldots})$,
the plain execution has two events $\action_{\mathtt{BR}}$ and $\action_{\mathtt{GF}}$, and the
definitions of $\gettagsbrl$ and $\tagppo$ (cell G12 in \cref{fig:to}) imply
$\tup{\action_{\mathtt{BR}}, \tagnrw} \arr{\ppo} \tup{\action_{\mathtt{GF}}, \taggf}$.



Recall that, for an execution $\exec$, $\exec.\Write$ represents write subevents
(stamps \tagcwrite, \tagcas, \tagnlw[], and \tagnrw[]), while $\exec.\Read$
represents read subevents (stamps \tagcread, \tagcas, \tagnlr[], \tagnarr[], and
\tagnrr[]).
Recall also that we note \eg
$\exec.\Write_x \defeq \setpred{\saction \in \exec.\Write}{\loc(\saction) =
  \set{x}}$ to constrain a set to subevents on a specific location $x$.
For the \brl library, we additionally define
$\exec.\Write^\node \defeq \setpred{\tup{\action,
    \tagcwrite}}{\nodefun{\threadfun{\action}} = \node} \cup \exec.\tagnrw$ as
the set of write subevents occurring on node $\node$. This includes CPU writes
on the node, as well as broadcast writes towards $\node$ from all threads. We
also note $\exec.\Write_x^\node \defeq \exec.\Write_x \cap \exec.\Write^\node$
as expected. Similarly,
$\exec.\Read^\node \defeq \setpred{\saction}{\saction \in \exec.\Read \land
  \nodefun{\threadfun{\saction}} = \node}$ covers reads occurring on $\node$,
either by a CPU read or as part of a broadcast.
We now work towards a definition of consistency for shared variables.
\begin{definition}  
  For an execution $\exec = \tup{E, \po, \gettagsbrl, \_, \_}$, we define
  the following:
  \begin{itemize}
  \item The \emph{value-read} function $\vr : \exec.\Read \rightarrow \Val$ that
    associates each read subevent with the value returned, if available, \ie if
    $\action = \tup{\_, \_, \tup{\brlread, \_, v}}$, then $\vr(\action) = v$.
  \item The \emph{value-written} function $\vw : \exec.\Write \rightarrow \Val$
    that associates each write subevent with a value $\exec$, \ie if
    $\action = \tup{\_, \_, \tup{\brlwrite, (\_, v), \_}}$, then
    $\vw(\action) = v$.
  \item A \emph{reads-from} relation, $\rf \defeq \bigcup_{\node} \rf^{\node}$,
    where each
    $\rf^{\node} \subseteq \exec.\Write^\node \times \exec.\Read^\node$ is a
    relation on subevents of the same location and node with matching values,
    \ie if $\tup{\saction_1, \saction_2} \in \rf^{\node}$ then
    $\loc(\saction_1) = \loc(\saction_2)$ and
    $\vw(\saction_1) = \vr(\saction_2)$.
  \item A \emph{modification-order} relation
    $\mo \defeq \bigcup_{x,\node} \mo_x^{\node}$ describing the order in which
    writes on $x$ on node $\node$ reach memory.
\end{itemize}
\end{definition}

We define {\em well-formedness} for $\rf$ and $\mo$ as follows. For each remote,
a broadcast writes the corresponding read value: if
$\saction_1 = \tup{\action, \tagnlr} \in \exec.\SEvents$ and
$\saction_2 = \tup{\action, \tagnrw} \in \exec.\SEvents$, then
$\vr(\saction_1) = \vw(\saction_2)$. Each $\rf^{\node}$ is functional on its
range, \ie every read in $\exec.\Read^\node$ is related to at most one write in
$\exec.\Write^\node$. If a read is not related to a write, it reads the initial
value of zero, \ie if
$\saction_2 \in \exec.\Read^\node$ and $\saction_2 \not\in \img{\rf^\node}$ then
$\vr(\saction_2) = 0$. Finally, each $\mo_x^{\node}$ is a strict total order on
$\exec.\Write_x^{\node}$.

We further define the \emph{reads-from-internal} relation as
$\rfi \defeq [\tagcwrite] ; (\po \cap \rf) ; [\tagcread]$ (which corresponds to
CPU reads and writes using the same TSO store buffer), and the
\emph{reads-from-external} relation as $\rfe \defeq \rf \setminus \rfi$.
Moreover, given an execution $\exec$ and well-formed \rf and \mo, we derive
additional relations.
\begin{align*}
  \fr^\node & \defeq \setpred{\tup{r,w} \in \exec.\Read^\node \times \exec.\Write^\node} {\begin{matrix}
    \loc(r) = \loc(w) \ \land \\
    \left(\tup{r,w} \in (\inv{(\rf^\node)}; \mo^\node) \lor r \not\in
    \img{\rf^\node} \right)
  \end{matrix}}
  \\
  \pf & \defeq \setpred{\tup{\tup{\action_1, \tagnlr},\tup{\action_2, \tagwait}}}
        {\exists \wid. \ \tup{\action_1,\action_2} \in \po\rst{\wid}}
        \qquad \quad \fr \defeq \bigcup_{\node} \fr^{\node}
  \\
  \iso & \defeq \setpred{\tup{\tup{\action, \tagnlr},\tup{\action, \tagnrw}}} {\action = \tup{\_,
         \_, \tup{\brlbr, (\_, \_, \set{\ldots,\node,\ldots}), \_}} \in E}
\end{align*}

The \emph{polls-from} relation $\pf$ states that a $\brlwait$ operation
synchronises with the NIC local read subevents of previous broadcasts that use
the same work identifier. The \emph{reads-before} relation $\fr$ states that a
read $r$ executes before a specific write $w$ on the same node and location.
This is either because $r$ reads the initial value of~$0$, or because $r$ reads
from a write that is $\mo$-before $w$. Finally, the
\emph{internal-synchronisation-order} relation $\iso$ states that, within a
broadcast, for each remote node the reading part occurs before the writing part.

We can then define the consistency predicate $\brl.\cons$ as follows.
\begin{definition}[\brl-consistency]
  \label{def:brl-consistency}
  $\tup{E, \po, \gettags, \so, \hb}$ is \brl-consistent if:
  \begin{itemize}
  \item $\gettags = \gettagsbrl$;
  \item there exists well-formed \vr, \vw, \rf, and \mo, such that \\
    $[\tagcread] ; (\inv{\po} \cap \fr) ; [\tagcwrite] = \emptyset$ and
    $\so = \iso \cup \rfe \cup \pf \cup \fr \cup \mo$.
  \end{itemize}
\end{definition}

It is straightforward to check that this consistency predicate
satisfies monotonicity and decomposability.  For CPU reads and writes,
we ask that \fr does not contradict the program order.  \Eg, a program
$\brlwrite(x,1) ; \brlread(x)$ must return $1$ and cannot return $0$,
even if the semantics of TSO allows for the read to finish before the
write.


%% file: proofs.tex



\section{Correctness Proofs}
\label{sec:proofs}

Correctness proofs of the \framework framework can be found in~\cite{POPL-LOCO}.
We recall the main definitions and results in \Cref{sec:proofs-mowgli} before
proving the soundness of the weak lock~(\cref{sec:lklw-library}), strong
lock~(\cref{sec:lkls-library}), node lock~(\cref{sec:lkln-library}), and \strl
libraries~(\cref{sec:strl-library}).



\subsection{Background: \framework Definitions and Results}
\label{sec:proofs-mowgli}

\framework assumes a type $\Val$ of values, a type $\Loc \suq \Val$ of
locations, and a type $\Method$ of methods. The syntax of sequential programs is
given by the following grammar:
\begin{align*}
  v, v_i \in \Val \qquad \quad & m \in \Method  \qquad\quad \progf \in \Val \rightarrow \Progp \qquad\quad k \in \mathbb{N}^+
  \\
  \Progp \ni \progp & ::= v
  \mid m(v_1,\ldots,v_k)
  \mid \LetF{\progp}{\progf}
  \mid \code{loop} \ \progp
  \mid \code{break}_k \ v
\end{align*}

\framework assumes top-level concurrency, \ie there is a fixed set of threads
$\Threads \defeq \{1, 2, \dots, T\}$, and a concurrent program is given by a
tuple $\progps = \tup{\progp_1,\ldots,\progp_T}$, where each thread $t$
corresponds to a program $\progp_t \in \Progp$.

The semantics of a program is given by an execution, which is a graph over
events. Recall that events are defined in~\Cref{def:app-events}. The first two components $\tup{E, \po}$ of an execution form a \emph{plain execution}:
\begin{definition}
  We say that $\tup{E, \po}$ is a \emph{plain execution} iff $E \suq \Events$,
  $\po \subseteq E \times E$, and
  $\po = \bigcup_{\threadt \in \Threads} \po\rst{t}$ where every $\po\rst{t}$
  (\ie $\po$ restricted to the events of thread $\threadt$) is a total order.
\end{definition}

We write $\emptyset_G \defeq \tup{\emptyset, \emptyset}$ for the empty execution
and $\set{\ev}_G \defeq \tup{\set{\ev}, \emptyset}$ for the execution with a
single event $\ev$. Given two executions, $G_1 = \tup{E_1, \po_1}$ and
$G_2 = \tup{E_2 , \po_2}$, with disjoint sets of events (\ie
$E_1 \cap E_2 = \emptyset$), we define their sequential composition $G_1 ; G_2$
and parallel composition $G_1 {\parallel} G_2$ as follows:
\begin{align*}
  G_1 ; G_2 & \defeq \tup{E_1 \cup E_2, \po_1 \cup \po_2 \cup (E_1 \times E_2 )} & G_1 {\parallel} G_2 & \defeq \tup{E_1 \cup E_2 , \po_1 \cup \po_2} 
\end{align*}

The plain semantics of a program $\progp$ executed by a thread $t$ is given by
$\interpt{\progp}$, which is a set of pairs of the form $\tup{r, G}$, where $r$
is the output and $G$ is a plain execution. This set represents all conceivable
unfoldings of the program into method calls, even those that will be rejected by
the semantics of the corresponding libraries. Each output is 
a pair
$\tup{v, k}$, where $v$ is a value and $k$ a break number, indicating the
program terminates by requesting to exit $k$ nested loops and returning the
value $v$.
\begin{align*}
  \interpt{v} \defeq & \set{\tup{ \tup{v,0}, \emptyset_G }} \qquad \qquad \interpt{\texttt{break}_k \ v} \defeq \set{\tup{ \tup{v,k}, \emptyset_G }}
  \\
  \interpt{m(\vect{v})} \defeq & \setpred{\tup{ \tup{v',0}, \set{\tup{\threadt,\aident,\tup{m,\vect{v},v'}}}_G }}{v' \in \Val \ \land\  \aident \in \EId}
  \\
  \interpt{\LetF{\progp}{\progf}} \defeq &
  \begin{array}[t]{@{}l@{}}
    \setpred{\tup{r, G_1;G_2}}{\tup{\tup{v,0}, G_1} \in \interpt{\progp} \ \land \ \tup{r, G_2} \in \interpt{\progf \ v}} \\[2pt]
    \ \cup \setpred{\tup{\tup{v,k}, G_1}}{ \tup{\tup{v,k}, G_1} \in \interpt{\progp} \ \land \ k \neq 0}\end{array} \\
  \interpt{\texttt{loop} \ \progp} \defeq & \bigcup_{j \in \mathbb{N}} \setpred{\tup{\tup{v,k}, G_0;\ldots;G_j}}{\begin{matrix}
    (\forall 0 \leq i < j. \ \tup{\tup{\_,0}, G_i} \in \interpt{\progp})
    \ \land \ \tup{\tup{v,k+1}, G_j} \in \interpt{\progp} 
  \end{matrix}}
\end{align*}

We lift the plain semantics to the level of concurrent programs and define
\[
  \interp{\progps} \defeq \setpred{\tup{\tup{v_1,\ldots,v_T},\parallel_{t \in \Threads} G_\threadt}}
{\begin{matrix}
  \forall t \in \Threads. \tup{\tup{v_\threadt,0},G_\threadt} \in
  \interpt{\progp_t}
\end{matrix}}
\]
Concurrent programs only properly terminate if each thread terminates with a
break number of $0$. In which case, the output of the concurrent program is the
parallel composition of the values and plain executions of the different
threads.

Then, we can define executions~(\cref{def:execs}),
libraries~(\cref{subsec:rdmasync_prelim}), and consistent
executions~(\cref{def:lambda-cons}).

Given a concurrent program $\progps$ using libraries $\Lambda$, we note
$\outcome_\Lambda(\progps)$ the set of all output values of its
$\Lambda$-consistent executions.
$$\outcome_\Lambda(\progps) \defeq \setpred{\vect{v}}{ \exists \tup{E, \po,
    \gettags, \so, \hb} \ \Lambda\text{-consistent.\ }
  \tup{\vect{v},\tup{E,\po}} \in \interp{\progps} }$$



Then, an implementation for a library $L$ is a function
$I : (\Threads \times L.M \times \Val^*) \rightarrow \Progp$ associating every
method call of the library $L$ to a sequential program.
\begin{definition}
  We say that $I$ is \emph{well defined} for a library $L$ using $\Lambda$ iff
  for all $\threadt \in \Threads$, $m \in L.M$ and $\vect{v} \in \Val^*$, we
  have:
\begin{enumerate}
\item $L \not\in \Lambda$, and $I(\threadt,m,\vect{v})$ only calls methods of
  the libraries of $\Lambda$.
\item $\tup{\tup{-,k+1}, -} \not\in \interpt{I(\threadt,m,\vect{v})}$, \ie the
  implementation of a method call $m(\vect{v})$ cannot return with a non-zero
  break number, and thus cannot cause a loop containing a call to $m(\vect{v})$
  to break inappropriately.
\item if $\tup{\tup{v,0}, \tup{E, \po}} \in \interpt{I(\threadt,m,\vect{v})}$
  then $E \neq \emptyset$, \ie if an implementation successfully executes, it
  must contain at least one method call.
\end{enumerate}
\end{definition}

We note $\loc(I)$ the set of all locations that can be accessed by the
implementation of $I$:
$\loc(I) \defeq \bigcup_{\threadt,m,\vect{v}} \bigcup_{(\_,\tup{E,\_}) \in
  \interpt{I(\threadt,m,\vect{v})}} \loc(E)$.
We then define a function $\impli{\_}$ to map an implementation $I$ to a
concurrent program as follows.
\begin{align*}
  \implti{v} & \defeq v  & 
  \implti{m(v_1,\ldots,v_k)} &
    \defeq \begin{cases}
      I(\threadt,m, \tup{v_1,\ldots,v_k}) & \text{if } m \in L.M \\
      m(v_1,\ldots,v_k) & \text{otherwise} \\
    \end{cases} \\
  \implti{\texttt{loop} \ \progp} & \defeq
                                    \texttt{loop} \ \implti{\progp} & 
  \implti{\LetF{\progp}{\progf}} & \defeq
                                   \LetF{\implti{\progp}}{(\lambda v. \implti{\progf \ v})} \\
  \implti{\texttt{break}_k \ v} & \defeq
                                  \texttt{break}_k \ v & 
  \impli{\tup{\progp_1,\ldots,\progp_T}} & \defeq \tup{\impl{\progp_1}_{1,L},\ldots,\impl{\progp_T}_{T,L}} 
\end{align*}

Using these definitions, we arrive at a notion of a sound
implementation, which holds whenever the implementation is a
refinement of the library specification.
\begin{definition}
  \label{def:sound}
  We say that $I$ is a \emph{sound implementation} of $L$ using
  $\Lambda$ if, for any program $\progps$ such that
  $\loc(I) \cap \loc(\progps) = \emptyset$, we have that
  $\outcome_\Lambda(\impli{\progps}) \suq
  \outcome_{\Lambda\uplus\set{L}}(\progps)$.
\end{definition}

As soundness is difficult to prove directly, \framework develops a modular proof
technique using an {\em abstraction function} mapping the implementation to its
abstract library specification.
For $f : A \rightarrow B$ and $r \suq A \times A$, we note
$f(r) \defeq \setpred{\tup{f(x),f(y)}}{\tup{x,y} \in r}$.
\begin{definition}
  \label{def:abstraction}
  Suppose $I$ is a well-defined implementation of a library $L$ using $\Lambda$,
  and that $G = \tup{E, \po}$ and $G'= \tup{E', \po'}$ are plain executions
  using methods of $\Lambda$ and $L$ respectively. We say that a surjective
  function $f : E \rightarrow E'$ abstracts $G$ to $G'$, denoted
  $\absf{L}{I}{G}{G'}$, iff
  \begin{itemize}
  \item $E\rst{L} = \emptyset$ (\ie $G$ contains no calls to the abstract
    library $L$) and $E'\rst{L} = E' $ (\ie $G'$ only contains calls to the
    abstract library $L$);
  \item $f(\po) \suq (\po')^*$ and
    $\forall \action_1,\action_2, \ \tup{f(\action_1),f(\action_2)} \in \po'
    \implies \tup{\action_1,\action_2} \in \po$; and
  \item if $\action' = \tup{\threadt,\aident,\tup{m,\vect{v},v'}} \in E'$ then
    $\tup{\tup{v',0}, G\rst{f^{-1}(\action')}} \in
    \interpt{I(\threadt,m,\vect{v})}$
  \end{itemize}
\end{definition}

\begin{lemma}
  \label{lem:find-abstraction}
  Given $\progps$ on library $L$ and a well-defined implementation $I$ of $L$,
  if $\tup{\vect{v}, G} \in \interp{\impli{\progps}}$ then there exists
  $\tup{\vect{v}, G'} \in \interp{\progps}$ and $f$ such that
  $\absf{L}{I}{G}{G'}$.
\end{lemma}

\begin{definition}
  \label[definition]{def:locally-sound}
  We say that a well defined implementation $I$ of a library $L$ is
  \emph{locally sound} iff, whenever we have a $\Lambda$-consistent execution
  $\exec = \tup{E, \po, \gettags, \so, \hb}$ and
  $\absf{L}{I}{\tup{E, \po}}{\tup{E', \po'}}$, then there exists $\gettags'$,
  $\so'$, and a concretisation function
  $g : \tup{E',\po',\gettags'}.\SEvents \rightarrow \exec.\SEvents$ such that:
  \begin{itemize}
  \item $g(\tup{\action', \tagt'}) = \tup{\action, \tagt}$ implies
    $f(\action) = \action'$ and
    \begin{itemize}
    \item For all $\tagt_0$ such that $\tup{\tagt_0, \tagt'} \in \tagppo$, there
      exists $\tup{\action_1, \tagt_1} \in \exec.\SEvents$ such that
      $f(\action_1) = \action'$, $\tup{\tagt_0, \tagt_1} \in \tagppo$, and
      $\tup{\tup{\action_1, \tagt_1}, \tup{\action, \tagt}} \in \hb^*$;
    \item For all $\tagt_0$ such that $\tup{\tagt', \tagt_0} \in \tagppo$, there
      exists $\tup{\action_2, \tagt_2} \in \exec.\SEvents$ such that
      $f(\action_2) = \action'$, $\tup{\tagt_2, \tagt_0} \in \tagppo$, and
      $\tup{\tup{\action, \tagt}, \tup{\action_2, \tagt_2}} \in \hb^*$.
    \end{itemize}
  \item $g(\so') \suq \hb$;
  \item For all $\hb'$ transitive such that $(\ppo' \cup \so')^+ \suq \hb'$ and
    $g(\hb') \suq \hb$, we have \\
    $\tup{E', \po', \gettags', \so', \hb'} \in L.\cons$, where
    $\ppo' \defeq \tup{E', \po', \gettags'}.\ppo$.
  \end{itemize}
\end{definition}

\begin{theorem}
  \label{thm:sound-sound}
  If a well-defined implementation is locally sound, then it is sound.
\end{theorem}

We show the local soundness of the different implementations given in this
paper. Thus, from the theorem above, these implementations are sound.



\subsection{\lklw Library}
\label{sec:lklw-library}

\lklwlocsound*

\begin{proof}
  We assume an $\set{\rdmasync,\brl}$-consistent execution
  $\exec = \tup{E, \po, \gettags, \so, \hb}$ which is abstracted via $f$ to
  $\tup{E',\po'}$ that uses (only) the \lklw library, \ie
  $\absf{\lklw}{\impllklw}{\tup{E, \po}}{\tup{E', \po'}}$ holds. We need to
  provide $\gettags'$, $\so'$, and
  $g : \tup{E',\po',\gettags'}.\SEvents \rightarrow \exec.\SEvents$ respecting
  some conditions. From $\tup{E',\po'}$, we simply take
  $\gettags' = \gettagslklw$.

  Since $\exec$ is $\set{\rdmasync,\brl}$-consistent, it means
  $(\ppo \cup \so\rst{\rdmasync} \cup \so\rst{\brl}) \suq \hb$, $\hb$ is
  transitive and irreflexive, and the two restrictions of $\exec$ are
  respectively \rdmasync-consistent and \brl-consistent.

  \rdmasync-consistency implies there is some well-formed \vr, \vw, \rf, \mo,
  \ro, and \rao such that $\ib$ is irreflexive,
  $\forall \action. \gettags\rst{\rdmasync}(\action) \in \gettagsrl(\action)$, and
  $\so\rst{\rdmasync} = \iso \cup \rfe \cup \pfget \cup \ro \cup \fr \cup \mo
  \cup \rao \cup ([\tagnrw]; \inv{\iso} ; \rao) \cup ([\Inst];\ib)$.

  \brl-consistency implies there is some well-formed $\vr''$, $\vw''$, $\rf''$,
  and $\mo''$, such that $\gettags\rst{\brl} = \gettagsbrl$,
  $[\tagcread] ; (\inv{\po\rst{\brl}} \cap \fr'') ; [\tagcwrite] = \emptyset$,
  and $\so\rst{\brl} = \iso'' \cup \rfe'' \cup \pf'' \cup \fr'' \cup \mo''$. (We
  will use double apostrophes for references to the \brl library.)

  We define $g$ as follows.
  \begin{itemize}
  \item For an event $\action' = (\threadt, \_, (\lklwa, (x), ()))$, we choose
    $g(\action', \tagmf) = (\action_{r}, \tagcread)$ with
    $\action_{r} = (\threadt, \_, (\brlread, (x_{\threadt'}), (v))) \in
    \inv{f}(\action')$ the last event of the implementation (reading a shared
    variable owned by some thread $\threadt'$).
  \item For an event $\action' = (\threadt, \_, (\lklwr, (x), ()))$, we choose
    $g(\action', \tagcwrite) = (\action_{w}, \tagcwrite)$ with
    $\action_{w} = (\threadt, \_, (\brlwrite, (x_\threadt, v+1), ())) \in
    \inv{f}(\action')$ the second event of the implementation.
  \end{itemize}
  First, let us show that $g$ preserves \tagppo (first property of local
  soundness). For \lklwr this is trivial using the identity function. For
  \lklwa, the stamp \tagcread is similar to \tagmf w.r.t. later stamps, so
  $(\action_2, \tagt_2) = (\action_{r}, \tagcread)$ is enough. For an earlier
  stamp $\tagt_0$ such that $(\tagt_0, \tagmf) \in \tagppo$, we take
  $(\action_1, \tagt_1) = ((\threadt, \_, (\rlrfaa, (\ldots,\wid), ())),
  \tagnlw)$ the first event of the implementation, and with
  $\action_{wt} = (\threadt, \_, (\rlwait, (\wid), ()))$ the second event we
  have
  $(\action_1, \tagt_1) \arr{\pfget} (\action_{wt}, \tagwait) \arr{\ppo}
  (\action_{r}, \tagcread)$ (thus included in \hb) with
  $(\tagt_0, \tagnlw) \in \tagppo$.

  Now we need to pick a suitable $\so'$ such that $g(\so') \suq \hb$ and
  $\tup{E',\po',\gettags',\so',\_}$ is \lklw-consistent. We can assume that
  $\tup{E',\po'}$ respects locks, as otherwise $\so' = \emptyset$ is enough.
  Thus, for each location $x$ we need to define a total order $\lo'_x$ on
  $A'_x \defeq \setpred{\action'}{\action' \in E'_x \land \methodfun{\action'} =
    \lklwa}$. Each event $\action' \in A'_x$ can be associated to its first
  subevent of the form
  $((\threadt', \_, (\rlrfaa, (p^{\threadt'}_x,x_a,1,\wid), ())), \tagnarr)$,
  with $\node = \nodefun{x}$. From \rdmasync-consistency, $\rao$ induces a total
  ordering on these subevents, and we simply keep the same ordering for $A'_x$.
  As such, we define \linebreak
  $\so' = \bigcup_x \setpred{\tup{\action'_1, \tagcwrite}, \tup{\action'_2,
      \tagmf}} {(\action'_1, \action'_2) \in \inv{(\imm{\po'_x})} ; \lo'_x}$ as
  expected, and we have that $\tup{E',\po',\gettags',\so',\_}$ is
  \lklw-consistent.

  Thus, the rest of the proof is to show that $g(\so') \suq \hb$, \ie that the
  synchronisations promised by the \lklw library are enforced in the
  implementation. We can assume
  $(\action'_0, \tagmf) \arr{\lo'_x} (\action'_2, \tagmf)$ and
  $(\action'_0, \tagmf) \arr{\imm{\po'_x}} (\action'_1, \tagcwrite)$, with
  $\action'_0$ running $\lklwa(x)$ by thread $\threadt_1$, $\action'_1$ running
  $\lklwr(x)$ by thread $\threadt_1$, and $\action'_2$ running $\lklwa(x)$ by
  thread $\threadt_2$. We also note
  $(\action_1, \tagcwrite) = g(\action'_1, \tagcwrite)$ and
  $(\action_2, \tagcread) = g(\action'_2, \tagmf)$. Our goal is then to show
  $(\action_1, \tagcwrite) \arr{\hb} (\action_2, \tagcread)$.

  We proceed by induction on the ordering $\lo'_x$. The base case is for
  $(\action'_0, \tagmf) \arr{\imm{\lo'_x}} (\action'_2, \tagmf)$. This base case
  trivially implies the general case by transitivity, since the program respects
  locks (\ie intermediate acquires are being released) and
  $(\tagcread, \tagcwrite) \in \tagppo$.

  Let
  $\action_0^{faa} = (\threadt_1, \_, (\rlrfaa, (p^{\threadt_1}_x,x_a,1,\wid), ()))$
  be the FAA in the implementation of $\action'_0$ and
  $\action_2^{faa} = (\threadt_2, \_, (\rlrfaa, (p^{\threadt_2}_x,x_a,1,\wid),
  ()))$ in the implementation of $\action'_2$. By definition we have
  $(\action_0^{faa}, \tagnarr) \arr{\imm{(\rao\rst{E_{x_a}})}} (\action_2^{faa},
  \tagnarr)$, since any remote RMW in $E_{x_a}$ is from an implementation of some
  $\lklwa(x)$ event. From the semantics of \rdmasync we have
  $(\action_0^{faa}, \tagnrw) \arr{\hb} (\action_2^{faa}, \tagnarr)$ (from the
  $([\tagnrw[]]; \inv{\iso} ; \rao)$ component), and thus we necessarily have
  $(\action_0^{faa}, \tagnrw) \arr{\rf} (\action_2^{faa}, \tagnarr)$, \ie the
  second FAA reads the modified value of the first. This is because
  $\action_2^{faa}$ cannot read from an earlier write (or the initial value of
  $0$) as that would imply an \fr dependency and an \hb cycle; and cannot read
  ($\rfe \subseteq \hb$) from a later write, as any later write is \hb after
  $\action_2^{faa}$ (via \rao and \ppo).

  There is some value $v_0 = \vr((\action_0^{faa}, \tagnarr))$ read by the first
  FAA operation. By well-formedness of \vr, \vw, and \rf, we have
  $\vr((\action_2^{faa}, \tagnarr)) = \vw((\action_0^{faa}, \tagnarr)) = v_0 +
  1$, \ie the following $\lklwa(x)$ gets the next ticket. More generally, it is
  clear every $\lklwa(x)$ gets a different ticket. We also have
  $\vw((\action_0^{faa}, \tagnlw)) = v_0$, \ie $p^{\threadt_1}_x$ is modified to
  contain $v_0$. Respectively $p^{\threadt_2}_x$ is modified to contain $v_0+1$.

  Let $\action_0^{r}$ be the third event of the implementation of $\action'_0$
  reading $p^{\threadt_1}_x$. We necessarily have
  $(\action_0^{faa}, \tagnlw) \arr{\rf} (\action_0^{r}, \tagcread)$. This is
  because $\action_0^{r}$ cannot read from the future (it would create an
  $\rf;\ippo$ cycle in \ib) and the second event $\rlwait(\wid)$ makes sure all
  previous modifications of $p^{\threadt_1}_x$ are available (ignoring the last
  one would be an $\fr;\hb$ cycle). Thus, in the implementation of $\action'_0$,
  the meta-variable $\code{v}$ corresponds to the value $v_0$. More generally,
  in any implementation of $\lklwa(x)$, $\code{v}$ corresponds to the ticket
  obtained (\eg $v_0+1$ for $\action'_2$).

  The implementation of $\action'_1$ (running $\lklwr(x)$) also reads
  $p^{\threadt_1}_x$. For the same reason, $\code{v}$ corresponds to the ticket
  of the previous $\lklwa(x)$, \ie $v_0$ in our case. Since the program respects
  locks, every $\lklwr(x)$ handles a different ticket, and $\action'_1$ is the
  only one handling ticket $v_0$ for $x$.

  The second event in the implementation of $\action'_1$ is
  $\action_1 = (\threadt_1, \_, (\brlwrite, (x_{\threadt_1}, v_0+1), ()))$
  modifying $x_{\threadt_1}$. (There is also a broadcast propagating the new
  value across the network.) By well-formedness we have
  $\vw''((\action_1, \tagcwrite)) = v_0+1$. The last event in the implementation
  of $\action'_2$ is of the form
  $\action_2 = (\threadt_2, \_, (\brlread, (x_{\threadt_2}), (v_0+1)))$
  returning a value of $v_0 + 1$, and by well-formedness
  $\vr''(\action_2, \tagcread) = v_0+1$. The read is necessarily on
  $x_{\threadt_2}$ as other $x_{\threadt}$ shared variables are never modified
  to contain $v_0+1$. Now, by well-formedness of $\rf''$ we can create a
  dependency between $\action_1$ and $\action_2$.

  If $\nodefun{\threadt_1} = \nodefun{\threadt_2}$ (\ie the two commands are on
  the same node, perhaps even the same thread), then we have
  $(\action_1, \tagcwrite) \arr{\rf''} (\action_2, \tagcread)$ as
  $(\action_1, \tagcwrite)$ is the only element of
  $\exec.\Write^{\nodefun{\threadt_1}}$ writing $v_0+1$. If they are different
  threads or $\action_2 \arr{\po} \action_1$, then $\rfe'' \subseteq \hb$ is
  enough. Otherwise $\threadt_1 = \threadt_2$ with
  $\action_1 \arr{\po} \action_2$ and the $\rlrfaa$/$\rlwait$ in-between
  $\action_1$ and $\action_2$ forces a sequence of dependencies
  $\ppo;\pfget;\ppo \subseteq \hb$.

  Else $\nodefun{\threadt_1} \neq \nodefun{\threadt_2}$ and $\action_2$ reads
  from an element of $\exec.\Write^{\nodefun{\threadt_2}}$, which is a subevent
  of a broadcast reading from $\action_1$. (Technically, this could be from a
  delayed broadcast of a previous $\lklwr(x)$ by thread $\threadt_1$, not
  necessarily the broadcast immediately after $\action_1$.) Thus we similarly
  have
  $((\action_1, \tagcwrite), (\action_2, \tagcread)) \in \rfe'';\iso'';\rfe''
  \subseteq \hb$.

\end{proof}



\subsection{\lkls Library}
\label{sec:lkls-library}

\begin{align*}
\impllkls(\threadt, \lklsa, (x)) & \defeq \lklwa(x)
&&&
\impllkls(\threadt, \lklsr, (x)) & \defeq \brlgf(\Nodes) ; \lklwr(x)
\end{align*}

\lklslocsound*
\begin{proof}
  This is very straightforward from the semantics of the different libraries.

  If an execution is lock-well-formed (\Cref{def:locklib-respect}) with respect
  to strong locks, the implementation is clearly lock-well-formed with respect
  to weak locks.

  A strong acquire $\lklsa(x)$ should behave as stamp \tagmf, which is the case
  of the implementation $\lklwa(x)$. A strong release $\lklsr(x)$ should behave
  as a global fence (stamps \taggf) and synchronise with later acquires. In the
  implementation, the first call executes a global fence (stamps \taggf, see
  \Cref{sec:app-brl}), while the latter call is a weak release that synchronises
  with later acquires (\Cref{def:locklib-weak-cons}). The two components execute
  in order according to $\ppo \subseteq \hb$ (cell L2 in \Cref{fig:to}).
\end{proof}



\subsection{\lkln Library}
\label{sec:lkln-library}

\lklnlocsound*
\begin{proof}

  We assume an $\set{\rdmasync}$-consistent execution
  $\exec = \tup{E, \po, \gettags, \so, \hb}$ which is abstracted via $f$ to
  $\tup{E',\po'}$ that uses (only) the \lkln library, \ie
  $\absf{\lkln}{\impllkln}{\tup{E, \po}}{\tup{E', \po'}}$ holds. We need to
  provide $\gettags'$, $\so'$, and
  $g : \tup{E',\po',\gettags'}.\SEvents \rightarrow \exec.\SEvents$ respecting
  some conditions. From $\tup{E',\po'}$, we simply take
  $\gettags' = \gettagslkln$.

  Since $\exec$ is $\set{\rdmasync}$-consistent, it means
  $(\ppo \cup \so) \suq \hb$, $\hb$ is transitive and irreflexive, and $\exec$
  is \rdmasync-consistent. Thus there is some well-formed \vr, \vw, \rf, \mo,
  \ro, and \rao such that $\ib$ is irreflexive,
  $\forall \action. \gettags(\action) \in \gettagsrl(\action)$, and
  $\so = \iso \cup \rfe \cup \pfget \cup \ro \cup \fr \cup \mo \cup \rao \cup
  ([\tagnrw[]]; \inv{\iso} ; \rao) \cup ([\Inst];\ib)$.

  As an intermediate result: for each thread $\threadt$ we have
  $\mo_{p^{\threadt}_x} \subseteq \po$, \ie the modifications of the temporary
  location $p^{\threadt}_x$ happen in program order. Since \hb containing \mo is
  acyclic, it is enough to show that whenever
  $\saction_1, \saction_2 \in \exec.\Write_{p^{\threadt}_x}$ and
  $(\saction_1,\saction_2) \in \po$ then we have
  $(\saction_1,\saction_2) \in \hb$. If $\saction_1$ has a stamp $\tagcwrite$ we
  immediately have $(\saction_1,\saction_2) \in \ppo \subseteq \hb$. From the
  implementation, in the other cases $\saction_1$ has a stamp $\tagnlw$ from
  either a \rlrfaa or \rlget operation. In each case, $\saction_1$ is
  immediately followed by some $(\action, \tagwait)$ forcing the write to
  finish. Thus we have $(\saction_1,\saction_2) \in \pfget;\ppo \subseteq \hb$.

  We now define $g$ as follows.
  \begin{itemize}
  \item For an event $\action' = (\threadt, \_, (\lklna, (x), ()))$, we choose
    $g(\action', \tagmf) = (\action_{r}, \tagcread)$ with
    $\action_{r} = (\threadt, \_, (\rlread, (p_x^\threadt), (v))) \in
    \inv{f}(\action')$ the last read event before breaking the loop, and
    penultimate event of the implementation.
  \item For an event $\action' = (\threadt, \_, (\lklnr, (x), ()))$, we choose:
    $g(\action', \tagnf) = (\action_{rf}, \tagnf)$ with
    $\action_{rf} = (\threadt, \_, (\rlrf, (\nodefun{x}), ())) \in
    \inv{f}(\action')$ the first event of the implementation; and
    $g(\action', \tagnrw) = (\action_{put}, \tagnrw)$ with
    $\action_{put} = (\threadt, \_, (\rlput, (x_r, p_x^\threadt, \_), ())) \in
    \inv{f}(\action')$ the second event of the implementation.
  \end{itemize}
  First, let us show that $g$ preserves \tagppo (first property of local
  soundness). For \lklnr this is trivial as $g$ maps to the same stamps. For
  \lklna, the stamp \tagcread is similar to \tagmf w.r.t. later stamps, so
  $(\action_2, \tagt_2) = (\action_{r}, \tagcread)$ is enough. For an earlier
  stamp $\tagt_0$ such that $(\tagt_0, \tagmf) \in \tagppo$, we take
  $(\action_1, \tagt_1) = ((\threadt, \_, (\rlrfaa, (\ldots,\wid), ())),
  \tagnlw)$ the first event of the implementation, and with
  $\action_{wt} = (\threadt, \_, (\rlwait, (\wid), ()))$ the second event we
  have
  $(\action_1, \tagt_1) \arr{\pfget} (\action_{wt}, \tagwait) \arr{\ppo}
  (\action_{r}, \tagcread)$ (thus included in \hb) with
  $(\tagt_0, \tagnlw) \in \tagppo$.

  Now we need to pick a suitable $\so'$ such that $g(\so') \suq \hb$ and
  $\tup{E',\po',\gettags',\so',\_}$ is \lkln-consistent. We can assume that
  $\tup{E',\po'}$ respects locks, as otherwise $\so' = \emptyset$ is enough.
  Thus, for each location $x$ we need to define a total order $\lo'_x$ on
  $A'_x \defeq \setpred{\action'}{\action' \in E'_x \land \methodfun{\action'} =
    \lklna}$. Each event $\action' \in A'_x$ can be associated to its first
  subevent of the form
  $((\threadt', \_, (\rlrfaa, (p^{\threadt'}_x,x_a,1,\wid), ())), \tagnarr)$,
  with $\node = \nodefun{x}$. From \rdmasync-consistency, $\rao$ induces a total
  ordering on these subevents, and we simply keep the same ordering for $A'_x$.
  As such, we define
  \begin{align*}
    \so' = & \setpred{\tup{\action', \tagnf[\nodefun{\loc(\action')}]}, \tup{\action', \tagnrw[\nodefun{\loc(\action')}]}}{\methodfun{\action'} = \lklnr} \\
           & \bigcup_{x \in \Loc} \setpred{\tup{\action'_1, \tagnrw[\nodefun{\loc(\action'_1)}]}, \tup{\action'_2, \tagmf}} {(\action'_1, \action'_2) \in \inv{(\imm{\po'_x})} ; \lo'_x}
  \end{align*}
  as expected, and we have that $\tup{E',\po',\gettags',\so',\_}$ is
  \lkln-consistent.

  Thus, the rest of the proof is to show that $g(\so') \suq \hb$, \ie that the
  synchronisations promised by the \lkln library are enforced in the
  implementation. The easy case is for the internal synchronisation. For
  $(\tup{\action', \tagnf}, \tup{\action', \tagnrw}) \in \so'$, we clearly have
  $(g(\tup{\action', \tagnf}), g(\tup{\action', \tagnrw})) \in \ppo \subseteq
  \hb$.

  For the main case, we can assume
  $(\action'_0, \tagmf) \arr{\lo'_x} (\action'_2, \tagmf)$ and
  $(\action'_0, \tagmf) \arr{\imm{\po'_x}} (\action'_1, \tagnrw)$, with
  $\nodefun{x} = \node$, $\action'_0$ running $\lklna(x)$ by thread
  $\threadt_1$, $\action'_1$ running $\lklnr(x)$ by thread $\threadt_1$, and
  $\action'_2$ running $\lklna(x)$ by thread $\threadt_2$. We also note
  $(\action_1, \tagnrw) = g(\action'_1, \tagnrw)$ and
  $(\action_2, \tagcread) = g(\action'_2, \tagmf)$. Our goal is then to show
  $(\action_1, \tagnrw) \arr{\hb} (\action_2, \tagcread)$.

  We proceed by induction on the ordering $\lo'_x$. The base case is for
  $(\action'_0, \tagmf) \arr{\imm{\lo'_x}} (\action'_2, \tagmf)$. This base case
  trivially implies the general case by transitivity, since the program respects
  locks (\ie intermediate acquires are being released) and
  $(\tagcread, \tagnrw) \in \tagppo$.

  Let
  $\action_0^{faa} = (\threadt_1, \_, (\rlrfaa, (p^{\threadt_1}_x,x_a,1,\wid),
  ()))$ be the FAA in the implementation of $\action'_0$ and
  $\action_2^{faa} = (\threadt_2, \_, (\rlrfaa, (p^{\threadt_2}_x,x_a,1,\wid),
  ()))$ in the implementation of $\action'_2$. By definition we have
  $(\action_0^{faa}, \tagnarr) \arr{\imm{(\rao\rst{E_{x_a}})}} (\action_2^{faa},
  \tagnarr)$, since any remote RMW in $E_{x_a}$ is from an implementation of some
  $\lklwa(x)$ event. From the semantics of \rdmasync we have
  $(\action_0^{faa}, \tagnrw) \arr{\hb} (\action_2^{faa}, \tagnarr)$ (from the
  $([\tagnrw[]]; \inv{\iso} ; \rao)$ component), and thus we necessarily have
  $(\action_0^{faa}, \tagnrw) \arr{\rf} (\action_2^{faa}, \tagnarr)$, \ie the
  second FAA reads the modified value of the first. This is because
  $\action_2^{faa}$ cannot read from an earlier write (or the initial value of
  $0$) as that would imply an \fr dependency and an \hb cycle; and cannot read
  ($\rfe \subseteq \hb$) from a later write, as any later write is \hb after
  $\action_2^{faa}$ (via \rao and \ppo).

  There is some value $v_0 = \vr((\action_0^{faa}, \tagnarr))$ read by the first
  FAA operation. By well-formedness of \vr, \vw, and \rf, we have
  $\vr((\action_2^{faa}, \tagnarr)) = \vw((\action_0^{faa}, \tagnarr)) = v_0 +
  1$, \ie the following $\lklwa(x)$ gets the next ticket. More generally, it is
  clear every $\lklwa(x)$ gets a different ticket. We also have
  $\vw((\action_0^{faa}, \tagnlw)) = v_0$, \ie $p^{\threadt_1}_x$ is modified to
  contain $v_0$. Respectively $p^{\threadt_2}_x$ is modified to contain $v_0+1$.

  Let $\action_0^{r}$ be the third event of the implementation of $\action'_0$
  reading $p^{\threadt_1}_x$. We necessarily have
  $(\action_0^{faa}, \tagnlw) \arr{\rf} (\action_0^{r}, \tagcread)$. This is
  because $\action_0^{r}$ cannot read from the future (it would create an
  $\rf;\ippo$ cycle in \ib) and the second event $\rlwait(\wid)$ makes sure all
  previous modifications of $p^{\threadt_1}_x$ are available (ignoring the last
  one would be an $\fr;\hb$ cycle since $\mo_{p^{\threadt_1}_x} \subseteq \po$).
  Thus, in the implementation of $\action'_0$, the meta-variable $\code{v}$
  corresponds to the value $v_0$. More generally, in any implementation of
  $\lklwa(x)$, $\code{v}$ corresponds to the ticket obtained (\eg $v_0+1$ for
  $\action'_2$). So the last event $\action_0^{w}$ of the implementation of
  $\action'_0$ modifies $p^{\threadt_1}_x$ to $v_0+1$.

  The implementation of $\action'_1$ (running $\lklwr(x)$) has an operation
  \rlput (event $\action_1$) reading $p^{\threadt_1}_x$ to send to $x_r$. We
  necessarily have $(\action_0^{w}, \tagcwrite) \arr{\rf} (\action_1, \tagnlr)$,
  since the write is available ($(\tagcwrite, \tagnlr) \in \tagppo$) and later
  write on $p^{\threadt_1}_x$ from later \rlrfaa are not finished
  ($(\tagnlr, \tagnlw) \in \tagppo$ and $\nodefun{x_r} = \nodefun{x_a}$). Thus
  $\vw((\action_1, \tagnrw)) = \vr((\action_1, \tagnlr)) = v_0+1$. More
  generally, each $\lklwr(x)$ modifies $x_r$ to contain the next value after the
  ticket obtained by the previous $\lklwa(x)$ operation. Since each $\lklwa(x)$
  handles a different ticket, this is the only modification of $x_r$ to contain
  $v_0+1$.

  The penultimate event in the implementation of $\action'_2$ (causing the loop
  break) is of the form
  $\action_2 = (\threadt_2, \_, (\rlread, (p^{\threadt_2}_x), (v_0+1)))$
  returning a value of $v_0 + 1$, and by well-formedness
  $\vr((\action_2, \tagcread)) = v_0+1$. If we note
  $\action_2^{get} = (\threadt_2, \_, (\rlget, (p^{\threadt_2}_x, x_r, \wid),
  ()))$ the last \rlget event preceding $\action_2$ in the implementation, we
  clearly have $(\action_2^{get}, \tagnlw) \arr{\rf} (\action_2, \tagcread)$ (as
  previously, the intermediate \rlwait makes the write available), and
  $\vr((\action_2^{get}, \tagnrr)) = \vw((\action_2^{get}, \tagnlw)) =
  \vr((\action_2, \tagcread)) = v_0+1$.

  By well-formedness of \rf, we also have
  $(\action_1, \tagnrw) \arr{\rf} (\action_2^{get}, \tagnrr)$ from the only
  write of $v_0+1$ on $x_r$. Finally, we have
  $((\action_1, \tagnrw), (\action_2, \tagcread)) \in \rfe;\iso;\rfe \subseteq
  \hb$.
\end{proof}



\subsection{\strl Library}
\label{sec:strl-library}

\strllocsound*

\begin{proof}
  We assume an $\set{\rdmasync,\lkln}$-consistent execution
  $\exec = \tup{E, \po, \gettags, \so, \hb}$ which is abstracted via $f$ to
  $\tup{E',\po'}$ that uses (only) the \strl library, \ie
  $\absf{\strl}{\implstrl}{\tup{E, \po}}{\tup{E', \po'}}$ holds. We need to
  provide $\gettags'$, $\so'$, and
  $g : \tup{E',\po',\gettags'}.\SEvents \rightarrow \exec.\SEvents$ respecting
  some conditions. From $\tup{E',\po'}$, we simply take
  $\gettags' = \gettagsstrl$.

  Since $\exec$ is $\set{\rdmasync,\lkln}$-consistent, it means
  $(\ppo \cup \so\rst{\rdmasync} \cup \so\rst{\lkln}) \suq \hb$, $\hb$ is
  transitive and irreflexive, and the two restrictions of $\exec$ are
  respectively \rdmasync-consistent and \lkln-consistent.

  \rdmasync-consistency implies there is some well-formed \vr, \vw, \rf, \mo,
  \ro, and \rao such that $\ib$ is irreflexive,
  $\forall \action. \gettags\rst{\rdmasync}(\action) \in \gettagsrl(\action)$, and
  $\so\rst{\rdmasync} = \iso \cup \rfe \cup \pfget \cup \ro \cup \fr \cup \mo
  \cup \rao \cup ([\tagnrw[]]; \inv{\iso} ; \rao) \cup ([\Inst];\ib)$.

  \implstrl respects locks, as every operation is implemented to contain an
  $\lklna$ (first) and a $\lklnr$ operation (later) on the same lock location.
  As such $\tup{E\rst{\lkln}, \po\rst{\lkln}}$ respects locks. So
  \lkln-consistency implies $\gettags\rst{\lkln} = \gettagslkln$ and for each
  lock location $l$ there is a total order $\lo_l$ on
  $\setpred{\action}{\action \in E_l \land \methodfun{\action} = \lklna}$ for the
  acquiring of location $l$ such that:
  \begin{align*}
    \so\rst{\lkln} = & \setpred{\tup{\action, \tagnf[\nodefun{\loc(\action)}]}, \tup{\action, \tagnrw[\nodefun{\loc(\action)}]}}{\methodfun{\action} = \lklnr} \\
                     & \bigcup_{l \in \Loc} \setpred{\tup{\action_1, \tagnrw[\nodefun{\loc(\action_1)}]}, \tup{\action_2, \tagmf}} {(\action_1, \action_2) \in \inv{(\imm{\po_l})} ; \lo_l}
  \end{align*}

  We define $g$ to map to the first subevent of the implementation. For an event
  $\action' \in E'$, we choose $g(\action', \tagmf) = (\action, \tagmf)$ with
  $\action = (\threadt, \_, (\lklna, \_, \_)) \in \inv{f}(\action')$ the first
  event of the implementation. This $g$ clearly preserves \tagppo (first
  property of local soundness), as it maps subevents to subevents using the same
  stamp.

  Now we need to pick a suitable $\so'$ such that $g(\so') \suq \hb$ and
  $\tup{E',\po',\gettags',\so',\_}$ is \strl-consistent. \Ie, we need
  well-formed $\vr'$, $\vw'$, $\rf'$, and $\mo'$ such that $g(\po')$, $g(\rf')$,
  $g(\mo')$, and $g(\fr')$ are all included in $\hb$. We immediately have
  $g(\po') \in \ppo \subseteq \hb$ since $(\tagmf, \tagmf) \in \tagppo$. For the
  other relations, we can consider each location $x$ independently. Let us note
  $\node = \nodefun{x} = \nodefun{l_x}$. All the relevant operations acquire the
  lock $l_x$, as such we can use $\lo_{l_x}$ to order them.

  We define $\mo'_x$ and $\rf'_x$ as follows:
  $$\mo_x' \defeq \setpred{(\saction'_1, \saction'_2)}{\saction'_1, \saction'_2
    \in \exec'.\Write \ \land \ (g(\saction'_1), g(\saction'_2)) \in \lo_{l_x}}$$
  $$\rf_x' \defeq \setpred{(\saction'_1, \saction'_2)}{
    \begin{matrix}
      \saction'_1 \in \exec'.\Write \ \land \ \saction'_2 \in \exec'.\Read \ \land \ (g(\saction'_1), g(\saction'_2)) \in \lo_{l_x} \ \land \ \\
      \forall \saction'_0. (\saction'_1, \saction'_0) \in \mo'_x \implies (g(\saction'_0), g(\saction'_2)) \not\in \lo_{l_x}
    \end{matrix}
  }$$ with the slight abuse of notation of writing
  $((\action_1, \tagt_1), (\action_2, \tagt_2)) \in \lo_{l}$ to mean
  $(\action_1, \action_2) \in \lo_{l}$. \Ie, the location $x$ is modified in the
  order of the acquires, and reads read from the latest previous write.

  We define $\vr'$ and $\vw'$ from the values of $\vr$ and $\vw$ on the RDMA
  subevent (on $x$) of the implementation. \Eg, for $\action'$ running
  $\strlfaa(x,v)$, there is an event
  $\action = (\_, \_, (\rlrfaa, (\_, x, v, \_), ())) \in \inv{f}(\action')$ and
  we note $\vr'((\action', \tagmf)) = \vr((\action, \tagnarr))$ and
  $\vw'((\action', \tagmf)) = \vw((\action, \tagnrw))$.

  We can easily see that $g(\rf')$, $g(\mo')$, and $g(\fr')$ are all included in
  $\hb$ by design. This comes from the fact that
  $(\action_1, \action_2) \in \lo_{l_x}$ implies
  $((\action_1, \tagmf), (\action_2, \tagmf)) \in \ppo; \so \subseteq \hb$
  (since $E$ respects nodes and the release operation exists) and for $\fr'$
  because $\lo_{l_x}$ is total on the acquiring of the lock $l_x$ (Thus if
  $(g(\saction'_0), g(\saction'_2)) \not\in \lo_{l_x}$ and
  $\saction'_0 \neq \saction'_2$ then
  $(g(\saction'_2), g(\saction'_0)) \in \lo_{l_x}$).

  The remaining part of the proof is to show that $\vr'$, $\vw'$, and $\rf'$ are
  well-formed.

  Firstly, let's consider $\vw'$. For RMW operations, the value is correct from
  the well-formedness of $\vw$. For an event $\action'$ running $\strlw(x,v)$,
  the implementation contains
  $\rlwrite(p_x^\threadt,v) ; \rlput(x,p_x^\threadt,\_)$ (let's call them
  $\action_1$ and $\action_2$), and we need to show
  $\vw'((\action', \tagmf)) = v$. By definition
  $\vw'((\action', \tagmf)) = \vw((\action_2, \tagnrw)) = \vr((\action_2,
  \tagnlr))$ and $\vw((\action_1, \tagcwrite)) = v$. To conclude, it is enough
  to show $((\action_1, \tagcwrite), (\action_2, \tagnlr)) \in \rf$. Clearly
  $\action_1$ is finished when we run $\action_2$ (\ie,
  $((\action_2, \tagnlr), (\action_1, \tagcwrite)) \in \fr$ would create an \hb
  cycle). It is less obvious that $\action_2$ cannot read from a later
  $\rlwrite(p_x^\threadt,v')$ (let's call it event $\action_3$) of a later
  operation $\strlw(x,v')$ by the same thread. This is because this later
  operation would need to acquire the lock $l_x$. By the semantics of $\lkln$,
  this creates a synchronisation (since $\action_1$ is also towards node
  $\node$), and we have
  $((\action_2, \tagnlr) , (\action_3, \tagcwrite)) \in \ppo; \so\rst{\lkln} ;
  \ppo \subseteq \hb$. As such, reading from $\action_3$ would create an \hb
  cycle and is not possible.

  Secondly, for $\vr'$ and non-$\strlw$ operations, we need to show that the
  value returned (\ie by $\action_r$ running $\rlread(\temp_\threadt)$) is the value
  read by the RDMA operation $\action$ running $m(\temp_\threadt, x, \ldots, \wid)$
  with $m\in\set{\strlr,\strlcas,\strlfaa}$. For this, we simply show
  $((\action, \tagnlw), (\action_r, \tagcread)) \in \rf$. From the in-between
  $\rlwait$ operation, we have
  $((\action, \tagnlw), (\action_r, \tagcread)) \in \pfget; \ppo \subseteq \hb$.
  Thus, $\action_r$ cannot ignore $\action$ (\ie $\fr$ would create an \hb cycle),
  and cannot read from a later operations (it would create an \ib cycle).

  Finally, we are left with checking that $\rf'$ is well-formed. We need to show
  that whenever $(\saction'_1,\saction'_2) \in \rf'$, with
  $\saction'_i = (\action'_i, \tagmf)$, we have
  $\vw'(\saction'_1) = \vr'(\saction'_2)$. (Technically, also that
  $(\_, \saction'_2) \not\in \rf'$ implies $\vr'(\saction'_2) = 0$, but this
  follows from a similar reasoning.) Let $\action_i$ be the RDMA operation in
  the implementation of $\action'_i$, by definition we have
  $\vw'(\saction'_1) = \vw((\action_1, \tagnrw))$ and
  $\vr'(\saction'_2) = \vr((\action_2, \tagt_2))$ (with
  $\tagt_2 \in \set{\tagnarr,\tagnrr}$ depending on the case). Our sufficient
  goal is then to show that we necessarily have
  $((\action_1, \tagnrw), (\action_2, \tagt_2)) \in \rf$. By definition of
  $\rf'$, we have $(g(\saction'_1), g(\saction'_2)) \in \lo_{l_x}$ as well as
  $\forall \saction'_0 \in \exec'.\Write. \ \left( (\saction'_1, \saction'_0)
    \in \mo'_x \implies (g(\saction'_0), g(\saction'_2)) \not\in \lo_{l_x}
  \right)$.

  The first point implies
  $((\action_1, \tagnrw), (\action_2, \tagt_2)) \in \ppo;\so\rst{\lkln};\ppo
  \subseteq \hb$ by the semantics of locks. This makes
  $((\action_2, \tagt_2), (\action_1, \tagnrw)) \in \fr$ impossible (\hb cycle),
  and $\action_2$ reads from either $\action_1$ or a later write: there exists
  an RDMA operation $\action_3$ (in the implementation of some $\action'_3$)
  such that $((\action_3, \tagnrw), (\action_2, \tagt_2)) \in \rf$ with
  $(\action_1, \tagnrw) \arr{\mo^*} (\action_3, \tagnrw)$. Note that
  $\action_2 \neq \action_3$ or it would create a $\rfe;\iso$ cycle in $\hb$;
  \ie an event cannot read from itself. Thus we need to show
  $\action_1 = \action_3$, and by contradiction let us assume
  $(\action_1, \tagnrw) \arr{\mo} (\action_3, \tagnrw)$.

  We then show $(\action'_1, \tagmf) \arr{\mo'} (\action'_3, \tagmf)$. Since
  $\lo_{l_x}$ is a total order, we have either
  $(g(\saction'_1), g(\saction'_3)) \in \lo_{l_x}$ or
  $(g(\saction'_3), g(\saction'_1)) \in \lo_{l_x}$. To show the first, we assume
  the second by contradiction, \ie that $\action'_3$ acquires first. Given the
  implementation $\implstrl$, there is a $\lklnr(x)$ event $\action_3^r$ such
  that $g(\action'_3) \arr{\po} \action_3 \arr{\po} \action_3^r$.
  Thus from the semantics of \lkln we have an \hb cycle
  $(\action_3, \tagnrw) \arr{\ppo} (\action_3^r, \tagnrw) \arr{\so\rst{\lkln}}
  g(\action'_1) \arr{\ppo} (\action_1, \tagnrw) \arr{\mo} (\action_3, \tagnrw)$
  providing a contradiction, and we necessarily have
  $(\action'_1, \tagmf) \arr{\mo'} (\action'_3, \tagmf)$.

  Now, from the definition of $\rf'$, the fact $\lo_{l_x}$ is a total order, and
  that $\action_3 \neq \action_2$, we have
  $(g(\saction'_2), g(\saction'_3)) \in \lo_{l_x}$. Similarly to previously,
  this implies
  $((\action_2, \tagt_2), (\action_3, \tagnrw)) \in
  \ppo;\ppo;\so\rst{\lkln};\ppo \subseteq \hb$ by the semantics of locks (using
  both the \tagnf and \tagnlw stamps of the release). This contradicts
  $((\action_3, \tagnrw), (\action_2, \tagt_2)) \in \rfe \subseteq \hb$. Thus
  $\action_1 = \action_3$,
  $((\action_1, \tagnrw), (\action_2, \tagt_2)) \in \rf$, and $\rf'$ is
  well-formed.

\end{proof}




%% file: rdmatsormw.tex

\section{Declarative Semantics of \rdmatsormw à la \framework}
\label{sec:rdmatsormw}

In this appendix, we first (\cref{sec:rdmatsormw-sem}) present the declarative
semantics of \rdmatsormw in a format similar to that of \rdmatso
in~\cite{POPL-LOCO}, but extended with remote RMW operations similarly to the
semantics of \rdmawaitrmw given in~\cref{sec:rdmasync}. It is slightly
different from the one in \cref{sec:rmw}, as we use the stamps and subevents
system of \framework.

We then (\cref{sec:rdmatsormw-implem}) provide a definition of the
implementation of \rdmawaitrmw into \rdmatsormw. Finally
(\cref{sec:rdmatsormw-soudness}), we give a proof of the soundness of this
implementation, similarly to~\cite{POPL-LOCO}.



\subsection{Semantics}
\label{sec:rdmatsormw-sem}

Our definition of \rdmatsormw is closer to an independent language than a
library. We do not need a relation \hb to represent the potential rest of the
program, as a program \emph{cannot} combine instructions from \rdmatsormw and
other libraries presented in this paper.

We use the following 13 methods:
\begin{align*}
m(\vect{v}) & ::=
\rtsowrite(x,v)
\mid \rtsoread(x)
\mid \rtsocas(x,v_1,v_2)
\mid \rtsomf() \\
& \quad
\mid \rtsoget(x,y)
\mid \rtsoput(x,y)
\mid \rtsopoll(\node)
\mid \rtsorf(\node) \\
& \quad
\mid \rtsorcas(x,y,v_1,v_2)
\mid \rtsorfaa(x,y,v) \\
& \quad
\mid \rtsoadd(x,v)
\mid \rtsorm(x,v)
\mid \rtsoempty(x)
\end{align*}

\begin{multicols}{2}
\begin{itemize}
\item $\rtsowrite : \Loc \times \Val \rightarrow ()$
\item $\rtsoread : \Loc \rightarrow \Val$
\item $\rtsocas : \Loc \times \Val \times \Val \rightarrow \Val$
\item $\rtsomf : () \rightarrow ()$
\item $\rtsoget : \Loc \times \Loc \rightarrow \Val$
\item $\rtsoput : \Loc \times \Loc \rightarrow \Val$
\item $\rtsopoll : \Nodes \rightarrow \Val$
\item $\rtsorf : \Nodes \rightarrow ()$
\item $\rtsorcas : \Loc \times \Loc \times \Val^2 \rightarrow \Val$
\item $\rtsorfaa : \Loc \times \Loc \times \Val \rightarrow \Val$
\item $\rtsoadd : \Loc \times \Val \rightarrow ()$
\item $\rtsorm : \Loc \times \Val \rightarrow ()$
\item $\rtsoempty : \Loc \rightarrow \mathbb{B}$
\end{itemize}
\end{multicols}

This version is based on~\cite{POPL-LOCO} extended with remote RMW. Compared to
\rdmatso from~\cite{OOPSLA-24}, we slightly extend the language so that RDMA
operations return an arbitrary unique identifier, and polling also returns the
same identifier of the operation being polled. In addition, we also assume basic
set operations $\rtsoadd$, $\rtsorm$, and $\rtsoempty$ to store these new
identifiers, where the locations used for sets do not overlap with locations
used for other operations.

\paragraph{Consistency predicate} An execution of an \rdmatsormw program is of
the form $\exec = \tup{E, \po, \gettags, \so}$. Note that
$\hb = (\ppo \cup \so)^+$ does not have the flexibility of containing additional
external constraints.

We say that a stamping function $\gettagsrtso$ is valid if:
\begin{itemize}
\item Polls have stamp \tagwait:
  $\gettagsrtso((\_, \_, (\rtsopoll, \_, \_))) = \set{\tagwait}$.
\item Auxiliary set operations have stamp \tagmf:
  $\gettagsrtso((\_, \_, (\rtsoadd, \_, \_))) =$ \\
  $\gettagsrtso((\_, \_, (\rtsorm, \_, \_))) = \gettagsrtso((\_, \_,
  (\rtsoempty, \_, \_))) = \set{\tagmf}$.
\item Other events follow the validity constraints of \rdmawaitrmw (\cf
  \Cref{sec:rdmasync}). E.g., events calling \rtsowrite have stamp \tagcwrite,
  while events calling \rtsoget towards node $\node$ have stamps \tagnrr and
  \tagnlw. We also define \loc on subevents similarly to \rdmawaitrmw.
\end{itemize}

We mark set operations with \tagmf to simplify the consistency conditions, as we
do not want to explicitly integrate them in the read ($\Read$) and write
($\Write$) subevents.

Given $\exec = \tup{E, \po, \gettagsrtso, \so}$, we say that \vr, \vw, \rf, \mo, \ro, \pf, and \rao are well-formed if:

\begin{itemize}
\item \vr, \vw, \rf, \mo, \ro, and \rao are well-formed, as in \rdmawaitrmw;
\item Let
  $P_\node \defeq \setpred{(\action, \tagwait)}{\action = (\_, \_, (\rtsopoll,
    (\node), \_)) \in E}$ be the set of poll (sub)events towards node
  $\node$. Let
  $C_\node \defeq \setpred{(\action, \tagnlw)}{\methodfun{\action} \in
    \set{\rtsoget,\rtsorcas,\rtsorfaa}} \cup \setpred{(\action,
    \tagnrw)}{\methodfun{\action} = \rtsoput}$ be the set of (final writes of) remote
  operations towards node $\node$ that need polling. Note that, for remote RMW
  operations, polling only synchronises with the \tagnlw part and not with the
  (potential) \tagnrw part.

  Then $\pf \suq \bigcup_{\node \in \Nodes} C_\node \times P_\node$ is the
  \emph{polls-from} relation, relating earlier NIC writes to later polls.
  Moreover:
  \begin{itemize}
  \item $\pf \suq \po$ (we can only poll previous operations of the same
    thread);
  \item $\pf$ is functional on its domain (every NIC write can be polled at most
    once);
  \item $\pf$ is total and functional on its range (every $\rtsopoll$ polls from
    exactly one NIC write);
  \item $\rtsopoll$ events poll-from the oldest non-polled remote operation towards the given node:\\
    for each node $\node$, if $w_1,w_2 \in C_\node$ and
    $w_1 \arr{\po} w_2 \arr{\pf} p_2$, then there exists $p_1$ such that
    $w_1 \arr{\pf} p_1 \arr{\po} p_2$;
  \item and a $\rtsopoll$ returns the unique identifier of the polled
    operation:\\ if
    $((\_, \_, (\_, \_, v_1)), \_) \arr{\pf} ((\_, \_, (\rtsopoll, \_, v_2)),
    \tagwait)$ then $v_1 = v_2$.
  \end{itemize}
\end{itemize}

We use the derived relations $\fr$, $\fri$, $\rfe$, $\rfi$, $\ippo$, and $\iso$
as defined for \rdmawaitrmw. We can then define $\ib$ as follows:
$$\ib \defeq (\ippo \ \cup \ \iso \ \cup \ \rf \ \cup \ \pf \ \cup \ \ro
\ \cup \ \fri \cup \set{(\action, \tagnrw), (\action, \tagnlw)})^+$$

The last new component states that, for remote RMW operations, the remote
write part \emph{starts} before the local write part. As mentioned previously,
this does not imply that they finish in order, and this component is not
included in \so. Since it does not prevent any behaviour, we remove this
component in the definition \rdmawaitrmw (\cref{sec:rdmasync}), but we keep it
here as it simplifies the soundness proof (\cref{thm:wait-to-tso}) and the
equivalence proof with the operational semantics~(\Cref{sec:annot-sem-a2d}).

\begin{definition}[\rdmatsormw-consistency]
  \label{def:rtsormw-consistency}
  $\exec = \tup{E, \po, \gettags, \so}$ is \rdmatsormw-consistent if:
  \begin{itemize}
  \item $(\ppo \cup \so)^+$ is irreflexive;
  \item $\tup{E, \po}$ respects nodes (as in \rdmawaitrmw);
  \item $\gettags$ is valid;
  \item there exists well-formed \vr, \vw, \rf, \mo, \ro, \pf, and \rao such that
    $\ib$ is irreflexive and \\
    $\so = \iso \cup \rfe \cup [\tagnlw[]];\pf \cup \ro \cup \fr \cup
    \mo \cup \rao \cup ([\tagnrw[]]; \inv{\iso} ; \rao) \cup ([\Inst];\ib)$;
  \item identifiers for RDMA operations are unique: if $\action_1$ and
    $\action_2$ are both of the form $(\_, \_, (m, \_, v))$ with
    $m \in \set{\rtsoput,\rtsoget,\rtsorcas,\rtsorfaa}$ then
    $\action_1 = \action_2$;
  \item and the set operations are (per-thread) sound: if \rtsoempty returns
    $\true$, then every value added to the set was subsequently removed. I.e.,
    if $\action_1 = (\threadt, \_, (\rtsoadd, (x,v), \_)$,
    $\action_3 = (\threadt, \_, (\rtsoempty, (x), \true))$, and
    $\action_1 \arr{\po} \action_3$, then there exists
    $\action_2 = (\threadt, \_, (\rtsorm, (x,v), \_)$ such that
    $\action_1 \arr{\po} \action_2 \arr{\po} \action_3$.
  \end{itemize}
\end{definition}



\subsection{Implementation Function}
\label{sec:rdmatsormw-implem}

In \Cref{fig:implwait} we define the implementation $\implwait$ from a full
program using only the \rdmawaitrmw library into a program using only
\rdmatsormw. We assume threads use disjoint work identifiers $\wid \in \Wid$,
otherwise it is straightforward to rename them.

For each location $x$ of \rdmawaitrmw, we also use a location $x$ for \rdmatsormw. For
each work identifier $\wid$ of \rdmawaitrmw, we use new \rdmatsormw locations
$\set{\wid^1,\ldots,\wid^N}$ where $N \defeq \cardinal{\Nodes}$ is the number of
nodes. Each location $\wid^\node$ is used as a set containing the identifiers of
ongoing operations towards node $\node$.

Most \rdmawaitrmw operations ($\rlwrite$, $\rlread$, $\rlcas$, $\rlmf$, and
$\rlrf$) are directly translated into their \rdmatsormw counterparts. An
operation $\rlget(x,y,\wid)$ towards node $\node$ is translated into a similar
$\rtsoget(x,y)$ whose output is added to the set $\wid^\node$; We proceed
similarly for other RDMA operations. Finally, a $\rlwait(\wid)$ operation needs
to poll until all relevant operations are finished, \ie the sets
$\set{\wid^1,\ldots,\wid^N}$ are all empty. Whenever we poll, we obtain the
identifier of a finished operation, and we remove it from \emph{all} sets where
it might be held. We remove it from $\wid^\node$ but also from any other set
$\wid_k^\node$ tracking a different group of operations, as otherwise a later
call to $\rlwait(\wid_k)$ would hang and never return.

\begin{figure}
  $$\texttt{For a thread } \threadt \texttt{ using work identifiers } \set{\wid_1,\ldots,\wid_K} \texttt{:}$$
\begin{minipage}{.48\textwidth}
\begin{align*}
  & \implwait(\threadt, \rlwrite, (x,v)) \defeq \rtsowrite(x,v) \\
  & \implwait(\threadt, \rlread, (x)) \defeq \rtsoread(x) \\
  & \implwait(\threadt, \rlcas, (x,v_1,v_2)) \defeq \rtsocas(x,v_1,v_2) \\
  & \implwait(\threadt, \rlmf, ()) \defeq \rtsomf() \\
  & \implwait(\threadt, \rlrf, (\node)) \defeq \rtsorf(\node) \\
  & \\
  & \implwait(\threadt, \rlget, (x,y,\wid)) \defeq \\
  & \LetC{v}{\rtsoget(x,y)}{\rtsoadd(\wid^{\nodefun{y}}, v)}
  & \\
  & \implwait(\threadt, \rlput, (x,y,\wid)) \defeq \\
  & \LetC{v}{\rtsoput(x,y)}{\rtsoadd(\wid^{\nodefun{x}}, v)}
\end{align*}
\end{minipage}
\qquad
\begin{minipage}{.45\textwidth}
\begin{align*}
  & \implwait(\threadt, \rlwait, (\wid)) \defeq \\
  & \texttt{For } \node \texttt{ in } 1,\ldots,N \texttt{ do \{} \\
  & \quad \texttt{While } (\rtsoempty(\wid^\node) \neq \true) \texttt { do \{} \\
  & \qquad \LetC{v}{\rtsopoll(\node)}{} \\
  & \qquad \texttt{For } k \texttt{ in } 1,\ldots,K \texttt{ do \{} \\
  & \qquad \quad \rtsorm(\wid_k^\node, v) \quad \} \ \} \ \} \\
  & \\
  & \implwait(\threadt, \rlrcas, (x,y,v_1,v_2,\wid)) \defeq \\
  & \LetC{v}{\rtsorcas(x,y,v_1,v_2)}{\rtsoadd(\wid^{\nodefun{y}}, v)}
  & \\
  & \implwait(\threadt, \rlrfaa, (x,y,v',\wid)) \defeq \\
  & \LetC{v}{\rtsorfaa(x,y,v')}{\rtsoadd(\wid^{\nodefun{y}}, v)}
\end{align*}
\end{minipage}
\caption{Implementation \implwait of \rdmawaitrmw into \rdmatsormw}
\label{fig:implwait}
\end{figure}



\subsection{Soundness}
\label{sec:rdmatsormw-soudness}

We do not prove that the implementation above is locally sound as it does not
apply for this case. Instead, we assume a full program using only the
\rdmawaitrmw library and compile it into \rdmatsormw.

\begin{theorem}
  \label{thm:wait-to-tso}
  Let $\progps$ be a program using only the $\rdmawaitrmw$ library. Then we have \\
  $\outcome_{\rdmatsormw}(\impliw{\progps}) \suq
  \outcome_{\{\rdmawaitrmw\}}(\progps)$, where:
  \begin{align*}
   \outcome_{\{\rdmawaitrmw\}}(\progps) &= \setpred{\vect{v}}{ \exists \tup{E, \po,
      \gettags, \so, \hb} \ \{\rdmawaitrmw\}\text{-consistent.\ }
    \tup{\vect{v},\tup{E,\po}} \in \interp{\progps}} \\
   \outcome_{\rdmatsormw}(\impliw{\progps}) &= \setpred{\vect{v}}{ \exists \tup{E,
      \po, \gettags, \so} \ \rdmatsormw\text{-consistent.\ }
    \tup{\vect{v},\tup{E,\po}} \in \interp{\impliw{\progps}} }
  \end{align*}
\end{theorem}

\begin{proof}
  By definition, we are given $\exec = \tup{E, \po, \gettags, \so}$
  $\rdmatsormw$-consistent (\Cref{def:rtsormw-consistency}) such that
  $\tup{\vect{v},\tup{E,\po}} \in \interp{\impliw{\progps}}$. Among others, it
  means $\tup{E, \po}$ respects nodes and there exists well-formed $\vr$, $\vw$,
  $\rf$, $\mo$, $\ro$, $\pf$, and $\rao$ such that $\ib$ is irreflexive,
  $\gettags$ is valid,
  $\so = \iso \cup \rfe \cup [\cup_\node \ \tagnlw];\pf \cup \ro \cup \fr \cup
  \mo \cup \rao \cup ([\tagnrw[]]; \inv{\iso} ; \rao) \cup ([\Inst];\ib)$, and
  $\hb \defeq (\ppo \cup \so)^+$ is irreflexive.

  From \Cref{lem:find-abstraction}, since $\progps$ uses only $\rdmawaitrmw$,
  there is $E',\po',f$ such that
  $\tup{\vect{v}, \tup{E', \po'}} \in \interp{\progps}$ and
  $\absf{\rdmawaitrmw}{\implwait}{\tup{E, \po}}{\tup{E', \po'}}$. Note that this
  clearly implies $\tup{E, \po}$ also respects nodes, as the implementation
  $\implwait$ keeps the same locations. Our objective is to find $\gettags'$,
  $\so'$, and $\hb'$ such that $\exec' = \tup{E', \po', \gettags', \so', \hb'}$
  is $\{\rdmawaitrmw\}$-consistent (\Cref{def:lambda-cons,def:rl-consistency}).
  To choose a valid function $\gettags'$, most values are forced. For remote
  compare-and-swap, we make the same choice as $\gettags$. \Ie for each \rlrcas
  we assert it succeeds iff the corresponding \rtsorcas in its implementation
  succeeds. We will also pick $\hb' \defeq (\ppo' \cup \so')^+$ since there is
  no external constraints. Thus, we only need to carefully pick $\so'$ and show
  it works.

  While our objective is not exactly local soundness (\Cref{def:locally-sound}),
  we still use a concretisation function
  $g : \tup{E',\po',\gettags'}.\SEvents \rightarrow \exec.\SEvents$ to then
  define $\so'$.

  \begin{itemize}
  \item For $\action' = (\threadt, \_, (\rlwrite,(x,v), ()))$, from the
    definition of the implementation $\implwait$ and the abstraction $f$, there
    is some event
    $\action = (\threadt, \_, (\rtsowrite, (x,v), ())) \in \inv{f}(\action')$.
    We define $g(\action', \tagcwrite) = (\action, \tagcwrite)$. For events
    calling $\rlread$, $\rlcas$, $\rlmf$, and $\rlrf$, we proceed similarly and
    let $g$ map each subevent to their counterpart in the implementation.
  \item For $\action' = (\threadt, \_, (\rlget,(x,y,\wid), ()))$, there
    is some event
    $\action = (\threadt, \_, (\rtsoget, (x,y), (v))) \in \inv{f}(\action')$.
    We define $g(\action', \tagnrr[\node(y)]) = (\action, \tagnrr[\node(y)])$
    and $g(\action', \tagnlw[\node(y)]) = (\action, \tagnlw[\node(y)])$. We
    proceed similarly for \rlput, \rlrcas, and \rlrfaa events.
  \item Finally for $\action' = (\threadt, \_, (\rlwait,(\wid), ()))$, there is
    in $\inv{f}(\action')$ some last event (in $\po$ order) of the form
    $\action = (\threadt, \_, (\rtsoempty, (\wid^N), \true))$ confirming the set
    $\wid^N$ tracking operations towards the last node $N$ is empty. We define
    $g(\action', \tagwait) = (\action, \tagmf)$.
  \end{itemize}

  We can see that $g(\tup{\action', \tagt'}) = \tup{\action, \tagt}$ implies
  that $f(\action) = \action'$ and that $\tagt$ is more restrictive than
  $\tagt'$.

  Each subevent in $\exec'.\Read$ (resp. $\exec'.\Write$) is mapped through $g$
  to a subevent in $\exec.\Read$ (resp. $\exec.\Write$) using the same stamp and
  location. Thus it is straightforward to define $\vr'$, $\vw'$, $\rf'$, $\mo'$,
  $\ro'$, and $\rao'$ by relying on their counterparts in $\exec$. \Eg
  $\vr'(\saction') \defeq \vr(g(\saction'))$ and
  $\rf' \defeq \setpred{(\saction_1',\saction_2')}{(g(\saction_1'),
    g(\saction_2')) \in \rf}$. The well-formedness of $\vr$, $\vw$, $\rf$,
  $\mo$, $\ro$, and $\rao$ trivially implies that of $\vr'$, $\vw'$, $\rf'$,
  $\mo'$, $\ro'$, and $\rao'$. From this, we can define all the expected derived
  relations, including $\pfget'$, $\pfput'$, and
  $\ib' \defeq (\ippo' \cup \iso' \cup \rf' \cup \pfget' \cup \pfput' \cup \ro'
  \cup \fri')^+$. We then define
  $\so' \defeq \iso' \cup \rfe' \cup \pfget' \cup \ro' \cup \fr' \cup \mo' \cup
  \rao' \cup ([\tagnrw[]]; \inv{\iso'} ; \rao') \cup ([\Inst];\ib')$, and as
  previously mentioned $\hb' \defeq (\ppo' \cup \so')^+$.

  To show $\{\rdmawaitrmw\}$-consistency, we are left to prove that $\ib'$ and
  $\hb'$ are irreflexive. For this, it is enough to show that $g(\ib') \suq \ib$
  and $g(\hb') \suq \hb \defeq (\ppo \cup \so)^+$ since we know both $\ib$ and
  $\hb$ to be irreflexive.

  For all subevent $\saction'$, $g(\saction')$ has a more restrictive stamp than
  $\saction'$ (in most cases it is the same stamp, but for \code{Wait} the stamp
  \tagmf is more restrictive than \tagwait); this implies that
  $g(\ppo') \subseteq \ppo$. Then, by definition, it is trivial to check that
  $g(\rf') \subseteq \rf$, $g(\mo') \subseteq \mo$, $g(\ro') \subseteq \ro$,
  $g(\ippo') \subseteq \ippo$, $g(\rfe') \subseteq \rfe$,
  $g(\iso') \subseteq \iso$, $g(\fr') \subseteq \fr$, and
  $g(\fri') \subseteq \fri$.

  To finish the proof, we need the following crucial pieces:
  $g(\pfput') \suq \ib$, $g(\pfget') \suq \ib$, and $g(\pfget') \suq \hb$. In
  fact, it is enough to show that $g(\pfput') \suq \ib^?;\pf;\ppo^+$ and
  $g(\pfget') \suq \pf;\ppo^+$. This is because $\pf;\ppo^+ \suq \ib$,
  $[\tagnlw[]];\pf;\ppo^+ \suq \hb$, and the domain of $g(\pfget')$ is included
  in $\cup_\node \ \exec.\tagnlw$ by definition.

  Let us start with $\pfget'$, assuming
  $((\action'_1, \tagnlw),(\action'_2, \tagwait)) \in \pfget'$. By definition
  they are of the form $\action'_1 = (\threadt, \_, (\_, (\ldots, \wid), ()))$
  and $\action'_2 = (\threadt, \_, (\rlwait, (\wid), ()))$, for some $\threadt$,
  $\wid$, and $\methodfun{\action'_1} \in \set{\rlget,\rlrcas,\rlrfaa}$, with
  $(\action'_1, \action'_2) \in \po'$ and $\node$ the remote node of this
  operation.
  By definition of the implementation and the abstraction, $\inv{f}(\action'_1)$
  contains two events $\action_1 = (\threadt, \_, (m, (\ldots), (v)))$ with a
  similar method $m \in \set{\rtsoget,\rtsorcas,\rtsorfaa}$ and
  $\action_a = (\threadt, \_, (\rtsoadd, (\wid^\node, v), ()))$, with
  $\action_1 \arr{\po} \action_a$. Meanwhile $\inv{f}(\action'_2)$ contains a
  last event $\action_2 = (\threadt, \_, (\rtsoempty, (\wid^N), \true))$ and an
  earlier event $\action_3 = (\threadt, \_, (\rtsoempty, (\wid^\node), \true))$,
  with $\action_3 \arr{\po^*} \action_2$, confirming operations towards $\node$
  are done (if $\node = N$ then $\action_2 = \action_3$).

  Since $f(\action_a) = \action'_1 \arr{\po'} \action'_2 = f(\action_3)$ and $f$
  is an abstraction, we have $\action_a \arr{\po} \action_3$, \ie the value $v$
  is added to $\wid^\node$ before the moment $\wid^\node$ is confirmed empty. By
  consistency (\Cref{def:rtsormw-consistency}), there is an in-between event
  $\action_4 = (\threadt, \_, (\rtsorm, (\wid^\node, v), ()))$ that removes this
  value, with $\action_a \arr{\po} \action_4 \arr{\po} \action_3$. From the
  definition of the implementation, such an event $\action_4$ is immediately
  preceded (with maybe other \rtsorm in-between) by an event
  $\action_p = (\threadt, \_, (\rtsopoll, (\node), (v)))$.
  Now we argue that we necessarily have
  $((\action_1, \tagnlw), (\action_p, \tagwait)) \in \pf$. From the
  well-formedness of \pf, we know that $(\action_p, \tagwait)$ has a preimage
  (\pf is total and functional on its range) and that this preimage outputs the
  value $v$. By consistency (\Cref{def:rtsormw-consistency}), $\action_1$, with
  $\methodfun{\action_1} \in \set{\rtsoget,\rtsorcas,\rtsorfaa}$, is the only RDMA
  operation with output $v$. Thus $(\action_1, \tagnlw)$ is the preimage of
  $(\action_p, \tagwait)$ by $\pf$.

  Finally we have
  $g(\action'_1, \tagnlw) = (\action_1, \tagnlw) \arr{\pf} (\action_p, \tagwait)
  \arr{\ppo} (\action_4, \tagmf) \arr{\ppo} (\action_3, \tagmf) \arr{\ppo^*}
  (\action_2, \tagmf) = g(\action'_2, \tagwait)$, which shows
  $g(\pfget') \suq \pf;\ppo^+$.

  For $\pfput'$ we have two cases. First, for a \rlput operation $\action'_1$,
  having $((\action'_1, \tagnrw),(\action'_2, \tagwait)) \in \pfput'$ similarly
  implies $g(\action'_1, \tagnrw) \arr{\pf;\ppo^+} g(\action'_2, \tagwait)$ for
  the same reasons. Second, for a successful remote RMW operation $\action'_1$,
  having $((\action'_1, \tagnrw),(\action'_2, \tagwait)) \in \pfput'$, with
  $g(\action'_1, \tagnrw) = (\action_1, \tagnrw)$ instead implies
  $(\action_1, \tagnlw) \arr{\pf;\ppo^+} g(\action'_2, \tagwait)$ for the same
  reasons. This is because the semantics of polls synchronises with the
  \emph{local write} part of the operation, not with the remote write part.
  However, we do have $g(\action'_1, \tagnrw) \arr{\ib} (\action_1, \tagnlw)$
  from the last component of \ib, and thus $g(\pfget') \suq \ib^?;\pf;\ppo^+$.

  Thus $\ib'$ and $\hb'$ are irreflexive, and $\exec'$ is
  $\{\rdmawaitrmw\}$-consistent.
  
\end{proof}




%% file: rmws.tex

\section{The \rdmatsormw Memory Model}
\label{sec:rmw}

In this section, we present an operational~(\cref{sec:op-sem}) and declarative
model~(\cref{sec:decl-sem}) for \rdmatsormw in the format of \rdmatso as defined
in~\cite{OOPSLA-24}, as well as an extension of their equivalence
proof~(\cref{sec:annot-sem-rules} onwards).

The declarative format used in \cref{sec:decl-sem} (based on~\cite{OOPSLA-24})
is slightly different from the one of \ref{sec:rdmatsormw-sem} (based
on~\cite{POPL-LOCO}), but they represent the same semantics.

\subsection{Operational Semantics}
\label{sec:op-sem}

\paragraph{Nodes and Threads}
We write $\Nodes = \set{1..N}$ for the set of node identifiers,
and $\Threads$ for the set of thread identifiers.
We write $\node$ (resp.\ $\threadt$) to range over nodes (resp.\ threads),
and given some node $\node$ we write $\neqn{\node}$ to range over the set of all other nodes $\Nodes \setminus \set{\node}$.
Each thread runs on a particular node, so we write $\node(\threadt)$ for the node the thread belongs to.

Note that the semantics of~\cite{OOPSLA-24} assumes that the remote node
$\neqn{\node}$ of an operation is different from the local node $\node$ (\ie
they ignore loopback). As we extend their operational model, we keep their
notations. However, as shown in~\cite{POPL-LOCO}, loopbacks are possible and
follow exactly the same semantics.

\paragraph{Memory}
Although all nodes can directly access all memory locations,
whether an operation is towards local or remote memory is pivotal to our semantics,
so we are always careful to note the node to which a memory location belongs.
We write $\Var_{\node}$ for the set of locations local to node $n$,
and $\Var = \biguplus_{\node}\Var_{\node}$ for the set of all locations.
We use $\Var_{\neqn{\node}} = \Var \setminus \Var_{\node}$ and write $x^{n}, y^{n}, z^{n}$ for values in $\Var_{\node}$, respectively $x^{\neqn{n}}, y^{\neqn{n}}, z^{\neqn{n}}$ for $\Var_{\neqn{\node}}$.
When the node in question is sufficiently clear,
we elide the superscript and instead simply write $x$ or $\neqn{x}$ for local or remote locations respectively.


\paragraph{Values and Expressions}
The language of expressions is standard and elided.
We write $v \in \Vals$ for values, with $\mathbb{N} \subseteq \Vals$,
and $e \in \Exps$ for expressions.
We write $\fv{e}$ for the set of memory locations referenced in $e$,
$e[v/x]$ for the expression obtained by substituting all references to location $x$ in $e$ with value $v$,
and $\db{e}$ for the evaluation of $e$ given it is \emph{closed},
that is, $\fv{e} = \emptyset$.
We use $e^{n}$ for expressions where $\fv{e^{n}} \subseteq \Var_{n}$

\paragraph{Commands and Programs}
Commands are described by the \comm{} grammar below.
CPU operations (\LComms) are assignment, assumption of the value of a location, memory fence, compare-and-swap, and poll, which awaits the earliest completion notification of a remote operations towards $\neqn{\node}$.

RDMA operations (\RComms) are either a `get' of the form $x \assign \neqn{y}$ which reads a remote location $\neqn{y}$ and writes its value to local location $x$,
a `put' ($\neqn{y} \assign x$) which does the reverse,
`remote-CAS' (resp.\ `remote-FAA') which executes a \emph{remote} compare-and-swap (resp.\ fetch-and-add),
and `remote fence' which ensures all prior RDMA operations towards $\neqn{n}$ complete before any later RDMA operations towards $\neqn{n}$ execute.
We note rRMW to cover both kind of remote read-modify-write operations, \ie \rcas and \rfaa.

Primitive operations (\PComms) are CPU or RDMA operations,
and commands (\Comms) are the no-op, primitive operations,
sequential composition (executes the first command, then the second),
non-deterministic choice (executes one command or the other),
and non-deterministic loop (executes the command some finite, possibly zero number of times).

A program $\prog$ consists of a map from threads to commands,
such that each $\threadt \in \Threads$ is mapped to a command on $\node(\threadt)$.
\[
\small
\begin{array}{@{}r @{\hspace{2pt}} l@{}}
\Comms \ni \comm ::= &
\cskip
\mid \pcomm
\mid \comm_1;\comm_2
\mid \comm_1 \choice \comm_2
\mid {\comm}^\kstar
\hspace{45pt}
\PComms \ni \pcomm ::= \lcomm \mid \rcomm
\vspace{-2pt}\ms\vspace{-2pt}
\LComms \!\ni\! \lcomm ::=
	& \x \!\assign\! \abexp^\node
	\mid \cassume(\x=\val)
	\mid  \cassume(\x\neq\val)
	\mid \cmfence
	\mid \x \!\assign\! \clcas(\y,\abexp_1,\abexp_2)
	\mid  \poll(\neqn\node)  \ms
\RComms \!\ni\! \rcomm ::=
	& \x \assign {\neqn{\y}}
	\mid \neqn{\y}  \assign \x
    \mid \x \assign \rcas(\neqn{y}, e_{1}, e_{2})
    \mid \x \assign \rfaa(\neqn{y}, e)
	\mid \cnfence{\neqn\node}
\end{array}
\]

\paragraph{Store Buffers}
To permit the weak behaviours of \TSO{} (\ie write-read reordering), we assign
each thread a \emph{store buffer} $\TSOmap(\threadt)$, which is a FIFO queue
containing pending writes to memory by that thread. When a thread performs a CPU
read, it reads the most recent entry for that location in its store buffer, if
there is one, instead of the value in memory. The write at the head of the queue
may be flushed to memory at any time, and $\mfence$ and $\CAS$ wait until the
store buffer is empty before executing.

\paragraph{Queue Pairs}
We follow the \emph{simplified} operational model described in \cite{OOPSLA-24},
and therefore consider a queue-pair structure comprising three FIFO queues:
\pipe, which contains pending or in-progress RDMA operations;
\wbR, the remote write buffer, which contains pending writes to the memory of the remote node;
and \wbL, the local write buffer, which contains pending writes to the memory of the local node.
The structure is shown in \cref{fig:sqpair_structure}.
Notice that under this simplified model,
the transition between local and remote node in \pipe{} is continuous --
we do not explicitly model the transition between local (yellow) and remote (pink) sides.

\input{simple-queue-pairs}

\paragraph{Remote Atomics}
To model the behaviour of RDMA atomic operations,
we assign each node a \emph{remote atomic lock} $\Atm(\node)$,
which is a boolean indicating whether an RDMA atomic is currently in progress \emph{towards} that node.

\paragraph{Transitions of the Operational Semantics}
We describe the rules governing the transitions between states,
which comprise a program $\prog$, global memory $\Mem$, store buffers $\TSOmap$, queue pairs $\NQPmap$ and remote atomic locks $\Atm$.

\input{example-trans}

Transitions take the form shown on the right,
which should be read as:
if $\phi$ is true, then it is allowed for the system to transition from the state described by $\prog\ldots\NQPmap$ to the state described by $\prog'\ldots\NQPmap'$.

In practice, however, writing each transition rule in such a way would be verbose and hard to understand,
as most transitions do not affect every part of the state.
We can separate \emph{program transitions} concerning $\prog$ from \emph{hardware transitions} concerning $\Mem, \TSOmap, \Atm, \NQPmap$.
In order to synchronise the two where necessary,
we assign labels to certain transitions and require that a labelled program transition only occur if it is matched by a hardware transition with the same label (or vice-versa).
Labels are of the form $\threadt:\lab$ where $\threadt$ is the thread executing at that step and $\lab \in \Labs$ is the label of the operation.
\emph{Silent transition}, which affect only the program (resp.\ only hardware) are written with the empty label, $\epsilon$,
and may be taken independently.

\cref{fig:opsem-combined} shows the top-level rules of the operational semantics which govern this separation.
We can henceforth consider the program and hardware transitions separately.
\input{opsem-combined}

\paragraph{Program Transitions}
\cref{fig:prog_transitions} shows the program and command transitions (middle),
labels (above) and expression rewriting rules (below).
The transitions for non-remote commands are familiar from \TSO{}.
Notice that the transitions for get and put simply transition to $\cskip$ with the relevant label;
we know that this means there will be some relevant transition in the hardware.
The transition to $\cskip$ allows the program to continue executing,
which we expect,
as remote operations are handled asynchronously by the NIC.\@

This is similarly the case for the rules for remote-CAS and remote-FAA.
The expressions involved are required to be closed,
similarly to the rules for local write and CAS;\@
the expressions must be evaluated before the transition.
It only makes sense to use a value, not an expression, in the label,
since the corresponding hardware transition will only be concerned with values.

\input{prog-trans}

\paragraph{Hardware Domains}
The upper section of \cref{fig:hardware_transitions} shows the hardware domains --
that is, the states we are interested in \emph{other} than the program.
We have already described memory, store buffers, remote atomic locks and queue pairs,
but note that the structures $\TSOmap$, $\Atm$, and $\NQPmap$ are specifically maps from threads, nodes, and both, respectively, to the particular structures.

A remote atomic lock is a boolean, $\bot$ (available) or $\top$ (unavailable).
A store buffer is a sequence of CPU writes and RDMA operations.
A queue pair is a tuple of three sequences $\pipe$, $\wbR$, and $\wbL$,
where $\pipe$ may contain any of the operations described below except for a confirmation notification,
$\wbR$ may contain NIC remote writes and NIC remote atomic writes,
and $\wbL$ may contain NIC local writes and confirmation notifications.

\begin{itemize}
  \item $y^{\neqn{n}} \assign x^{n}$ denotes a put operation where the value of local memory location $x$ is yet to be read (NIC local read);
  \item $y^{\neqn{n}} \assign v$ denotes a NIC remote write of value $v$ to remote location $y$, which occurs as the latter part of a put;
  \item $\doneE$ denotes the acknowledgement message returned by a put;
  \item $x^{n} \assign y^{\neqn{n}}$ denotes a get operation where the value of the remote location $y$ is yet to be read (pending NIC remote read)
  \item $x^{n} \assign v$ denotes a NIC local write of value $v$ to local location $x$, which occurs as the latter part of a get or rRMW;\@
  \item $\rcas(z^{n},x^{\neqn{n}},v,v')$ denotes a remote CAS towards remote location $x$, with expected value $v$, update value $v'$, and returning to local location $z$;
  \item $\rfaa(z^{n},x^{\neqn{n}},v)$ similarly denotes a remote FAA towards $x$ and returning to $z$, with increment value $v$;
  \item $\y^{\neqn{n}} \assign_{A} v$ denotes a NIC remote write specifically in the case of an rRMW -- it is necessary for this to be disambiguated from the NIC remote write of a put, as we will see later;
  \item $\cnfence{\neqn{n}}$ denotes a remote fence towards node $\neqn{n}$;
  \item $\cn$ denotes a confirmation of a successful NIC remote write.
\end{itemize}

\paragraph{Hardware Transitions}
All remote commands enter the queue-pair pipe via the thread's store buffer.
When the program takes a transition step labelled with a remote CAS or FAA,
the hardware takes a transition with a matching label,
which adds that operation to the store buffer.
The seventh transition rule allows remote commands at the head of the store buffer to enter the pipe of a queue pair, determined by their target node.

So far, we have seen that when an rRMW appears in the program,
we can expect there to be a hardware transition which adds it to the store buffer,
and later another hardware transition which removes it from the head of the store buffer and adds it to the suitable queue pair.

The final hardware transition introduces the queue-pair transitions,
indicated by $\pipearrow$\footnote{SQP stands for simplified queue pair. We only considered the simplified three-buffer queue pair, so this disambiguation is technically unnecessary, but we maintain the notation for consistency with \cite{OOPSLA-24}}.
When a particular queue pair takes a transition step,
involving memory and the global remote atomic lock,
the hardware takes a suitable corresponding transition.
The queue-pair transitions merely involve a particular subset of the hardware states,
so the relationship is straightforward.
This separation is purely made for clarity and simplification of the queue-pair transition rules.

\input{hw-dom-trans}

\paragraph{Queue-Pair Transitions}
From \cref{fig:sqpair_structure},
recall that remote operations enter the main $\pipe$ of the queue pair,
then are suitably processed until they exit the pipe,
possibly adding a write to $\wbL$ or $\wbR$ (or both).
Note that the pipe grows to the left,
so throughout, $\alpha$ contains operations which are later in the program,
while $\beta$ contains earlier operations which have not yet completed.

The rules for remote fence, put, and get share a simple structure,
where the premise for a transition either requires the operation to be at the head of the pipe $\simqp.\pipe = \alpha \cdot (\text{operation})$,
or allows it to be in the middle of the pipe $\simqp.\pipe = \alpha \cdot (\text{operation}) \cdot \beta$, with some stipulation as to the operations allowed in $\beta$.
In the prior case, the operation never executes before another, earlier operation;
in the latter, it can execute before any operation in $\beta$ which was issued before it.
There may also be some requirement that buffers $\wbL$ or $\wbR$ contain no writes,
due to PCIe guarantees:
$\wbL \in \queueset{\cn}$ ($\wbL$ contains only confirmation notifications) or $\wbR = \epsilon$ (there are no operations in $\wbR$).
Consider, for example, the first step of a put,
which is a NIC local read described by rule 2.
The value of location $x$ is read from memory,
so long as $\wbL$ has no pending writes and there are no other NIC local reads earlier in the pipe.

\input{qp-trans}

We can then describe the rules for rfence, put, and get at a high level:
\begin{description}
  \item[Remote fence (rule 1)]
        An rfence may be removed from the pipe once it reaches the head (there are no earlier operations remaining to be processed).
        In combination with the fact that no other transition rule allows a step to be taken when there is an rfence later in the pipe,
        this enforces the behaviour that all remote operations prior to an rfence complete before it, and all later ones after it.
  \item[Put (rules 2-5)]
        Rule 2: a NIC local read is performed, replacing the location $x$ with its value in memory.
        Rule 3: the NIC remote write is sent to $\wbR$,
        and an acknowledgement created in the pipe.
        Rule 4: the remote write is committed to memory once it reaches the head of the queue.
        Rule 5: the acknowledgement in the pipe is converted to a confirmation notification in $\wbL$, so that it can be polled.
  \item[Get (rules 6-8)]
        Rule 6: a NIC remote read replaces the location $\neqn{y}$ with its value in memory.
        Rule 7: the NIC local write is sent to $\wbL$, with a confirmation notification for the purpose of polling.
        Rule 8: the local write is committed to memory once there are no pending earlier writes in the queue.
\end{description}

Now, consider the rules for rRMWs.
These rules are more complicated due to the need to check and update the remote atomic lock for the target node,
which we see as $\Atm(\node(\neqn{x})) = \bot$ (the remote atomic lock for the target node is available),
and $\Atm' = \Atm\upvar{\node(\neqn{x})}{\top}$ (update the remote atomic lock for the target node to indicate it is busy).
We also have distinct rules for success and failure of RCAS,
depending on whether the remote memory location holds the expected value ($\Mem(\neqn{x}) = v$ or $\Mem(\neqn{x}) \neq v$).

The rules can then be interpreted as follows:
\begin{description}
  \item[(Rule 9)]
        A failed RCAS read -- the remote memory location does not hold the expected value.
        This read can only occur when the remote atomic lock is available,
        otherwise it would violate the atomicity guarantee.
        The value of $\neqn{x}$ is read,
        and a NIC local write is added to the pipe to return that value to $z$.
        This is then handled by the same rules as for a get.
        The remote atomic lock is not obtained,
        since the remote location will not be written to.
  \item[(Rule 10)]
        A successful RCAS read -- the remote location contains the expected value.
        Once again, this requires that the remote atomic lock be available,
        and it is also obtained to ensure atomicity until the remote location is written to.
        A NIC local write is added to the pipe (similarly to 9),
        and a NIC remote atomic write to update the remote location is also added.
  \item[(Rule 11)]
        The remote read of an RFAA -- this is unconditionally successful.
        It is very similar to a successful RCAS, but the value for the NIC remote atomic write is calculated by adding $v$ to the value of $\neqn{x}$ in memory.
  \item[(Rule 12)]
        A NIC remote atomic write in the pipe is processed into $\wbR$ similarly to a regular NIC remote write.
  \item[(Rule 13)]
        A NIC remote atomic write is committed to memory,
        and the remote atomic lock is released.
\end{description}

\subsection{Declarative Semantics}
\label{sec:decl-sem}

A declarative semantics, in contrast to an operational one,
describes only the events that occur in a system,
not the state of the system itself.
An execution is represented by a \emph{graph},
with various relations over events.
For example, given an event $r$,
we say that it ``reads-from'' event $w$
if $r$ reads the value written to memory by $w$.
We write $(w,r) \in \rf$ in this case.
We then constrain these relations suitably to only allow execution graphs which make sense in the context of a program:
considering $\rf$ again, we would naturally only allow $(w,r) \in \rf$ if the values of the read and write match.

Then, we know that an execution of a program is allowed if the graph of the execution is consistent.
Contrast this with the operational semantics:
there, our guarantee comes from the individual transition rules;
here, it is due to the overall structure of the graph.

\paragraph{Events and Executions}
An \emph{execution} is a graph comprising a set of events and several relations over events;
events are represented as graph nodes,
and the relations are edges.
An event has a unique identifier $\aident$,
is created by a thread $\threadt \in \Threads$,
and has an event label $\lab \in \Alab$ which describes the event.

\begin{definition}[Labels and events]
\label{def:app-labels}
\label{def:app-actions}
\label{def:app-events}
Each event label is associated to a node $\node$.
The set of \emph{event labels of node $\node$} is denoted by
$\lab \in \Alab_{\node}$,
where $\lab$ is a tuple with one of the following forms:
\vspace{-2.4em}
\begin{multicols}{2}
\raggedcolumns
\vspace*{\fill}
\begin{itemize}[itemsep=12pt]
\item (CPU) local read: $\lab=\lR(\x^\node,\valr)$
\item (CPU) local write: $\lab=\lW(\x^\node,\valw)$
\item (CPU) CAS: $\lab=\RMW(\x^\node,\valr,\valw)$
\item (CPU) memory fence: $\lab=\lF$
\item (CPU) poll: $\lab= \lP(\neqn{\node})$
\end{itemize}
\vspace*{\fill}

\columnbreak

\vspace*{\fill}
\begin{itemize}
\item NIC local read: $\lab=\nlR(\x^\node,\valr,\neqn{\node})$
\item NIC remote write: $\lab=\nrW(\y^{\neqn\node},\valw)$
\item NIC remote read: $\lab=\nrR(\y^{\neqn\node},\valr)$
\item NIC local write: $\lab=\nlW(\x^\node,\valw,\neqn{\node})$
\item NIC fence: $\lab=\nF(\neqn{\node})$
\item NIC atomic remote read:\\ $\lab=\narR(\y^{\neqn\node},\valr)$
\item NIC atomic remote write:\\ $\lab=\narW(\y^{\neqn\node},\valw)$
\end{itemize}
\end{multicols}
\vspace{-0.4em}

The set of event labels are defined $\Alab \defeq \bigcup_{\node} \Alab_{\node}$.

An \emph{event}, $\action \in \Events$, is a triple $(\aident,\threadt,\lab)$, where $\aident \in \mathbb{N}$,
$\threadt \in \Threads$ and $\lab\in\Alab_{\node(\threadt)}$.
\end{definition}

We distinguish between events associated with the CPU (left) or NIC (right),
with the prefix $\mathtt{n}$ used for all NIC event labels.
Note that a put, get, or rRMW is modelled by multiple events:
a put $\neqn{x} \assign y$ comprises a NIC local read event of type $\nlR$ (on $y$) followed by a NIC remote write event $\nrW$ (on $\neqn{x}$);
conversely a get $x \assign \neqn{y}$ comprises events of type $\nrR$ (on $\neqn{y}$) and $\nlW$ (on $x$).
A successful rRMW (either successful RCAS or RFAA) is modelled by three events of type $\narR$, $\narW$ and $\nlW$,
while a failed rRMW (RCAS only) is modelled by only $\narR$ and $\nlW$.

For a given label $\lab$, we write $\typ(\lab)$, $\loc(\lab)$, $\valr(\lab)$, $\valw(\lab)$,
$\node(\lab)$ and $\neqn\node(\lab)$
for the type, location, value read or written, and local or remote node,
where applicable.
\begin{samepage}
For example, consider $\lab = \nlR(x^{\node}, \valr, \neqn{\node})$:
\vspace{-2.4em}
\begin{multicols}{2}
\raggedcolumns
\vspace*{\fill}
\begin{itemize}
  \item $\typ(\nlR(x^{\node}, \valr, \neqn{\node})) = \nlR$
  \item $\loc(\nlR(x^{\node}, \valr, \neqn{\node})) = x$
  \item $\valr(\nlR(x^{\node}, \valr, \neqn{\node})) = \valr$
\end{itemize}
\vspace*{\fill}

\columnbreak

\vspace*{\fill}
\begin{itemize}
  \item $\valw(\nlR(x^{\node}, \valr, \neqn{\node}))$ is undefined
  \item $\node(\nlR(x^{\node}, \valr, \neqn{\node})) = \node$
  \item $\neqn{\node}(\nlR(x^{\node}, \valr, \neqn{\node})) = \neqn{\node}$
\end{itemize}
\end{multicols}
\vspace{-0.4em}
\end{samepage}

We write $\aident(\action)$, $\threadt(\action)$, $\lab(\action)$ for the relevant constituents of an event tuple $\action = (\aident,\threadt,\lab)$.
We lift the functions on event labels to functions on events,
for example $\typ(\action) \defeq \typ(\lab(\action))$.

\paragraph{Difference with \framework}
The labels of this declarative semantics (à la \cite{OOPSLA-24}) roughly
corresponds to the stamps of \framework (à la \cite{POPL-LOCO}) in the main
paper. \Eg a NIC local read has a label $\nlR$ here and corresponds to the stamp
(family) $\tagnlr[]$ in~\cref{fig:to}. However, there is one major discrepancy.
The declarative semantics of this section distinguishes between NIC remote
writes performed by put operations (label $\nrW$) and performed by rRMW
operations (label $\narW$), while they both correspond to the single stamp
$\tagnrw$.

The main reason is that this semantics, by decomposing operations into multiple
events, creates a \po ordering between the remote write and local write parts of
a (successful) remote RMW. As such, we cannot enforce a \ppo ordering between the
two parts ($\narW$ and $\nlW$) as they might not finish in order, but we can
enforce a \ppo ordering between the remote write of a put and later local writes
($\nrW$ and $\nlW$), making the semantics more straightforward. With \framework,
each operation generates a \emph{single} event, and there is no \po ordering
between subevents of the same event. Thus we can add a dependency between
$\tagnrw$ and $\tagnlw$ (cell G10 in~\cref{fig:to}), and it will not create an
internal dependency within the same rRMW operation.

A secondary reason is that the two labels ($\nrW$ and $\narW$) correspond to
different behaviours of the operational semantics. Making the distinction
renders the equivalence proof more tractable.

\paragraph{Issue and Observation Points}
Some types of events do not occur instantaneously:
for example, a local write event $\lW$ first enters the store before,
before later being committed to memory.
We therefore distinguish between the point at which an event is \emph{issued} by the CPU or NIC,
and the point at which it is \emph{observed},
when its effect becomes visible in memory.
An event is \emph{instantaneous} if it either has no visible effect on memory,
or if it affects memory immediately,
as is the case for a local CAS operation.
For instantaneous events, the issue and observation points coincide.

\paragraph{Notation}
Once again, we follow and extend the notation of \cite{OOPSLA-24}.
For a set $A$ and relations $r,r_{1},r_{2}$, we write:
\vspace{-0.6em}
\begin{description}[itemindent=0pt, itemsep=2pt]
  \item[$\bm{r^{+}}$] for the transitive closure of $r$;
  \item[$\bm{r^{-1}}$] for the inverse of $r$;
  \item[$\bm{r\rst{A}}$] $\defeq r \cap (A \times A)$ for the restriction of $r$ to set $A$;
  \item[$\bm{[A]}$] $\defeq \set{(a,a) \st a \in A}$ for the identity relation
  \item[$\bm{r_{1};r_{2}}$] {$\defeq \set{(a,b) \st \exists c. (a,c) \in r_{1} \land (c,b) \in r_{2}}$} for relational composition;
  \item[$\bm{\imm{r}}$] $\defeq r \setminus (r;r)$ for the immediate edges in $r$, when $r$ is a strict partial order.
\end{description}
For a set of events $\events$, location $x$ and label type $\lX$, we also define:
\vspace{-0.6em}
\begin{description}[itemindent=0pt, itemsep=2pt]
  \item[{$\events_{x}$}] $\defeq \set{\action \in \events \st \loc(\action) = x}$, the events towards $x$;
  \item[{$\events$}$.\lX$] $\defeq \set{\action \in \events \st \typ(\action) = \lX}$, the events of type $\lX$;
  \item[{$\events$}$.\Read$] $\defeq \events.\lR \cup \events.\RMW \cup \events.\nlR \cup \events.\nrR \cup \events.\narR$, the set of reads;
  \item[{$\events$}$.\Write$] $\defeq \events.\lW \cup \events.\RMW \cup \events.\nlW \cup \events.\nrW \cup \events.\narW$, the set of writes;
  \item[{$\events$}$.\Inst$] $\defeq \events \setminus (\events.\lW \cup \events.\nlW \cup \events.\nrW \cup \events.\narW)$, the set of instantaneous events.
\end{description}
Finally, we define the following relations:
\par\vspace{0.5\baselineskip}
\begin{tabular}{@{} l r c l @{}}
\textbf{Same-location:} & $\sloc$ & $\defeq$ & $\set{(\action,\action') \in \Events^2 \st \loc(\action) = \loc(\action')}$ \\
\textbf{Same-thread:} & $\sthd$ & $\defeq$ & $\set{(\action,\action') \in \Events^2 \st \threadt(\action) = \threadt(\action')}$ \\
\textbf{Same-queue-pair:} & $\sqp$ & $\defeq$ & $\set{(\action,\action') \in \Events^2 \st \threadt(\action) = \threadt(\action') \land \neqn\node(\action) = \neqn\node(\action')}$ \\
\end{tabular}
\par\vspace{0.5\baselineskip}%
\noindent
Note that these relations are all symmetric,
and $\sqp \suq \sthd$.
Given events $\events$, we write $\events.\sloc$ for $\sloc\rst{\events}$,
likewise for $\events.\sthd$ and $\events.\sqp$.

\begin{samepage}
\begin{definition}[Pre-executions]
\label{def:pre-executions}
A \emph{pre-execution} is a tuple $\EG=\tup{\events,\po,\rf,\mo, \pf, \ro, \rao}$, where:
\begin{itemize}
  \item $\events \suq \Events$ is the set of events and includes a set of \emph{initialisation} events, $\events^0 \subseteq \events$, comprising a single write with label $\lW(\x, 0)$ for each $\x \in \Var$.
  \item $\po \suq \events \times \events $ is the `\emph{program order}' relation defined as a disjoint union of strict total orders,
each ordering the events of one thread, with $\events^0 \times (\events \setminus \events^0) \suq \po$.
  \item $\rf \suq \events.\Write \times \events.\Read$ is the `\emph{reads-from}' relation on events of the same location with
matching values; \ie $(a, b) \in \rf \Rightarrow (a, b) \in \sloc \land \valw (a) \!=\! \valr (b)$.
Moreover, $\rf$ is total and functional on its range: every read in $\events.\Read$ is related to exactly one write in $\events.\Write$.
  \item $\mo \defeq \bigcup_{\x \in \Var} \mo_\x$ is the `\emph{modification-order}', where each $\mo_\x$
is a strict total order on $\events.\Write_\x$ with $\events^0_\x \times (\events.\Write_\x \setminus \events^0_\x) \suq \mo_\x$
describing the order in which writes on $\x$ reach the memory.
  \item $\pf \suq (\events.\nlW \cup \events.\nrW) \times \events.\lP$ is the `\emph{polls-from}' relation, relating earlier (in program-order) NIC writes to later poll operations on the \emph{same queue pair};
  \ie $\pf \suq \po \cap \sqp$.
  Moreover, $\pf$ is functional on its domain (every NIC write can be be polled at most once),
  and $\pf$ is total and functional on its range (every poll in $\events.\lP$ polls from exactly one NIC write).
  \item $\ro \suq \events.\sqp$ is the `\emph{NIC flush order}', such that
  for all $(a, b) \in \events.\sqp$, if $a \in \events.\nlR, b \in \events.\nlW$, then $(a, b) \!\in\! \ro \cup \inv\ro$, and
  if $a \in (\events.\nrR \cup \events.\narR), b \in (\events.\nrW \cup \events.\narW)$, then $(a, b) \in \ro \cup \inv\ro$.
  \item $\rao \defeq \bigcup_{n \in \Nodes} \rao_{n}$ is the `\emph{remote-atomic-order}',
        where each $\rao_{n}$ is a strict total order on $\set{e \st e \in \events.\narR \land \neqn{\node}(e) = \node}$ describing the order in which remote atomics towards $\node$ are executed.
\end{itemize}\vspace{-5pt}
\end{definition}
\end{samepage}

The definitions of $\po$, $\rf$ and $\mo$ are familiar from \TSO{}, while $\pf$
and $\ro$ are introduced in \cite{OOPSLA-24}. As mentioned previously, $\ro$
represents the PCIe guarantee that a NIC local read flushes pending NIC remote
writes on the same queue pair, and likewise for NIC local reads/writes. We
introduce $\rao$, which totally orders NIC remote atomic reads towards a given
node and help enforce the rRMW atomicity guarantee.

\paragraph{Derived Relations}
Given a pre-execution $\tup{\events, \po, \rf, \mo, \pf, \ro, \rao}$, we define the following \emph{derived} relations:
\begin{itemize}
  \item $\fr \defeq (\rf^{-1};\mo) \setminus [\events]$
        is the \emph{reads-before} relation,
        relating each read $r$ to writes that are $\mo$-after the write from which $r$ reads.
  \item $\rfi \defeq [\lW];(\rf\cap\sthd);[\lR]$ is the $\rf$-\emph{buffer} relation,
        which includes $\rf$ edges only for CPU operations on the same thread,
        which thus share a store buffer;
        therefore when $w \arr{\rfi} r$,
        it may be that the write $w$ is not yet visible (committed to memory) when it is read by $r$,
        since CPU reads check the store buffer.
  \item $\rfe \defeq \rf \setminus \rfi$ is the $\rfi$-complement:
        if $w \arr{\rfe} r$, then $r$ only occurs after $w$ is observable.
  \item $\fri \defeq [\lR];(\fr\cap\sthd);[\lW]$ is the $\fr$-\emph{buffer}
        relation, analogously.
  \item $\ar \defeq [\narW];(\imm{\po}^{-1})$ is the \emph{atomic-write-to-read} relation,
        connecting the remote write of a successful rRMW to their corresponding read.
\end{itemize}
Note that these derived relations contain no additional information.
We introduce them for ease and brevity of notation.

\paragraph{Preserved Program Order}
We identify which events in $\po$ are \emph{issued} in order,
and which are \emph{observed} in order.
The observation point of an event is no earlier than its issue point,
so two events in $\po$ are only observed in order if they are issued in order.
Furthermore, when the $\po$-earlier event is instantaneous,
the events are observed in order if and only if they are issued in order.

\input{reordering}

We therefore define two relations:
$\ippo$, the \emph{issue-preserved-program-order} relation,
and $\oppo$, the \emph{observation-preserved-program-order} relation,
where $\oppo \subseteq \ippo \subseteq \po$.
The tables in \cref{fig:ippo-oppo} show these relations.
Each row indicates the $\po$-earlier event, while each column indicates that which is $\po$-later.
A cell labelled \checkyes{} indicates the event pair is in $\ippo$ (resp.\ $\oppo$) and must be issued (resp.\ observed) in program order,
while \checkno{} indicates they are not in $\ippo/\oppo$ and may be issued/observed out of program order.
The label \sqp{} indicates that the events are in $\ippo/\oppo$ if they are events on the same queue pair.

We can observe high-level reordering rules by looking at each quadrant of the two tables,
which partition the event pairs by their categorisation as CPU or NIC events.
The top left quadrant contains pairs of CPU events.
Observe that CPU events are always issued in program order,
and only an earlier CPU write may be observed out of order,
as all other CPU events are instantaneous.
The bottom left quadrants shows that an earlier NIC event may always be issued or observed after a later CPU event,
matching our intuition that NIC events execute concurrently,
as if in a separate thread;
conversely the top right shows that earlier CPU events always complete before later NIC events.
In the bottom right quadrant, we can see that a pair of NIC events are only ordered if they are on the same queue pair.

The relations $\ippo$ and $\oppo$ differ in only six cells.
A CPU write may be buffered and hence not observed by a later CPU read or poll (B1 and B5).
Other CPU writes and CAS or fence operations go via the store buffer,
so earlier writes will be observed first.
Similarly, a remote fence may be observed before an earlier NIC remote (atomic) write (resp.\ local),
if that write is buffered in $\wbR$ (resp.\ $\wbL$) (G12 and I12, resp. K12).
Finally, a $\po$-later $\nlW$ may be observed before a $\po$-earlier $\narW$ (I11).
This occurs specifically in the case where both are created by the same rRMW,
because the writes are sent to $\wbL$ and $\wbR$ respectively and may be committed in either order.

\begin{definition}[Executions]
\label{def:well-formed}
\label{def:executions}
A pre-execution $\EG \!=\! \tup{\events, \po, \rf, \mo, \pf, \ro, \rao}$ is \emph{well-formed} if
the following hold for all $w, r, w_1, w_2, p_2$:
\begin{enumerate}
\item Poll events poll-from the oldest non-polled remote operation on the same queue pair:\\
if $w_1 \in \EG.\nlW \cup \EG.\nrW$ and $w_1 \arr{\po \cap \sqp} w_2 \arr{\pf} p_2$, then there exists $p_1$
  such that $w_1 \arr{\pf} p_1 \arr{\po} p_2$.
\item Each \textput (\resp \textget) operation corresponds to  two events: a read and a write with the read immediately preceding the write in $\po$:
\begin{enumerate*}
	\item if $r \!\in\! \EG.\nlR$ (\resp $r \!\in\! \EG.\nrR$), then  $(r,w) \!\in\! \imm{\po}$ for some $w \!\in\! \EG.\nrW$ ($w \!\in\! \EG.\nlW$); and
	\item if $w \!\in\! \EG.\nrW$ then $(r,w) \!\in\! \imm{\po}$ for some $r \!\in\! \EG.\nlR$. The case $w \in \EG.\nlW$ is handled by (6) below.
\end{enumerate*}
\item Read and write events of a \textput (\resp \textget) have matching values:\\
if $(r,w) \in \imm{\EG.\po}$, $\typ(r) \in \set{\nlR,\nrR}$ and $\typ(w) \in \set{\nlW,\nrW}$, then $\valr(r) = \valw(w)$.
\item Each rRMW operation corresponds to either
        an atomic remote read followed by a local write, or
        an atomic remote read, followed by an atomic remote write, followed by a local write:
        \begin{enumerate*}
          \item if $r \!\in\! \EG.\narR$ then $(r, w_{1}) \!\in\! \imm{\po}$ for some $w_{1} \!\in\! \EG.\narW \cup \EG.\nlW$,
          and if $w_{1} \!\in\! \EG.\narW$ then $(w_{1}, w_{2}) \!\in\! \imm{\po}$ for some $w_{2} \!\in\! \EG.\nlW$, and
          \item if $w_{1} \!\in\! \EG.\narW$ then $(r,w_{1}) \!\in\! \imm{\po}$ for some $r \in \narR$, and $(w_{1},w_{2}) \!\in\! \imm{\po}$ for some $w_{2} \in \nlW$.
        \end{enumerate*}
        The case for $w_{2} \in \nlW$ is handled by (6) below.
  \item Remote atomic read and local write events of an rRMW have matching values:
        if $(r, w) \in \EG.\imm{\po}$, $\typ(r) = \narR$ and $\typ(w) = \nlW$, then $\valr(r) = \valw(w)$;
        and if $(r, w_{1}), (w_{1}, w_{2}) \in \EG.\imm{po}$, $\typ(r) = \narR$, $\typ(w_{1}) = \narW$ and $\typ(w_{2}) = \nlW$, then $\valr(r) = \valw(w_{2})$.
  \item[6.] (2) and (4) auxiliary in the case of $w \in \nlW$.
        If $w \in \EG.\nlW$ then either:
        \begin{enumerate}[nosep, left=1em, topsep=0pt, partopsep=0pt]
          \item $(r,w) \in \imm{\po}$ for some $r \in \EG.\nrR$ or
          \item $(r,w) \in \imm{\po}$ for some $r \in \EG.\narR$ or
          \item $(r,w'),(w',w) \in \imm{\po}$ for some $r \in \EG.\narR$ and $w' \in \EG.\narW$.
        \end{enumerate}
\end{enumerate}
An \emph{execution} is a pre-execution (\cref{def:pre-executions}) that is well-formed.

\end{definition}
Given an execution $G$, we write $G.\events$, $G.\mo$, $G.\ippo$ and so forth to project the components and derived relations of $G$.
When the execution is question is clear,
we simply write $\events$, $\mo$ or similar.
\begin{definition}[\rdmatso-consistency]
\label{def:consistency}
An execution $\tup{\events, \po, \rf, \mo, \pf, \ro, \rao}$ is
\emph{\rdmatso-consistent} iff
\begin{enumerate*}
  \item $\ib$ is irreflexive; and
  \item $\ob$ is irreflexive,
\end{enumerate*}
where:
\begin{align*}
  \ib  & \defeq
  \big(
           \ippo
           \cup \rf
           \cup \pf
           \cup \ro
           \cup \fri
           \cup (\ob;[\Inst])
   \big)^{+}
         \tag{\text{`issued-before'}} \\
  \ob  & \defeq
  \big(
           \oppo
           \cup \rfe
           \cup ([\nlW]; \pf)
           \cup \ro
           \cup \fr
           \cup \mo
           \cup \rao
           \cup (\ar;\rao)
           \cup ([\Inst];\ib)
   \big)^{+}
         \tag{\text{`observed-before'}}
\end{align*}\vspace*{-15pt}
\end{definition}

These relations extend $\ippo$ and $\oppo$ respectively to describe the issue and observation orders across threads and nodes.
They are required to be irreflexive,
i.e.\ an event cannot be issued or observed before itself.

The remaining components of $\ib$ are
\begin{enumerate*}[label=(\alph*)]
  \item $\rf$: if $w \arr{\rf} r$ then $w$ was at least issued (if not observed) before $r$ -- recall that if the read and write are both CPU events on the same thread, $w$ may not be observable;
  \item $\pf$: similarly $w \arr{\pf} p$ only if $w$ was issued before $p$;
  \item $\ro$: NIC events arrive in $\wbL$/$\wbR$ in the order they are issued;
  \item $\fri$: if $r \arr{\fri} w$, then $r$ must be issued before $w$,
  otherwise $r$ would read from $w$ or an $\mo$-later $w'$;
  \item $\ob;[\Inst]$: in general, an event is observed no earlier than it is
    issued, and for an instantaneous event, the two points coincide. Thus
    $e \arr{\ob} e'$ implies $e \arr{\ib} e'$ when $e'$ is instantaneous. As
    noted in~\cite{OOPSLA-24}, this last component is optional and does not
    modify the semantics.
\end{enumerate*}

On the other hand, for $\ob$ we have
\begin{enumerate*}[label=(\alph*)]
  \item $\rfe$: if $w \arr{\rfe} r$ then $w$ was committed to memory before $r$,
  since $r$ cannot read from the store buffer of another thread;
  \item $[\nlW];\pf$: NIC local writes cannot be polled until they are committed to memory;
  \item $\ro$: NIC events are observed in the same order they arrive in $\wbL$/$\wbR$;
  \item $\fr$: if $r \arr{\fr} w$, then $w$ was not observed before $r$,
  otherwise it would have been committed to memory before $r$;
  \item $\mo$: if $w \arr{\mo} w'$, then $w$ was observed in memory before $w'$;
  \item $\rao$: remote atomic reads are (issued and) completed in the defined order;
  \item $\ar;\rao$: if $w \arr{\ar} r \arr{\rao} r'$, then we have that $r$ and $w$ are the read and write of the same rRMW operation,
  thus $w$ must be observed before the $\rao$-later $r'$ to ensure atomicity.
  \item $[\Inst];\ib$: by a similar logic to above, we know that the $\ib$-earlier instantaneous event is also observed earlier, since its issue and observation points coincide.
\end{enumerate*}

\paragraph{Semantics of a Program}
Given a program $\prog$, we can generate an event graph $(\events,\po)$,
by a standard process, which we describe below.
We then choose any $\rf,\mo,\pf,\ro,\rao$ such that the execution is consistent.
The semantics of $\prog$ are the set of consistent executions of $\prog$.

\paragraph{Thread to Event Graph}
Given a thread identifier $\threadt \in \Threads$ and a sequence of labels $\lab_{1},\ldots,\lab_{n} \in \Alab$,
we define the \emph{event graphs of $\threadt$} as $(\set{e_{1},\ldots,e_{n}}, \po) \in G^{\threadt}(\lab_{1},\ldots,\lab_{n})$ where:
\begin{enumerate*}[label=(\alph*)]
  \item $\lab(e_{i}) = \lab_{i}$ for all $1 \leq i \leq n$;
  \item $\aident(e_{i}) \neq \aident(e_{j})$ for all $1 \leq i < j \leq n$;
  \item $\threadt(e_{i}) = \threadt$ for all $1 \leq i \leq n$;
  \item $\po = \set{(e_{i},e_{j}) \st 1 \leq i < j \leq n}$.
\end{enumerate*}

\paragraph{Initial Event Graph}
Given a set of locations $\Var$, we define $G_{init} = (\events_{0}, \emptyset)$,
such that for each $x \in \Var$ there is exactly one $e \in \events_{0}$ with $\lab(e) = \lW(x, 0)$,
and every event in $\events_{0}$ has a unique identifier.
We call $\events_{0}$ the set of \emph{initialisation events}.

\paragraph{Sequential Composition}
For two event graphs $G_{1}$ and $G_{2}$,
we define their \emph{sequential} composition $G_{1};G_{2} = (\events, \po)$ where
\begin{align*}
  \events &\defeq G_{1}.\events \uplus G_{2}.\events \\
  \po &\defeq G_{1}.\po \cup G_{2}.\po \cup (G_{1}.\events \times G_{2}.\events)
\end{align*}
Note that all events in $G_{2}$ are ordered $\po$-after every event in $G_{1}$.
Sequential composition is defined only where the set of events of each graph are disjoint,
i.e.\ $G_{1}.\events \cap G_{2}.\events = \emptyset$.

\paragraph{Parallel Composition}
We define \emph{parallel} composition by $G_{1} \parallel G_{2} = (\events, \po)$ where
\begin{align*}
  \events &\defeq G_{1}.\events \uplus G_{2}.\events \\
  \po &\defeq G_{1}.\po \cup G_{2}.\po
\end{align*}
Note that the events of each graph are not $\po$-ordered with respect to one another.
We also require that the event sets be disjoint.
As this operation is commutative and associative,
it is straightforward to lift it to sets of graphs,
which we denote by $\parallel\SAG$, where $\SAG$ is a set of event graphs.

\paragraph{Program to Event Graph}
\label{par:prog-to-graph}
A program $\prog$ \emph{generates} $G$ if $G = G_{init};(\parallel_{t\in\Threads}G_{t})$ and there is a set of sequences $s_{\threadt} \in S$ such that $\prog(\threadt) \actseq s_{t}$ and $G_{\threadt} \in G^{\threadt}(s_{\threadt})$ for all $\threadt \in \Threads$.

The operation $C \actseq s$ relates a sequential program $C$ to a sequence of labels $s$ it generates.
The definition is standard and show in \cref{fig:action-sequence}.
Note that RDMA operations generate multiple events,
and for local and remote CAS operations,
we distinguish between success and failure cases.

\input{event-sequence}

\begin{theorem}
  \label{thm:equiv}
  The operational and declarative semantics of \rdmatsormw{} are equivalent.
\end{theorem}
\begin{proof}
  See \Cref{sec:annot-sem-rules} onwards, extending the proof
  of~\cite{OOPSLA-24}.
\end{proof}


%% file: simple-queue-pairs.tex

\begin{figure}[t]
  \centering
  \definecolor{myyellow}{rgb}{1,1,0}
  \begin{tikzpicture}[scale=0.8, every node/.style={scale=0.75}]
    \draw[preaction={fill, green}, pattern={Lines[angle=90,distance=4pt]}]
    (0,-.1) -- (1.2,-.1)
    -- (1.2,0.4)
    node[pos=0.7,label=above left:{SBuff}]{}
    -- (0,0.4)--(0,-.1);
    \fill [myyellow]
    (1.3,0.2) -- (4.7,0.2) -- (4.7,0.7) -- (1.3,0.7)
    (4.7,-1.2)-- (3.1,-1.2) -- (3.1,-1.7) -- (4.7,-1.7)
    ;
    \fill [purple!30!white]
    (5.7,0.2) -- (5.7,0.7)
    -- (7.5,0.7) arc(90:0:1.2)
    -- (7.5,0.7) arc(90:0:1.2) -- +(0:-0.5)
       arc(0:90:0.7)
    (5.7,-1.2) -- (5.7,-1.7)
    -- (7.5,-1.7) arc(-90:0:1.2)
    -- (7.5,-1.7) arc(-90:0:1.2) -- +(0:-0.5)
       arc(0:-90:0.7)
    ;
    \shade [shading=axis, left color=myyellow, right color=purple!30!white]
    (4.6,0.2) -- (5.8,0.2) -- (5.8,0.7) -- (4.6,0.7)
    (5.8,-1.2)-- (4.6,-1.2) -- (4.6,-1.7) -- (5.8,-1.7)
    ;
    \draw
    (1.3,0.2) -- (4,0.2) -- (7.5,0.2)
    arc(90:-90:0.7)  (7.5,-1.2)--(3.1,-1.2)
    (1.3,0.2) -- (1.3,0.7) -- (7.5,0.7)
    arc(90:-90:1.2)
    (7.5,-1.7)--(3.1,-1.7) -- (3.1,-1.2)
    (7.5,0.2) ++ (-90:0.7) ++(0:0.7) -- +(0:0.5)
    ;
    \coordinate  (wlr) at  ($(7.5,0.2) + (-90:0.7) +(0:0.7) + (0:0.6) + (90:0.25)$);
    \draw[preaction={fill, purple!30!white}, pattern={Lines[angle=90,distance=4pt]}]
     (wlr) -- ($(wlr)+(1.2,0)$) -- ($(wlr)+(1.2,-0.5)$)-- ($(wlr)+(0,-0.5)$)--(wlr);
    \draw[preaction={fill, myyellow}, pattern={Lines[angle=90,distance=4pt]}]
    (1.9,-1.7) -- (3.1,-1.7) -- (3.1,-1.2)  node[pos=0.7,label={[xshift=-18pt,yshift=-30pt]$\wbL$}]{}
 -- (1.9,-1.2)--(1.9,-1.7);
     \path
     (4,0.9) node {$\pipe$}
     (9.5,-1) node {$\wbR$}
     ;
     \draw [draw=none](5.7,1.3) -- (5.7,1.5) ;
     \draw [fill=orange!80!white] (1.3,0) rectangle (1.8,-1.7) node [pos=0.5,rotate=90]{Memory};
     \draw [fill=orange!80!white] (10.6,0.3) rectangle (10.1,-1.4) node [pos=0.5,rotate=90]{Memory};
     \path[ultra thick,blue,->]
     (0.2,.15) edge (1,.15)
     (1.5,.45) edge (2.3,.45)
     (2.3,.45) node[right] {\small Queue grows this way}
     (6.1,.45) edge (6.8,.45)
     ($(wlr)+(.2,-.25)$) edge ($(wlr)+(1,-.25)$)
     (6.8,-1.45) edge (6.1,-1.45)
     (2.9,-1.45) edge (2.1,-1.45);
  \end{tikzpicture}
  \hrule
  \caption{Simple queue-pair structure.}
  \label{fig:sqpair_structure}
\end{figure}
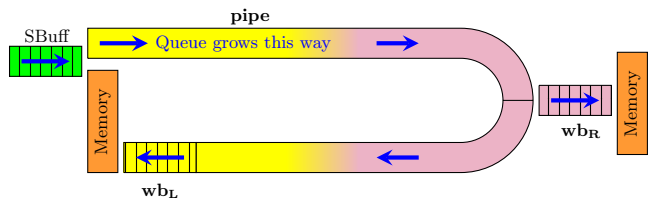


%% file: example-trans.tex
\begin{wrapfigure}{r}{0.4\textwidth}  
  \vspace{-2.5em}
\begin{mathpar}
\inferrule*{\phi}
{\prog, \Mem, \TSOmap, \Atm, \NQPmap \trp \prog', \Mem', \TSOmap', \Atm', \NQPmap'}
\end{mathpar}
\label{fig:example_transition}
  \vspace{-2.5em}
\end{wrapfigure}

%% file: opsem-combined.tex
\begin{figure}
  \small
\begin{mathpar}
\inferrule*{
	\prog \arrprog{\empL} \prog'
}{
	\prog, \Mem, \TSOmap, \Atm, \NQPmap \trp \prog'\!, \Mem, \TSOmap, \Atm, \NQPmap
}
\quad
\inferrule*{
	\Mem, \TSOmap, \Atm, \NQPmap \arreps \Mem', \TSOmap', \Atm', \NQPmap'
}{
	\prog, \Mem, \TSOmap, \Atm, \NQPmap \trp \prog, \Mem'\!, \TSOmap'\!, \Atm'\!, \NQPmap'\!
}
\\
\inferrule*{
  \prog \arrprog{\lb} \prog'
  \quad
  \Mem, \TSOmap, \Atm, \NQPmap \arrlab{\lb} \Mem'\!, \TSOmap'\!, \Atm'\!, \NQPmap'\!
}
{\prog,\Mem,\TSOmap,\Atm,\NQPmap \trp \prog',\Mem',\TSOmap',\Atm',\NQPmap'}
\end{mathpar}
\hrule
\caption{\rdmatso operational semantics with the program and hardware transitions given in \cref{fig:prog_transitions} and \cref{fig:hardware_transitions}}
\label{fig:opsem-combined}
\end{figure}

%% file: prog-trans.tex
\begin{figure}
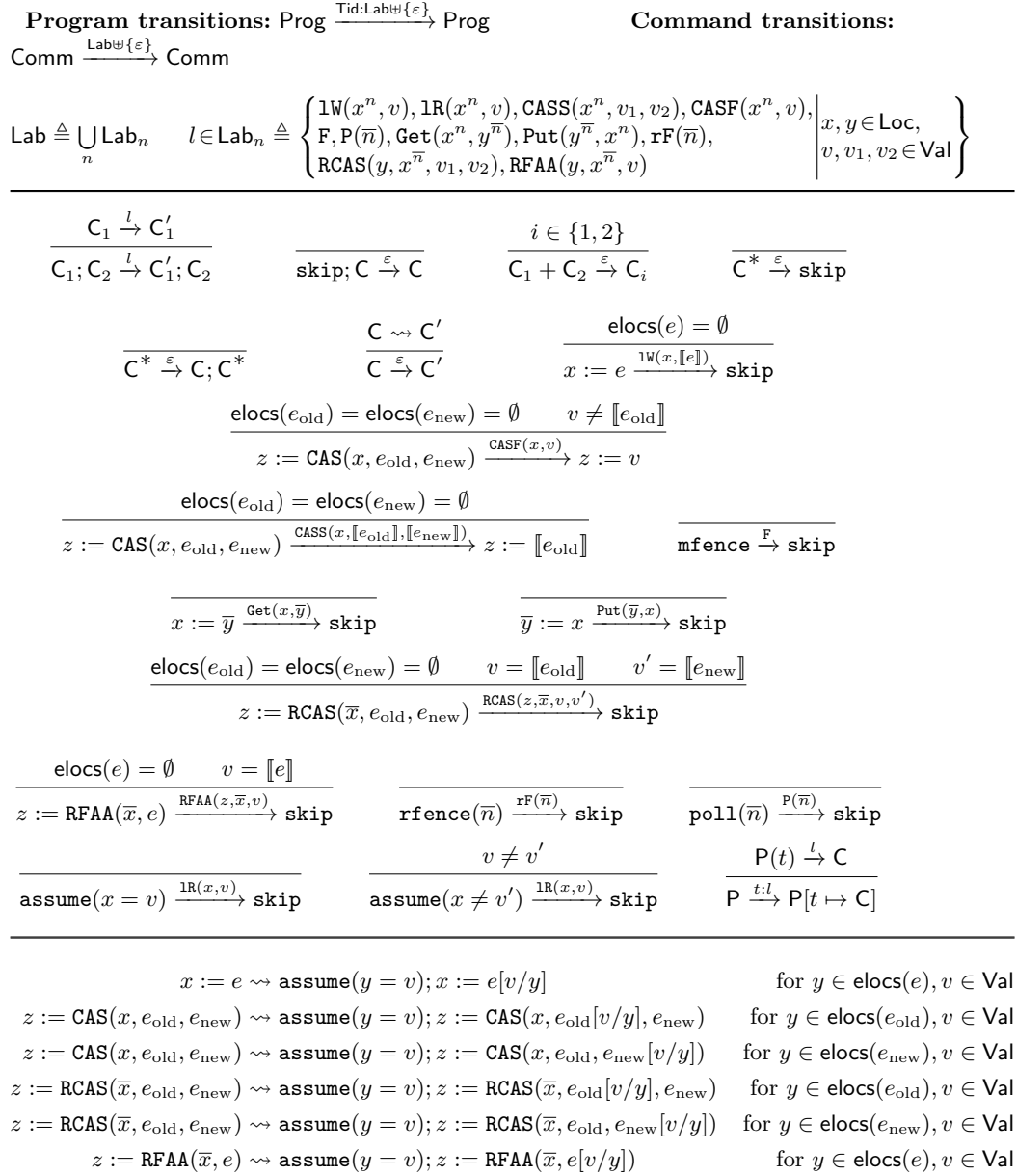

\footnotesize
\textbf{\;\;Program transitions:} $\Progs  \arrcomm{\!\Threads: \Labs \uplus \{\empL\} \!} \Progs$
\hfill
\textbf{\;\;Command transitions:} $\Comms \arrcomm{\Labs \uplus \{\empL\}\!} \Comms$
\;\;\\
\small
\[
\begin{array}{@{} l @{\hspace{15pt}} l @{}}
	\Labs \defeq \bigcup\limits_{\node} \Labs_\node
&
	\lb \!\in\! \Labs_\node \defeq
	\left\{
	\begin{array}{@{} l | l @{}}
		\begin{array}{@{} l @{}}
			\slW(x^\node,v), \slR(x^\node,v), \CASSucc(x^\node,v_1, v_2), \CASFail(x^\node,v), \\
			\slF, \slP(\neqn\node), \srget(\x^\node, \y^{\neqn \node}), \srput(\y^{\neqn \node},\x^\node),
\snF(\neqn\node), \\
      \RCAS(\y, x^{\neqn\node}, \val_{1}, \val_{2}),
      \RFAA(\y, x^{\neqn\node}, \val)
		\end{array}
		&
		\begin{array}{@{} l @{}}
			x, y \!\in\! \Var, \\
			v,v_1,v_2 \!\in\! \Vals
		\end{array}
	\end{array}
	\right\}
\end{array}
\]
\hrule
\hrule\vspace{-3pt}
\begin{mathpar}
\inferrule*{\comm[]_1 \arrcomm{\lb} {\comm[]}'_1}
{\comm[]_1;\comm[]_2 \arrcomm{\lb} {\comm[]}'_1;\comm[]_2}
\and
\inferrule*{ }
{\cskip;\comm[] \arrcomm{\empL} \comm[]}
\and
\inferrule*{i \in \set{1,2}}
{\comm[]_1+\comm[]_2 \arrcomm{\empL} \comm[]_i}
\and
\inferrule*{ }
{{\comm[]}^\kstar \arrcomm{\empL} \cskip}
\and
\inferrule*{ }
{{\comm[]}^\kstar \arrcomm{\empL} \comm[];{\comm[]}^\kstar}
\vsep
\and
\inferrule*{\comm[]\leadsto {\comm[]}'}
{\comm[] \arrcomm{\empL} {\comm[]}'}
\and
\inferrule*{\fv{\abexp} = \emptyset}
{\x \assign \abexp \arrcomm{\slW(x,\db{\abexp})} \cskip}
\and
\inferrule*{\fv{\abexp_\old} = \fv{\abexp_\new} = \emptyset \\
  v \neq \db{\abexp_\old}}
{\z \assign \clcas(\x,\abexp_\old,\abexp_\new)
  \arrcomm{\CASFail(x,v)} \z \assign v}
\vsep
\and
\inferrule*{\fv{\abexp_\old} = \fv{\abexp_\new} = \emptyset}
{\z \assign \clcas(\x,\abexp_\old,\abexp_\new)
 \arrcomm{\CASSucc(x,\db{\abexp_\old},\db{\abexp_\new})} \z \assign \db{\abexp_\old}}
\and
\inferrule*{ }
{\mfence \arrcomm{\slF}  \cskip}
\and
\inferrule*{ }
{\x \assign \neqn\y \arrcomm{\srget(\x,\neqn\y)} \cskip}
\vsep
\and
\inferrule*{ }
{\neqn\y \assign \x \arrcomm{\srput(\neqn\y,\x)} \cskip}
\and
\inferrule*{\fv{\abexp_\old} = \fv{\abexp_\new} = \emptyset \\
  v = \db{\abexp_\old} \\
  v' = \db{\abexp_\new}}
{\z \assign \RCAS(\neqn{\x},\abexp_\old,\abexp_\new)
  \arrcomm{\RCAS(\z, \neqn{\x}, v, v')} \cskip}
\and
\inferrule*{\fv{\abexp} = \emptyset \\
    v = \db{\abexp}}
{\z \assign \RFAA(\neqn{\x},\abexp)
  \arrcomm{\RFAA(\z, \neqn{\x}, v)} \cskip}
\and
\inferrule*{ }
{\cnfence{\neqn\node} \arrcomm{\snF(\neqn\node)} \cskip}
\and
\inferrule*{ }
{\poll(\neqn\node) \arrcomm{\slP(\neqn\node)} \cskip}
\vsep\vspace{-2pt}
\and
\inferrule*{ }
{\cassume(x = v) \arrcomm{\slR(x,v)}  \cskip}
\and
\inferrule*{v \neq v'}
{\cassume(x \neq v') \arrcomm{\slR(x,v)}  \cskip}
\and
\inferrule*{
	\prog(\threadt) \arrcomm{\lb} \comm[]
}
{
	\prog \arrprog{\lb} \prog\upvar{\threadt}{\comm[]}
}
\end{mathpar}
\hrule
\hrule
\begin{align*}
\x  \assign  \abexp &\leadsto \cassume(\y=v);\x  \assign  \abexp[v/\y] & \text{for\ } \y \in \fv{\abexp}, \val\in\Vals \\
\z  \assign  \clcas(\x,\abexp_\old,\abexp_\new) &\leadsto
\cassume(y=v); \z  \assign  \clcas(\x,\abexp_\old[v/y],\abexp_\new) & \text{for\ } y \in \fv{\abexp_\old}, \val\in\Vals \\
\z  \assign  \clcas(\x,\abexp_{\old},\abexp_\new) &\leadsto
\cassume(y=v); \z  \assign  \clcas(\x,\abexp_\old,\abexp_\new[v/y]) & \text{for\ } y \in \fv{\abexp_\new} , \val\in\Vals \\
\z  \assign  \RCAS(\neqn{x},\abexp_\old,\abexp_\new) &\leadsto
\cassume(y=v); \z  \assign  \RCAS(\neqn{x},\abexp_\old[v/y],\abexp_\new) & \text{for\ } y \in \fv{\abexp_\old}, \val\in\Vals \\
\z  \assign  \RCAS(\neqn{x},\abexp_{\old},\abexp_\new) &\leadsto
\cassume(y=v); \z  \assign  \RCAS(\neqn{x},\abexp_\old,\abexp_\new[v/y]) & \text{for\ } y \in \fv{\abexp_\new} , \val\in\Vals \\
\z  \assign  \RFAA(\neqn{x},\abexp) &\leadsto
\cassume(y=v); \z  \assign  \RFAA(\neqn{x},\abexp[v/y]) & \text{for\ } y \in \fv{\abexp}, \val\in\Vals
\end{align*}
\hrule\hrule
\caption{The \rdmatso program and command transitions}\label{fig:prog_transitions}
\end{figure}

%% file: hw-dom-trans.tex
\begin{figure}[p]
\thispagestyle{empty}
  \noindent
  \begin{adjustbox}{center,minipage=\textwidth}
  \centering
\footnotesize
\[
\begin{array}{@{} c @{}}

  \Mem \!\in\! \Mems  \defeq \Var \to \Vals
  \qquad

  \TSOmap  \!\in\! \TSOmaps \defeq \lambda \threadt \in \Threads. \STSOs_{\node(\threadt)} \\

  \Atm \!\in\! \RAMap \defeq \lambda \node. \set{\bot, \top}

  \qquad

  \NQPmap \!\in\! \SQPmaps \defeq \lambda \threadt. \big(\lambda \neqn{\node(\threadt)}. \SQPairs_\node^{\neqn\node}\big)
  \\

  \stso \!\in\! \STSOs_n \!\!\defeq \!\!
  \queueset{
    \x^\node \!\!\assign\! v, \!
    \y^{\neqn \node} \!\!\assign\! \x^\node, \!
    \x^\node \!\!\assign\! \y^{\neqn\node}, \!
    \RCAS(\z^\node,\x^{\neqn{\node}},v,v'), \!
    \RFAA(\z^\node,\x^{\neqn{\node}},v), \!
    \cnfence{\neqn\node}}\\

  \simqp \in \SQPairs_\node^{\neqn\node} \defeq
  \Pipes_\node^{\neqn\node} \times
  \WBRs_\node^{\neqn\node} \times
  \WBLs_\node^{\neqn\node}
  \\

  \wbL \!\in\! \WBLs_\node^{\neqn\node} \defeq
  \queueset{\cn,\x^{\node} \assign v}
  \qquad

  \wbR \!\in\! \WBRs_\node^{\neqn\node} \defeq\!
  \queueset{\y^{\neqn\node} \assign v,y^{\neqn\node} \assign_A v} \\

  \pipe \in \Pipes_\node^{\neqn\node} \defeq
  \queueset{
  \begin{array}{l}
  \y^{\neqn\node} \assign \x^\node,
  \y^{\neqn\node} \assign v,
  \y^{\neqn\node} \assign_A v,
  \doneE,
  \x^\node \assign \y^{\neqn\node},
  \x^\node \assign v, \\
  \RCAS(\z^\node,\x^{\neqn{\node}},v,v'), \!
  \RFAA(\z^\node,\x^{\neqn{\node}},v), \!
  \cnfence{\neqn\node}
  \end{array}
  }
\end{array}
\]
%
\hrule
\hrule
\hrule
\begin{mathpar}
\inferrule*{\TSOmap'\! = \TSOmap\upvar{\threadt}
  {(x \assign v) \cdot \TSOmap(\threadt)}}
{
  \Mem, \TSOmap, \Atm, \NQPmap
  \arrlab{\slW(x,v)\!}
  \Mem, \TSOmap'\!, \Atm, \NQPmap
}
\quad
\inferrule*{(\Mem \updmem \TSOmap(\threadt))(x) = v}
{
  \Mem, \TSOmap, \Atm, \NQPmap
  \arrlab{\slR(x,v)}
  \Mem, \TSOmap, \Atm, \NQPmap
}
\vspace{-4pt}
\and
\inferrule*{\TSOmap(\threadt) = \varepsilon \\ \Mem(x) = v_1}
{
  \Mem, \TSOmap, \Atm, \NQPmap
  \arrlab{\CASSucc(x,v_1,v_2)\!}
  \Mem\upvar{x}{v_2}, \TSOmap, \Atm, \NQPmap
}
\vspace{-4pt}
\and
\inferrule*{\TSOmap(\threadt) = \varepsilon \\ \Mem(x) = v}
{
  \Mem, \TSOmap, \Atm, \NQPmap
  \arrlab{\CASFail(x,v)\!}
  \Mem, \TSOmap, \Atm, \NQPmap
}
\and
\inferrule*{\TSOmap(\threadt) = \varepsilon}
{
  \Mem, \TSOmap, \Atm, \NQPmap
  \arrlab{\slF}
  \Mem, \TSOmap, \Atm, \NQPmap
}
\and
\inferrule*{\TSOmap(\threadt) = \stso \cdot (\x \assign v)}
{
  \Mem, \TSOmap, \Atm, \NQPmap
  \arreps
    \Mem\upvar{x}{v}, \TSOmap\upvar{\threadt}{\stso}, \Atm, \NQPmap
}
\vspace{-4pt}
\and
\inferrule*{\TSOmap(\threadt) \!=\! \stso \!\cdot\! \rcomm \\
  \rcomm \!\in\! \set{\x  \assign \y^{\neqn\node}, \y^{\neqn\node} \assign \x, \RCAS(\z, \neqn{\x}, \val, \val'), \RFAA(\z, \neqn{\x}, \val), \cnfence{\neqn\node}}\\
  \NQPmap(\threadt)(\neqn\node) \!=\! \simqp\\
  \simqp' = \simqp\upvar{\pipe}{\rcomm \cdot \simqp.\pipe}
}
{
  \Mem, \TSOmap, \Atm, \NQPmap
  \arreps
    \Mem, \TSOmap \upvar{\threadt}{\stso}, \Atm,
    \NQPmap\upvar{\threadt}{\NQPmap(\threadt)\upvar{\neqn\node}{\simqp'}}
}
\vspace{-4pt}
\and
\inferrule*{\TSOmap' = \TSOmap\upvar{\threadt}
  {(\x \assign \neqn\y) \cdot \TSOmap(\threadt)}}
{
  \Mem, \TSOmap, \Atm, \NQPmap
  \arrlab{\srget(\x,\neqn\y)}
  \Mem, \TSOmap', \Atm, \NQPmap
}
\and
\inferrule*{\TSOmap' = \TSOmap\upvar{\threadt}
  {(\neqn\y \assign \x) \cdot \TSOmap(\threadt)}}
{
  \Mem, \TSOmap, \Atm, \NQPmap
  \arrlab{\srput(\neqn\y,\x)}
  \Mem, \TSOmap', \Atm, \NQPmap
}
\\
\and
\inferrule*{\TSOmap' = \TSOmap\upvar{\threadt}{\RCAS(\z, \neqn{\x}, \val, \val') \cdot \TSOmap(\threadt)}} { \Mem, \TSOmap, \Atm, \NQPmap \arrlab{\RCAS(\z, \neqn{\x}, \val, \val')} \Mem, \TSOmap', \Atm, \NQPmap}%
\and
\inferrule*{\TSOmap' = \TSOmap\upvar{\threadt}{\RFAA(\z, \neqn{\x}, \val) \cdot \TSOmap(\threadt)}} { \Mem, \TSOmap, \Atm, \NQPmap \arrlab{\RFAA(\z, \neqn{\x}, \val)} \Mem, \TSOmap', \Atm, \NQPmap}%
\and
\inferrule*{\TSOmap' = \TSOmap\upvar{\threadt}
  {(\cnfence{\neqn\node}) \cdot \TSOmap(\threadt)}}
{
  \Mem, \TSOmap, \Atm, \NQPmap
  \arrlab{\snF(\neqn\node)}
  \Mem, \TSOmap', \Atm, \NQPmap
}
\vspace{-4pt}
\and
\inferrule*{
  \NQPmap(\threadt)(\neqn\node) \!=\! \simqp
  \\
  \simqp.\wbL \!=\! \alpha \cdot \cn
  \\
  \simqp' \!=\! \simqp\upvar{\wbL}{\alpha}
}
{
  \Mem, \TSOmap, \Atm, \NQPmap
  \arrlab{\slP(\neqn\node)\!\!}
    \Mem, \TSOmap, \Atm,  \NQPmap\upvar{\threadt}{\NQPmap(\threadt)\upvar{\neqn\node}{\simqp'}}
}
\vspace{-4pt}
\and
\inferrule*{
  \Mem, \Atm, \NQPmap(\threadt)(\neqn\node) \pipearrow \Mem', \Atm', \simqp
  \quad (\cref{fig:qp_transitions})
}
 {
   \Mem, \TSOmap, \Atm, \NQPmap
   \arrlab{\empL\!\!}
   \Mem'\!, \TSOmap, \Atm', \NQPmap\upvar{\threadt}{\NQPmap(\threadt)\upvar{\neqn\node}{\simqp}}
  }
\end{mathpar}
\vspace{5pt}
%
%
with
\quad
$
  (\Mem\updmem\alpha)(\x) \defeq
  \begin{cases}
    \val
    & \text{if } \alpha = \beta \cdot (\x \assign \val) \cdot - \land \forall \val'.\, \x \assign \val' \not\in \beta \\
    \Mem(\x)
    & \text{if } \forall \val.\, \x \assign \val \not\in \alpha \\
  \end{cases}
$
\hfill
\phantom{a}
\hrule
\hrule
\hrule
\caption{\rdmatso simplified hardware domains (above) and hardware transitions
  (below)}\label{fig:hardware_transitions}
  \end{adjustbox}
\end{figure}

%% file: qp-trans.tex
\begin{figure}[p]
\scriptsize
\begin{mathpar}
\namedinferrule{\simqp.\pipe = \alpha \cdot (\cnfence{\neqn\node})}
{
  \Mem, \Atm, \simqp \pipearrow
    \Mem, \Atm, \simqp\upvar{\pipe}{\alpha}
}{rfence}
\and
\namedinferrule{
  {
  \begin{array}{@{} c @{}}
    \simqp.\pipe = \alpha \cdot (\neqn{\y} \assign \x) \cdot \beta
    \quad\wbL \!\in\! \queueset{\cn}\\
    \beta \in \queueset{\y' \assign \val', \y' \assign_A \val',
    \neqn{\y'} \assign \val', \x' \assign \neqn{\y'},
    \RCAS(z, \neqn{x}, v, v'), \RFAA(z, \neqn{x}, v), \doneE}
  \end{array}
  }
}{
  \Mem, \Atm, \simqp \pipearrow
  \Mem, \Atm, \simqp\upvar{\pipe}{\alpha \cdot (\neqn{\y} \assign \Mem(\x))
    \cdot \beta}
}{local-read}
\and
\namedinferrule{
  {
  \begin{array}{@{} c @{}}
    \simqp.\pipe = \alpha \cdot (\neqn{\y} \assign \val) \cdot \beta\\
    \beta \in \queueset{\x' \assign \neqn{\y'}, \x' \assign \val', \doneE}\\
    \simqp'\! = \simqp\upvar{\pipe}{\alpha \cdot \doneE \cdot \beta}\upvar{\wbR}{
    (\neqn{\y} \assign \val) \cdot \simqp.\wbR}
  \end{array}
  }
}{
  \Mem, \Atm, \simqp
  \pipearrow
  \Mem, \Atm, \simqp'
}{send-write}
\and
\namedinferrule{
  \simqp.\wbR = \alpha \cdot (\neqn{\y} \assign \val)
}{
  \Mem, \Atm, \simqp
  \pipearrow
  \Mem\upvar{\neqn{\y}}{\val}, \Atm, \simqp\upvar{\wbR}{\alpha}
}{remote-write}
\and
\namedinferrule{
  {
  \begin{array}{@{} c @{}}
    \simqp.\pipe = \alpha \cdot \doneE
    \\
    \simqp'\! = \simqp\upvar{\pipe}{\alpha}\upvar{\wbL}{\cn \cdot \simqp.\wbL}
  \end{array}
  }
}{
  \Mem, \Atm, \simqp
  \pipearrow
  \Mem, \Atm, \simqp'
}{ack}
\and
\namedinferrule{
  {
  \begin{array}{@{} c @{}}
    \simqp.\pipe \!=\! \alpha \!\cdot\! (\x \assign \neqn{\y}) \!\cdot\! \beta
    \qquad
    \beta \!\in\! \queueset{\x' \assign \neqn{\y'}, \x' \assign \val',\doneE}
    \\
    \simqp.\wbR\! =\! \varepsilon
    \quad
    \simqp'\! \!=\! \simqp\upvar{\pipe}{\alpha \cdot (\x \assign \Mem(\neqn{\y})) \!\cdot\! \beta}
   \end{array}
   }
}{
  \Mem, \Atm, \simqp
  \pipearrow
    \Mem, \Atm, \simqp'
}{remote-read}
\and
\namedinferrule{
  {
  \begin{array}{@{} c @{}}
    \simqp.\pipe = \alpha \cdot (\x \assign \val)
    \qquad
    \\
    \simqp'\! = \simqp\upvar{\pipe}{\alpha}\upvar{\wbL}{\cn \cdot (\x \assign \val) \cdot \simqp.\wbL}
  \end{array}
   }
}{
  \Mem, \Atm, \simqp \pipearrow \Mem, \Atm, \simqp'
}{send-read}
\and
\namedinferrule{
  {
  \begin{array}{@{} c @{}}
    \simqp.\wbL = \alpha \cdot (\x \assign \val) \cdot \beta
    \quad \beta \in \queueset{\cn} \\
    \simqp'\! = \simqp\upvar{\wbL}{\alpha \cdot \beta}
  \end{array}
  }
}{
  \Mem, \Atm, \simqp \pipearrow \Mem\upvar{\x}{\val}, \Atm, \simqp'
}{local-write}
\and
\namedinferrule{
  {
  \begin{array}{@{} c @{}}
    \simqp.\pipe = \alpha \cdot \RCAS(\z, \neqn{\x}, \val, \val') \cdot \beta
    \quad \simqp.\wbR = \epsilon \\
    \Atm(\node(\neqn{x})) = \bot
    \quad \Mem(\neqn{\x}) \neq \val
    \quad \beta \in \queueset{\x' \assign \neqn{\y'}, \x' \assign \val', \doneE}\\
    \simqp' = \simqp\upvar{\pipe}{\alpha \cdot (z \assign \Mem(\neqn{x})) \cdot \beta}
  \end{array}
   }
}{
  \Mem, \Atm, \simqp \pipearrow \Mem, \Atm, \simqp'
}{nCAS-F}
\and
\namedinferrule{
  {
  \begin{array}{@{} c @{}}
    \simqp.\pipe = \alpha \cdot \RCAS(\z, \neqn{\x}, \val, \val') \cdot \beta
    \quad \simqp.\wbR = \epsilon \\
    \Mem(\neqn{\x}) = \val
    \quad \beta \in \queueset{\x' \assign \neqn{\y'}, \x' \assign \val', \doneE} \\
    \Atm(\node(\neqn{x})) = \bot
    \quad \Atm' = \Atm\upvar{\node(\neqn{x})}{\top} \\
  \end{array}
   }
}{
  \Mem, \Atm, \simqp \pipearrow \Mem, \Atm', \simqp\upvar{\pipe}{\alpha \cdot (z \assign v) \cdot (\neqn{x} \assign_A v') \cdot \beta}
}{nCAS-S}%
\and
\namedinferrule{
  {
  \begin{array}{@{} c @{}}
    \simqp.\pipe = \alpha \cdot \RFAA(\z, \neqn{\x}, \val) \cdot \beta
    \quad \simqp.\wbR = \epsilon \\
    \Mem(\neqn{\x}) + \val = \val'
    \quad \beta \in \queueset{\x' \assign \neqn{\y'}, \x' \assign \val', \doneE} \\
    \Atm(\node(\neqn{x})) = \bot
    \quad \Atm' = \Atm\upvar{\node(\neqn{x})}{\top} \\
  \end{array}
   }
}{
  \Mem, \Atm, \simqp \pipearrow \Mem, \Atm', \simqp\upvar{\pipe}{\alpha \cdot (z \assign v) \cdot (\neqn{x} \assign_A v') \cdot \beta}
}{nFAA}
\and
\namedinferrule{
  {
  \begin{array}{@{} c @{}}
    \simqp.\pipe = \alpha \cdot (\neqn{x} \assign_A v) \cdot \beta \\
    \wbR' =  (\neqn{x} \assign_A \val) \cdot \simqp.\wbR \\
    \quad \beta \in \queueset{\x' \assign \neqn{\y'}, \x' \assign \val', \doneE} \\
    \simqp' = \simqp\upvar{\pipe}{\alpha\cdot\beta}\upvar{\wbR}{\wbR'}
  \end{array}
   }
}{
  \Mem, \Atm, \simqp \pipearrow \Mem, \Atm, \simqp'
}{nRMW-1}%
\and
\namedinferrule{
  {
  \begin{array}{@{} c @{}}
    \simqp.\wbR = \alpha \cdot (\neqn{x} \assign_A v) \\
    \quad \Atm' = \Atm\upvar{\node(\neqn{x})}{\bot} \\
    \simqp' = \simqp\upvar{\wbR}{\alpha}
  \end{array}
   }
}{
  \Mem, \Atm, \simqp \pipearrow \Mem\upvar{\neqn{x}}{v}, \Atm', \simqp'
}{nRMW-2}%
\end{mathpar}
\caption{Queue-pair transitions of the simplified \rdmatso operational semantics}\label{fig:qp_transitions}
\end{figure}

%% file: reordering.tex


\newcommand{\csqp}{sqp\xspace}

\setcounter{rowcounter}{1}
\setcounter{colcounter}{1}
\definecolor{cellname}{rgb}{0.5,0.5,0.5}
\definecolor{highlightcoloralt}{HTML}{FFFF66}


\begin{figure}
%
%
\begin{center}
\begin{tabular}{c|c|c|c|c|c|c|c|c?c|c|c|c|c|c|c|}
\multicolumn{3}{c}{} & \multicolumn{12}{c}{Later in Program Order} \\
  \hhline{~|-|-|-|-|-|-|-|-|-|-|-|-|-|-|-|}
\multicolumn{1}{c|}{} & \multicolumn{3}{c|}{\multirow{3}{*}{\LARGE \ippo}} & \multicolumn{5}{c?}{CPU} & \multicolumn{7}{c|}{NIC} \\
  \hhline{~|~~~|-|-|-|-|-|-|-|-|-|-|-|-|}
\multicolumn{1}{c|}{} & \multicolumn{3}{c|}{} & \colheader & \colheader & \colheader & \colheader
   & \colheader & \colheader & \colheader & \colheader & \colheader & \colheader & \colheader & \colheader\\
  \hhline{~|~~~|-|-|-|-|-|-|-|-|-|-|-|-|}
\multicolumn{1}{c|}{} & \multicolumn{3}{c|}{}
                      & \lR   & \lW   & \RMW  & \lF   & \lP   & \nlR  & \nrW  & \narR & \narW & \nrR  & \nlW  & \nF\\
  \hhline{~|-|-|-|-|-|-|-|-|-|-|-|-|-|-|-|}
\multirow{10}{*}{\rotatebox{90}{Earlier in Program Order}} &
\multirow{5}{*}{\rotatebox{90}{CPU}}
   & \rowheader{\lR}  & \cyes & \cyes & \cyes & \cyes & \cyes & \cyes & \cyes & \cyes & \cyes & \cyes & \cyes & \cyes\\
\hhline{~|~|-|-|-|-|-|-|-|-|-|-|-|-|-|-|}
&  & \rowheader{\lW}  & \cyes & \cyes & \cyes & \cyes & \cyes & \cyes & \cyes & \cyes & \cyes & \cyes & \cyes & \cyes\\
\hhline{~|~|-|-|-|-|-|-|-|-|-|-|-|-|-|-|}
&  & \rowheader{\RMW} & \cyes & \cyes & \cyes & \cyes & \cyes & \cyes & \cyes & \cyes & \cyes & \cyes & \cyes & \cyes\\
\hhline{~|~|-|-|-|-|-|-|-|-|-|-|-|-|-|-|}
&  & \rowheader{\lF}  & \cyes & \cyes & \cyes & \cyes & \cyes & \cyes & \cyes & \cyes & \cyes & \cyes & \cyes & \cyes\\
\hhline{~|~|-|-|-|-|-|-|-|-|-|-|-|-|-|-|}
&  & \rowheader{\lP}  & \cyes & \cyes & \cyes & \cyes & \cyes & \cyes & \cyes & \cyes & \cyes & \cyes & \cyes & \cyes\\
\hhline{~|-|-|-|-|-|-|-|-|-|-|-|-|-|-|-|}
\hhline{~|-|-|-|-|-|-|-|-|-|-|-|-|-|-|-|}
& \multirow{5}{*}{\rotatebox{90}{NIC}}
   & \rowheader{\nlR} & \cno  & \cno  & \cno  & \cno  & \cno  & \csqp  & \csqp  & \csqp  & \csqp  & \csqp  & \csqp  & \csqp \\
\hhline{~|~|-|-|-|-|-|-|-|-|-|-|-|-|-|-|}
&  & \rowheader{\nrW} & \cno  & \cno  & \cno  & \cno  & \cno  & \cno  & \csqp  & \csqp  & \csqp  & \csqp  & \csqp  & \csqp \\
\hhline{~|~|-|-|-|-|-|-|-|-|-|-|-|-|-|-|}
& & \rowheader{\narR}& \cno  & \cno  & \cno  & \cno  & \cno  & \cno  & \csqp  & \csqp  & \csqp  & \csqp  & \csqp  & \csqp \\ \hhline{~|~|-|-|-|-|-|-|-|-|-|-|-|-|-|-|}
& & \rowheader{\narW}& \cno  & \cno  & \cno  & \cno  & \cno  & \cno  & \csqp  & \csqp  & \csqp  & \csqp  & \csqp  & \csqp \\ \hhline{~|~|-|-|-|-|-|-|-|-|-|-|-|-|-|-|}
&  & \rowheader{\nrR} & \cno  & \cno  & \cno  & \cno  & \cno  & \cno  & \cno  & \cno  & \cno  & \cno  & \csqp  & \csqp \\
\hhline{~|~|-|-|-|-|-|-|-|-|-|-|-|-|-|-|}
&  & \rowheader{\nlW} & \cno  & \cno  & \cno  & \cno  & \cno  & \cno  & \cno  & \cno  & \cno  & \cno  & \csqp  & \csqp \\
\hhline{~|~|-|-|-|-|-|-|-|-|-|-|-|-|-|-|}
&  & \rowheader{\nF}  & \cno  & \cno  & \cno  & \cno  & \cno  & \csqp  & \csqp  & \csqp  & \csqp  & \csqp  & \csqp  & \csqp \\
\hhline{~|-|-|-|-|-|-|-|-|-|-|-|-|-|-|-|}
\end{tabular}
\end{center}

\setcounter{rowcounter}{1}
\setcounter{colcounter}{1}
\vspace{3pt}
\begin{center}
\begin{tabular}{c|c|c|c|c|c|c|c|c?c|c|c|c|c|c|c|}
\multicolumn{3}{c}{} & \multicolumn{12}{c}{Later in Program Order} \\
  \hhline{~|-|-|-|-|-|-|-|-|-|-|-|-|-|-|-|}
\multicolumn{1}{c|}{} & \multicolumn{3}{c|}{\multirow{3}{*}{\LARGE \oppo}} & \multicolumn{5}{c?}{CPU} & \multicolumn{7}{c|}{NIC} \\
  \hhline{~|~~~|-|-|-|-|-|-|-|-|-|-|-|-|}
\multicolumn{1}{c|}{} & \multicolumn{3}{c|}{} & \colheader & \colheader & \colheader & \colheader
   & \colheader & \colheader & \colheader & \colheader & \colheader & \colheader & \colheader & \colheader \\
  \hhline{~|~~~|-|-|-|-|-|-|-|-|-|-|-|-|}
\multicolumn{1}{c|}{} & \multicolumn{3}{c|}{}
                      & \lR   & \lW   & \RMW  & \lF   & \lP   & \nlR  & \nrW  & \narR & \narW & \nrR  & \nlW  & \nF\\
  \hhline{~|-|-|-|-|-|-|-|-|-|-|-|-|-|-|-|}
\multirow{12}{*}{\rotatebox{90}{Earlier in Program Order}} &
\multirow{5}{*}{\rotatebox{90}{CPU}}
   & \rowheader{\lR}  & \cyes & \cyes & \cyes & \cyes & \cyes & \cyes & \cyes & \cyes & \cyes & \cyes & \cyes & \cyes\\
\hhline{~|~|-|-|-|-|-|-|-|-|-|-|-|-|-|-|}
&  & \rowheader{\lW}  & \cno  & \cyes & \cyes & \cyes & \cno  & \cyes & \cyes & \cyes & \cyes & \cyes & \cyes & \cyes\\
\hhline{~|~|-|-|-|-|-|-|-|-|-|-|-|-|-|-|}
&  & \rowheader{\RMW} & \cyes & \cyes & \cyes & \cyes & \cyes & \cyes & \cyes & \cyes & \cyes & \cyes & \cyes & \cyes\\
\hhline{~|~|-|-|-|-|-|-|-|-|-|-|-|-|-|-|}
&  & \rowheader{\lF}  & \cyes & \cyes & \cyes & \cyes & \cyes & \cyes & \cyes & \cyes & \cyes & \cyes & \cyes & \cyes\\
\hhline{~|~|-|-|-|-|-|-|-|-|-|-|-|-|-|-|}
&  & \rowheader{\lP}  & \cyes & \cyes & \cyes & \cyes & \cyes & \cyes & \cyes & \cyes & \cyes & \cyes & \cyes & \cyes\\
\hhline{~|-|-|-|-|-|-|-|-|-|-|-|-|-|-|-|}
\hhline{~|-|-|-|-|-|-|-|-|-|-|-|-|-|-|-|}
& \multirow{5}{*}{\rotatebox{90}{NIC}}
   & \rowheader{\nlR} & \cno  & \cno  & \cno  & \cno  & \cno  & \csqp  & \csqp  & \csqp  & \csqp  & \csqp  & \csqp  & \csqp \\
\hhline{~|~|-|-|-|-|-|-|-|-|-|-|-|-|-|-|}
&  & \rowheader{\nrW} & \cno  & \cno  & \cno  & \cno  & \cno  & \cno  & \csqp  & \csqp  & \csqp  & \csqp  & \csqp  & \cno \\
\hhline{~|~|-|-|-|-|-|-|-|-|-|-|-|-|-|-|}
& & \rowheader{\narR}& \cno  & \cno  & \cno  & \cno  & \cno  & \cno  & \csqp  & \csqp  & \csqp  & \csqp  & \csqp  & \csqp \\
\hhline{~|~|-|-|-|-|-|-|-|-|-|-|-|-|-|-|}
& & \rowheader{\narW}& \cno  & \cno  & \cno  & \cno  & \cno  & \cno  & \csqp  & \csqp  & \csqp  & \csqp  & \cno  & \cno \\
\hhline{~|~|-|-|-|-|-|-|-|-|-|-|-|-|-|-|}
&  & \rowheader{\nrR} & \cno  & \cno  & \cno  & \cno  & \cno  & \cno  & \cno  & \cno  & \cno  & \cno  & \csqp  & \csqp \\
\hhline{~|~|-|-|-|-|-|-|-|-|-|-|-|-|-|-|}
&  & \rowheader{\nlW} & \cno  & \cno  & \cno  & \cno  & \cno  & \cno  & \cno  & \cno  & \cno  & \cno  & \csqp  & \cno \\
\hhline{~|~|-|-|-|-|-|-|-|-|-|-|-|-|-|-|}
&  & \rowheader{\nF}  & \cno  & \cno  & \cno  & \cno  & \cno  & \csqp  & \csqp  & \csqp  & \csqp  & \csqp  & \csqp  & \csqp \\
\hhline{~|-|-|-|-|-|-|-|-|-|-|-|-|-|-|-|}
\end{tabular}
\end{center}

\caption{The \rdmatso ordering constraints on $\ippo$ (above) and $\oppo$ (below), 
 where \cyes denotes that instructions are ordered (and cannot be reordered), \cno denotes they are not ordered (and may be reordered), and \csqp denotes they are ordered iff they are on the same queue pair.}
\label{fig:ippo-oppo}
\end{figure}


%% file: event-sequence.tex
\begin{figure*}
\begin{mathpar}
\inferrule*{
  C \leadsto C' \\
  C' \actseq s
}{
  C \actseq s
}
\and
\inferrule*{
  C_1 \actseq s_1 \\
  C_2 \actseq s_2
}{
  C_1;C_2 \actseq s_1,s_2
}
\and
\inferrule*{
  \fv{\abexp} = \emptyset
}{
  \x \assign \abexp \actseq \lW(\x,\db{\abexp})
}
\and
\inferrule*{
  \fv{\abexp_\old} = \fv{\abexp_\new} = \emptyset \\\\
  \z \assign \db{\abexp_\old} \actseq s
}{
  \z \assign \clcas(\x,\abexp_\old,\abexp_\new) \actseq
  \RMW(\x,\db{\abexp_\old},\db{\abexp_\new}),s
}
\and
\inferrule*{
  \fv{\abexp_\old} = \fv{\abexp_\new} = \emptyset \\\\
  \val \neq \db{\abexp_\old} \\
  \z \assign v \actseq s
}{
  \z \assign \clcas(\x,\abexp_\old,\abexp_\new) \actseq
  \lF,\lR(\x,\val),s
}
\and
\inferrule*{
}{
  \mfence \actseq \lF
}
\and
\inferrule*{
}{
  \cassume(\x = \val) \actseq \lR(\x,\val)
}
\and
\inferrule*{
  \val'\neq \val
}{
  \cassume(\x \neq \val) \actseq \lR(\x,\val')
}
\and
\inferrule*{
}{
  \x \assign y^{\neqn\node} \actseq
  \nrR(y^{\neqn\node},\val),\nlW(\x,\val,\neqn{\node})
}
\and
\inferrule*{
}{
  y^{\neqn\node} \assign \x \actseq
  \nlR(\x,\val,\neqn{\node}),\nrW(y^{\neqn\node},\val)
}
\and
\inferrule*{
}{
  \cnfence{\neqn\node} \actseq \nF(\neqn\node)
}
\and
\inferrule*{
}{
  \poll(\neqn\node) \actseq \lP(\neqn\node)
}
\and
\inferrule*{
}{
  \cskip \actseq \epsilon
}
\and
\inferrule*{
  \fv{e_{old}} = \fv{e_{new}} = \emptyset \\
  \val \neq \db{e_{old}}
}{
  z \assign \rcas(x^{\neqn\node}, e_{old}, e_{new}) \actseq
  \narR(x^{\neqn\node}, \val), \nlW(z, \val, \neqn\node)
}
\and
\inferrule*{
  \fv{e_{old}} = \fv{e_{new}} = \emptyset
}{
  z \assign \rcas(x^{\neqn\node}, e_{old}, e_{new}) \actseq
  \narR(x^{\neqn\node}, \db{e_{old}}), \narW(x^{\neqn\node}, \db{e_{new}}), \nlW(z, \db{e_{old}}, \neqn\node)
}
\and
\inferrule*{
  \fv{e} = \emptyset \\
  \val' = \val + \db{e}
}{
  z \assign \rfaa(x^{\neqn\node}, e) \actseq
  \narR(x^{\neqn\node}, \val), \narW(x^{\neqn\node}, \val'), \nlW(z, \val, \neqn\node)
}
\end{mathpar}
\caption{Label Sequences Construction}
\label{fig:action-sequence}
\end{figure*}

%% file: op-decl-proof.tex

\newcommand{\PushTSO}{\ensuremath{\mathtt{Push}}\xspace}
\newcommand{\PushNIC}{\ensuremath{\mathtt{NIC}}\xspace}
\newcommand{\lE}{\ensuremath{\mathcal{E}}\xspace}
\newcommand{\lCN}{\ensuremath{\mathsf{CN}}\xspace}

\newcommand{\aupdmem}{\blacktriangleleft}

\newcommand{\actrtc}[2][\threadt]{\ctrtc[#1:#2]}

\newcommand{\diff}{\bowtie}
\newcommand{\nodup}{\ensuremath{\mathsf{nodup}}\xspace}
\newcommand{\relevant}{\ensuremath{\mathbf{Relevant}}\xspace}

\newcommand{\succpi}{\succ_{\pi}}
\renewcommand{\precpi}{\prec_{\pi}} 

\newcommand{\wfpipe}{\ensuremath{\mathsf{wfpipe}}\xspace}

\newcommand{\backComp}{\ensuremath{\mathsf{backComp}}\xspace}

\newcommand{\wfrdCPU}{\ensuremath{\mathsf{wfrdCPU}}\xspace}
\newcommand{\wfrdNIC}{\ensuremath{\mathsf{wfrdNIC}}\xspace}

\newcommand{\mktso}{\ensuremath{\mathsf{mksbuff}}\xspace}
\newcommand{\mkatm}{\ensuremath{\mathsf{chkatm}}\xspace}
\newcommand{\mkpipe}{\ensuremath{\mathsf{mkpipe}}\xspace}
\newcommand{\mkwbR}{\ensuremath{\mathsf{mkwbR}}\xspace}
\newcommand{\mkwbL}{\ensuremath{\mathsf{mkwbL}}\xspace}

\newcommand{\generates}[3]{\ensuremath{#1 \text{ generates } #2 \text{ in } #3}}

\newcommand{\extend}{\ensuremath{\mathsf{extend}}\xspace}




\input{annotated-semantics}


\clearpage
\input{annot-path}


\clearpage
\input{annot-to-decl}


\clearpage
\input{decl-to-annot}


\clearpage
\input{op-vs-annot}



%% file: annotated-semantics.tex


\subsection{Annotated Labels and Inference Rules}
\label{sec:annot-sem-rules}


On top of the 12 labels presented in \Cref{sec:decl-sem}, we create six
new labels: $\rput(\neqn\y,\x)$, $\rget(\x,\neqn\y)$, $\rcas(\z,\neqn{\x},\val,\upd)$, $\rfaa(\z,\neqn{x},\upd)$, $\nlEX(\neqn{\node})$, and
$\nrEX(\neqn{\node})$. These labels can also be used to create events (when
bundled with an event identifier and a thread identifier).


We note $\ActionsExt$ the extended set of all events, including the six new
labels.

Recall that $\Read = \lR \cup \RMW \cup \nlR \cup \nrR \cup \narR \subseteq \ActionsExt$
and $\Write = \lW \cup \RMW \cup \nlW \cup \nrW \cup \narW \subseteq \ActionsExt$. We also
note $\nEX = \nlEX \cup \nrEX$ and $\rrmw = \rcas \cup \rfaa$.

\begin{figure}[h]
  $\lambda \in \aLabels$
\begin{align*}
  \lambda \defeq\;
  &\mid \lR  \tup{r,w}    &\text{where}\;& r \in \lR, w \in \Write, \locveq(r,w) \\
  &\mid \lW  \tup{w}      &\text{where}\;& w \in \lW \\
  &\mid \RMW \tup{u,w}    &\text{where}\;& u \in \RMW, w \in \Write, \locveq(u,w) \\
  &\mid \lF  \tup{f}      &\text{where}\;& f \in \lF \\
  &\mid \PushTSO\tup{a}   &\text{where}\;& a \in (\rput \cup \rget \cup \rcas \cup \rfaa \cup \nF) \\ 
  &\mid \PushNIC\tup{a}   &\text{where}\;& a \in (\rput \cup \rget \cup \rcas \cup \rfaa \cup \nF) \\ 
  &\mid \nlR \tup{r,w,a,w'}&\text{where}\;& r \in \nlR, w \in \Write, a \in \rput, w' \in \nrW, \locveq(r,w), \\
  &                       &               & \loc_r(a) = \loc(r), \loc_w(a) = \loc(w'), \valr(r) = \valw(w') \\
  &\mid \nrR \tup{r,w,a,w'}&\text{where}\;& r \in \nrR, w \in \Write, a \in \rget, w' \in \nlW, \locveq(r,w), \\
  &                       &               & \loc_r(a) = \loc(r), \loc_w(a) = \loc(w'), \valr(r) = \valw(w') \\
  &\mid \narR \tup{r,w,a,w',w''}&\text{where}\;& r \in \narR, w \in \Write, a \in \rRMW, w' \in \nlW, w'' \in \narW, \\
  &                       &               & \locveq(r,w), \loc_{r}(a) = \loc(r) = \loc(w''), \\
  &                       &               & \loc_{w}(a) = \loc(w'), \valr(r) = \valw(w'), \\
  &                       &               & a \in \rcas \implies \valr(r) = \vale(a) \land \valw(w'') = \valu(a) \\
  &                       &               & a \in \rfaa \implies \valw(w'') = \valr(r) + \val(a) \\
  &\mid \naF \tup{r, w, a, w'}&\text{where}\;& r \in \narR, w \in \Write, a \in \rrmw, w' \in \nlW, \locveq(r,w), \\
  &                       &               & \loc_{r}(a) = \loc(r), \loc_{w}(a) = \loc(w'), \\
  &                       &               & \valr(r) = \valw(w^{l}), \valr(r) \neq \vale(a) \\
  &\mid \nlW \tup{w,e}    &\text{where}\;& w \in \nlW, e \in \nlEX, \sameqp(w,e) \\
  &\mid \nrW \tup{w,e}    &\text{where}\;& w \in \nrW, e \in \nrEX, \sameqp(w,e) \\
  &\mid \narW \tup{w}  &\text{where}\;& w \in \narW \\
  &\mid \lCN \tup{e}      &\text{where}\;& e \in \nrEX \\
  &\mid \lP  \tup{p,e}    &\text{where}\;& p \in \lP, e \in \nEX, \sameqp(p,e) \\
  &\mid \nF  \tup{f}      &\text{where}\;& f \in \nF \\
  &\mid \flB \tup{w}      &\text{where}\;& w \in \Write \\
  &\mid \lE\tup{\threadt} &\text{where}\;& \threadt \in \Threads \\
\end{align*}
\vspace{-3em}
\begin{align*}
  \locveq(r,w) \ \ &\defeq \ \ \loc(r) = \loc(w) \wedge \valr(r) = \valw(w) \\
  \sameqp(e,e') \ \ &\defeq \ \ \threadt(e) = \threadt(e') \land \neqn\node(e) =
                      \neqn\node(e')
\end{align*}
\caption{Annotated Labels}
\label{fig:alabels}
\end{figure}

For annotated labels, we reuse most names from labels, but they are different
entities. For instance we note $r \in \lR$ for an event with label $\lR$, while
$\lambda = \lR\tup{\ldots}$ is an annotated label.

We use $\typ(\lambda)$ to denote the type of the annotated label (\lR, \lW,
\RMW, \lF, \PushTSO, \PushNIC, \nlR, \nrR, \nlW, \nrW, \lCN, \lP, \nF, \flB,
\lE). We use
$r(\lambda), w(\lambda), u(\lambda), a(\lambda), f(\lambda), p(\lambda),
e(\lambda), \ldots$ to access the elements of a $\lambda\in\aLabels$ where
applicable. Also, we note $\threadt(\lambda)$ for the thread of the first
argument of $\lambda$.

The annotated program transitions (\Cref{fig:annotated-prog-transitions}) use an
additional annotated label $\RMWF\tup{r,w}$ with $r \in \lR$ and $w \in \Write$
to represent a failed CAS operation. This case is then translated into two
labels (a memory fence and a local read) when creating a path in
\cref{sec:annot-sem-path}. Also, note that the annotated domains (\eg the store
buffers and the queue pairs) contain events, not annotated labels.


\input{annotated-semantics-program}


\input{annotated-semantics-hardware}


\begin{figure}
\begin{mathpar}
\scriptsize
\inferrule{
  \pipe = \alpha \cdot f \\
  f = (\aident, \threadt, \nF(\node))
}{
  \Mem,\Atm, \tup{\pipe,\wbR,\wbL} \pipealab{\nF\tup{f}}
  \Mem,\Atm, \tup{\alpha,\wbR,\wbL}
}
\and
\inferrule{
  \pipe = \alpha \cdot a \cdot \beta \\
  a = (\aident_a, \threadt, \rput(\neqn{\y}, \x)) \\
  \Mem(\x) = w \\
  r = (\aident_r, \threadt, \nlR(\x, \valw(w), \node(\neqn{\y}))) \\
  w' = (\aident_{w'}, \threadt, \nrW(\neqn{\y}, \valw(w))) \\
  \beta \in \queue{(\nrW \cup \narW \cup \rget \cup \nlW \cup \rcas \cup \rfaa \cup \nrEX)} \\
  \wbL \in \queue{\nEX}
}{
  \Mem,\Atm, \tup{\pipe,\wbR,\wbL} \pipealab{\nlR\tup{r,w,a,w'}}
  \Mem,\Atm, \tup{\alpha \cdot w' \cdot \beta,\wbR,\wbL}
}
\and
\inferrule{
  \pipe = \alpha \cdot w \cdot \beta \\
  w = (\aident_w, \threadt, \nrW(\neqn{\y}, \val)) \\
  e = (\aident_e, \threadt, \nrEX(n(\neqn{\y}))) \\
  \beta \in \queue{(\rget \cup \nlW \cup \nrEX)}
}{
  \Mem,\Atm, \tup{\pipe,\wbR,\wbL} \pipealab{\nrW\tup{w,e}}
  \Mem,\Atm, \tup{\alpha \cdot e \cdot \beta, w \cdot \wbR, \wbL}
}
\and
\inferrule{
  \wbR = \alpha \cdot w \\
  w \in \nrW
}{
  \Mem,\Atm, \tup{\pipe,\wbR,\wbL}
  \pipealab{\flB\tup{w}}
  \Mem\upvar{\loc(w)}{w},\Atm,\tup{\pipe,\alpha,\wbL}
}
\and
\inferrule{
  \pipe = \alpha \cdot e \\
  e \in \nrEX
}{
  \Mem,\Atm, \tup{\pipe,\wbR,\wbL} \pipealab{\lCN\tup{e}}
  \Mem,\Atm, \tup{\alpha, \wbR, e \cdot \wbL}
}
\and
\inferrule{
  \pipe = \alpha \cdot a \cdot \beta \\
  a = (\aident_a, \threadt, \rget(\x, \neqn{\y})) \\
  \Mem(\neqn{\y}) = w \\
  r = (\aident_r, \threadt, \nrR(\neqn{\y}, \valw(w))) \\
  w' = (\aident_{w'}, \threadt, \nlW(\x, \valw(w), \node(\neqn{\y}))) \\
  \beta \in \queue{(\rget \cup \nlW \cup \nrEX)} \\
  \wbR = \varepsilon
}{
  \Mem,\Atm, \tup{\pipe,\wbR,\wbL} \pipealab{\nrR\tup{r,w,a,w'}}
  \Mem,\Atm, \tup{\alpha \cdot w' \cdot \beta,\wbR,\wbL}
}
\and
\inferrule{
  \pipe = \alpha \cdot w \\
  w = (\aident_w, \threadt, \nlW(\x, \val, n)) \\
  e = (\aident_e, \threadt, \nlEX(n))
}{
  \Mem,\Atm, \tup{\pipe,\wbR,\wbL} \pipealab{\nlW\tup{w,e}}
  \Mem,\Atm, \tup{\alpha, \wbR, e \cdot w \cdot \wbL}
}
\and
\inferrule{
  \wbL = \alpha \cdot w \cdot \beta \\
  w \in \nlW \\
  \beta \in \queue{\nEX}
}{
  \Mem,\Atm,\tup{\pipe,\wbR,\wbL}
  \pipealab{\flB\tup{w}}
  \Mem\upvar{\loc(w)}{w},\Atm,\tup{\pipe,\wbR,\alpha \cdot \beta}
}
\and
\inferrule{
  \pipe = \alpha \cdot a \cdot \beta \quad
  \wbR = \varepsilon \quad
  \Mem(\neqn{\x}) = w \\
  \valw(w) \neq v \\
  \Atm(\node(\neqn{x})) = \bot \\
  a = (\aident_a, \threadt, \RCAS(\z, \neqn{\x}, \val, \upd)) \\
  r = (\aident_r, \threadt, \narR(\neqn{\x}, \valw(w))) \\
  w' = (\aident_{w'}, \threadt, \nlW(\z, \valw(w), \node(\neqn{x}))) \\
  \beta \in \queue{(\rget \cup \nlW \cup \nrEX)}
}{
  \Mem,\Atm, \tup{\pipe,\wbR,\wbL} \pipealab{\naF\tup{r,w,a,w'}}
  \Mem,\Atm, \tup{\alpha \cdot w' \cdot \beta,\wbR,\wbL}
}
\and
\inferrule{
  \pipe = \alpha \cdot a \cdot \beta \quad
  \wbR = \varepsilon \quad
  \Mem(\neqn{\x}) = w \\
  \valw(w) = v \\
  \Atm(\node(\neqn{x})) = \bot \\
  a = (\aident_a, \threadt, \RCAS(\z, \neqn{\x}, \val, \upd)) \\
  r = (\aident_r, \threadt, \narR(\neqn{\x}, \valw(w))) \\
  w'' = (\aident_{w''}, \threadt, \narW(\neqn{\x}, u)) \\
  w' = (\aident_{w'}, \threadt, \nlW(\z, \valw(w), \node(\neqn{x}))) \\
  \beta \in \queue{(\rget \cup \nlW \cup \nrEX)}
}{
  \Mem,\Atm, \tup{\pipe,\wbR,\wbL} \pipealab{\narR\tup{r,w,a,w',w''}}
  \Mem,\Atm\upvar{\node(\neqn{x})}{\top}, \tup{\alpha \cdot w' \cdot w'' \cdot \beta,\wbR,\wbL}
}
\and
\inferrule{
  \pipe = \alpha \cdot a \cdot \beta \quad
  \wbR = \varepsilon \quad
  \Mem(\neqn{\x}) = w \\
  \valw(w) + v = u \\
  \Atm(\node(\neqn{x})) = \bot \\
  a = (\aident_a, \threadt, \RFAA(\z, \neqn{\x}, \val)) \\
  r = (\aident_r, \threadt, \narR(\neqn{\x}, \valw(w))) \\
  w'' = (\aident_{w''}, \threadt, \narW(\neqn{\x}, u))) \\
  w' = (\aident_{w'}, \threadt, \nlW(\z, \valw(w))) \\
  \beta \in \queue{(\rget \cup \nlW \cup \nrEX)}
}{
  \Mem,\Atm, \tup{\pipe,\wbR,\wbL} \pipealab{\narR\tup{r,w,a,w',w''}}
  \Mem,\Atm\upvar{\node(\neqn{x})}{\top}, \tup{\alpha \cdot w' \cdot w'' \cdot \beta,\wbR,\wbL}
}
\and
\inferrule{
  \pipe = \alpha \cdot w \cdot \beta \\
  \beta \in \queue{(\rget \cup \nlW \cup \nrEX)} \\
  w = (\aident_w, \threadt, \narW(\neqn{x}, \val))
}{
  \Mem,\Atm, \tup{\pipe,\wbR,\wbL} \pipealab{\narW\tup{w}}
  \Mem,\Atm, \tup{\alpha\cdot\beta, w \cdot \wbR, \wbL}
}
\and
\inferrule{
  \wbR = \alpha \cdot w \\
  w = (\aident_w, \threadt, \narW(\neqn{x}, \val))
}{
  \Mem, \Atm, \tup{\pipe, \wbR, \wbL} \pipealab{\flB\tup{w}}
  \Mem\upvar{\loc(w)}{w}, \Atm\upvar{\node(\neqn{x})}{\bot}, \tup{\pipe, \alpha, \wbL}
}
\end{mathpar}
\caption{Annotated 3 Buffers NIC Semantics}
\label{fig:annot-nic-sem-3}
\end{figure}


\paragraph{initialisation}
Given a program \Program, let
$$
\begin{array}{ll}
  \Mem_0\in\AMems & \text{ s.t. } \forall \x\in\Var.\;\Mem_0(\x)=init_{\x}
                    \text{ with }\lab(init_x)\defeq\lW(\x,0) \\
  \tso_0\in\ATSOs & \tso_0 \defeq\varepsilon \\
  \ATSOmap_0\in\ATSOmaps & \ATSOmap_0 \defeq \lambda\threadt.\tso_0 \\
  \Atm_0\in\RAMap & \Atm_0 \defeq \lambda\threadt.\bot \\
  \qp_0\in\AQPairs & \qp_0 \defeq\tup{\varepsilon,\varepsilon,\varepsilon} \\
  \ANQPmap_0\in\ANQPmaps & \ANQPmap_0 \defeq \lambda \threadt. \lambda \node. \qp_0 \end{array}
$$


%% file: annotated-semantics-program.tex

\begin{figure}
\small
\textbf{\;\;Program transitions:} $\Progs \amtr{\aLabels \uplus \set{\RMWF}} \Progs$ \\
\textbf{\;\;Command transitions:} $\Comms \amtr{\aLabels \uplus \set{\RMWF}} \Comms$ \\
\begin{mathpar}
\inferrule*{\comm[]_1 \amtr{\lambda} {\comm[]}'_1}
{\comm[]_1;\comm[]_2 \amtr{\lambda} {\comm[]}'_1;\comm[]_2}
\and
\inferrule*{ }
{\cskip;\comm[] \amtr{\lE\tup{\threadt}} \comm[]}
\and
\inferrule*{i \in \set{1,2}}
{\comm[]_1+\comm[]_2 \amtr{\lE\tup{\threadt}} \comm[]_i}
\and
\inferrule*{ }
{{\comm[]}^\kstar \amtr{\lE\tup{\threadt}} \cskip}
\and
\inferrule*{ }
{{\comm[]}^\kstar \amtr{\lE\tup{\threadt}} \comm[];{\comm[]}^\kstar}
\and
\inferrule*{\comm[]\leadsto {\comm[]}'}
{\comm[] \amtr{\lE\tup{\threadt}} {\comm[]}'}
\and
\inferrule*{\fv{\abexp} = \emptyset \\
  w = (\aident, \threadt, \lW(\x, \db{\abexp}))}
{\x \assign \abexp \amtr{\lW\tup{w}} \cskip}
\and
\inferrule*{\fv{\abexp_\old} = \fv{\abexp_\new} = \emptyset \\
  v \neq \db{\abexp_\old} \\
  r = (\aident, \threadt, \lR(x, v))
}
{\z \assign \clcas(\x,\abexp_\old,\abexp_\new)
  \amtr{\RMWF\tup{r,w}}
  \z \assign v}
\and
\inferrule*{\fv{\abexp_\old} = \fv{\abexp_\new} = \emptyset \\
  u = (\aident, \threadt, \RMW(x, \db{\abexp_\old}, \db{\abexp_\new}))
}
{\z \assign \clcas(\x,\abexp_\old,\abexp_\new)
  \amtr{\RMW\tup{u,w}} \z \assign \db{\abexp_\old}}
\and
\inferrule*{f = (\aident, \threadt, \lF)}
{\mfence \amtr{\lF\tup{f}} \cskip}
\and
\inferrule*{a = (\aident, \threadt, \rget(\x, \neqn\y))}
{\x \assign \neqn\y  \amtr{\PushTSO\tup{a}} \cskip}
\and
\inferrule*{a = (\aident, \threadt, \rput(\neqn\y, \x))}
{\neqn\y \assign \x \amtr{\PushTSO\tup{a}} \cskip}
\and
\inferrule*{a = (\aident, \threadt, \nF(\neqn\node))}
{\cnfence{\neqn\node} \amtr{\PushTSO\tup{a}} \cskip}
\and
\inferrule*{\fv{\abexp_\old} = \fv{\abexp_\new} = \emptyset \\
  v = \db{\abexp_\old} \\
  u = \db{\abexp_\new} \\
  a = (\aident, \threadt, \rcas(z, \neqn{x}, v, u))
}
{\z \assign \rcas(\x,\abexp_\old,\abexp_\new)
  \amtr{\PushTSO\tup{a}}
  \cskip}
\and
\inferrule*{\fv{\abexp} = \emptyset \\
  u = \db{\abexp} \\
  a = (\aident, \threadt, \rfaa(z, \neqn{x}, u))
}
{\z \assign \rfaa(\x,\abexp)
  \amtr{\PushTSO\tup{a}}
  \cskip}
\and
%
%
%
\inferrule*{p = (\aident, \threadt, \lP(n))}
{\poll(\node) \amtr{\lP\tup{p, e}} \cskip}
\and
\inferrule*{r = (\aident, \threadt, \lR(x, v))}
{\cassume(x = v) \amtr{\lR\tup{r,w}} \cskip}
\and
\inferrule*{v \neq v' \\ r = (\aident, \threadt, \lR(x, v'))}
{\cassume(x \neq v) \amtr{\lR\tup{r,w}} \cskip}
\and
\inferrule*{
  \prog(\threadt(\lambda)) \amtr{\lambda} \comm[]
}
{
  \prog \amtr{\lambda} \prog\upvar{\threadt(\lambda)}{\comm[]}
}
\end{mathpar}
\caption{\rdmatso program and command transitions for the annotated semantics}
\label{fig:annotated-prog-transitions}
\end{figure}


%% file: annotated-semantics-hardware.tex

\begin{figure}
\small
\[
\begin{array}{@{} c @{}}
  \Mem \in \AMems \defeq \setpred{m \in \Var\to\Write} {\forall x \in \Var.\loc(m[x])=x}
  \qquad
  \ATSOmap \in \ATSOmaps \defeq \Threads \to \ASTSOs
  \\
  \Atm \!\in\! \RAMap \defeq \lambda \node. \set{\bot, \top}
  \qquad
  \ANQPmap \in \ANQPmaps \defeq \Threads \to (\Nodes \to \AQPairs)
  \\
  \tso \in \ATSOs \defeq
  \queue{(\lW \cup \rget \cup \rput \cup \nF \cup\; \rcas \cup \rfaa)}
  \qquad
  \qpbis \in \AQPairs \defeq \APipes \times \AWBRs \times \AWBLs
  \\
  \pipe \in \APipes \defeq \queue{(\rget \cup \rput \cup \nF \cup \nrW \cup\; \narW \cup \nrEX \cup \nlW \cup\; \rcas \cup \rfaa)}
  \\
  \wbR \in \AWBRs \defeq \queue{(\nrW, \narW)}
  \qquad
  \wbL \in \AWBLs \defeq \queue{(\nlW \cup \nlEX \cup \nrEX)}
\end{array}
\]
\hrule
\hrule
\begin{mathpar}
\inferrule*{\TSOmap' = \TSOmap\upvar{\threadt(w)}
  {w \cdot \TSOmap(\threadt(w))}}
{
  \Mem, \TSOmap, \Atm, \NQPmap
  \amtr{\lW\tup{w}}
  \Mem, \TSOmap', \Atm, \NQPmap
}
\and
\inferrule*{(\Mem \aupdmem \TSOmap(\threadt(r)))(\loc(r)) = w \\
  \valr(r) = \valw(w)}
{
  \Mem, \TSOmap, \Atm, \NQPmap
  \amtr{\lR\tup{r,w}}
  \Mem, \TSOmap, \Atm, \NQPmap
}
\and
\inferrule*{\TSOmap(\threadt(u)) = \varepsilon \\
  \Mem(\loc(u)) = w \\
  \valr(u) = \valw(w)}
{
  \Mem, \TSOmap, \Atm, \NQPmap
  \amtr{\RMW\tup{u,w}}
  \Mem\upvar{x}{u}, \TSOmap, \Atm, \NQPmap
}
\and
\inferrule*{\TSOmap(\threadt(f)) = \varepsilon}
{
  \Mem, \TSOmap, \Atm, \NQPmap
  \amtr{\lF\tup{f}}
  \Mem, \TSOmap, \Atm, \NQPmap
}
\and
\inferrule*{\TSOmap' = \TSOmap\upvar{\threadt(a)}
  {a \cdot \TSOmap(\threadt(a))}}
{
  \Mem, \TSOmap, \Atm, \NQPmap
  \amtr{\PushTSO\tup{a}}
  \Mem, \TSOmap', \Atm, \NQPmap
}
\and
\inferrule*{\TSOmap(\threadt(w)) = \stso \cdot w \\ w \in \lW}
{
  \Mem, \TSOmap, \Atm, \NQPmap
  \amtr{\flB\tup{w}}
  \Mem\upvar{x}{w}, \TSOmap\upvar{\threadt(w)}{\stso}, \Atm, \NQPmap
}
\and
\inferrule*{\TSOmap(\threadt(a)) = \stso \cdot a \\
  a \notin \lW \\
  \NQPmap(\threadt(a))(\node(a)) = \qp\\
  \qp' = \qp\upvar{\pipe}{a \cdot \qp.\pipe}
}{
  \Mem, \TSOmap, \Atm, \NQPmap
  \amtr{\PushNIC\tup{a}}
    \Mem, \TSOmap \upvar{\threadt(a)}{\stso}, \Atm,
    \NQPmap\upvar{\threadt(a)}{\NQPmap(\threadt(a))\upvar{\node(a)}{\qp'}}
}
\and
\inferrule*{
  \NQPmap(\threadt(p))(\node(p)) = \qp \\
  \qp.\wbL = \alpha \cdot e \\
  e \in \nEX \\
  \qp' = \qp\upvar{\wbL}{\alpha}
}{
  \Mem, \TSOmap, \Atm, \NQPmap
  \amtr{\lP\tup{p,e}}
  \Mem, \TSOmap, \Atm,  \NQPmap\upvar{\threadt(p)}{\NQPmap(\threadt(p))\upvar{\node(p)}{\qp'}}
}
\and
\inferrule*{
  \Mem, \Atm, \NQPmap(\threadt(\lambda))(\neqn\node) \pipealab{\lambda} \Mem', \Atm', \qp
}{
  \Mem, \TSOmap, \Atm, \NQPmap
  \amtr{\lambda}
  \Mem', \TSOmap, \Atm', \NQPmap\upvar{\threadt(\lambda)}{\NQPmap(\threadt(\lambda))\upvar{\neqn\node}{\qp}}
}
\end{mathpar}
\vspace{8pt}
\hrule
$$\text{with} \quad(\Mem \aupdmem \alpha)(\x) = \begin{cases}
\Mem[\x]                  & \alpha = \varepsilon \\
w                         & \alpha = w \cdot \beta \land w \in \Write
                            \land \loc(w) = \x \\
(\Mem \aupdmem \beta)(\x) & \alpha = e \cdot \beta \land
                            (e \not\in \Write \lor \loc(e) \neq \x)
\end{cases}
$$
\hrule
\hrule
\caption{\rdmatso hardware domains and hardware transitions for the annotated semantics}
\label{fig:annotated-hardware-transitions}
\end{figure}


%% file: annot-path.tex


\subsection{Paths, Gluing, and Other Definitions}
\label{sec:annot-sem-path}

We define a path as:
$\pi \in \paths \defeq \queue{(\aLabels \setminus \lE\tup{\threadt})}$

We define Annotated Operational Semantics Gluing with the following rules.

\begin{mathpar}
\inferrule*{
  \prog \amtr{\lE\tup{\threadt}} \prog'
}{
  \prog, \Mem, \TSOmap, \Atm, \NQPmap, \pi
  \trp
  \prog', \Mem, \TSOmap, \Atm, \NQPmap, \pi
}
\and
\inferrule*{
  \prog \amtr{\lambda} \prog' \\
  \Mem, \TSOmap, \Atm, \NQPmap \amtr{\lambda} \Mem', \TSOmap', \Atm, \NQPmap' \\
  \lambda \in \bigpar{\lR \cup \lW \cup \RMW \cup \lF \cup \PushTSO \cup \lP} \\
  \fresh(\lambda,\pi)
}{
  \prog, \Mem, \TSOmap, \Atm, \NQPmap, \pi
  \trp
  \prog', \Mem', \TSOmap', \Atm, \NQPmap', \lambda \cdot \pi
}
\and
\inferrule*{
  \Mem, \TSOmap, \Atm, \NQPmap \amtr{\lambda} \Mem', \TSOmap', \Atm', \NQPmap' \\
  \lambda \in \bigpar{\PushNIC \cup \nlR \cup \nrR \cup \nlW \cup \nrW
    \cup \naF \cup \narR \cup \narW \cup \lCN \cup \nF \cup \flB} \\
  \fresh(\lambda,\pi)
}{
  \prog, \Mem, \TSOmap, \Atm, \NQPmap, \pi
  \trp
  \prog, \Mem', \TSOmap', \Atm', \NQPmap', \lambda \cdot \pi
}
\and
\inferrule*{
  \prog \amtr{\RMWF\tup{r,w}} \prog' \\
  \lambda_1 = \lF\tup{(\aident,\threadt(r),\lF)} \\
  \lambda_2 = \lR\tup{r,w} \\
  \Mem, \TSOmap, \Atm, \NQPmap \amtr{\lambda_1}
  \Mem, \TSOmap, \Atm, \NQPmap \amtr{\lambda_2} \Mem, \TSOmap, \Atm, \NQPmap \\
  \fresh(\lambda_1,\pi) \\
  \fresh(\lambda_2,\pi)
}{
  \prog, \Mem, \TSOmap, \Atm, \NQPmap, \pi
  \trp
  \prog', \Mem, \TSOmap, \Atm, \NQPmap, \lambda_2 \cdot \lambda_1 \cdot \pi
}
\end{mathpar}


Two annotated labels are non-conflicting ($\lambda_1 \diff \lambda_2$) if they
are of a different type or if their relevant arguments are disjoints.
An annotated label is fresh if it does not conflict with any previous annotated
label.

$$\relevant : \aLabels \rightarrow 2^{\ActionsExt}$$
\begin{minipage}[t]{0.5\textwidth}
\vspace{-1em}
\begin{align*}
  \relevant(\lR \tup{r,\_})     \ \defeq& \ \{r\} \\
  \relevant(\lW \tup{w})        \ \defeq& \ \{w\} \\
  \relevant(\RMW \tup{u,\_})    \ \defeq& \ \{u\} \\
  \relevant(\lF \tup{f})        \ \defeq& \ \{f\} \\
  \relevant(\PushTSO\tup{a})    \ \defeq& \ \{a\} \\
  \relevant(\PushNIC\tup{a})    \ \defeq& \ \{a\} \\
  \relevant(\nlR \tup{r,\_,a,w'})\ \defeq& \ \{r,a,w'\} \\
  \relevant(\nrR \tup{r,\_,a,w'})\ \defeq& \ \{r,a,w'\} \\
  \relevant(\naF \tup{r,\_,a,w'})\ \defeq& \ \{r,a,w'\} \\
\end{align*}
\end{minipage}
\begin{minipage}[t]{0.5\textwidth}
\vspace{-1em}
\begin{align*}
  \relevant(\narR \tup{r,\_,a,w'',w'})\ \defeq& \ \{r,a,w',w''\} \\
  \relevant(\narW \tup{w})    \ \defeq& \ \{w\} \\
  \relevant(\nlW \tup{w,e})     \ \defeq& \ \{w,e\} \\
  \relevant(\nrW \tup{w,e})     \ \defeq& \ \{w,e\} \\
  \relevant(\lCN \tup{e})       \ \defeq& \ \{e\} \\
  \relevant(\lP \tup{p,e})      \ \defeq& \ \{p,e\} \\
  \relevant(\nF \tup{f})        \ \defeq& \ \{f\} \\
  \relevant(\flB \tup{w})       \ \defeq& \ \{w\} \\
  \relevant(\lE\tup{\_})        \ \defeq& \ \{\} \\
\end{align*}
\end{minipage}
\begin{align*}
\lambda_1 \diff \lambda_2 \ \defeq \
  & \ \typ(\lambda_1) \neq \typ(\lambda_2) \ \lor \relevant(\lambda_1) \cap \relevant(\lambda_2) = \emptyset \\
\fresh(\lambda,\pi) \ \defeq \
  & \ \forall \lambda'\in\pi, \ \lambda \diff \lambda' \\
\nodup(\pi) \ \defeq \
  & \ \forall \pi_2,\lambda,\pi_1. \ \pi = \pi_2\cdot\lambda\cdot\pi_1 \implies \fresh(\lambda,\pi_1)
\end{align*}

$\relevant(\lambda)$ are the arguments that are important to consider to avoid
duplicating events. The excluded events are the write operations we lookup when
reading. For instance:
\begin{itemize}
\item Having both $\lR\tup{r_1,w}$ and $\lR\tup{r_2,w}$ during an execution is
  fine, since $w$ can be looked up any number of time.
\item Having both $\nlR\tup{r_1,w_1,a,e_1}$ and $\nlR\tup{r_2,w_2,a,e_2}$ during
  an execution is problematic, since it means the \textput operation $a$ is
  being run twice.
\end{itemize}


\begin{minipage}{\textwidth}
\paragraph{Completeness}
\begin{align*}
\complete(\pi) \defeq
  &\ \forall a,w',e,r,w,f,w''. \ \\
  &\ \phantom{~\land~} \ \lW\tup{w}\in\pi \implies \flB\tup{w}\in\pi \\
  &\ \land \ \PushTSO\tup{a}\in\pi \implies \PushNIC\tup{a}\in\pi \\
  &\ \land \ \PushNIC\tup{f}\in\pi \land f \in \nF \implies \nF\tup{f}\in\pi \\
  &\ \land \ \PushNIC\tup{a}\in\pi \land a \in \rput \implies
    \exists r,w,w'. \ \nlR\tup{r,w,a,w'}\in\pi \\
  &\ \land \ \PushNIC\tup{a}\in\pi \land a \in \rget \implies
    \exists r,w,w'. \ \nrR\tup{r,w,a,w'}\in\pi \\
  &\ \land \ \PushNIC\tup{a}\in\pi \land a \in \rfaa \implies
    \exists r,w,w'. \ \narR\tup{r,w,a,w',w''}\in\pi \\
  &\ \land \ \PushNIC\tup{a}\in\pi \land a \in \rcas \implies
    \bigpar{
    \begin{array}{l}
      \phantom{\lor}\ \exists r,w,w'. \ \naF\tup{r,w,a,w'}\in\pi \\
      \lor\ \exists r,w,w',w''. \ \narR\tup{r,w,a,w',w''}\in\pi
    \end{array}
    } \\
  &\ \land \ \nlR\tup{r,w,a,w'}\in\pi \implies \exists e. \ \nrW\tup{w',e}\in\pi \\
  &\ \land \ \nrR\tup{r,w,a,w'}\in\pi \implies \exists e. \ \nlW\tup{w',e}\in\pi \\
  &\ \land \ \narR\tup{r,w,a,w',w''}\in\pi \implies
    \narW\tup{w''}\in\pi \\
  &\ \land \ \narR\tup{r,w,a,w',w''}\in\pi \implies
    \exists e. \ \nlW\tup{w', e}\in\pi \\
  &\ \land \ \naF\tup{r,w,a,w'}\in\pi \implies
    \exists e. \ \nlW\tup{w', e}\in\pi \\
  &\ \land \ \nlW\tup{w,e}\in\pi \implies \flB\tup{w}\in\pi \\
  &\ \land \ \nrW\tup{w,e}\in\pi \implies
    \flB\tup{w}\in\pi \land \lCN\tup{e} \in \pi \\
  &\ \land \ \narW\tup{w}\in\pi \implies
    \flB\tup{w}\in\pi
\end{align*}
\end{minipage}

Informal: every pending operation is done and (most) buffers are empty. Note
that some \nEX (i.e., completion notifications) might still be in \wbL.




For a path $\pi$ without duplicate (e.g. if $\nodup(\pi)$ holds), we define the
total ordering of its annotated labels as follows. Note that the early part of
the path is on the right.
$$\lambda_1 \precpi \lambda_2 \ \defeq \ \exists \pi_1,\pi_2,\pi_3 \ \text{s.t.}
\ \pi = \pi_3 \cdot \lambda_2 \cdot \pi_2 \cdot \lambda_1 \cdot \pi_1$$





\begin{minipage}{\textwidth}
\paragraph{Backward Completeness} (with ordering)
\begin{align*}
\backComp(\pi) \defeq
  &\ \forall a,w',e,r,w,f,p,w''. \ \\
  &\ \phantom{~\land~} \ \flB\tup{w}\in\pi \implies
     \bigpar{
     \begin{array}{l}
       \phantom{\lor} \ \lW\tup{w} \precpi \flB\tup{w} \\
       \lor \ \exists e. \nlW\tup{w,e} \precpi \flB\tup{w} \\
       \lor \ \exists e. \nrW\tup{w,e} \precpi \flB\tup{w} \\
       \lor \ \exists e. \narW\tup{w} \precpi \flB\tup{w}
     \end{array} } \\
  &\ \land \ \PushNIC\tup{a}\in\pi \implies \PushTSO\tup{a} \precpi \PushNIC\tup{a} \\
  &\ \land \ \nF\tup{f}\in\pi \implies \PushNIC\tup{f} \precpi \nF\tup{f} \\
  &\ \land \ \nlR\tup{r,w,a,w'}\in\pi \implies
    \PushNIC\tup{a} \precpi \nlR\tup{r,w,a,w'} \\
  &\ \land \ \nrR\tup{r,w,a,w'}\in\pi \implies
    \PushNIC\tup{a} \precpi \nrR\tup{r,w,a,w'} \\
  &\ \land \ \naF\tup{r,w,a,w'}\in\pi \implies
    \PushNIC\tup{a} \precpi \naF\tup{r,w,a,w'} \\
  &\ \land \ \narR\tup{r,w,a,w',w''}\in\pi \implies
    \PushNIC\tup{a} \precpi \narR\tup{r,w,a,w',w''} \\
  &\ \land \ \nrW\tup{w',e}\in\pi \implies \exists r,w,a. \ \nlR\tup{r,w,a,w'} \precpi \nrW\tup{w',e} \\
  &\ \land \ \nlW\tup{w',e}\in\pi \implies
    \bigpar{
    \begin{array}{l}
      \phantom{\lor} \ \exists r,w,a. \ \nrR\tup{r,w,a,w'} \precpi \nlW\tup{w',e} \\
      \lor \ \exists r,w,a. \ \naF\tup{r,w,a,w'} \precpi \nlW\tup{w',e} \\
      \lor \ \exists r,w,a,w''. \bigpar{
      \begin{array}{l}
        \narR\tup{r,w,a,w',w''} \\
        \precpi \narW\tup{w''} \precpi \nlW\tup{w',e}
      \end{array}
      }
    \end{array}} \\
  &\ \land \ \narW\tup{w'}\in\pi \implies
    \exists r,w,a,w''. \narR\tup{r,w,a,w',w''} \precpi \narW\tup{w'} \\
  &\ \land \ \lCN\tup{e}\in\pi \implies
      \phantom{\lor} \ \exists w. \ \nrW\tup{w,e} \precpi \lCN\tup{e} \\
  &\ \land \ \lP\tup{p,e}\in\pi \implies \ \bigpar{
    \begin{array}{l}
      \phantom{\lor} \ \exists w. \ \nlW\tup{w,e} \precpi \flB\tup{w} \precpi \lP\tup{p,e} \\
      \lor \ \lCN\tup{e} \precpi \lP\tup{p,e}
    \end{array} }
\end{align*}
\end{minipage}


\bigskip
\paragraph{Poll Order}
$$
\pollOrder(\pi) \ \defeq \ \forall e_1,e_2. \ \bigpar{
  \begin{array}{l}
    \phantom{\land} \ \sameqp(e_1,e_2) \\
    \land \ \lambda_1 \in \{\nlW\tup{\_,e_1}, \lCN\tup{e_1}\} \\
    \land \ \lambda_2 \in \{\nlW\tup{\_,e_2}, \lCN\tup{e_2}\} \\
    \land \ \lambda_1 \precpi \lambda_2 \\
    \land \ \lP\tup{\_,e_2}\in\pi
  \end{array}} \implies \lP\tup{\_,e_1} \precpi \lP\tup{\_,e_2}  \\
$$


\begin{minipage}{\textwidth}
\paragraph{Flush Order}
\begin{align*}
  & \bufFlushOrd(\pi) \ \defeq \ \\
  & \phantom{~\land~} \ \forall w_1,w_2 \in \lW. \ \bigpar{
    \begin{array}{l}
      \threadt(w_1)=\threadt(w_2) \implies \\
      \bigpar{ \flB\tup{w_2}\in\pi \land \lW\tup{w_1} \precpi \lW\tup{w_2} }
        \iff \flB\tup{w_1} \precpi \flB\tup{w_2}
    \end{array}
    } \\
  & \land \ \forall a_1,a_2 \in (\rget \cup \rput \cup \nF \cup \rcas \cup \rfaa). \\
  & \phantom{\land \ \forall} \bigpar{
    \begin{array}{l}
      \threadt(a_1)=\threadt(a_2) \implies \\
      \bigpar{ \PushNIC\tup{a_2}\in\pi \land
               \PushTSO\tup{a_1} \precpi \PushTSO\tup{a_2} }
        \iff \PushNIC\tup{a_1} \precpi \PushNIC\tup{a_2}
    \end{array}
    } \\
  & \land \ \forall a_1 \in (\rget \cup \rput \cup \nF \cup \rcas \cup \rfaa), w_2 \in \lW. \\
  & \phantom{\land \ \forall} \bigpar{
    \begin{array}{l}
      \threadt(a_1)=\threadt(w_2) \implies \\
      \phantom{\land} \bigpar{ \flB\tup{w_2}\in\pi \land
                               \PushTSO\tup{a_1} \precpi \lW\tup{w_2} }
        \iff \PushNIC\tup{a_1} \precpi \flB\tup{w_2} \\
      \land \bigpar{ \PushNIC\tup{a_1}\in\pi \land
                     \lW\tup{w_2} \precpi \PushTSO\tup{a_1} }
        \iff \flB\tup{w_2} \precpi \PushNIC\tup{a_1}
    \end{array}
    } \\
  & \land \ \forall w_1,w_2 \in \nlW. \ \bigpar{
    \begin{array}{l}
      \sameqp(w_1,w_2) \implies \\
      \bigpar{ \flB\tup{w_2}\in\pi \land \nlW\tup{w_1} \precpi \nlW\tup{w_2} }
        \iff \flB\tup{w_1} \precpi \flB\tup{w_2}
    \end{array}
    } \\
  & \land \ \forall w_1,w_2 \in \nrW. \ \bigpar{
    \begin{array}{l}
      \sameqp(w_1,w_2) \implies \\
      \bigpar{ \flB\tup{w_2}\in\pi \land \nrW\tup{w_1} \precpi \nrW\tup{w_2} }
        \iff \flB\tup{w_1} \precpi \flB\tup{w_2}
    \end{array}
    } \\
  & \land \ \forall w_1,w_2 \in \narW. \ \bigpar{
    \begin{array}{l}
      \sameqp(w_1,w_2) \implies \\
      \bigpar{ \flB\tup{w_2}\in\pi \land \narW\tup{w_1} \precpi \narW\tup{w_2} }
        \iff \flB\tup{w_1} \precpi \flB\tup{w_2}
    \end{array}
    } \\
  & \land \ \forall w_1 \in \nrW, w_2 \in \narW. \ \bigpar{
    \begin{array}{l}
      \sameqp(w_1,w_2) \implies \\
      \bigpar{ \flB\tup{w_2}\in\pi \land \nrW\tup{w_1} \precpi \narW\tup{w_2} }
        \iff \flB\tup{w_1} \precpi \flB\tup{w_2}
    \end{array}
    } \\
  & \land \ \forall w_1 \in \narW, w_2 \in \nrW. \ \bigpar{
    \begin{array}{l}
      \sameqp(w_1,w_2) \implies \\
      \bigpar{ \flB\tup{w_2}\in\pi \land \narW\tup{w_1} \precpi \nrW\tup{w_2} }
        \iff \flB\tup{w_1} \precpi \flB\tup{w_2}
    \end{array}
    } \\
  & \land \ \forall w \in \lW, f \in \lF. \ \lW\tup{w} \precpi \lF\tup{f}
    \land \threadt(w) = \threadt(f) \implies \flB\tup{w} \precpi \lF\tup{f} \\
  & \land \ \forall w \in \lW, u \in \RMW. \ \lW\tup{w} \precpi \RMW\tup{u,\_}
    \land \threadt(w) = \threadt(u) \implies \flB\tup{w} \precpi \RMW\tup{u,\_} \\
  & \land \ \forall w \in \nlW, r \in \nlR. \\
  & \phantom{\land \ \forall} \bigpar{ \nlW\tup{w,\_} \precpi
    \nlR\tup{r,\_,\_,\_} \land \sameqp(w,r) } \implies
    \flB\tup{w} \precpi \nlR\tup{r,\_,\_,\_} \\
  & \land \ \forall w \in \nrW, r \in \nrR. \\
  & \phantom{\land \ \forall} \bigpar{ \nrW\tup{w,\_} \precpi
    \nrR\tup{r,\_,\_,\_} \land \sameqp(w,r) } \implies
    \flB\tup{w} \precpi \nrR\tup{r,\_,\_,\_} \\
  & \land \ \forall w \in \nrW, r \in \narR. \\
  & \phantom{\land \ \forall} \bigpar{ \nrW\tup{w,\_} \precpi
    \naF\tup{r,\_,\_,\_} \land \sameqp(w,r) } \implies
    \flB\tup{w} \precpi \naF\tup{r,\_,\_,\_} \\
  & \land \ \forall w \in \nrW, r \in \narR. \\
  & \phantom{\land \ \forall} \bigpar{ \nrW\tup{w,\_} \precpi
    \narR\tup{r,\_,\_,\_,\_} \land \sameqp(w,r) } \implies
    \flB\tup{w} \precpi \narR\tup{r,\_,\_,\_,\_} \\
  & \land \ \forall w \in \narW, r \in \nrR. \\
  & \phantom{\land \ \forall} \bigpar{ \narW\tup{w} \precpi
    \nrR\tup{r,\_,\_,\_} \land \sameqp(w,r) } \implies
    \flB\tup{w} \precpi \nrR\tup{r,\_,\_,\_} \\
  & \land \ \forall w \in \narW, r \in \narR. \\
  & \phantom{\land \ \forall} \bigpar{ \narW\tup{w} \precpi
    \naF\tup{r,\_,\_,\_} \land \sameqp(w,r) } \implies
    \flB\tup{w} \precpi \naF\tup{r,\_,\_,\_} \\
  & \land \ \forall w \in \narW, r \in \narR. \\
  & \phantom{\land \ \forall} \bigpar{ \narW\tup{w} \precpi
    \narR\tup{r,\_,\_,\_,\_} \land \sameqp(w,r) } \implies
    \flB\tup{w} \precpi \narR\tup{r,\_,\_,\_,\_}
\end{align*}
\end{minipage}


\begin{minipage}{\textwidth}
\paragraph{NIC Order}
\begin{align*}
  & \nicActOrder(\pi) \ \defeq \ \forall a_1,a_2. \
    \PushNIC\tup{a_1} \precpi \PushNIC\tup{a_2}
    \land \sameqp(a_1,a_2) \implies \\
  & \phantom{\land~} \ \bigpar{ a_1 \in \nF \land a_2 \in \rget
    \land \nrR\tup{\_,\_,a_2,\_} \in \pi
    \implies \nF\tup{a_1} \precpi \nrR\tup{\_,\_,a_2,\_} } \\
  & \land \ \bigpar{ a_1 \in \nF \land a_2 \in \rput
    \land \nlR\tup{\_,\_,a_2,\_} \in \pi
    \implies \nF\tup{a_1} \precpi \nlR\tup{\_,\_,a_2,\_} } \\
  & \land \ \bigpar{ a_1 \in \nF \land a_2 \in \RCAS
    \land \naF\tup{\_,\_,a_2,\_} \in \pi
    \implies \nF\tup{a_1} \precpi \naF\tup{\_,\_,a_2,\_,} } \\
  & \land \ \bigpar{ a_1 \in \nF \land a_2 \in \rrmw
    \land \narR\tup{\_,\_,a_2,\_,\_} \in \pi
    \implies \nF\tup{a_1} \precpi \narR\tup{\_,\_,a_2,\_,\_} } \\
  & \land \ \bigpar{ a_1 \in \nF \land a_2 \in \nF
    \land \nF\tup{a_2} \in \pi
    \implies \nF\tup{a_1} \precpi \nF\tup{a_2} } \\
  & \land \ \bigpar{ a_1 \in \rget \land a_2 \in \nF
    \land \nF\tup{a_2} \in \pi \implies
    \nrR\tup{\_,\_,a_1,w_1} \precpi \nlW\tup{w_1,\_} \precpi \nF\tup{a_2} } \\
  & \land \ \bigpar{
    \begin{array}{l}
      a_1 \in \rput \land a_2 \in \nF
      \land \nF\tup{a_2} \in \pi \\
      \implies \nlR\tup{\_,\_,a_1,w_1} \precpi
      \nrW\tup{w_1,e_1} \precpi \lCN\tup{e_1} \precpi \nF\tup{a_2}
    \end{array}} \\
  & \land \ \bigpar{
    \begin{array}{l}
      a_1 \in \RCAS \land a_2 \in \nF
      \land \nF\tup{a_2} \in \pi \\
      \implies \narR\tup{\_,\_,a_1,w_1,w_{2}} \precpi \narW\tup{w_{2}}
      \precpi \nlW\tup{w_1,\_} \precpi \nF\tup{a_2} \\
      \phantom{\implies}\llap{$\lor$\ }\ \naF\tup{\_,\_,a_1,w_1} \precpi
      \nlW\tup{w_1,\_} \precpi \nF\tup{a_2}
    \end{array}} \\
  & \land \ \bigpar{
    \begin{array}{l}
      a_1 \in \RFAA \land a_2 \in \nF
      \land \nF\tup{a_2} \in \pi \\
      \implies \narR\tup{\_,\_,a_1,w_1,w_{2}} \precpi \narW\tup{w_{2}}
      \precpi \nlW\tup{w_1,\_} \precpi \nF\tup{a_2} \\
    \end{array}} \\
  & \land \ \bigpar{
    \begin{array}{l}
      a_1 \in \rget \land a_2 \in \rget \land
      \nrR\tup{\_,\_,a_2,w_2} \precpi \nlW\tup{w_2,\_} \\
      \implies \nrR\tup{\_,\_,a_1,w_1} \precpi
      \nlW\tup{w_1,\_} \precpi \nlW\tup{w_2,\_}
    \end{array}} \\
  & \land \ \bigpar{
    \begin{array}{l}
      a_1 \in \rget \land a_2 \in \rput \land \nlR\tup{\_,\_,a_2,w_2}
      \precpi \nrW\tup{w_2,e_2} \precpi \lCN\tup{e_2} \\
      \implies \nrR\tup{\_,\_,a_1,w_1} \precpi \nlW\tup{w_1,\_}
      \precpi \lCN\tup{e_2}
    \end{array}} \\
  & \land \ \bigpar{
    \begin{array}{l}
      a_1 \in \rget \land a_2 \in \RCAS \land \bigpar{
      \begin{array}{l}
      \narR\tup{\_,\_,a_2,w_2,\_}
      \precpi \nlW\tup{w_2,\_} \\
      \lor \ \naF\tup{\_,\_,a_2,w_2}
      \precpi \nlW\tup{w_2,\_}
      \end{array}} \\
      \implies \nrR\tup{\_,\_,a_1,w_1} \precpi \nlW\tup{w_1,\_}
      \precpi \nlW\tup{w_{2},\_}
    \end{array}} \\
  & \land \ \bigpar{
    \begin{array}{l}
      a_1 \in \rget \land a_2 \in \RFAA \land
      \narR\tup{\_,\_,a_2,w_2,\_}
      \precpi \nlW\tup{w_2,\_} \\
      \implies \nrR\tup{\_,\_,a_1,w_1} \precpi \nlW\tup{w_1,\_}
      \precpi \nlW\tup{w_{2},\_}
    \end{array}} \\
  & \land \ \bigpar{
    \begin{array}{l}
      a_1 \in \rput \land a_2 \in \rget
      \land \nrR\tup{\_,\_,a_2,\_} \in \pi \\
      \implies \nlR\tup{\_,\_,a_1,w_1}
      \precpi \nrW\tup{w_1,\_} \precpi \nrR\tup{\_,\_,a_2,\_}
    \end{array}} \\
  & \land \ \bigpar{
    \begin{array}{l}
      a_1 \in \rput \land a_2 \in \rget
      \land \nrR\tup{\_,\_,a_2,w_2} \precpi \nlW\tup{w_2,\_} \\
      \implies \nlR\tup{\_,\_,a_1,w_1} \precpi \nrW\tup{w_1,e_1}
      \precpi \lCN\tup{e_1} \precpi \nlW\tup{w_2,\_}
    \end{array}} \\
  & \land \ \bigpar{
    \begin{array}{l}
      a_1 \in \rput \land a_2 \in \RCAS \land \naF\tup{\_,\_,a_2,w_2} \in \pi \\
      \implies \nlR\tup{\_,\_,a_1,w_1} \precpi \nrW\tup{w_1,e_{1}}
      \precpi \naF\tup{\_,\_,a_2,w_2}
    \end{array}} \\
  & \land \ \bigpar{
    \begin{array}{l}
      a_1 \in \rput \land a_2 \in \RCAS \cup \rfaa \land \narR\tup{\_,\_,a_2,w_2,\_} \in \pi \\
      \implies \nlR\tup{\_,\_,a_1,w_1} \precpi \nrW\tup{w_1,e_{1}}
      \precpi \narR\tup{\_,\_,a_2,w_2,\_}
    \end{array}} \\
  & \land \ \bigpar{ a_1 \in \rput \land a_2 \in \rput
    \land \nlR\tup{\_,\_,a_2,\_} \in \pi \implies
    \nlR\tup{\_,\_,a_1,\_} \precpi \nlR\tup{\_,\_,a_2,\_}} \\
  & \land \ \bigpar{
    \begin{array}{l}
      a_1 \in \rput \land a_2 \in \rput \land
      \nlR\tup{\_,\_,a_2,w_2} \precpi \nrW\tup{w_2,\_} \\
      \implies \nlR\tup{\_,\_,a_1,w_1} \precpi \nrW\tup{w_1,\_}
      \precpi \nrW\tup{w_2,\_}
    \end{array}} \\
  & \land \ \bigpar{
    \begin{array}{l}
      a_1 \in \rput \land a_2 \in \rput \land \nlR\tup{\_,\_,a_2,w_2}
      \precpi \nrW\tup{w_2,e_2} \precpi \lCN\tup{e_2} \\
      \implies \nlR\tup{\_,\_,a_1,w_1} \precpi \nrW\tup{w_1,e_1}
      \precpi \lCN\tup{e_1} \precpi \lCN\tup{e_2}
    \end{array}}
\end{align*}
\end{minipage}

\begin{align*}
  & \land \ \bigpar{
    \begin{array}{l}
      a_{1} \in (\rcas \cup \rfaa) \land a_{2} \in \rget \land \nrR\tup{\_,\_,a_{2},\_}
      \in \pi \\
      \implies \bigpar{
      \begin{array}{l}
        \phantom{\lor\ } a_{1} \in \RCAS \land \naF\tup{\_,\_,a_{1},\_}
        \precpi \nrR\tup{\_,\_,a_{2},w_{2}} \\
        \lor\ \narR\tup{\_,\_,a_{1},w_{1},\_}
        \precpi \narW\tup{w_{1}}
        \precpi \nrR\tup{\_,\_,a_{2},w_{2}}
      \end{array}}
    \end{array}} \\
  & \land \ \bigpar{
    \begin{array}{l}
      a_{1} \in (\rcas \cup \rfaa) \land a_{2} \in \rget \land \nrR\tup{\_,\_,a_{2},w_{2}}
      \precpi \nlW\tup{w_{2},\_} \\
      \implies \bigpar{
      \begin{array}{l}
        \phantom{\lor\ } a_{1} \in \rcas \land \naF\tup{\_,\_,a_{1},w_{1}}
        \precpi \nlW\tup{w_{1},\_} \precpi \nlW\tup{w_{2},\_} \\
        \lor\ \narR\tup{\_,\_,a_{1},w_{1},\_}
        \precpi \nlW\tup{w_{1},\_} \precpi \nlW\tup{w_{2},\_} \\
      \end{array}}
    \end{array}} \\
  & \land \ \bigpar{
    \begin{array}{l}
      a_{1} \in (\rcas \cup \rfaa) \land a_{2} \in \rput \land \nlR\tup{\_,\_,a_{2},w_{2}}
      \precpi \nrW\tup{w_{2},\_} \\
      \implies \bigpar{
      \begin{array}{l}
        \phantom{\lor\ } a_{1} \in \rcas \land \naF\tup{\_,\_,a_{1},w_{1}}
        \precpi \nrW\tup{w_{2},\_} \\
        \lor\ \narR\tup{\_,\_,a_{1},w_{1},\_}
        \precpi \narW\tup{w_{1}} \precpi \nrW\tup{w_{2},\_}
      \end{array}}
    \end{array}} \\
  & \land \ \bigpar{
    \begin{array}{l}
      a_{1} \in (\rcas \cup \rfaa) \land a_{2} \in \rput \land \nlR\tup{\_,\_,a_{2},w_{2}}
      \precpi \nrW\tup{w_{2},e_{2}} \precpi \lCN\tup{e_{2}} \\
      \implies \bigpar{
      \begin{array}{l}
        \phantom{\lor\ } a_{1} \in \rcas \land \naF\tup{\_,\_,a_{1},w_{1}}
        \precpi \nlW\tup{w_{1},\_} \precpi \lCN\tup{e_{2}} \\
        \lor\ \narR\tup{\_,\_,a_{1},w_{1},\_}
        \precpi \nlW\tup{w_{1},\_} \precpi \lCN\tup{e_{2}} \\
      \end{array}}
    \end{array}} \\
  & \land \ \bigpar{
    \begin{array}{l}
      a_{1} \in (\rcas \cup \rfaa) \land a_{2} \in \rcas \land \naF\tup{\_,\_,a_{2},\_}
      \in \pi \\
      \implies \bigpar{
      \begin{array}{l}
        \phantom{\lor\ } a_{1} \in \rcas \land \naF\tup{\_,\_,a_{1},\_}
        \precpi \naF\tup{\_,\_,a_{2},\_} \\
        \lor\ \narR\tup{\_,\_,a_{1},\_,w_{1}} \precpi \narW\tup{w_{1}}
        \precpi \naF\tup{\_,\_,a_{2},\_}
      \end{array}}
    \end{array}} \\
  & \land \ \bigpar{
    \begin{array}{l}
      a_{1}, a_{2} \in (\rcas \cup \rfaa) \land \narR\tup{\_,\_,a_{2},\_,\_} \in \pi \\
      \implies \bigpar{
      \begin{array}{l}
        \phantom{\lor\ } a_{1} \in \rcas \land \naF\tup{\_,\_,a_{1},w_{1}}
        \precpi \narR\tup{\_,\_,a_{2},\_,\_} \\
        \lor\ \narR\tup{\_,\_,a_{1},\_,w_{1}} \precpi \narW\tup{w_{1}}
        \precpi \narR\tup{\_,\_,a_{2},\_,\_}
      \end{array}}
    \end{array}} \\
  & \land \ \bigpar{
    \begin{array}{l}
      a_{1}, a_{2} \in (\rcas \cup \rfaa) \land \bigpar{
      \begin{array}{l}
        \phantom{\lor\ } a_{1} \in \rcas \land \naF\tup{\_,\_,a_{2},w_{2}} \precpi \nlW\tup{w_{2},\_} \\
        \lor\ \narR\tup{\_,\_,a_{2},w_{2},\_} \precpi \nlW\tup{w_{2},\_}
      \end{array}} \\
      \implies \bigpar{
      \begin{array}{l}
        \phantom{\lor\ } a_{1} \in \rcas \land \naF\tup{\_,\_,a_{1},w_{1}}
        \precpi \nlW\tup{w_{1},\_} \precpi \nlW\tup{w_{2},\_} \\
        \lor\ \narR\tup{\_,\_,a_{1},w_{1},\_}
        \precpi \nlW\tup{w_{1},\_} \precpi \nlW\tup{w_{2},\_} \\
      \end{array}}
    \end{array}}
\end{align*}


\bigskip
\begin{minipage}{\textwidth}
\paragraph{NIC Atomicity}
\begin{align*}
  \remoteAtomic&(\pi) \ \defeq \ \forall a_{1},a_{2},r,w. \\
  & \bigpar{
  \begin{array}{l}
    \phantom{\land \ } \lambda_{1} = \narR\tup{r_{1},\_,a_{1},\_,w} \\
    \land \ \lambda_{2} \in \set{\naF\tup{\_,\_,a_{2},\_}, \narR\tup{\_,\_,a_{2},\_,\_}} \\
    \land \ a_{1},a_{2} \in \rrmw \
    \land \ \neqn{\node}(a_{1}) = \neqn{\node}(a_{2}) \
    \land \ \lambda_{1} \precpi \lambda_{2} \\
  \end{array}
  }
  \implies \flB\tup{w} \precpi \lambda_{2}
\end{align*}
\end{minipage}


\bigskip
\begin{minipage}{\textwidth}
\paragraph{Read Order}
\begin{align*}
    \wfrd(\pi) \ \defeq \
  & \phantom{~\land~} \ \forall \pi_2,r,w,\pi_1. \
    \pi = \pi_2 \cdot \lR\tup{r,w} \cdot \pi_1
    \implies \wfrdCPU(r,w,\pi_1) \\
  & \land \ \forall \pi_2,u,w,\pi_1. \
    \pi = \pi_2 \cdot \RMW\tup{u,w} \cdot \pi_1
    \implies \wfrdCPU(u,w,\pi_1) \\
  & \land \ \forall \pi_2,r,w,\pi_1. \
    \pi = \pi_2 \cdot \nlR\tup{r,w,\_,\_} \cdot \pi_1
    \implies \wfrdNIC(r,w,\pi_1) \\
  & \land \ \forall \pi_2,r,w,\pi_1. \
    \pi = \pi_2 \cdot \nrR\tup{r,w,\_,\_} \cdot \pi_1
    \implies \wfrdNIC(r,w,\pi_1) \\
  & \land \ \forall \pi_2,r,w,\pi_1. \
    \pi = \pi_2 \cdot \naF\tup{r,w,\_,\_} \cdot \pi_1
    \implies \wfrdNIC(r,w,\pi_1) \\
  & \land \ \forall \pi_2,r,w,\pi_1. \
    \pi = \pi_2 \cdot \narR\tup{r,w,\_,\_,\_} \cdot \pi_1
    \implies \wfrdNIC(r,w,\pi_1)
\end{align*}
\end{minipage}

\begin{align*}
  \wfrdCPU(r,w,\pi) \ \defeq \
  & \phantom{\lor~} \bigpar{
    \begin{array}{l}
      \exists \pi_2,\lambda,\pi_1. \ \pi = \pi_2 \cdot \lambda \cdot \pi_1 \\
      \land \ \lambda \in \{\flB\tup{w}, \RMW\tup{w,\_}\} \\
      \land \ \setpred{\flB\tup{w'},\RMW\tup{w',\_} \in \pi_2}
              {\loc(w')=\loc(r)} = \emptyset \\
      \land \ \setpred{w'}{
      \begin{array}{l}
        \lW\tup{w'} \in \pi \land \flB\tup{w'} \notin \pi \ \land \\
        \loc(w')=\loc(r) \land \threadt(w')=\threadt(r)
      \end{array}} = \emptyset
    \end{array}
  } \\
  & \lor \bigpar{
    \begin{array}{l}
      \exists \pi_2,\lambda,\pi_1. \ \pi = \pi_2 \cdot \lambda \cdot \pi_1 \\
      \land \ \lambda = \lW\tup{w} \land \threadt(w)=\threadt(r)
      \land \flB\tup{w} \notin \pi_2 \\
      \land \ \setpred{\lW\tup{w'} \in \pi_2}{\loc(w')=\loc(r)
      \land \threadt(w')=\threadt(r)} = \emptyset
    \end{array}
  } \\
  & \lor \bigpar{
    \begin{array}{l}
      w = init_{\loc(w)} \land \\
      \setpred{
      \begin{array}{l}
        \flB\tup{w'}, \RMW\tup{w',\_} \in \pi, \\
        \lW\tup{w''} \in \pi
      \end{array}
      }{
      \begin{array}{l}
        \loc(w')=\loc(r) \ \land \\
        \loc(w'')=\loc(r) \land \threadt(w'')=\threadt(r)
      \end{array}
      } = \emptyset
    \end{array}
  }
\end{align*}

\begin{align*}
  \wfrdNIC(r,w,\pi) \ \defeq \
  & \phantom{\lor~} \bigpar{
    \begin{array}{l}
      \exists \pi_2,\lambda,\pi_1. \ \pi = \pi_2 \cdot \lambda \cdot \pi_1 \\
      \land \ \lambda \in \{\flB\tup{w}, \RMW\tup{w,\_}\} \\
      \land \ \setpred{\flB\tup{w'},\RMW\tup{w',\_} \in \pi_2}
              {\loc(w')=\loc(r)} = \emptyset
    \end{array}
  } \\
  & \lor \bigpar{
    \begin{array}{l}
      w = init_{\loc(w)} \land \\
      \setpred{\flB\tup{w'}, \RMW\tup{w',\_} \in \pi}
      {\loc(w')=\loc(r)} = \emptyset
    \end{array}
  }
\end{align*}


\bigskip
{\bf Well-formed path.}

\begin{align*}
  \wfp(\pi) \defeq \
  & \phantom{~\land~} \ \nodup(\pi) \\
  & \land \ \backComp(\pi) \\
  & \land \ \bufFlushOrd(\pi) \\
  & \land \ \pollOrder(\pi) \\
  & \land \ \nicActOrder(\pi) \\
  & \land \ \remoteAtomic(\pi) \\
  & \land \ \wfrd(\pi) \\
\end{align*}


\begin{definition}
\begin{align*}
\wf(\Mem, \TSOmap, \Atm, \NQPmap, \pi) \ \defeq
  & \phantom{~\land~} \ \wfp(\pi)\\
  & \land \ \forall \x \in \Var. \ \Mem(\x)=\pathread(\pi,\x) \\
  & \land \ \forall \threadt \in \Threads.
    \TSOmap(\threadt) = \mktso(\varepsilon,\threadt,\pi) \\
  & \land \ \forall \node \in \Nodes.
    \Atm(\node) = \mkatm(\node,\pi) \\
  & \land \ \forall \threadt \in \Threads.
    \forall \neqn\node \in (\Nodes \setminus \set{\node(\threadt)}).
    \bigpar{
    \begin{array}[c]{l}
      \NQPmap(\threadt)(\neqn\node).\pipe =
      \mkpipe(\varepsilon,\threadt,\neqn\node,\pi)\\
      \NQPmap(\threadt)(\neqn\node).\wbR =
      \mkwbR(\varepsilon,\threadt,\neqn\node,\pi)\\
      \NQPmap(\threadt)(\neqn\node).\wbL =
      \mkwbL(\varepsilon,\threadt,\neqn\node,\pi)\\
    \end{array}}
\end{align*}
\end{definition}

Where, the functions $\pathread$, $\mktso$, $\mkatm$, $\mkpipe$, $\mkwbR$, and $\mkwbL$
are defined below.
\begin{align*}
\pathread(\lambda{\cdot}\pi,\x)\defeq
  & \begin{cases}
      w & \lambda \in \{\flB\tup{w}, \RMW\tup{w,\_}\} \land \loc(w) = x \\
      \pathread(\pi,\x) & \text{otherwise}
    \end{cases}\\
  \pathread(\varepsilon,\x)\defeq & \ init_{x}
\end{align*}

\begin{align*}
\mktso(\tso,\threadt,\varepsilon)\defeq
  & \ \tso\\
\mktso(\tso,\threadt,\pi{\cdot}\lambda) \defeq
  & \begin{cases}
      \mktso(w{\cdot}\tso,\threadt, \pi)
      & \lambda=\lW\tup{w}
        \land\threadt(w)=\threadt \land \flB\tup{w}\notin\pi\\
      \mktso(a{\cdot}\tso,\threadt,\pi)
      & \lambda=\PushTSO\tup{a}
        \land \PushNIC\tup{a} \notin \pi
        \land \threadt(a)=t \\
      \mktso(\tso,\threadt,\pi)
      & \text{otherwise}
    \end{cases}
\end{align*}

\begin{align*}
\mkatm(\node,\pi) \defeq
  & \begin{cases}
      \bot
    & \forall w. \bigpar{
      \begin{array}{l}
        \narR\tup{\_,\_,a,\_,w} \in \pi  \\
        \land \ \neqn{\node}(a) = n
      \end{array}
      } \implies \flB\tup{w} \in \pi \\
      \top
      & \text{otherwise}
    \end{cases}
\end{align*}

\begin{adjustbox}{center,minipage=0.82\paperwidth}
\begin{align*}
\mkpipe(\pipe,\threadt,\neqn\node,\varepsilon)\defeq
  & \ \pipe\\
\mkpipe(\pipe,\threadt,\neqn\node,\pi{\cdot}\lambda) \defeq
  & \begin{cases}
      \mkpipe(a{\cdot}\pipe,\threadt,\neqn\node,\pi)
      & \text{if} \bigpar{ \begin{array}{l}
          \threadt(\lambda) = \threadt
          \land \ \neqn\node(\lambda) = \neqn\node
          \land \ \lambda=\PushNIC\tup{a} \\
          \land \ \nlR\tup{\_,\_,a,\_} \not\in \pi
          \land \ \nrR\tup{\_,\_,a,\_} \not\in \pi \\
          \land \ \nF\tup{a} \not\in \pi
          \land \ \naF\tup{\_,\_,a,\_} \not\in \pi \\
          \land \ \narR\tup{\_,\_,a,\_,\_} \not\in \pi
        \end{array} } \\
      \mkpipe(w{\cdot}\pipe,\threadt,\neqn\node,\pi)
      & \text{if} \bigpar{ \begin{array}{l}
          \threadt(\lambda) = \threadt
          \land \ \neqn\node(\lambda) = \neqn\node
          \land \ \lambda=\PushNIC\tup{a} \\
          \land \ \nlR\tup{\_,\_,a,w} \in \pi
          \land \ \nrW\tup{w,\_} \not\in \pi
        \end{array} } \\
      \mkpipe(e{\cdot}\pipe,\threadt,\neqn\node,\pi)
      & \text{if} \bigpar{ \begin{array}{l}
          \threadt(\lambda) = \threadt
          \land \ \neqn\node(\lambda) = \neqn\node
          \land \ \lambda=\PushNIC\tup{a} \\
          \land \ \nlR\tup{\_,\_,a,w} \in \pi
          \land \ \nrW\tup{w,e} \in \pi \\
          \land \ \lCN\tup{e} \not\in \pi
        \end{array} } \\
      \mkpipe(w{\cdot}\pipe,\threadt,\neqn\node,\pi)
      & \text{if} \bigpar{ \begin{array}{l}
          \threadt(\lambda) = \threadt
          \land \ \neqn\node(\lambda) = \neqn\node
          \land \ \lambda=\PushNIC\tup{a} \\
          \land \ \nrR\tup{\_,\_,a,w} \in \pi
          \land \ \nlW\tup{w,\_} \not\in \pi
        \end{array} } \\
      \mkpipe(w{\cdot}\pipe,\threadt,\neqn\node,\pi)
      & \text{if} \bigpar{ \begin{array}{l}
          \threadt(\lambda) = \threadt
          \land \ \neqn\node(\lambda) = \neqn\node
          \land \ \lambda=\PushNIC\tup{a} \\
          \land \ \naF\tup{\_,\_,a,w} \in \pi
          \land \ \nlW\tup{w,\_} \not\in \pi
        \end{array} } \\
      \mkpipe(w{\cdot}w'{\cdot}\pipe,\threadt,\neqn\node,\pi)
      & \text{if} \bigpar{ \begin{array}{l}
          \threadt(\lambda) = \threadt
          \land \ \neqn\node(\lambda) = \neqn\node
          \land \ \lambda=\PushNIC\tup{a} \\
          \land \ \narR\tup{\_,\_,a,w,w'} \in \pi
          \land \ \narW\tup{w'} \not\in \pi
        \end{array} } \\
      \mkpipe(\pipe,\threadt,\neqn\node,\pi)
      & \text{otherwise}
    \end{cases}
\end{align*}
\end{adjustbox}

\begin{align*}
  \mkwbR(\wbR,\threadt,\neqn\node,\varepsilon)\defeq
  & \ \wbR \\
  \mkwbR(\wbR,\threadt,\neqn\node,\pi{\cdot}\lambda)\defeq
  & \begin{cases}
      \mkwbR(w{\cdot}\wbR,\threadt,\neqn\node,\pi)
      & \text{if} \bigpar{
        \begin{array}{l}
          \threadt(\lambda) = \threadt
          \land \neqn\node(\lambda) = \neqn\node
          \land \flB\tup{w} \notin \pi \\
          \land \ \lambda\in\set{\nrW\tup{w,\_}, \narW\tup{w}}
        \end{array}
        } \\
      \mkwbR(\wbR,\threadt,\neqn\node,\pi)
      & \text{otherwise}
    \end{cases}
\end{align*}

\begin{adjustbox}{center,minipage=0.82\paperwidth}
\begin{align*}
  \mkwbL(\wbL,\threadt,\neqn\node,\varepsilon)\defeq
  & \ \wbL \\
  \mkwbL(\wbL,\threadt,\neqn\node,\pi{\cdot}\lambda)\defeq
  & \begin{cases}
      \mkwbL(e{\cdot}w{\cdot}\wbL,\threadt,\neqn\node,\pi)
      & \text{if} \bigpar{ \begin{array}{l}
          \threadt(\lambda) = \threadt
          \land \ \neqn\node(\lambda) = \neqn\node
          \land \ \lambda=\nlW\tup{w,e} \\
          \land \ \flB\tup{w} \notin \pi \land \lP\tup{\_,e} \notin \pi
        \end{array}} \\
      \mkwbL(e{\cdot}\wbL,\threadt,\neqn\node,\pi)
      & \text{if} \bigpar{ \begin{array}{l}
          \threadt(\lambda) = \threadt
          \land \ \neqn\node(\lambda) = \neqn\node
          \land \ \lambda=\nlW\tup{w,e} \\
          \land \ \flB\tup{w} \in \pi \land \lP\tup{\_,e} \notin \pi
        \end{array}} \\
      \mkwbL(e{\cdot}\wbL,\threadt,\neqn\node,\pi)
      & \text{if} \bigpar{ \begin{array}{l}
          \threadt(\lambda) = \threadt
          \land \ \neqn\node(\lambda) = \neqn\node
          \land \ \lambda=\lCN\tup{e} \\
          \land \ \lP\tup{\_,e} \notin \pi
        \end{array}} \\
      \mkwbL(\wbL,\threadt,\neqn\node,\pi)
      & \text{otherwise}
    \end{cases}
\end{align*}
\end{adjustbox}

\begin{theorem}
\label{thm:wfpi}
For all
$\prog, \prog', \Mem, \Mem', \TSOmap, \TSOmap', \Atm, \Atm', \NQPmap, \NQPmap', \pi, \pi'$:

\begin{itemize}
\item $\wf(\Mem_0, \TSOmap_0, \Atm_{0}, \NQPmap_0, \varepsilon)$;
\item if $\prog, \Mem, \TSOmap, \Atm, \NQPmap, \pi \trp \prog', \Mem', \TSOmap', \Atm', \NQPmap', \pi'$ and
  $\wf(\Mem,\TSOmap,\Atm,\NQPmap,\pi)$ then $\wf(\Mem',\TSOmap',\Atm',\NQPmap',\pi')$;
\item if
  $\prog, \Mem_0, \TSOmap_0, \Atm_{0}, \NQPmap_0, \varepsilon \trp^* (\lambda
  \threadt.\cskip), \Mem, \TSOmap_0, \Atm_{0}, \NQPmap, \pi$ such that forall
  $\threadt,\neqn\node$ we have
  $\NQPmap(\threadt)(\neqn\node) = \tup{\varepsilon, \varepsilon,
    \queue{\nEX}}$, then $\wf(\Mem,\TSOmap_0,\Atm_{0},\NQPmap,\pi)$ and $\complete(\pi)$.
\end{itemize}
\end{theorem}

The proof of the first part follows trivially from the definitions of $\Mem_0$,
$\TSOmap_0$, $\Atm_{0}$, and $\NQPmap_0$. The second part is proved by induction on the
structure of $\trp$. The last part follows from the previous two parts and
induction on the length of $\trp^*$, as well as how the definition of $\wf$ on
empty store buffers and queue pairs (regardless of \nEX in \wbL) implies
$\complete(\pi)$.


%% file: annot-to-decl.tex


\subsection{From Annotated Semantics to Declarative Semantics}
\label{sec:annot-sem-a2d}

We define $$\getG(\pi) \defeq
\begin{cases}
(\Actions,\po,\rf,\pf,\mo,\ro,\rao) & \text{if }\wfp(\pi) \land \complete(\pi)\\
\text{undefined}&\text{otherwise}
\end{cases}$$

with

$$ \Actions \defeq \Actions_0 \cup \set{\getA(\lambda)\st \lambda\in\pi}$$

Recall that $\Actions_0$ is the set of initialisation events
$\setpred{init_{\x}}{\x \in \Var}$, where $\lab(init_x) = \lW(\x,0)$

$$\getA : \aLabels \rightharpoonup \Actions$$ 
\begin{minipage}[t]{0.5\textwidth}
\vspace{-1em}
\begin{align*}
  \getA(\lR \tup{r,\_})       \ \defeq& \ r \\
  \getA(\lW \tup{w})          \ \defeq& \ w \\
  \getA(\RMW \tup{u,\_})      \ \defeq& \ u \\
  \getA(\lF \tup{f})          \ \defeq& \ f \\
  \getA(\nlR \tup{r,\_,\_,\_})\ \defeq& \ r \\
  \getA(\nrW \tup{w})      \ \defeq& \ w \\
  \getA(\nrR \tup{r,\_,\_,\_})\ \defeq& \ r \\
  \getA(\nlW \tup{w,\_})      \ \defeq& \ w \\
  \getA(\naF \tup{r,\_,\_,\_})\ \defeq& \ r \\
\end{align*}
\end{minipage}
\begin{minipage}[t]{0.5\textwidth}
\vspace{-1em}
\begin{align*}
  \getA(\narR \tup{r,\_,\_,\_,\_})\ \defeq& \ r \\
  \getA(\narW \tup{w})      \ \defeq& \ w \\
  \getA(\lP \tup{p,\_})       \ \defeq& \ p \\
  \getA(\nF \tup{f})          \ \defeq& \ f \\
  \getA(\flB \tup{w})         \ \defeq& \ w \\
  \getA(\PushTSO\tup{\_})     \quad & \ \text{is undefined} \\
  \getA(\PushNIC\tup{\_})     \quad & \ \text{is undefined} \\
  \getA(\lCN \tup{\_})        \quad & \ \text{is undefined} \\
  \getA(\lE\tup{\_})          \quad & \ \text{is undefined} \\
\end{align*}
\end{minipage}

We define $\getIL(\_, \pi)$ and $\getOL(\_, \pi)$ to perform the reverse
operation of $\getA$. In the case of write events, $\getIL(\_, \pi)$ retrieves
the first label sending the write to the buffer, while $\getOL(\_, \pi)$
retrieves the second label committing the write to memory.

$$\getIL(\_, \pi),\getOL(\_, \pi) : \set{\getA(\lambda)\st \lambda\in\pi}
\rightarrow \aLabels$$

For all $\lambda \in \pi$:
\begin{itemize}
\item if $\typ(\lambda) \in \set{\lR,\RMW,\lF,\lP,\nlR,\nrR,\narR,\naF,\nF}$, \\then
  $\getIL(\getA(\lambda), \pi) \defeq \getOL(\getA(\lambda), \pi) \defeq
  \lambda$; 
\item if $\typ(\lambda) \in \set{\lW,\nlW,\nrW,\narW}$, \\then
  $\getIL(\getA(\lambda), \pi) \defeq \lambda$ while
  $\getOL(\getA(\lambda), \pi) \defeq \flB\tup{\lambda}$.
\item if $\lambda = \flB\tup{w}$, then from $\backComp(\pi)$ there is
  $\lambda' \precpi \lambda$ such that $\typ(\lambda) \in \set{\lW,\nlW,\nrW,\narW}$
  and $\getA(\lambda') = \getA(\lambda) = w$. From the previous case, we have
  $\getIL(w, \pi) \defeq \lambda'$ and $\getOL(w, \pi) \defeq \lambda$.
\end{itemize}

From this we define two relations \IB and \OB on $\Actions$ total on all
meaningful events by copying the ordering in $\pi$.

$$\IB \defeq \setpred{(e_1,e_2)}{\getIL(e_1,\pi) \precpi \getIL(e_2,\pi)}
\cup \bigpar{ \Actions_0 \times (\Actions \setminus \Actions_0) }$$
$$\OB \defeq \setpred{(e_1,e_2)}{\getOL(e_1,\pi) \precpi \getOL(e_2,\pi)}
\cup \bigpar{ \Actions_0 \times (\Actions \setminus \Actions_0) }$$

From $\wfp(\pi)$, \IB and \OB are transitive and irreflexive. Note: we could
make \IB and \OB total by adding an arbitrary total order on $\Actions_0$.

$$\rf \defeq \setpred{(w,r)}{
  \begin{array}{l}
  \lR\tup{r,w}\in\pi
  \lor \nlR\tup{r,w,\_,\_}\in\pi
  \lor \nrR\tup{r,w,\_,\_}\in\pi
  \lor \RMW\tup{r,w}\in\pi \\
  \lor\ \narR\tup{r,w,\_,\_,\_}\in\pi
  \lor \naF\tup{r,w,\_,\_}\in\pi
  \end{array}
}$$

$$\pf \defeq
\setpred{(w,p)}{
  \begin{array}{l}
  \nlW\tup{w,e} \precpi \lP\tup{p,e} \\
  \lor\ \nrW\tup{w,e} \precpi \lP\tup{p,e}
  \end{array}
}
$$

$$\generates{\lambda}{e}{\pi} \ \defeq \bigpar{
\begin{array}{l}
  \phantom{\lor} \ \lambda \in \{\lR\tup{e,\_}, \lW\tup{e},
  \RMW\tup{e,\_}, \PushTSO\tup{e}, \lP\tup{e,\_}, \lF\tup{e}\} \\
  \lor \ \lambda = \PushTSO\tup{a} \land \bigpar{
  \begin{array}{l}
    \phantom{\lor} \ \lambda \precpi \nlR\tup{e,\_,a,\_} \\
    \lor \ \lambda \precpi \nlR\tup{\_,\_,a,e} \\
    \lor \ \lambda \precpi \nrR\tup{e,\_,a,\_} \\
    \lor \ \lambda \precpi \nrR\tup{\_,\_,a,e} \\
    \lor \ \lambda \precpi \naF\tup{e,\_,a,\_} \\
    \lor \ \lambda \precpi \naF\tup{\_,\_,a,e} \\
    \lor \ \lambda \precpi \narR\tup{e,\_,a,\_,\_} \\
    \lor \ \lambda \precpi \narR\tup{\_,\_,a,e,\_} \\
    \lor \ \lambda \precpi \narR\tup{\_,\_,a,\_,e}
  \end{array}}
\end{array}}
$$

$$\po \defeq \bigpar{
\begin{array}{l}
\Actions_0 \times (\Actions\setminus\Actions_0) \\ \cup
\setpred{(e_1,e_2)}{
\begin{array}{l}
  \phantom{\land} \ \lambda_1 \precpi \lambda_2
  \land \threadt(\lambda_1)=\threadt(\lambda_2) \\
  \land \ \generates{\lambda_1}{e_1}{\pi} \\
  \land \ \generates{\lambda_2}{e_2}{\pi}
\end{array}} \\ \cup
\setpred{(r,w)}{
  \begin{array}{l}
    \phantom{\lor} \ \nlR\tup{r,\_,\_,w} \in \pi \\
    \lor \ \nrR\tup{r,\_,\_,w} \in \pi \\
    \lor \ \naF\tup{r,\_,\_,w} \in \pi \\
    \lor \ \narR\tup{r,\_,\_,\_,w} \in \pi
  \end{array}} \\ \cup
  \setpred{(w_{1},w_{2})}{\narR\tup{\_,\_,\_,w_{2},w_{1}}} \in \pi
\end{array}}
$$

$$\mo \defeq
\setpred{(w_1,w_2)}{
\begin{array}{l}
\phantom{\land} \ w_1 = init_x \\
\land \ \bigpar{ \flB\tup{w_2} \in \pi \lor \RMW\tup{w_2,\_} \in \pi } \\
\land \ \loc(w_1) = x = \loc(w_2)
\end{array}}
\cup
\setpred{(w_1,w_2)}{
\begin{array}{l}
\phantom{\land} \ \lambda_1 \precpi \lambda_2 \\
\land \ \lambda_1 \in \{\flB\tup{w_1}, \RMW\tup{w_1,\_} \} \\
\land \ \lambda_2 \in \{\flB\tup{w_2}, \RMW\tup{w_2,\_} \} \\
\land \ \loc(w_1) = \loc(w_2)
\end{array}}
$$

$$\ro \defeq
\begin{array}{l}
\bigpar{
  \begin{array}{l}
    \phantom{\cup} \ \setpred{(r,w)}{\sameqp(r,w) \land
    \nlR\tup{r,\_,\_,\_} \precpi \nlW\tup{w,\_} \precpi \flB\tup{w}} \\
    \cup \ \setpred{(r,w)}{\exists \lambda_{r},\lambda_{w}. \sameqp(r,w) \land
    \lambda_{r} \precpi \lambda_{w} \precpi \flB\tup{w}} \\
    \cup \ \setpred{(w,r)}{\sameqp(w,r) \land
    \nlW\tup{w,\_} \precpi \flB\tup{w} \precpi \nlR\tup{r,\_,\_,\_}} \\
    \cup \ \setpred{(w,r)}{\exists \lambda_{r},\lambda_{w}. \sameqp(w,r) \land
    \lambda_{w} \precpi \flB\tup{w} \precpi \lambda_{r}} \\
  \end{array}} \\
  \text{where}\ \lambda_{r} \in \set{\nrR\tup{r',\ldots}, \naF\tup{r',\ldots}, \narR\tup{r',\ldots}} \\
  \phantom{\text{where}\ } \lambda_{w} \in \set{\nrW\tup{w,\_}, \narW\tup{w}}
\end{array}
$$

$$\rao \defeq \bigpar{
  \setpred{(r_{1},r_{2})}{\neqn{\node}(a_{1}) = \neqn{\node}(a_{2}) \land
    \bigpar{
      \begin{array}{l}
        \phantom{\land\ } \lambda_{1} \precpi \lambda_{2} \\
        \land\ \lambda_{1} \in \set{\naF\tup{r_{1},a_{1},\ldots}, \narR\tup{r_{1},a_{1},\ldots}} \\
        \land\ \lambda_{2} \in \set{\naF\tup{r_{2},a_{2},\ldots}, \narR\tup{r_{2},a_{2},\ldots}}
      \end{array}
    }
  }
}
$$

From an execution graph $E=\getG(\pi)$, we use the definitions of the paper to
define \oppo, \ippo, \rfi, \rfe, \fr, \fri, \ar, \ob, and \ib.


\begin{lemma}
  \label{lem:nlw-put-rrmw}
  $w \in \nlW \implies \exists r. \bigpar{
    \begin{array}{l}
      \phantom{\lor\ } r \in \nrR \land (r,w) \in \imm{\po} \\
      \lor\ r \in \narR \land (r,w) \in \imm{\po} \cup (\imm{\po})^{2}
    \end{array}
  }$
\end{lemma}
\begin{proof}
  By definition of \po, we can only have such $w \in \nlW$ if there is some $\lambda = \PushTSO\tup{a}$ which generates $w$ in $\pi$.
  Then we can consider the cases of $a$ such that $\PushTSO\tup{a}$ generates some $w \in \nlW$.
  Either:
  \begin{itemize}
    \item $a \in \rput$, then there is some $r \in \nrR$ with $(r,w) \in \imm{\po}$
    \item $a \in \rcas \cup \rfaa$, then there is some $r \in \narR$ with either $(r,w) \in \imm{\po}$ (in the case of a failed \rcas) or $(r,w) \in (\imm{\po})^{2}$ (in the case of a successful \rcas\ or \rfaa)
  \end{itemize}
\end{proof}

\begin{theorem}
  $\getG(\pi)$ is well-formed.
\end{theorem}
\begin{proof}
  We need to check the conditions of a pre-execution (\cref{def:pre-executions})
  and of well-formedness (\cref{def:well-formed}). For the pre-execution
  conditions:
  \begin{itemize}
  \item Checking $\Actions^0 \times (\Actions \setminus \Actions^0) \suq \po$:

    by definition.

  \item Checking $\po$ is a union of strict partial orders each on one thread:

    If $\threadt(\action_1) \neq \threadt(\action_2)$, then
    $(\action_1,\action_2) \not\in \po$ and $(\action_2,\action_1) \not\in \po$
    by definition. If $\threadt(\action_1) = \threadt(\action_2)$, then either
    $(\action_1,\action_2) \in \po$ or $(\action_2,\action_1) \in \po$. This
    comes from the second case of the definition of $\po$: if there is
    $\lambda_1$ and $\lambda_2$ such that
    $\generates{\lambda_i}{\action_i}{\pi}$, then either
    $\lambda_1 \precpi \lambda_2$ or $\lambda_2 \precpi \lambda_1$.

  \item Checking that \rf is functional on its range:

    If $r \in \Read \subseteq \set{\getA(\lambda)\st \lambda\in\pi}$,
    then we have either $\lR\tup{r,\_}$, $\nlR\tup{r,\_,\_,\_}$,
    $\nrR\tup{r,\_,\_,\_}$, $\naF\tup{r,\_,\_,\_}$, or $\narR\tup{r,\_,\_,\_,\_}$ in $\pi$, and $r$ have at least one antecedent.

    If $(w,r) \in \rf$, let us assume $r \in \nlR$, then by definition
    $\nlR\tup{r,w,\_,\_} \in \pi$. Since $\nodup(\pi)$, for all $w' \neq w$, we
    have $\nlR\tup{r,w',\_,\_} \notin \pi$, and syntactically we cannot write
    $\lR\tup{r,\_}$ or $\nrR\tup{r,\_,\_,\_}$, so $(w',r) \notin \rf$.
    Similarly, $r\in\lR$, $r\in\nrR$, $r\in\naF$ or $r\in\narR$ only have one antecedent.

  \item Checking that \rf relates events on the same location with matching
    values:

    By syntactic definition of the annotated labels $\lR$, $\nlR$, $\nrR$, $\naF$ and $\narR$,
    e.g., $\lR\tup{r,w} \implies \locveq(r,w)$.

  \item Checking that $\mo$ is a union of strict total orders for writes on each
    variables:

    By definition of $\mo$, given that we have $\complete(\pi)$, e.g., if
    $\lW\tup{w} \in \pi$ then $\flB\tup{w} \in \pi$.

  \item Checking that $\pf \suq \po \cap \sqp$:

    If $(w,p) \in \pf$ with $w \in \nlW$ (resp.\ $\nrW$), then we have
    $\nlW\tup{w,e} \precpi \lP\tup{p,e}$. There is $\lambda$ such that
    $\generates{\lambda}{w}{\pi}$, and we have
    $\lambda \precpi \nlW\tup{w,e} \precpi \lP\tup{p,e}$. Also,
    $\threadt(p) = \threadt(w)$ and
    $\neqn\node(p) = \neqn\node(e) = \neqn\node(w)$, so we have $(w,p) \in \po$
    and $(w,p) \in \sqp$.

  \item Checking that $\pf$ is functional on its domain:

    If $(w,p) \in \pf$ with $w \in \nlW$ (resp.\ $\nrW$), then we have
    $\nlW\tup{w,e} \precpi \lP\tup{p,e}$. From $\nodup(\pi)$, for all
    $p' \neq p$ we have $\lP\tup{p,e} \notin \pi$, so $w$ has at most one image.

  \item Checking that $\pf$ is total and functional on its range:

    If $p \in \Actions$, then there is $e \in \nlEX$ (resp.\ $\nrEX$) such that
    $\lP\tup{p,e} \in \pi$. From $\backComp(\pi)$ there is $w \in \nlW$ (resp.
    $\nrW$) such that $\nlW\tup{w,e} \precpi \lP\tup{p,e}$, and so
    $(w,p) \in \pf$. From $\nodup(\pi)$, $e$ cannot be used in another $\nlW$
    (resp.\ $\nrW$) annotated label, and $p$ has exactly one antecedent.

  \item Checking that for all $(a, b) \in \sqp$, $a \in \nrR \cup \naF \cup \narR$, $b \in \nrW \cup \narW$, (resp.\ $\nlR/\nlW$) then $(a, b) \in \ro \cup \inv\ro$:

    By definition of $\ro$, given that $\bufFlushOrd(\pi)$ forbids such
    interleavings as
    $\nrW\tup{w,\_} \precpi \nrR\tup{r,\_,\_,\_} \precpi \flB\tup{w}$ (resp.
    $\nlW$ and $\nlR$) when $\sameqp(r,w)$.

    \item Checking that $\rao$ is a union of strict total orders for remote atomic reads:

          By definition of $\rao$.

  \end{itemize}

  For the well-formedness conditions:

  \begin{itemize}

  \item[(1)] Let us assume $(w_1,w_2) \in \po \cap \sqp$ and
    $(w_2,p_2) \in \pf$. The three events are on the same thread and queue pair.

    If $w_1 \in \nlW$, then by $\complete(\pi)$ there is a chain
    $\PushTSO\tup{a_1} \precpi \PushNIC\tup{a_1} \precpi \nR
    \precpi \nlW\tup{w_1,e_1}$ for some $\nR \in \set{\nrR\tup{\_,\_,a_{1},w_{1}}, \naF\tup{\_,\_,a_{1},w_{1}}, \narR\tup{\_,\_,a_{1},w_{1},\_}}$; if $w_1 \in \nrW$, there is instead a chain
    $\PushTSO\tup{a_1} \precpi \PushNIC\tup{a_1} \precpi \nlR\tup{\_,\_,a_1,w_1}
    \precpi \nrW\tup{w_1,e_1} \precpi \lCN\tup{e_1}$. Similarly there is a chain
    for $w_2$. By $(w_1,w_2) \in \po$ we have
    $\PushTSO\tup{a_1} \precpi \PushTSO\tup{a_2}$, and by $\bufFlushOrd(\pi)$ we
    have $\PushNIC\tup{a_1} \precpi \PushNIC\tup{a_2}$.

    Let us call $\lambda_1$ the last annotated label on the chain for $w_1$,
    i.e., either $\nlW\tup{w_1,e_1}$ or $\lCN\tup{e_1}$. Similarly, $\lambda_2$
    is the last annotated label on the chain for $w_2$. There are four cases to
    consider, but in all four $\nicActOrder(\pi)$ implies
    $\lambda_1 \precpi \lambda_2$.

    Then, from $\pollOrder(\pi)$, there is $p_1$ such that
    $\lP\tup{p_1,e_1} \precpi \lP\tup{p_2,e_2}$. By definitions, we have both
    $(w_1,p_1) \in \pf$ and $(p_1,p_2) \in \po$.
    
  \item[(2)] If $r \in \nlR$, then there is $w \in \nrW$ (taken from
    $\nlR\tup{r,\_,\_,w}$) such that $(r,w) \in \imm{\po}$. This is by the
    last case of definition of $\po$, since there is $\lambda_a$ such that we
    have both
    $\generates{\lambda_a}{r}{\pi}$ and $\generates{\lambda_a}{w}{\pi}$. \\
    Similarly for $\nrR$/$\nlW$ and $\nrW$/$\nlR$.
  \item [(3)] If $(r,w) \in \imm{\po}$, $\typ(r) \in \set{\nlR,\nrR}$, and
    $\typ(w) \in \set{\nlW,\nrW}$, then $(r,w) \in \po$ comes from the third case
    of the definition of $\po$, and we have either $\nlR\tup{r,\_,\_,w}$ or
    $\nrR\tup{r,\_,\_,w}$ in $\pi$. In both cases, we have $\valr(r) = \valw(w)$
    by syntactic definition of the annotated labels.

  \item [(4)]
    \begin{enumerate*}
      \item [(a)] If $r \in \narR$, then either:
        There is $\naF\tup{r,\_,\_,w} \in \pi$ such that $w \in \nlW$ and $(r,w) \in \imm{\po}$.
        This follows from the second case definition of $\po$.
        There is $\narR\tup{r,\_,\_,w_{2},w_{1}} \in \pi$ such that $w_{1} \in \narW$, $w_{2} \in \nlW$, and $(r,w_{1}),(w_{1},w_{2}) \in \imm{\po}$.
        This follows from the second and third cases of the definition of $\po$ since there is $\lambda_{a}$ which generates $r$, $w_{1}$ and $w_{2}$ in $\pi$.
      \item [(b)] If $w \in \narW$ then $(r,w),(w,w') \in \imm{\po}$ with $r \in \narR$ and $w' \in \nlW$ comes from the second case definition of $\po$.
    \end{enumerate*}

  \item [(5)] If $(r, w) \in \EG.\imm{\po}$, $\typ(r) = \narR$ and $\typ(w) = \nlW$,
    then $(r,w)$ comes from the second case definition of $\po$ and we have $\naF\tup{r,\_,\_,w} \in \pi$.
    Then $\valr(r) = \valw(w)$ by the syntax of annotated labels.
    If $(r, w_{1}), (w_{1}, w_{2}) \in \EG.\imm{po}$, $\typ(r) = \narR$, $\typ(w_{1}) = \narW$ and $\typ(w_{2}) = \nlW$,
    then $(r,w_{1})$ comes from the second case definition of $\po$ and $(w_{1},w_{2})$ from the third case,
    so we have $\narR\tup{r,\_,\_,w_{2},w_{1}} \in \pi$.
    Then $\valr(r) = \valw(w_{2})$ by the syntax of annotated labels.
  \item [(6)] Comes from \cref{lem:nlw-put-rrmw}.
  \end{itemize}
\end{proof}

\begin{lemma}
  \label{lem:OBIB}
  $\OB ; [\Inst] \subseteq \IB$ and $[\Inst] ; \IB \subseteq \OB$.
\end{lemma}
\begin{proof}
  If $(e_1,e_2) \in \OB ; [\Inst]$, then
  $\getOL(e_1,\pi) \precpi \getOL(e_2,\pi) = \getIL(e_2,\pi)$.
  \begin{itemize}
  \item If $e_1 \in \Inst$, then $\getOL(e_1,\pi) = \getIL(e_1,\pi)$, so we have
    $\getIL(e_1,\pi) \precpi \getIL(e_2,\pi)$ and $(e_1,e_2) \in \IB$.
  \item If $e_1 \in \set{\lW,\nlW,\nrW,\narW}$, there is $\lambda$ such that
    $\typ(\lambda) \in \set{\lW,\nlW,\nrW,\narW}$, $\getA(\lambda) = e_1$, and
    $\getIL(e_1,\pi) = \lambda \precpi \flB\tup{e_1} = \getOL(e_1,\pi)$. By
    transitivity we again have $\getIL(e_1,\pi) \precpi \getIL(e_2,\pi)$ and
    $(e_1,e_2) \in \IB$.
  \end{itemize}

  With a similar reasoning, we can see that $[\Inst] ; \IB \subseteq \OB$.

\end{proof}

\begin{theorem}
  $\getG(\pi)$ is consistent.
\end{theorem}
\begin{proof}
  From Definition~\ref{def:consistency}, we need to check that both $\ib$
  and $\ob$ are irreflexive. Since $\IB$ and $\OB$ are irreflexive, it
  is enough to show that $\ib \subseteq \IB$ and $\ob \subseteq \OB$.

  The explicit definition using limits is the following (where
  $\rfe \defeq (\rf \setminus \rfi)$ includes $(\rf \cap \sqp)$ since we assume
  the PCIe guarantees hold):
  \begin{align*}
    \ib^0 & \defeq \bigpar{\ippo \ \cup \ \rf \ \cup \ \pf \ \cup \
            \fri \ \cup \ \ro}^+ \\
    \ob^0 & \defeq \bigpar{\oppo \ \cup \ \rfe \ \cup \ [\nlW];\pf \
            \cup \ \fr \ \cup \ \ro \ \cup \ \mo \ \cup \ \rao \ \cup \ \ar;\rao}^+ \\
    \ib^{n+1} & \defeq \bigpar{\ib^n \cup \ob^n;[\Inst]}^+ \\
    \ob^{n+1} & \defeq \bigpar{\ob^n \cup [\Inst];\ib^n}^+ \\
    \ib & \defeq \lim_{n \rightarrow \infty} \ib^n \\
    \ob & \defeq \lim_{n \rightarrow \infty} \ob^n
  \end{align*}

  It is then enough to show that $\ib^0 \subseteq \IB$ and
  $\ob^0 \subseteq \OB$. Using Lemma~\ref{lem:OBIB} above, we can check the
  induction case:
  \begin{align*}
    \ib^{n+1} &= \bigpar{\ib^n \cup \ob^n;[\Inst]}^+ \subseteq \bigpar{\ib^n \cup
                \OB;[\Inst]}^+ \subseteq \bigpar{\IB \cup \IB}^+ = \IB \\
    \ob^{n+1} &= \bigpar{\ob^n \cup [\Inst];\ib^n}^+ \subseteq \bigpar{\ob^n \cup
                [\Inst];\IB}^+ \subseteq \bigpar{\OB \cup \OB}^+ = \OB
  \end{align*}

  Since \IB and \OB are transitive, we need to check the components of $\ib^0$
  and $\ob^0$. There are twelve cases to verify.
  \begin{itemize}
  \item Checking $\ippo \subseteq \IB$.\\
    Let $\CPUactions = \set{\lR,\lW,\RMW,\lF,\lP}$ and
    $\NICactions = \set{\nlR,\nrR,\narR,\naF,\nlW,\nrW,\narW,\nF}$.
    $[\CPUactions];\po \subseteq \IB$ by definition of $\po$ and $\IB$:
    $\CPUactions$ are the events for which the same annotated label is used to
    define $\po$ and $\IB$, i.e.,
    $\forall e\in \CPUactions, \ \generates{\getIL(e, \pi)}{e}{\pi}$.\\
    To check that $[\NICactions];\ippo;[\NICactions] \subseteq \IB$, there are 36
    cases to consider. They are all trivially satisfied by $\nicActOrder(\pi)$ and $\backComp(\pi)$.
  \item Checking $\oppo \subseteq \OB$.\\
    From above we have $[\Inst];\oppo \subseteq
    [\Inst];\ippo \subseteq [\Inst];\IB \subseteq \OB$. \\
    $[\lW];\po;[\Actions \setminus (\lR \cup \lP)] \subseteq \OB$ by using
    $\bufFlushOrd(\pi)$. \\
    For the remaining cases:
    \begin{itemize}
    \item[(G7)] $[\nrW];(\po\cap\sqp);[\nrW] \subseteq \OB$ comes from
      $\nicActOrder(\pi)$ (i.e., $\nrW\tup{\ldots} \precpi \nrW\tup{\ldots}$)
      and $\bufFlushOrd(\pi)$ (i.e.,
      $\flB\tup{\ldots} \precpi \flB\tup{\ldots}$).
    \item[(G8)] $[\nrW];(\po\cap\sqp);[\narR] \subseteq \OB$ comes from
      $\nicActOrder(\pi)$ (i.e., $\nrW\tup{\ldots} \precpi \narR\tup{\ldots}$)
      and $\bufFlushOrd(\pi)$ (i.e., $\flB\tup{\ldots} \precpi \narR\tup{\ldots}$).
    \item[(G9)] If $e_1 \in \nrW$, $e_3 \in \narW$, and
      $(e_1,e_3) \in (\po\cap\sqp)$, then from \cref{def:executions} there is
      $e_2 \in \narR$ such that $(e_2, e_3) \in \imm{\po}$ and thus
      $(e_1,e_2) \in (\po\cap\sqp)$. From case G8 above, we have
      $(e_1,e_2) \in \OB$. From $\backComp(\pi)$, we have
      $(e_2,e_3) \in [\Inst];\IB \subseteq \OB$. Thus
      $[\nrW];(\po\cap\sqp);[\narW] \subseteq \OB$.
    \item[(G10)] $[\nrW];(\po\cap\sqp);[\nrR] \subseteq \OB$ comes from
      $\nicActOrder(\pi)$ (i.e., $\nrW\tup{\ldots} \precpi \nrR\tup{\ldots}$)
      and $\bufFlushOrd(\pi)$ (i.e.,
      $\nrW\tup{\ldots} \precpi \flB\tup{\ldots} \precpi \nrR\tup{\ldots}$).
    \item[(G11)] If $e_1 \in \nrW$, $e_3 \in \nlW$, and
      $(e_1,e_3) \in (\po\cap\sqp)$, then from \cref{def:executions} there is
      $e_2 \in (\narR \cup \nrR)$ such that $(e_2, e_3) \in \imm{\po}^{\set{1,2}}$ and thus
      $(e_1,e_2) \in (\po\cap\sqp)$.
      Then $(e_{1},e_{2}) \subseteq \OB$ comes from cases G9 and G10 respectively.
      From $\backComp(\pi)$, we have
      $(e_2,e_3) \in [\Inst];\IB \subseteq \OB$. Thus
      $[\nrW];(\po\cap\sqp);[\nlW] \subseteq \OB$.
    \item[(I7)] $[\narW];(\po\cap\sqp);[\nrW] \subseteq \OB$ comes from
      $\nicActOrder(\pi)$ (i.e., $\narW\tup{\ldots} \precpi \nrW\tup{\ldots}$)
      and $\bufFlushOrd(\pi)$ (i.e.,
      $\flB\tup{\ldots} \precpi \flB\tup{\ldots})$.
    \item[(I8)] $[\narW];(\po\cap\sqp);[\narR] \subseteq \OB$ follows from \cref{def:executions}, $\nicActOrder(\pi)$ and $\bufFlushOrd(\pi)$ by similar reasoning to I7.
    \item[(I9)] $[\narW];(\po\cap\sqp);[\narW] \subseteq \OB$ follows similarly to I7.
    \item[(I10)] $[\narW];(\po\cap\sqp);[\nrR] \subseteq \OB$ follows similarly to I7.
    \item[(K11)] $[\nlW];(\po\cap\sqp);[\nlW] \subseteq \OB$ comes from
      $\nicActOrder(\pi)$ (i.e., $\nlW\tup{\ldots} \precpi \nlW\tup{\ldots}$)
      and $\bufFlushOrd(\pi)$ (i.e.,
      $\flB\tup{\ldots} \precpi \flB\tup{\ldots}$).
    \end{itemize}
  \item Checking $\rfe \subseteq \OB$.\\
    If $(w,r) \in \rfe$, there is $\pi_1$ and $\pi_2$ such that
    $\pi = \pi_2 \cdot \getOL(r,\pi) \cdot \pi_1$, and we use $\wfrd(\pi)$.
    \begin{itemize}
    \item If $r \in \lR$, we have $\wfrdCPU(r,w,\pi_1)$. The definition allow
      for three different cases. In the first case,
      $\lambda \in \{\flB\tup{w},\RMW\tup{w,\_}\}$ is in $\pi_1$; we have
      $\lambda = \getOL(w,\pi) \precpi \getOL(r,\pi)$ and so $(w,r) \in \OB$. In
      the second case, we have $\lambda = \lW\tup{w}$ and
      $\threadt(w) = \threadt(r)$; so
      $(w,r) \in [\lW] ; (\rf \cap \sthd) ; [\lR] = \rfi$, which contradicts
      $(w,r) \in \rfe = \rf \setminus \rfi$. In the third case, $w = init_x$ for
      some location $x$, so
      $(w,r) \in \Actions_0 \times (\Actions \setminus \Actions_0) \subseteq
      \OB$.
    \item If $r \in \RMW$, similarly to above, except the second case of
      $\wfrdCPU(r,w,\pi_1)$ is not possible because of $\bufFlushOrd(\pi)$:
      $\flB\tup{w} \notin \pi_1$ while $\RMW$ acts as a memory fence.
    \item If $r \in \nlR$, we have $\wfrdNIC(r,w,\pi_1)$, with two
      possibilities. In the first case,
      $\lambda \in \{\flB\tup{w},\RMW\tup{w,\_}\}$ is in $\pi_1$; we have
      $\lambda = \getOL(w,\pi) \precpi \getOL(r,\pi)$ and so $(w,r) \in \OB$. In
      the second case, $w = init_x$ for some location $x$, so
      $(w,r) \in \Actions_0 \times (\Actions \setminus \Actions_0) \subseteq
      \OB$.
    \item If $r \in \nrR$ or $\narR$, similarly to above.
    \end{itemize}
  \item Checking $\rf \subseteq \IB$.\\
    From above we have
    $\rfe = \rfe;[\Inst] \subseteq \OB;[\Inst] \subseteq \IB$. \\
    If $(w,r) \in \rfi \subseteq [\lW];\rf;[\lR]$, then there is
    $\lR\tup{r,w} \in \pi$. There is $\pi_1$ and $\pi_2$ such that
    $\pi = \pi_2 \cdot \lR\tup{r,w} \cdot \pi_1$. So by $\wfrd(\pi)$ we have
    $\wfrdCPU(r,w,\pi_1)$ which implies $\lW\tup{w} \precpi \lR\tup{r,w}$ and
    $(w,r) \in \IB$.
  \item Checking $[\nlW];\pf \subseteq \OB$.\\
    If $(w,p) \in \pf$ with $w \in \nlW$, then there exists $e$ such that
    $\nlW\tup{w,e} \precpi \lP\tup{p,e}$. From $\backComp(\pi)$, we have
    $\nlW\tup{w,e} \precpi \flB\tup{w} \precpi \lP\tup{p,e}$ and so
    $(w,p) \in \OB$.
  \item Checking $\pf \subseteq \IB$.\\
    If $(w,p) \in \pf$, then there exists $e$ such that either
    $\nlW\tup{w,e} \precpi \lP\tup{p,e}$ or
    $\nrW\tup{w,e} \precpi \lP\tup{p,e}$. In both cases we immediately have
    $(w,p) \in \IB$.
  \item Checking $\fri \subseteq \IB$.\\
    If $(r,w') \in \fri$ then $r \in \lR$, $w' \in \lW$,
    $\threadt(r) = \threadt(w')$, and there exists $w$ such that $(w,r) \in \rf$
    and $(w,w') \in \mo$. There is $\pi_4$ and $\pi_3$ such that
    $\pi = \pi_4 \cdot \lR\tup{r,w} \cdot \pi_3$. So by $\wfrd(\pi)$ we have
    $\wfrdCPU(r,w,\pi_3)$, and there is three cases to consider.
    \begin{itemize}
    \item In the first case, $\pi_3 = \pi_2 \cdot \lambda_w \cdot \pi_1$, with
      $\lambda_w \in \{\flB\tup{w}, \RMW\tup{w,\_}\}$, and
      $\flB\tup{w'} \notin \pi_2$. Since $(w,w') \in \mo$ we have
      $\flB\tup{w'} \notin \pi_1$, an so $\flB\tup{w'} \notin \pi_3$. The last
      condition of the first case then gives us $\lW\tup{w'} \notin \pi_3$,
      which implies $(r,w') \in \IB$.
    \item In the second case, $\pi_3 = \pi_2 \cdot \lambda_w \cdot \pi_1$, with
      $\lambda_w = \lW\tup{w}$, $\thread(w) = \thread(r)$, and
      $\flB\tup{w} \notin \pi_3$. Then $w$ and $w'$ are on the same thread, and
      by $\bufFlushOrd(\pi)$ and $(w,w') \in \mo$ we have
      $\lW\tup{w} \precpi \lW\tup{w'}$ and $\lW\tup{w'} \notin \pi_1$. The last
      condition of the second case gives us $\lW\tup{w'} \notin \pi_2$, so
      $\lW\tup{w'} \notin \pi_3$ and $(r,w') \in \IB$.
    \item In the last case, $w = init_x$ for some location $x$, and we
      immediately get $\lW\tup{w'} \notin \pi_3$, which implies
      $(r,w') \in \IB$.
    \end{itemize}
  \item Checking $\fr \subseteq \OB$.\\
    If $(r,w') \in \fr$, then there exists $w$ such that $(w,r) \in \rf$ and
    $(w,w') \in \mo$. By definition of $\rf$, there is $\pi_4$ and $\pi_3$ such
    that $\pi = \pi_4 \cdot \lambda_r \cdot \pi_3$, with
    $\lambda_r \in$ $\{\lR\tup{r,w},$ $\RMW\tup{r,w},$ $\nlR\tup{r,w,\_,\_},$ $
      \nrR\tup{r,w,\_,\_},$ $\naF\tup{r,w,\_,\_},$ $\narR\tup{r,w,\_,\_,\_}\}$. So by $\wfrd(\pi)$ we have either
    $\wfrdNIC(r,w,\pi_3)$ or $\wfrdCPU(r,w,\pi_3)$, and there are five cases to
    consider.
    \begin{itemize}
    \item In the first case of $\wfrdNIC(r,w,\pi_3)$,
      $\pi_3 = \pi_2 \cdot \getOL(w,\pi) \cdot \pi_1$, and
      $\getOL(w',\pi) \notin \pi_2$. Since $(w,w') \in \mo$ we have
      $\getOL(w',\pi) \notin \pi_1$, and thus $\getOL(w',\pi) \notin \pi_3$. So
      $\getOL(w',\pi) \in \pi_4$ and $(r,w') \in \OB$.
    \item In the last case $\wfrdNIC(r,w,\pi_3)$, $w = init_x$ for some location
      $x$, and we immediately have $\getOL(w',\pi) \notin \pi_3$, which implies
      $(r,w') \in \OB$.
    \item For the first case of $\wfrdCPU(r,w,\pi_3)$, same reasoning as for the
      first case of $\wfrdNIC$.
    \item For the second case of $\wfrdCPU(r,w,\pi_3)$,
      $\pi_3 = \pi_2 \cdot \getIL(w,\pi) \cdot \pi_1$, with
      $\thread(w) = \thread(r)$, and $\getOL(w,\pi) \notin \pi_3$. So
      $\getOL(w,\pi) \in \pi_4$, and since $(w,w') \in \mo$ we have
      $\getOL(w',\pi) \in \pi_4$ as well, and $(r,w') \in \OB$.
    \item For the last case of $\wfrdCPU(r,w,\pi_3)$, same reasoning as for the
      last case of $\wfrdNIC$.
    \end{itemize}
  \item Checking $\ro \subseteq \IB$.\\
    By definition of $\ro$.
  \item Checking $\ro \subseteq \OB$.\\
    By definition of $\ro$.
  \item Checking $\mo \subseteq \OB$.\\
    By definition of $\mo$, as what matters are the $init_x$, $\flB\tup{w}$, and
    $\RMW\tup{w,\_}$ events.
  \item Checking $\rao \subseteq \OB$.\\
    By definition of $\rao$.
  \item Checking $\ar;\rao \subseteq \OB$.\\
    If $(w,r_{1}) \in \ar$ then $\narR\tup{r_{1},a_{1},\_,\_,w} \in \pi$ for some $a_{1} \in \rrmw$,
    and if $(r_{1},r_{2}) \in \rao$ then $\narR\tup{r_{1},\_,a_{1},\_,w} \precpi \lambda_{r}$ for some $\lambda_{r} \in \set{\naF\tup{r_{2},\_,a_{2},\_}, \narR\tup{r_{2},\_,a_{2},\_,\_}}$,
    with $\neqn{\node}(a_{1}) = \neqn{\node}(a_{2})$.
    Then using $\remoteAtomic(\pi)$ we have that $\flB\tup{w} \precpi \lambda_{r}$.
  \end{itemize}

\end{proof}


%% file: decl-to-annot.tex


\subsection{From Declarative Semantics to Annotated Semantics}
\label{sec:annot-sem-d2a}

From a program $\Program$ and a well-formed consistent execution graph
$\EG = (\Actions,\po,\rf,\pf,\mo,\ro,\rao)$, where $(\Actions,\po)$ is generated by \Program, we want to reconstruct an annotated semantics execution.


\begin{theorem}
  \label{thm:extend-ibob}
  $\ib$ and $\ob$ can be extended into total relations $\IB$ and $\OB$ on
  $\Actions$ such that:
  \begin{itemize}
  \item $\IB$ and $\OB$ are irreflexive and transitive;
  \item $\OB ; [\Inst] \subseteq \IB$ and $[\Inst] ; \IB \subseteq \OB$.
  \end{itemize}
\end{theorem}
\begin{proof}
  We show that if $\ib$ is not already total we can extend it (and maybe $\ob$)
  into a strictly bigger relation satisfying the constraints of the theorem. Let
  us assume that there is $(a,b) \in \Actions^2$ such that $(a,b) \notin \ib$
  and $(b,a) \notin \ib$. We arbitrarily decide to include $(a,b)$ in our
  relation and we define $\ib' = (\ib \cup \set{(a,b)})^+$ and
  $\ob' = (\ob \cup [\Inst];\ib')^+$.

  Clearly $\ib'$ and $\ob'$ are transitive, $\ib'$ is irreflexive, and
  $[\Inst];\ib' \subseteq \ob'$. We need to prove the following two facts:
  $\ob'$ is still irreflexive; and $\ob';[\Inst] \subseteq \ib'$.

  First, let us check that $(\ob \cup [\Inst];\ib')^+$ is irreflexive. Since
  $\ob$ and $([\Inst];\ib')$ are both transitive and irreflexive, a cycle would
  only be possible by alternating between the two components, so it is enough to
  show that $(\ob ; ([\Inst] ; \ib'))^+$ is irreflexive. But
  $(\ob ; ([\Inst] ; \ib'))^+ = ((\ob ; [\Inst]) ; \ib')^+ \subseteq (\ib ;
  \ib')^+ \subseteq \ib'$ is irreflexive. Thus $\ob'$ is irreflexive.

  Then, we need to check that $\ob';[\Inst] \subseteq \ib'$. Using some
  rewriting,
  $\ob' = (\ob \cup [\Inst];\ib')^+ = \ob \cup (\ob^* ;
  ([\Inst];\ib'))^+;\ob^*$. We know $\ob ; [\Inst] \subseteq \ib'$, which also
  implies $\ob^* ; [\Inst] \subseteq \ib'^*$. So
  $\ob';[\Inst] = \ob;[\Inst] \cup ((\ob^* ; [\Inst]);\ib')^+;(\ob^*;[\Inst])
  \subseteq \ib' \cup (\ib'^*;\ib')^+;\ib'^* \subseteq \ib'$.

  Once \ib is a total relation on $\Actions$, we can similarly freely extend \ob
  into a total relation.
 \end{proof}

We use Theorem~\ref{thm:extend-ibob} above to extend \ib and \ob into total
relations \IB and \OB.

Since $(\Actions, \po)$ is derived from $\prog$,
by \Cref{par:prog-to-graph} we have that for all $\threadt \in \Threads$
there are $s_{\threadt}$ and $G_{\threadt}$ such that $G_{\threadt} \in G^{\threadt}(s_{t})$, $\prog(\threadt) \actseq s_{t}$ and $(\Actions, \po) = G_{init};(\parallel_{\threadt\in\Threads}G_{\threadt})$.
We consider each premise of the form $C \actseq s$,
where $C$ is a primitive command,
to generate new events and annotated labels.
\begin{itemize}
\item If $s=r\in\lR$, from well-formedness conditions, there is $w$ such
  that $(w,r) \in \rf$ and $\locveq(r,w)$. We create an annotated label
  $\lR\tup{r,w}$.
\item If $s=u,s'$ where $u\in\RMW$, from well-formedness conditions,
  there is $w$ such that $(w,u) \in \rf$ and $\locveq(u,w)$. We create an annotated label $\RMW\tup{u,w}$, then process $s'$.
  \item If $s=f,r,s'$ where $f\in\lF$, $r\in\lR$, and $w\in\lW$, from well-formedness conditions, there is $w'$ such that $(w',r) \in \rf$ and $\locveq(r,w')$. We create annotated labels $\lF\tup{f}$, $\lR\tup{r,w'}$, $\lW\tup{w}$ and $\flB\tup{w}$, then process $s'$.
\item If $s=w\in\lW$, we create annotated labels $\lW\tup{w}$ and $\flB\tup{w}$.
\item If $s=f\in\lF$, we create annotated labels $\lF\tup{f}$.
\item If $s=r,w$ where $r \in \nlR$ and $w \in \nrW$, we create two events $a \in \rput$ and $e \in \nrEX$,
  and the annotated labels $\PushTSO\tup{a}$, $\PushNIC\tup{a}$,
  $\nlR\tup{r,w',a,w}$ (where $(w',r) \in \rf$), $\nrW\tup{w,e}$, $\flB\tup{w}$,
  and $\lCN\tup{e}$. If there is $p$ such that $(w,p) \in \pf$, we also create
  an annotated label $\lP\tup{p,e}$. To simplify later definition, we also
  extend $\po$ such that the event $a$ is placed just before $r$, and $e$ just
  after $w$. I.e., let
  $\po' = \po \cup \setpred{(e',a)}{(e',r)\in\po} \cup
  \setpred{(a,e')}{(r,e')\in\po^*}$ and redefine
  $\po = \po' \cup \setpred{(e',e)}{(e',w)\in\po'^*} \cup
  \setpred{(e,e')}{(w,e')\in\po'}$. \\
  Note: from well-formedness conditions, every $\nlR$ and every $\nrW$ are part
  of such a pair.
\item If $s=r,w$ where $r\in\nrR$ and $w\in\nlW$, we similarly create $a \in \rget$, $e \in \nlEX$,
  $\PushTSO\tup{a}$, $\PushNIC\tup{a}$, $\nrR\tup{\ldots}$, $\nlW\tup{\ldots}$,
  $\flB\tup{\ldots}$, and potentially $\lP\tup{\ldots}$.
  \item If $s=r,w$ where $r\in\narR$ and $w\in\nlW$, we have $C$ of the form $z\assign\rcas(\neqn{x},e,e')$,
  so we use the values $\db{e}$ and $\db{e'}$ to create $a \in \RCAS$,
  $\PushTSO\tup{a}$, $\PushNIC\tup{a}$, $\naF\tup{\ldots}$, $\nlW\tup{\ldots}$,
  $\flB\tup{\ldots}$, and potentially $\lP\tup{\ldots}$.
  \item If $s=r,w_{1},w_{2}$ where $r \in \narR$, $w_{1} \in \narW$, $w_{2} \in \nlW$,
  we have $C$ either of the form $z\assign\rfaa(\neqn{x},e)$ or $z\assign\rcas(\neqn{x},e_{1},e_{2})$,
  so we create $a \in \rfaa$ or $a \in \rcas$ accordingly, and
  $\PushTSO\tup{a}$, $\PushNIC\tup{a}$, $\narR\tup{\ldots}$, $\narW\tup{w_{1}}$, $\nlW\tup{w_{2},\ldots}$,
  $\flB\tup{w_{1}}$, $\flB\tup{w_{2}}$ and potentially $\lP\tup{\ldots}$.
\item If $s=f\in\nF$, we create the annotated labels $\PushTSO\tup{f}$,
  $\PushNIC\tup{f}$, and $\nF\tup{f}$.
\item We ignore $s=p\in\lP$, as this is already handled by our earlier cases.
\end{itemize}

Then, we use \IB and \OB to reconstruct a partial path from these annotated
labels. We define a path $\pi_0$ such that:
\begin{itemize}
\item $\pi_0 \in \queue{(\aLabels \setminus (\PushTSO \cup \PushNIC \cup \lCN))}$
\item $\getIL(e_1,\pi_0) \prec_{\pi_0} \getIL(e_2,\pi_0) \iff (e_1,e_2) \in \IB$
\item $\getOL(e_1,\pi_0) \prec_{\pi_0} \getOL(e_2,\pi_0) \iff (e_1,e_2) \in \OB$
\item $\forall w \in \set{\lW,\nlW,\nrW,\narW}$,
    $\getIL(w,\pi_0) \prec_{\pi_0} \getOL(w,\pi_0)$
\end{itemize}
This is possible from the properties of \IB and \OB. For pairs of annotated
labels not ordered by \IB or \OB, we decide to order
$\lW\tup{w}$/$\nlW\tup{w,\_}$/$\nrW\tup{w,\_}$/$\narW\tup{w}$ first and $\flB\tup{w}$ last.
Note that the annotated labels $\PushTSO\tup{\ldots}$, $\PushNIC\tup{\ldots}$,
and $\lCN\tup{\ldots}$ not covered by $\IB$/$\OB$ are not yet integrated in
$\pi_0$.

Then we extend $\pi_0$ to add annotated labels not considered by the declarative
semantics. We use the following extension function that introduces a new
annotated label as early as possible after a set of dependencies.

$$\extend(\pi, \lambda, S) \defeq
\begin{cases}
  \pi_2 \cdot \lambda \cdot \lambda' \cdot \pi_1
  & \text{if } \pi = \pi_2\cdot\lambda'\cdot\pi_1 \land
    \lambda' \in S \land \pi_2 \cap S = \emptyset \\
  \pi \cdot \lambda
  & \text{if } \pi \cap S = \emptyset
\end{cases}$$

We define a new function to recover the first annotated label corresponding to
an event:
$$\ActionsExt \defeq \Actions \cup (\rget \cup \rput \cup \RCAS \cup \rfaa \cup \nlEX \cup \nrEX)$$
$$\getCPU : \ActionsExt \rightharpoonup \aLabels$$
$$\getCPU(e) \defeq
\begin{cases}
  \getIL(e, \pi_0) & \text{if } e \in \CPUactions = \set{\lR,\lW,\RMW,\lF,\lP}\\
  \PushTSO\tup{e} & \text{if } e \in \set{\rput,\rget,\RCAS,\rfaa,\nF}\\
  \text{undefined} & \text{otherwise}
\end{cases}$$

And a similar function for events emptying a CPU buffer:
$$\getTSO : \ActionsExt \rightharpoonup \aLabels$$
$$\getTSO(e) \defeq
\begin{cases}
  \flB\tup{e} & \text{if } e \in \lW\\
  \PushNIC\tup{e} & \text{if } e \in \set{\rput,\rget,\RCAS,\rfaa,\nF}\\
  \text{undefined} & \text{otherwise}
\end{cases}$$

Let us consider $(a_1,\ldots,a_n) = \Actions \cap \set{\rput,\rget,\RCAS,\rfaa,\nF}$ in \po
order, i.e., if $i<j$ then $(a_j,a_i) \notin \po$. We extend $\pi_0$
successively until we get $\pi_n$:
\begin{itemize}
\item We introduce $\PushTSO$ as early as possible: \\
  Let
  $\pi' = \extend(\pi_{i-1}, \PushTSO\tup{a_i}, \setpred{\getCPU(e)}{(e,a_i) \in
    \po})$
\item We introduce $\PushNIC$ as early as possible: \\
  Let
  $\pi'' = \extend(\pi', \PushNIC\tup{a_i}, \set{\PushTSO\tup{a_i}} \cup
  \setpred{\getTSO(e)}{(e,a_i) \in \po})$
\item If $a_i \in \rput$, there is $e_i \in \nrEX$ such that
  $\nlR\tup{\_,\_,a_i,w} \prec_{\pi_0} \nrW\tup{w,e_i}$. We also introduce
  $\lCN$: Let
  $S = \set{\nrW\tup{w,e_i}} \cup \setpred{\nlW\tup{\_,e}}{(e,e_i) \in
    \po\cap\sqp} \cup \setpred{\lCN\tup{e}}{(e,e_i) \in \po\cap\sqp}$, we pose
  $\pi_i = \extend(\pi'', \lCN\tup{e_i}, S)$. \\
  Otherwise, i.e. $a_i \notin \rput$, we simply have $\pi_i = \pi''$
\end{itemize}

Finally, $\pi = \pi_n$ is our path for an annotated semantics reduction. We
clearly have $\complete(\pi)$ by definition. Our goal is then to prove that
$\wfp(\pi)$ holds. It is composed of seven properties. Note that we already have
the existence of the relevant annotated labels, and we need to show that the
ordering constraints are respected.


\bigskip
{\bf nodup}

$\nodup(\pi)$ directly comes from the definition of annotated labels. There is
no conflict in event usage.


\bigskip
{\bf backComp}

Here are a couple lemmas showing that the new annotated labels are not placed
too late and do not disturb the expected ordering.

\begin{lemma}
  \label{lem:extend-1}
  For all $a \in \set{\rput,\rget,\RCAS,\rfaa,\nF}$ and $b \in \Actions$, if
  $(a,b) \in \po^*$, then $\PushTSO\tup{a} \prec_{\pi} \getIL(b,\pi_0)$.
\end{lemma}
\begin{proof}
  We take an arbitrary $b$, and proceed for $a$ in $\po$ order, i.e., we can
  assume it holds for $e \in \set{\rput,\rget,\RCAS,\rfaa,\nF}$ such that $(e,a) \in \po$.
  By definition, $\PushTSO\tup{a}$ comes from an extension
  $\pi'' = \extend(\pi', \PushTSO\tup{a}, \setpred{\getCPU(e)}{(e,a) \in \po})$
  and has been placed either first—and the result is trivial—or just after some
  $\getCPU(e)$ with $(e,a) \in \po$. If $e \in \set{\rput,\rget,\RCAS,\rfaa,\nF}$, we have
  $\PushTSO\tup{e} \prec_{\pi''} \PushTSO\tup{a} \prec_{\pi''} \getIL(b,\pi_0)$
  by induction hypothesis. If $e \in \CPUactions = \set{\lR,\lW,\RMW,\lF,\lP}$,
  we have
  $\getIL(e,\pi_0) \prec_{\pi''} \PushTSO\tup{a} \prec_{\pi''} \getIL(b,\pi_0)$
  since $(e,b) \in \ippo \subseteq \IB$.
\end{proof}
\begin{lemma}
  \label{lem:extend-2}
  $\forall a \in \set{\rput,\rget,\RCAS,\rfaa,\nF}$, $\forall b \in \set{\nF,\nrR,\nlR,\narR,\lW}$, if
  $(a,b) \in \po^*$, then $\PushNIC\tup{a} \prec_{\pi} \getOL(b,\pi_0)$.
\end{lemma}
\begin{proof}
  We take an arbitrary $b \in \set{\nF,\nrR,\nlR,\narR}$, and proceed for $a$ in $\po$
  order, i.e., we can assume it holds for $e \in \set{\rput,\rget,\RCAS,\rfaa,\nF}$ such
  that $(e,a) \in \po$. By definition, $\PushNIC\tup{a}$ comes from an extension
  $\pi'' = \extend(\pi', \PushNIC\tup{a}, S)$, with
  $S = \set{\PushTSO\tup{a}} \cup \setpred{\getTSO(e)}{(e,a) \in \po}$, and has
  been placed just after some $\lambda \in S$.
  \begin{itemize}
  \item If $\lambda = \PushTSO\tup{a}$, then we have
    $\lambda \prec_{\pi''} \PushNIC\tup{a} \prec_{\pi''} \getOL(b,\pi_0)$ using
    Lemma~\ref{lem:extend-1} above, since $\getIL(b,\pi_0) = \getOL(b,\pi_0)$ or $\getIL(b,\pi_{0}) \prec_{\pi''} \getOL(b,\pi_{0})$.
  \item If $\lambda = \getTSO\tup{e} = \PushNIC\tup{e}$ for some
    $e \in \set{\rput,\rget,\RCAS,\rfaa,\nF}$, then we have
    $\lambda \prec_{\pi''} \PushNIC\tup{a} \prec_{\pi''} \getOL(b,\pi_0)$ by
    induction hypothesis.
  \item If $\lambda = \getTSO\tup{e} = \flB\tup{e}$ for some $e \in \lW$, then
    we have
    $\flB\tup{e} \prec_{\pi''} \PushNIC\tup{a} \prec_{\pi''} \getOL(b,\pi_0)$
    since $(e,b) \in \oppo \subseteq \OB$.
  \end{itemize}
\end{proof}
\begin{lemma}
  \label{lem:extend-3}
  Forall $w$,$e$,$p$, if $\nrW\tup{w,e} \in \pi$ and $\lP\tup{p,e} \in \pi$,
  then $\lCN\tup{e} \precpi \lP\tup{p,e}$.
\end{lemma}
\begin{proof}
  Once again, we proceed for $e$ in $\po$ order, i.e., we can assume the result
  holds for $e' \in \nrEX$ such that $(e',e) \in \po$. $\lCN\tup{e}$ is inserted
  in some operation $\pi'' = \extend(\pi', \lCN\tup{e}, S)$, with
  $S = \set{\nrW\tup{w,e}} \cup \setpred{\nlW\tup{\_,e'}}{(e',e) \in
    \po\cap\sqp} \cup \setpred{\lCN\tup{e'}}{(e',e) \in \po\cap\sqp}$. It is
  then placed just after some label $\lambda \in S$.
  \begin{itemize}
  \item If $\lambda = \nrW\tup{w,e}$, we have
    $\lambda \prec_{\pi''} \lCN\tup{e} \prec_{\pi''} \lP\tup{p,e}$ because
    $(w,p) \in \pf \subseteq \IB$.
  \item If $\lambda = \lCN\tup{e'}$ with $(e',e) \in \po\cap\sqp$, then there is
    some $w'$ such that $(w',w) \in \po\cap\sqp$ and $\nrW\tup{w',e'} \in \pi'$.
    From well-formedness condition number 1 (see
    Definition~\ref{def:well-formed}), there is some $p'$ such that
    $(w',p') \in \pf$ and $(p',p) \in \po$. By induction hypothesis, we have
    $\lCN\tup{e'} \prec_{\pi'} \lP\tup{p',e'}$, and from $(p',p) \in \IB$ we
    have $\lP\tup{p',e'} \prec_{\pi'} \lP\tup{p,e}$. In the end, we have the
    result $\lCN\tup{e'} \prec_{\pi''} \lCN\tup{e} \prec_{\pi''} \lP\tup{p,e}$.
  \item If $\lambda = \nlW\tup{w',e'}$ with $(e',e) \in \po\cap\sqp$, then we
    also have $(w',w) \in \po\cap\sqp$, so from well-formedness condition number
    1 (see Definition~\ref{def:well-formed}), there is some $p'$ such that
    $(w',p') \in \pf$ and $(p',p) \in \po$. We have
    $\nlW\tup{w',e'} \prec_{\pi''} \lCN\tup{e} \prec_{\pi''} \lP\tup{p',e'}
    \prec_{\pi''} \lP\tup{p,e}$.
  \end{itemize}
\end{proof}

We can then check that we have $\backComp(\pi)$:
\begin{itemize}
\item $\lW\tup{w} \precpi \flB\tup{w}$ comes from the third property when
  defining $\pi_0$; similarly for $\nlW$, $\nrW$ and $\narW$.
\item $\PushTSO\tup{a} \precpi \PushNIC\tup{a}$ comes from the extension
  process.
\item $\PushNIC\tup{f} \precpi \nF\tup{f}$ comes from Lemma~\ref{lem:extend-2};
  similarly for $\PushNIC\tup{a} \precpi \nlR/\nrR/\naF/\narR\tup{\ldots}$.
\item $\nlR\tup{r,w,a,w'} \precpi \nrW\tup{w',e}$ comes from
  $(r,w') \in \ippo \subseteq \IB$; similarly for $\nrR$/$\nlW$, $\naF/\nlW$, $\narR/\nlW$ and $\narR/\narW$.
\item $\nrW\tup{w,e} \precpi \lCN\tup{e}$ comes from the extension process
\item $\nlW\tup{w,e} \precpi \flB\tup{w} \precpi \lP\tup{p,e}$ comes from
  $(w,p) \in [\nlW];\pf \subseteq \OB$.
\item $\lCN\tup{e} \precpi \lP\tup{p,e}$ comes from Lemma~\ref{lem:extend-3}.
\end{itemize}
Thus we have $\backComp(\pi)$.


\bigskip
{\bf bufFlushOrd}

\begin{itemize}
\item
  $\lW\tup{w_1} \precpi \lW\tup{w_2} \iff \flB\tup{w_1} \precpi \flB\tup{w_2}$
  when $\threadt(w_1)=\threadt(w_2)$ comes the fact that
  $[\lW];\po;[\lW] \subseteq (\IB \cup \OB)$, so both sides are true if and
  only if $(w_1,w_2) \in \po$; similarly for $\nlW$ and $\nrW/\narW$ on the same
  queue pair.
\item When $\threadt(a_1)=\threadt(a_2)$,
  $\PushTSO\tup{a_1} \precpi \PushTSO\tup{a_2} \iff \PushNIC\tup{a_1} \precpi
  \PushNIC\tup{a_2} \iff (a_1,a_2) \in \po$ from the definition of the
  extension process (to define $\pi_n$).
\item For $a \in \set{\rput,\rget,\nF,\RCAS,\rfaa}$, $w \in \lW$, such that
  $\threadt(a)=\threadt(w)$:
  \begin{itemize}
  \item If $(w,a) \in \po$, then $\lW\tup{w} \precpi \PushTSO\tup{a}$ and
    $\flB\tup{w} \precpi \PushNIC\tup{a}$ from the definition of the extension
    process.
  \item If $(a,w) \in \po$, then $\PushTSO\tup{a} \precpi \lW\tup{w}$ and
    $\PushNIC\tup{a} \precpi \flB\tup{w}$ from Lemmas~\ref{lem:extend-1}
    and~\ref{lem:extend-2}.
  \end{itemize}
\item When $\threadt(w)=\threadt(f)$, $\lW\tup{w} \precpi \lF\tup{f}$ implies
  $(w,f) \in \po$ (since $[\lF];\po;[\lW] \subseteq \ippo \subseteq \IB$), which
  implies $\flB\tup{w} \precpi \lF\tup{f}$ (since
  $[\lW];\po;[\lF] \subseteq \oppo \subseteq \OB$); similarly for $\RMW$.
\item If $w \in \nlW$, $r \in \nlR$, and $\sameqp(w,r)$, then from
  the definition of pre-executions (see condition 6 of Definition~\ref{def:pre-executions}),
  either $(w,r) \in \ro$ or $(r,w) \in \ro$. If
  $\nlW\tup{w,\_} \precpi \nlR\tup{r,\_,\_,\_}$, then $(r,w) \notin \ro$ (since
  $\ro \subseteq \IB$) and $(w,r) \in \ro$. Thus,
  $\flB\tup{w} \precpi \nlR\tup{r,\_,\_,\_}$ (since $\ro \subseteq \OB$);
  similarly for $w\in\set{nrW,narW}$ and $r\in\set{\nrR,\narR}$.
\end{itemize}
Thus we have $\bufFlushOrd(\pi)$.


\bigskip
{\bf pollOrder}

\begin{lemma}
  \label{lem:extend-4}
  For all $e_1$, $e_2 \in \set{\nlEX, \nrEX}$, such that $\sameqp(e_1,e_2)$, let
  $\lambda_1 \in \{\nlW\tup{\_,e_1}, \lCN\tup{e_1}\}$,
  $\lambda_2 \in \{\nlW\tup{\_,e_2}, \lCN\tup{e_2}\}$, then
  $(e_1,e_2) \in \po \iff \lambda_1 \precpi \lambda_2$.
\end{lemma}
\begin{proof}
  By symmetry, we only need to show
  $(e_1,e_2) \in \po \implies \lambda_1 \precpi \lambda_2$. Once again, we
  proceed for $e_1$ in $\po$ order, i.e., we can assume the result holds for
  $e' \in \nEX$ such that $(e',e_1) \in \po$.
  \begin{itemize}
  \item If $\lambda_1 = \nlW\tup{w_1,e_1}$ and $\lambda_2 = \nlW\tup{w_2,e_2}$,
    then $(e_1,e_2) \in \po$ implies $(w_1,w_2) \in (\po \cap \sqp)$, so
    $(w_1,w_2) \in \ippo \subseteq \IB$ and $\lambda_1 \precpi \lambda_2$.
  \item If $\lambda_1 = \nlW\tup{w_1,e_1}$ and $\lambda_2 = \lCN\tup{e_2}$, then
    by definition of the extension process we have
    $\lambda_1 \precpi \lambda_2$.
  \item If $\lambda_1 = \lCN\tup{e_1}$ and $\lambda_2 = \nlW\tup{w_2,e_2}$, then
    $\lambda_1$ is inserted in some operation
    $\pi'' = \extend(\pi',$ $\lCN\tup{e_1}, S)$, with
    $S = \set{\nrW\tup{\_,e_1}} \cup \setpred{\nlW\tup{\_,e'}}{(e',e_1) \in
      \po\cap\sqp} \cup \setpred{\lCN\tup{e'}}{(e',e_1) \in \po\cap\sqp}$. It is
    then placed just after some label $\lambda \in S$.
    \begin{itemize}
    \item If $\lambda = \nrW\tup{w_1,e_1}$, we have
      $\lambda \prec_{\pi''} \lambda_1 \prec_{\pi''} \lambda_2$ because
      $(w_1,w_2) \in \ippo \subseteq \IB$.
    \item If $\lambda = \lCN\tup{e'}$ or $\lambda = \nlW\tup{\_,e'}$, with
      $(e',e_1) \in \po\cap\sqp$, then by induction hypothesis
      $\lambda \prec_{\pi''} \lambda_1 \prec_{\pi''} \lambda_2$.
    \end{itemize}
  \item If $\lambda_1 = \lCN\tup{e_1}$ and $\lambda_2 = \lCN\tup{e_2}$, then by
    definition of the extension process we have $\lambda_1 \precpi \lambda_2$.
  \end{itemize}
\end{proof}

Let us assume we have $e_1,e_2,p_2,\lambda_1,\lambda_2$ such that
$\sameqp(e_1,e_2)$, $\lambda_1 \in \{\nlW\tup{\_,e_1}, \lCN\tup{e_1}\}$,
$\lambda_2 \in \{\nlW\tup{\_,e_2}, \lCN\tup{e_2}\}$,
$\lambda_1 \precpi \lambda_2$, and $\lP\tup{p_2,e_2}\in\pi$.

From the creation of the events $e_1$ and $e_2$, there is some
$w_1,w_2 \in \set{\nlW,\nrW}$ such that $(w_i,e_i) \in \imm{\po}$. From
Lemma~\ref{lem:extend-4}, we have $(e_1,e_2) \in \po$ and thus
$(w_1,w_2) \in (\po \cap \sqp)$. By definition, we also have
$(w_2,p_2) \in \pf$. From well-formedness condition number 1 (see
Definition~\ref{def:well-formed}), there is some $p_1$ such that
$(w_1,p_1) \in \pf$ and $(p_1,p_2) \in \po$. Thus we have
$\lP\tup{p_1,e_1} \precpi \lP\tup{p_2,e_2}$ as required to prove
$\pollOrder(\pi)$.


\bigskip
{\bf nicActOrder}

Let $a_1$ and $a_2$ such that $\PushNIC\tup{a_1} \precpi \PushNIC\tup{a_2}$ and
$\sameqp(a_1,a_2)$. From the definition of the extension process, we have
$(a_1,a_2) \in \po$.
\begin{itemize}
\item If $a_1 \in \nF$ or $a_2 \in \nF$, then most of the required results hold
  by definition of $\ippo$. The only exception is
  $\lCN\tup{e} \precpi \nF\tup{a_2}$ which holds (by induction on $e$ in \po
  order) because all the dependencies of $\lCN\tup{e}$ are before $\nF\tup{a_2}$
  by $\ippo$.
\item If $\bigpar{ a_1 \in \rget \land a_2 \in \rget}$, the result holds by
  $\ippo$.
\item If $\bigpar{ a_1 \in \rget \land a_2 \in \rput}$, the result holds by
  Lemma~\ref{lem:extend-4}.
\item If $\bigpar{ a_{1} \in \rget \land a_{2} \in \RCAS \cup \rfaa }$, the results hold by $\ippo$.
\item If $\bigpar{ a_1 \in \rput \land a_2 \in \rget}$, the first result holds
  by $\ippo$, the second by Lemma~\ref{lem:extend-4}.
\item If $\bigpar{ a_1 \in \rput \land a_2 \in \rput}$, the first two results hold by $\ippo$, the last one by Lemma~\ref{lem:extend-4}.
\item If $\bigpar{a_{1} \in \rput \land a_{2} \in \RCAS \cup \rfaa}$, the results hold by \ippo.
\item If $\bigpar{a_{1} \in \RCAS \cup \rfaa \land a_{2} \in \rget}$, the results hold by \ippo.
\item If $\bigpar{a_{1} \in \RCAS \cup \rfaa \land a_{2} \in \rput}$, the first result holds by \ippo, the latter by Lemma~\ref{lem:extend-4}.
\item If $\bigpar{a_{1}, a_{2} \in \RCAS \cup \rfaa}$, the first result holds by \ippo, the latter by Lemma~\ref{lem:extend-4}.
\end{itemize}
Thus we have $\nicActOrder(\pi)$.


\bigskip
{\bf nicAtomicity}

For every $a_{1},a_{2} \in \rrmw$ where $\neqn{\node}(a_{1}) = \neqn{\node}(a_{2})$,
if $\narR\tup{r_{1},a_{1},\_,\_,w} \precpi \lambda_{r}$
where $\lambda_{r} \in \{\naF\tup{r_{2},\_,a_{2},\_},$ $\narR\tup{r_{2},\_,a_{2}\_,\_}\}$,
then from the extension process we have $(r_{1},w) \in \imm{\po}$,
and $w\in\narW$, so $(w,r_{1})\in\ar$.
Then we need to show that $(r_{1},r_{2})\in\rao$.
Suppose, for contradiction, that $(r_{1},r_{2})\not\in\rao$.
By definition of $\rao$, for each node $\node$, $\rao_{n}$ is a total order on $\set{e\in\narR\st\neqn{\node}(e)=n}$.
Thus we have either $(r_{1},r_{2})\in\rao$ or $(r_{2},r_{1})\in\rao$,
and by assumption the prior is not the case so $(r_{2},r_{1})\in\rao\subseteq\OB$.
However, since $\narR\tup{r_{1},\ldots} \precpi \lambda_{r}$,
we have $(r_{1},r_{2}) \in \OB$,
which is a contradiction, as $\OB$ is irreflexive.
Therefore reject our original assumption.
Thus $(r_{1},r_{2})\in\rao$,
then we have $(w,r_{2}) \in \ar;\rao \subseteq \OB$,
so $\flB\tup{w} \precpi \lambda_{r}$.
Thus we have $\remoteAtomic(\pi)$.


\bigskip
{\bf wfrd}

Let us assume we have $\pi = \pi_4 \cdot \lambda_r \cdot \pi_3$, with
$\lambda_r \in$ $\{\lR\tup{r,w},$ $\RMW\tup{r,w},$ $\nlR\tup{r,w,\_,\_},$ $\nrR\tup{r,w,\_,\_},$ $\naF\tup{r,w,\_,\_},$ $\narR\tup{r,w,\_,\_,\_}\}$. In all cases we have $(w,r) \in \rf$. Another important
fact is that $\forall w', (w,w') \in \mo \implies (r,w') \in \fr$.

\begin{itemize}
\item If $\lambda_r = \lR\tup{r,w}$, we need to show $\wfrdCPU(r,w,\pi_3)$.
  \begin{itemize}
  \item If $w = init_{\loc(w)}$, then we need to check that
    $\setpred{\flB\tup{w'}, \RMW\tup{w',\_} \in \pi_3}{\loc(w')=\loc(r)} =
    \emptyset$ and
    $\setpred{ \lW\tup{w''} \in \pi_3} {\loc(w'')=\loc(r) \land
      \threadt(w'')=\threadt(r)} = \emptyset$. For the first, such a $w'$ would
    imply $(r,w') \in \fr \subseteq \OB$, which contradicts the ordering with
    $\lambda_r$. For the second, such an $w''$ would imply
    $(r,w'') \in \fri \subseteq \IB$, and $\lambda_r \precpi \lW\tup{w''}$ which
    similarly contradicts the ordering with $\lambda_r$.
  \item If $w \in \lW$, $\threadt(w) = \threadt(r)$, and
    $\flB\tup{w} \notin \pi_3$. From $(w,r) \in \rfi \subseteq \IB$, we have
    $\lambda_w = \lW\tup{w} \precpi \lambda_r$, i.e.,
    $\pi_3 = \pi_2 \cdot \lambda_w \cdot \pi_1$. We need to show that
    $\setpred{\lW\tup{w'} \in \pi_2}{\loc(w')=\loc(r) \land
      \threadt(w')=\threadt(r)} = \emptyset$. Such a $w'$ would imply
    $(w,w') \in \po$ (from $[\lW];\po;[\lW] \subseteq \ippo \subseteq \IB$, and
    the execution graph forcing either $(w,w') \in \po$ or $(w',w) \in \po$),
    $(w,w') \in \mo$ (from $[\lW];\po;[\lW] \subseteq \oppo \subseteq \OB$, and
    well-formedness conditions forcing either $(w,w') \in \mo$ or
    $(w',w) \in \mo$), and $(r,w') \in \fri \subseteq \IB$ would contradicts the
    ordering with $\lambda_r$.
  \item Else we have $\lambda_w \in \pi_3$, with
    $\lambda_w\in \set{\flB\tup{w}, \RMW\tup{w,\_}}$. If $w \in \lW$ and
    $\threadt(w) = \threadt(r)$, this is the remaining subcase, else it comes from
    $(w,r) \in \rfe \subseteq \OB$. Thus we have
    $\pi_3 = \pi_2 \cdot \lambda_w \cdot \pi_1$, and we need to check two
    properties. First, we check that \\
    $\setpred{\flB\tup{w'},\RMW\tup{w',\_} \in \pi_2} {\loc(w')=\loc(r)} =
    \emptyset$. It holds because such a $w'$ would again imply
    $(r,w') \in \fr \subseteq \OB$ and contradict the ordering with $\lambda_r$.
    Second, we check that $\setpred{w'}{
      \begin{array}{l}
        \lW\tup{w'} \in \pi_3 \land \flB\tup{w'} \notin \pi_3 \ \land \\
        \loc(w')=\loc(r) \land \threadt(w')=\threadt(r) \end{array}}
    = \emptyset$. It holds because such a $w'$ would again imply
    $(w,w') \in \mo$, $(r,w') \in \fri \subseteq \IB$ and contradict the
    ordering with $\lambda_r$.
  \end{itemize}

\item If $\lambda_r = \RMW\tup{r,w}$, we similarly check that
  $\wfrdCPU(r,w,\pi_3)$ holds. The difference is that cases that previously
  contradicted $(\fri \subseteq \IB)$ now contradict $\bufFlushOrd(\pi)$ that
  forces the buffer of $\threadt(r)$ to be empty when performing $\lambda_r$.

\item If $\lambda_r = \nlR\tup{r,w,\_,\_}$, we need to show
  $\wfrdNIC(r,w,\pi_3)$.
  \begin{itemize}
  \item If $w = init_{\loc(w)}$, then we need to check that
    $\setpred{\flB\tup{w'}, \RMW\tup{w',\_} \in \pi_3}{\loc(w')=\loc(r)} =
    \emptyset$. Such a $w'$ would imply $(r,w') \in \fr \subseteq \OB$, which
    contradicts the ordering with $\lambda_r$.
  \item Else we have $\lambda_w \in \pi_3$, with
    $\lambda_w\in \set{\flB\tup{w}, \RMW\tup{w,\_}}$. This comes from
    $(w,r) \in \rfe \subseteq \OB$. Thus we have
    $\pi_3 = \pi_2 \cdot \lambda_w \cdot \pi_1$, and we need to check that \\
    $\setpred{\flB\tup{w'},\RMW\tup{w',\_} \in \pi_2} {\loc(w')=\loc(r)} =
    \emptyset$. It holds because such a $w'$ would again imply
    $(r,w') \in \fr \subseteq \OB$ and contradict the ordering with $\lambda_r$.
  \end{itemize}

\item If $\lambda_r = \nrR\tup{r,w,\_,\_}$, $\naF\tup{r,w,\_,\_}$ or $\narR\tup{r,w,\_,\_,\_}$, we similarly check that
  $\wfrdNIC(r,w,\pi_3)$ for the same reasons.
\end{itemize}
Thus we have $\wfrd(\pi)$.

\begin{theorem}
  Let $\EG$ be a well-formed consistent execution graph generated from a program
  $\Program$. Let $\pi$ be the path obtained from $\EG$ by the process defined
  above. Then there is $\Mem'$, $\NQPmap'$ (such that forall
  $\threadt,\neqn\node$ we have
  $\NQPmap'(\threadt)(\neqn\node) = \tup{\varepsilon, \varepsilon,
    \queue{\nEX}}$), and an equivalent path $\pi'$ (producing the same outcome
  as $\pi$) such that
  $\prog, \Mem_0, \TSOmap_0, \Atm_{0}, \NQPmap_0, \varepsilon \trp^* (\lambda
  \threadt.\cskip), \Mem', \TSOmap_0, \Atm_{0}, \NQPmap', \pi'$.
\end{theorem}
\begin{proof}
  From above, we have $\wf(\pi)$. This shows that the program configuration can
  perform the events described by the annotated labels of $\pi$. The remaining part of the proof is simply to
  check that the command rewritings used when deriving the execution graph from
  $\prog$ (see \Cref{fig:action-sequence}) can be used as $\lE$ transitions
  in the annotated semantics for $\prog$, which follows from the definitions.
\end{proof}


%% file: op-vs-annot.tex


\newcommand{\dbu}[2]{\db{#1}_{#2}}
\newcommand{\forget}[1]{\dbu{#1}{}}
\newcommand{\forgetmem}[1]{\dbu{#1}{\mathrm{M}}}
\newcommand{\forgetatm}[1]{\dbu{#1}{\mathrm{A}}}
\newcommand{\forgetop}[1]{\dbu{#1}{\mathrm{op}}}
\newcommand{\forgetopwbl}[1]{\dbu{#1}{\mathrm{opl}}}

\subsection{Operational Semantics and Annotated Semantics}
\label{sec:annot-sem-o2a}

We define forgetful functions from annotated configurations to operational
configurations. For memories, we replace the write event by the value written.
For labels within annotated configurations, we drop some arguments to recover
the data structure of the operational semantics.

$$\forgetmem{\cdot} : \AMems \to \Mems$$
$$\forgetmem{\Mem} \defeq \lambda x. \valw(\Mem(x))$$

$$\forgetop{\cdot} : \ActionsExt \rightharpoonup
\set{
  \begin{array}{l}
    \y^{\neqn\node} \assign \x^\node,
    \y^{\neqn\node} \assign v,
    \doneE,
    \x^\node \assign \y^{\neqn\node},
    \x^\node \assign v, \\
    \x \assign \RCAS(y^{n}, v, v'),
    \x \assign \RFAA(y^{n}, v),
    \cn,
    \cnfence{\neqn\node}
  \end{array}
}
$$

\begin{minipage}[t]{0.5\textwidth}
\vspace{-1em}
\begin{align*}
\forgetop{\lW(\x,\valw)}
  &\defeq  \x \assign \valw \\
\forgetop{\nrW(\neqn{\y},\valr)}
  &\defeq \neqn{\y} \assign \valr \\
\forgetop{\nlW(\x,\valw,\neqn{\node})}
  &\defeq \x \assign \valw \\
\forgetop{\nF(\neqn{\node})}
  &\defeq \cnfence{\neqn\node} \\
\forgetop{\rput(\neqn\y,\x)}
  &\defeq \neqn\y \assign \x \\
\forgetop{\rget(\x,\neqn\y)}
  &\defeq \x \assign \neqn\y \\
\forgetop{\RCAS(\z,\neqn\x,v,v')}
  &\defeq \z \assign \RCAS(\neqn\x,v,v') \\
\forgetop{\RFAA(\z,\neqn\x,v)}
  &\defeq \z \assign \RFAA(\neqn\x,v) \\
\forgetop{\nlEX(\neqn{\node})}
  &\defeq \cn \\
\forgetop{\nrEX(\neqn{\node})}
  &\defeq \doneE \\
\end{align*}
\end{minipage}
\begin{minipage}[t]{0.5\textwidth}
\vspace{-1em}
\begin{align*}
\forgetop{\lF}
  & \quad \text{is undefined} \\
\forgetop{\lP(\ldots)}
  & \quad \text{is undefined} \\
\forgetop{\lR(\ldots)}
  & \quad \text{is undefined} \\
\forgetop{\RMW(\ldots)}
  & \quad \text{is undefined} \\
\forgetop{\nlR(\ldots)}
  & \quad \text{is undefined} \\
\forgetop{\nrR(\ldots)}
  & \quad \text{is undefined} \\
\forgetop{\naF(\ldots)}
  & \quad \text{is undefined} \\
\forgetop{\narR(\ldots)}
  & \quad \text{is undefined} \\
\forgetop{\narW(\ldots)}
  & \quad \text{is undefined} \\
\end{align*}
\end{minipage}

$$\forgetopwbl{\cdot} : \ActionsExt \rightharpoonup
\set{
  \begin{array}{l}
    \y^{\neqn\node} \assign \x^\node,
    \y^{\neqn\node} \assign v,
    \x^\node \assign \y^{\neqn\node},
    \x^\node \assign v, \\
    \x \assign \RCAS(y^{n}, v, v'),
    \x \assign \RFAA(y^{n}, v),
    \cn,
    \cnfence{\neqn\node}
  \end{array}
}
$$
$$ \forgetopwbl{l} = \begin{cases}
\cn & \text{if } l = \nrEX(\neqn{\node}) \\
\forgetop{l} & \text{otherwise}
\end{cases} $$

The labels that cannot appear in a well-formed annotated configuration are not
mapped. For \textput operations, the operational semantics uses both $(\doneE)$ and
$(\cn)$ while the annotated semantics uses the label $\nrEX$, so the mapping is
different for labels in $\wbL$.

$\forgetop{\cdot}$ and $\forgetopwbl{\cdot}$ are extended to lists
in an obvious way.

\bigskip

We then extend this to configurations as expected. We overload notations to
simplify the formulas.

For $\qp = \tup{\pipe, \wbR, \wbL} \in \AQPairs$, we define
$\forget{\qp} \defeq \tup{\forgetop{\pipe}, \forgetop{\wbR},
  \forgetopwbl{\wbL}}$.

For $\ANQPmap \in \ANQPmaps$, we define
$\forget{\ANQPmap} \defeq \lambda \threadt. \lambda\neqn\node.
  \forget{\ANQPmap(\threadt)(\neqn\node)}$.

For $\ATSOmap \in \ATSOmaps$, we define
$\forget{\ATSOmap} \defeq \lambda \threadt. \forgetop{\ATSOmap(\threadt)}$.


\begin{theorem}
  \label{thm:annot-to-op}

  For all $\prog, \prog' \in \Progs$, $\AMem, \AMem' \in \AMems$,
  $\ATSOmap, \ATSOmap' \in \ATSOmaps$, $\Atm, \Atm' \in \RAMap$, $\ANQPmap, \ANQPmap' \in \ANQPmaps$,
  $\pi, \pi' \in \paths$, if
  $\prog, \AMem, \ATSOmap, \Atm, \ANQPmap, \pi \trp \prog', \AMem', \ATSOmap', \Atm',
  \ANQPmap', \pi'$ and $\wf(\AMem,\ATSOmap,\Atm,\ANQPmap,\pi)$, then
  $\prog, \forgetmem{\AMem}, \forget{\ATSOmap}, \Atm, \forget{\ANQPmap} \trp
  \prog', \forgetmem{\AMem'}, \forget{\ATSOmap'}, \Atm', \forget{\ANQPmap'}$.
\end{theorem}
\begin{proof}
  By straightforward induction on $\trp$.
\end{proof}

\bigskip

\begin{theorem}
  \label{thm:op-to-annot}
  For all $\AMem \in \AMems$, $\Mem'' \in \Mems$, $\ATSOmap \in \ATSOmaps$,
  $\TSOmap'' \in \TSOmaps$, $\Atm, \Atm' \in \RAMap$, $\ANQPmap \in \ANQPmaps$, $\NQPmap'' \in \NQPmaps$,
  and $\pi \in \paths$, if
  $\prog, \forgetmem{\AMem}, \forget{\ATSOmap}, \Atm, \forget{\ANQPmap} \trp \prog',
  \Mem'', \TSOmap'', \Atm', \NQPmap''$ and $\wf(\AMem,\ATSOmap,\Atm,\ANQPmap,\pi)$, then
  there exists $\AMem' \in \AMems$, $\ATSOmap' \in \ATSOmaps$,
  $\ANQPmap' \in \ANQPmaps$, and $\pi' \in \paths$ such that
  $\forgetmem{\AMem'} = \Mem''$, $\forget{\ATSOmap'} = \ATSOmap''$,
  $\forget{\ANQPmap'} = \ANQPmap''$, and
  $\prog, \AMem, \ATSOmap, \Atm, \ANQPmap, \pi \trp \prog', \AMem', \ATSOmap', \Atm',
  \ANQPmap', \pi'$.
\end{theorem}
\begin{proof}
  By straightforward induction on $\trp$. In some cases, the reduction enforces
  a specific annotated label $\lambda$ and we have $\pi' = \lambda \cdot \pi$;
  we then need $\wf(\AMem,\ATSOmap,\Atm,\ANQPmap,\pi)$ to check that $\lambda$ is
  fresh enough for $\pi$.
\end{proof}

\bigskip

\begin{theorem}[Operational and Annotated Semantics Equivalence]
  \label{thm:op-annot}
  For all program $\prog$.
  \begin{itemize}
  \item $\forgetmem{\AMem_0}$, $\forget{\ATSOmap_0}$, $\Atm_{0}$, and $\forget{\ANQPmap_0}$
    are the initialisation for the operational semantics;
  \item If
    $\prog, \AMem_0, \ATSOmap_0, \Atm_{0}, \ANQPmap_0, \varepsilon \trp^* \prog', \AMem',
    \ATSOmap', \Atm', \ANQPmap', \pi'$ then
    $\prog, \forgetmem{\AMem_0}, \forget{\ATSOmap_0}, \Atm_{0}, \forget{\ANQPmap_0} \trp^*
    \prog', \forgetmem{\AMem'}, \forget{\ATSOmap'}, \Atm', \forget{\ANQPmap'}$
  \item If
    $\prog, \forgetmem{\AMem_0}, \forget{\ATSOmap_0}, \Atm_{0}, \forget{\ANQPmap_0} \trp^*
    \prog', \Mem'', \TSOmap'', \Atm', \NQPmap''$ then there exists $\AMem' \in \AMems$,
    $\ATSOmap' \in \ATSOmaps$, $\ANQPmap' \in \ANQPmaps$, and $\pi' \in \paths$
    such that $\forgetmem{\AMem'} = \Mem''$, $\forget{\ATSOmap'} = \ATSOmap''$,
    $\forget{\ANQPmap'} = \ANQPmap''$, and
    $\prog, \AMem_0, \ATSOmap_0, \Atm_{0}, \ANQPmap_0, \varepsilon \trp^* \prog', \AMem',
    \ATSOmap', \Atm', \ANQPmap', \pi'$.
  \end{itemize}
\end{theorem}
\begin{proof}
  The first point comes from unfolding the definitions. The other two are proved
  by straightforward induction on $\trp^*$ and using
  Theorems~\ref{thm:annot-to-op} and~\ref{thm:op-to-annot}. The condition
  $\wf(\AMem,\ATSOmap,\Atm,\ANQPmap,\pi)$ is obtained by applying
  Theorem~\ref{thm:wfpi} when needed.
\end{proof}
